\documentclass[multphys,vecphys]{svmult2}

\usepackage{makeidx}     
\usepackage{graphicx}    
\usepackage{multicol}    

\usepackage{ckma4}
\usepackage{epsfig}
\usepackage{euscript}
\usepackage{rotating}
\usepackage{subfigure,array,float}
\usepackage{amsmath,amsfonts,amssymb}

\makeindex             

\begin{document}

\begin{titlepage}

\begin{figure}[p]
\centering
\includegraphics[height=240mm]{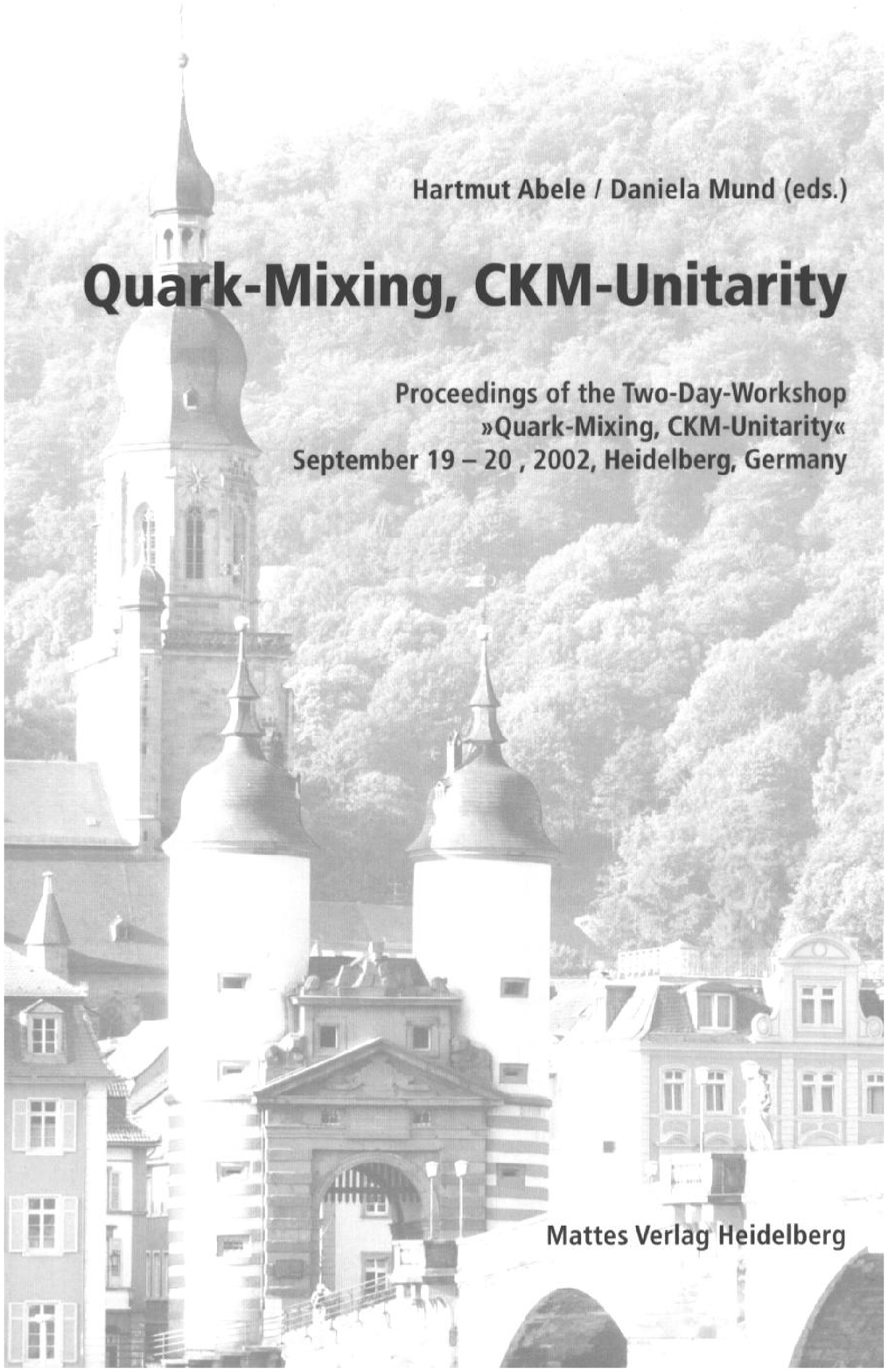}
%
%
\end{figure}
\end{titlepage}
\cleardoublepage

\frontmatter

%
%
%

\thispagestyle{empty}
\vspace*{3.5cm}
\begin{flushleft}
{\Huge{ {\emph{\scshape{Quark-Mixing,\\ \vspace{10pt} \   \ \ \ \ \ \ CKM-Unitarity}}}}}
\vspace{40pt}

{\LARGE { Proceedings of the two day workshop\\
\emph{\scshape{Quark-Mixing, CKM-Unitarity}}}
\vspace{6pt}\\Heidelberg, Germany 19 to 20 September
2002}\vspace{40pt}

\Large Edited by \\{\Large\bf{Hartmut Abele}} and {\Large\bf{Daniela Mund}}\\
\Large Universit\"at Heidelberg
\end{flushleft}

%
%
%

\thispagestyle{empty}
\vspace*{3.5cm}
\begin{flushleft}


\end{flushleft}

%
%
%

\thispagestyle{empty}
\vspace*{3.5cm}
\begin{flushleft}


{\large \bf Dedicated to \vspace{10pt} \\
 \huge Dirk Dubbers \vspace{10pt} \\
\large on the occasion of his 60th birthday}

\end{flushleft}

%
%
%

\thispagestyle{empty}
\vspace*{3.5cm}
\begin{flushleft}


\end{flushleft}

%
%
\setcounter{page}{7}
\preface

\vspace{-3cm}

The fundamental constituents of matter are quarks and leptons. The
quarks which are involved in the process of weak interaction mix
and this mixing is expressed in the so-called
``Cabibbo--Kobayashi--Maskawa" (CKM) matrix. The presently poorly
satisfied unitarity condition for the CKM matrix presents a puzzle
in which a deviation from unitarity may point towards new physics.

The two day workshop \emph{\scshape{Quark-Mixing, CKM-Unitarity}}
was held in Heidelberg (Germany) from 19 to 20 September 2002. The
workshop reviewed the information to date on the inputs for the
unitarity check from the experimental and theoretical side. The
Standard Model does not predict the content of the CKM matrix and
the value of individual matrix elements is determined from weak
decays of individual quarks. Especially the value of $V_{ud}$, the
first matrix element, is subject to scrutiny. $V_{ud}$ has been
derived from a series of experiments on superallowed nuclear
$\beta$-decay measurements, neutron $\beta$-decay and pion
$\beta$-decay. With the information from nuclear and neutron
$\beta$-decay for the first quark generation and from K and
hyperon-decays for the second generation, the unitarity-check
fails significantly for unknown reasons. This workshop is an
attempt to provide an opportunity for clarification of this
situation on the experimental and theoretical side.  \vspace{8pt}

Accordingly, these proceedings are devoted to these
topics:\vspace{4pt}

\hspace*{0.5cm}$\bullet$ Unitarity of the CKM matrix \\
\hspace*{1cm}$\bullet$ First quark flavor decays:
                                 nuclear $\beta$-decays,
                                 neutron $\beta$-decay,
                                 $\pi$--decay \\
 \hspace*{1cm}$\bullet$                           Radioactive beams \\
 \hspace*{1cm}$\bullet$                           Second quark flavor decays:
                                 kaon decays,
                                 hyperon decays \\
  \hspace*{1cm}$\bullet$                          Standard theory -- QED, electroweak and hadronic
  corrections\\
  \hspace*{1cm}$\bullet$                          New possibilities for experiments and
  facilities \\
  \hspace*{1cm}$\bullet$ T- and CP-violation\vspace{8pt}

With these proceedings, we present both a review of the
experimental and theoretical information on quark-mixing with
focus on the first quark generation. The papers present new
findings on these topics in the context of what is known so far.
Besides this, about half a dozen new neutron-decay instruments
being planned or under construction are presented. Better neutron
sources, in particular for high fluxes of cold and high densities
of ultra-cold neutrons will boost fundamental studies in these
fields. The workshop included invited talks and a panel
discussion. The results of the panel discussion are published in
``The European Physical Journal".

The editors wish to dedicate these proceedings to Prof. Dirk
Dubbers on the occasion of his 60th birthday. For many years he
has given advice and support to the ``Atom and Neutron Physics
Group" at the University of Heidelberg Institute of Physics.


We would like to thank all participants and the programme
committee members T. Bowles (LANL), W. Marciano (Brookhaven), A.
Serebrov (PNPI), D. Dubbers (Heidelberg) and O. Nachtmann
(Heidelberg). We would like to express our gratitude to C.
Kr\"amer and F. Schneyder, who have devoted a great deal of their
time and energy to making this meeting a success.

This workshop was sponsored by the Deutsche Forschungsgemeinschaft
(German Research Foundation).

\vspace{1cm}
\begin{flushright}\noindent
Heidelberg, September 2003\hfill {\it Hartmut Abele}\\
\hfill {\it Daniela Mund}\\
\end{flushright}

\tableofcontents

\mainmatter
%
%

\part{Thursday}



\def\Journal#1#2#3#4{{#1} {#2}, #3 (#4)}

\def\MPL{\em Mod. Phys. Lett.}
\def\NP{\em Nucl. Phys.}
\def\PL{\em Phys. Lett.}
\def\PRL{\em Phys. Rev. Lett.}
\def\PR{\em Phys. Rev.}
\def\PTP{\em Prog. Theor. Phys.}
\def\ZP{\em Z. Phys.}
\def\EPJ{\em Eur. Phys. J.}

\title*{Status of the Cabibbo-Kobayashi-Maskawa \protect\newline Quark-Mixing Matrix
 }
\toctitle{Status of the Cabibbo-Kobayashi-Maskawa \protect\newline Quark-Mixing
Matrix}
%
%
\titlerunning{Status of the Cabibbo-Kobayashi-Maskawa Quark Mixing Matrix
}
%
\author{B. Renk}
\authorrunning{B. Renk}
%
%
\institute{Johannes-Gutenberg Universit\"{a}t, Mainz,
Germany\\E-mail: Burkhard.Renk@uni-mainz.de}

\maketitle              

\begin{abstract}
This review, prepared for the 2002 Review of Particle Physics together
with F.J. Gilman and K. Kleinknecht, summarizes experimental inputs and
theoretical conclusions on the present status of the CKM mixing matrix,
and on the CP violating phase $\delta_{_{13}}$. Experimental data are
consistent with each other and with a phase of $\delta_{_{13}} = 59^o \pm 13^o$.
\end{abstract}

\section{The CKM Matrix}
In the Standard Model with $SU(2) \times U(1)$ as the gauge group
of electroweak interactions, both the quarks and leptons are
assigned to be left-handed doublets and right-handed singlets. The
quark mass eigenstates are not the same as the weak eigenstates,
and the matrix relating these bases was defined for six quarks and
given an explicit parametrization by Kobayashi and Maskawa
{\,}\cite{KobayashiMaskawa73} in 1973.  This generalizes the
four-quark case, where the matrix is described by a single
parameter, the Cabibbo angle{\,}\cite{Cabibbo63}.

By convention, the mixing is often expressed in terms of a
$3\times 3$ unitary matrix $V$ operating on the charge $-e/3$
quark mass eigenstates ($d$, $s$, and $b$):
\begin{equation}
\left(\begin{array}{ccc}
d ^{\,\prime}   \\
                s ^{\,\prime} \\
                b ^{\,\prime} \\
            \end{array}\right)
=
    \left(\begin{array}{ccc}
        V_{ud}&     V_{us}&     V_{ub}\\
        V_{cd}&     V_{cs}&     V_{cb}\\
        V_{td}&     V_{ts}&     V_{tb}\\
            \end{array}\right)
    \left(\begin{array}{ccc}
        d \\
                s \\
                b \\
            \end{array}\right) ~.
\end{equation}

The values of individual matrix elements can in principle
all be determined from weak decays of the relevant quarks,
or, in some cases, from deep inelastic neutrino scattering.

There are several parametrizations of the
Cabibbo-Kobayashi-Maskawa \protect\newline (CKM) matrix.  We advocate a
``standard'' parametrization{\,}\cite{StandardParametrization} of
V that utilizes angles $\theta_{12}$, $\theta_{23}$,
$\theta_{13}$, and a phase, $\delta_{_{13}}$
\begin{equation}
V = \left(\begin{array}{ccc}
    c_{_{12}} c_{_{13}}&
    s_{_{12}} c_{_{13}}&
    \: s_{_{13}} e^{-i\delta_{_{13}} } \\
    -s_{_{12}} c_{_{23}}
    -c_{_{12}} s_{_{23}} s_{_{13}} e^{i\delta_{_{13}} }&
    c_{_{12}} c_{_{23}}
        -s_{_{12}} s_{_{23}} s_{_{13}} e^{i\delta_{_{13}} }&
    s_{_{23}} c_{_{13}}                \\
    s_{_{12}} s_{_{23}} -c_{_{12}} c_{_{23}}
        s_{_{13}} e^{i\delta_{_{13}} }&
    -c_{_{12}} s_{_{23}}
           -s_{_{12}} c_{_{23}}
        s_{_{13}} e^{i\delta_{_{13}} } &
    c_{_{23}} c_{_{13}}                 \\
    \end{array}\right) ~,
\label{eq:CKMparametrization}
\end{equation}
with $c_{_{ij}} = \cos\theta_{ij}$ and $s_{ij} = \sin\theta_{ij}$
for the ``generation'' labels ${i,j = 1,2,3}$. This has distinct
advantages of interpretation, for the rotation angles are defined
and labelled in a way which relate to the mixing of two specific
generations and if one of these angles vanishes, so does the
mixing between those two generations; in the limit $\theta_{23} =
\theta_{13} = 0$ the third generation decouples, and the situation
reduces to the usual Cabibbo mixing of the first two generations
with $\theta_{12}$ identified as the Cabibbo
angle{\,}\cite{Cabibbo63}. The real angles $\theta_{12}$,
$\theta_{23}$, $\theta_{13}$ can all be made to lie in the first
quadrant by an appropriate redefinition of quark field phases.

The matrix elements in the first row and third column,
which have been directly measured in decay processes,
are all of a simple form, and, as $c_{_{13}}$ is known
to deviate from unity only in the sixth decimal place,
$V_{ud} = c_{_{12}}$, $V_{us} = s_{_{12}}$, $V_{ub}
= s_{_{13}}~e^{-i \delta_{_{13}} }$, $V_{cb} = s_{_{23}}$,
and $V_{tb} = c_{_{23}}$ to an excellent approximation.
The phase $\delta_{_{13}}$ lies in the range
$0 \leq \delta_{_{13}} < 2\pi$,
with non-zero values generally breaking $CP$ invariance
for the weak interactions.  The generalization to the
$n$~generation case contains $n(n-1)/ 2$~angles and
$(n-1)(n-2)/ 2$ phases.

\section{Brief summary of experimental results}
Most matrix elements can be measured in processes that occur at
the tree level. Further information, particularly on CKM matrix
elements involving the top quark, can be obtained from
flavor-changing processes that occur at the one-loop level.
 Derivation of values for $V_{td}$ and
$V_{ts}$ in this manner from, for example,
$B$~mixing or $b \rightarrow s \gamma$, require an
additional assumption that the top-quark loop,
rather than new physics, gives the dominant contribution
to the process in question.  Conversely, when we find agreement
between CKM matrix elements extracted from loop diagrams
and the values based on direct measurements plus
the assumption of three generations,
this can be used to place restrictions on new physics.

\subsection{Tree level processes }
A more detailed discussion of the experimental and theoretical
input used in the fits can be found in the review in
{\,}\cite{CKMPDG02renk}. From this we deduced the following values and
errors.

Nuclear beta decays {\,}\cite{HardyTowner98} and neutron decays{\,}\cite{polarizedneutrons}:
\begin{equation}
|V_{ud}| = 0.9734 \pm 0.0008 ~.
\label{eq:Vud}
\end{equation}

Analysis of $K_{e3}$ decays{\,}\cite{Leutwyler84}:
\begin{equation}
|V_{us}|  =  0.2196  \pm  0.0026 ~,
\label{eq:KdecaysVus}
\end{equation}

Neutrino and antineutrino production of charm{\,}\cite{Abramowicz82}:
\begin{equation}
|V_{cd}|  =  0.224 \pm  0.016 ~.
\label{eq:Vcd}
\end{equation}

Ratio of hadronic $W$ decays to leptonic decays{\,}\cite{SumWcs}:
\begin{equation}
|V_{cs}| = 0.996 \pm 0.013 ~.
\label{eq:SumWcs}
\end{equation}

Exclusive and inclusive b - decays to charm{\,}\cite{Vcb02}:
\begin{equation}
|V_{cb}| =  (41.2 \pm 2.0 ) \times 10^{-3}~.
\label{eq:Vcb02}
\end{equation}

Exclusive and inclusive b - decays to charmless
states{\,}\cite{Vub02}:
\begin{equation}
| V_{ub}|=  (3.6 \pm 0.7) \times 10^{-3}~.
\label{eq:Vub02}
\end{equation}

Fraction of decays
of the form $t \rightarrow b ~\ell^+ ~\nu_{\ell}$,
as opposed to semileptonic $t$ decays that involve
the light $s$ or $d$ quarks{\,}\cite{Vtblimit} :
\begin{equation}
{|V_{tb} |^2 \over |V_{td} |^2 + |V_{ts} |^2 + |V_{tb} |^2}
= 0.94_{-0.24}^{+0.31} ~.
\label{eq:Vtb}
\end{equation}

\subsection{Loop level processes }
Following the initial evidence{\,}\cite{NA31}, it is now
established that direct $CP$ violation in the weak transition from
a neutral ~$K$ to two pions exists, i.e., that the parameter
$\epsilon^\prime$ is non-zero{\,}\cite{epsilonprimenonzero}. While
theoretical uncertainties in hadronic matrix elements of
cancelling amplitudes presently preclude this measurement from
giving a significant constraint on the unitarity triangle, it
supports the assumption that the observed $CP$ violation is
related to a non-zero value of the CKM phase. This encourages
 the usage of one loop process, CP conserving and CP violating,
 to further constraint the CKM matrix. These inputs are summarized in the following.

Measurement of ${B_d}^0 - {\bar{B}_d}^0$ mixing with $\Delta M_{B_d} = 0.489 \pm 0.008$~ps${}^{-1}$
{\,}\cite{Mixing02} :
\begin{equation}
|{V_{tb}}^* \cdot V_{td}| = 0.0079 \pm 0.0015 ~,
\label{eq:BdMixingVtd}
\end{equation}

Ratio of $B_s$ to $B_d$ mass differences{\,}\cite{Mixing02}:
\begin{equation}
|V_{td}|/|V_{ts}| < 0.25 ~.
\label{eq:BsMixingVtd}
\end{equation}

The $CP$-violating parameter $\epsilon$ in the neutral~$K$ system and
theoretical predictions of the hadronic matrix
elements{\,}\cite{epsilonKQCD}{,\,}\cite{BKparameter}.

The non-vanishing asymmetry in the decays
$B_d (\bar{B}_d ) \rightarrow \psi K_S $ measured by
BaBar{\,}\cite{BabarSine2beta02} and Belle{\,}\cite{BelleSine2beta02}, when averaged yields
:
\begin{equation}
\sin 2\beta = 0.78 \pm 0.08 ~.
\label{eq:sine2beta}
\end{equation}

\section{Determination of the CKM matrix}

Using the eight tree-level constraints together with unitarity, and assuming only three generations,
the 90\% confidence limits on the magnitude of the elements
of the complete matrix are
\begin{equation}
        \left(\begin{array}{ccc}
            0.9741 \hbox{ to }\, 0.9756 &
        0.219  \hbox{ to }\, 0.226 &
        0.0025 \hbox{ to }\,  0.0048 \\
            0.219 \hbox{ to }\,  0.226 &
        0.9732 \hbox{ to }\, 0.9748 &
        0.038 \hbox{ to }\,  0.044 \\
            0.004 \hbox{ to }\, 0.014 &
        0.037 \hbox{ to }\, 0.044 &
        0.9990 \hbox{ to }\, 0.9993 \\
            \end{array}\right) \ .
\label{eq:90ConfidenceRange}
\end{equation}

The ranges shown are for the individual matrix elements.
The constraints of unitarity connect different elements, so
choosing a specific value for one element restricts the range
of others.  Using tree-level processes as
constraints only, the matrix elements
in Eq.~(\ref{eq:90ConfidenceRange}) correspond to
values of the sines of the angles of
$s_{_{12}} = 0.2229 \pm 0.0022 $,
$s_{_{23}} = 0.0412 \pm 0.0020$,
and $s_{_{13}} = 0.0036 \pm 0.0007$.

If we use the loop-level processes
as additional constraints, the sines of the angles remain
unaffected, and the CKM phase, sometimes referred to as
the angle $\gamma = \phi_3$ of the unitarity triangle,
is restricted to $\delta_{_{13}} = (1.02 \pm 0.22)$ radians
$= 59^o \pm 13^o$. \\

Direct and indirect information on the smallest matrix
elements of the CKM matrix is neatly
summarized in terms of the ``unitarity triangle,'' one of
six such triangles that correspond to the unitarity condition
applied to two different rows or columns of the CKM matrix.
Unitarity applied to the first and third columns yields
\begin{equation}
V_{ud} ~{V_{ub}}^{\!*} + V_{cd} ~{V_{cb}}^{\!*}
+ V_{td} ~{V_{tb}}^{\!*} = 0 ~.
\label{eq:FullUnitarityTriangle}
\end{equation}

The unitarity triangle is just a geometrical presentation
of this equation in the complex
plane{\,}\cite{UnitarityTriangle}, as in Figure 1(a). Setting
cosines of small angles to unity,
Eq.~(\ref{eq:FullUnitarityTriangle}) becomes
\begin{equation}
{V_{ub}}^{\!*} + V_{td} \approx s_{12} ~{V_{cb}}^{\!*} ~,
\label{eq:UnitarityTriangle}
\end{equation}
which is shown as the unitarity triangle.
\begin{figure}[ht]
\begin{center}
\includegraphics[width=0.6\textwidth]{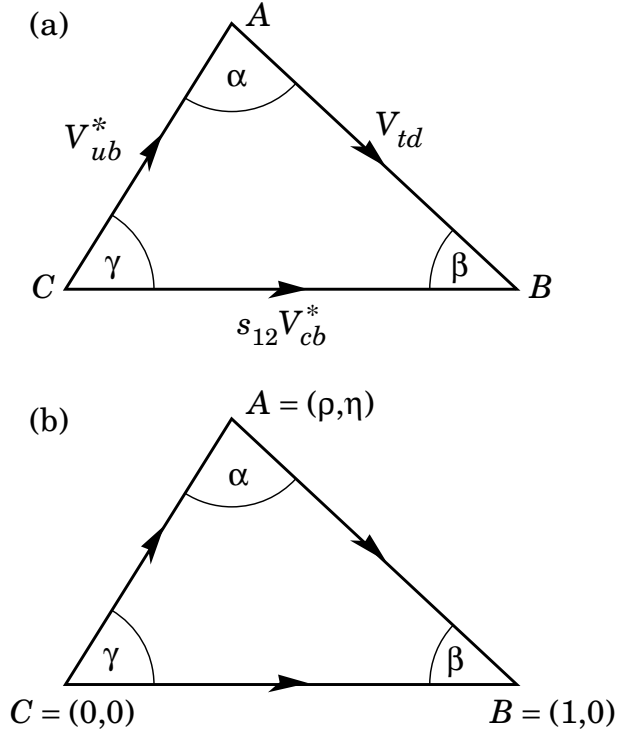}
\end{center}
\caption{(a)~Representation in the complex plane of the triangle
formed by the CKM matrix elements $V_{ud} ~{V_{ub}}^{\!*}$,
$V_{td} ~{V_{tb}}^{\!*} $, and $V_{cd} ~{V_{cb}}^{\!*}$.
(b)~Rescaled triangle with vertices A, B, and C at $(\bar{\rho},
\bar{\eta)}$, $(1,0)$, and $(0,0)$, respectively.}
\label{fig:UnitarityTriangle02}
\end{figure}
The angles $\alpha$, $\beta$ and $\gamma$ of the
triangle are also referred to as $\phi_2$, $\phi_1$,
and $\phi_3$, respectively, with $\beta$ and
$\gamma = \delta_{13}$ being the phases of the
CKM elements $V_{td}$ and $V_{ub}$ as per
\begin{equation}
{V_{td} = |{V_{td}}| e^{-i\beta }},
{V_{ub} = |{V_{ub}}| e^{-i\gamma}} ~.
\end{equation}

Rescaling the triangle so that the base is of unit length,
the coordinates of the vertices A, B, and C become respectively:
\begin{equation}
\bigl( \hbox{Re}(V_{ud}~V_{ub}^{\!*})/|V_{cd}~V_{cb}^{\!*}|,
~\hbox{Im}(V_{ud}~V_{ub}^{\!*})/|V_{cd}~V_{cb}^{\!*}| \bigr),
~(1,0),
~{\rm and}~(0,0) ~.
\label{eq:ScaledUnitarityTriangle}
\end{equation}

The coordinates of the apex of the rescaled unitarity triangle
take the simple form $( \bar{\rho}, \bar{\eta})$, with
$ \bar{\rho}= \rho (1-\lambda^2/2 )$ and
$\bar{\eta}=\eta (1-\lambda^2/2 )$ in the Wolfenstein
parametrization{,\,}\cite{Wolfenstein83}
as shown in Figure 1(b).

$CP$-violating processes involve the phase in the
CKM matrix, assuming that the observed $CP$ violation
is solely related to a nonzero value of this phase.
More specifically, a necessary and sufficient condition
for $CP$ violation with three generations can be formulated
in a parametrization-independent manner in terms of the
non-vanishing of $J$, the determinant of the commutator
of the mass matrices for the charge $2e/3$
and charge $-e/3$ quarks{\,}\cite{Jarlskog85}.
$CP$-violating amplitudes or differences of rates are
all proportional to the product of CKM factors in
this quantity, namely
$s_{_{12}} s_{_{13}} s_{_{23}} c_{_{12}}
c_{_{13}}^2 c_{_{23}} \sin{\delta_{13}}$.
This is just twice the area of the unitarity triangle.
\begin{figure}[p]
\begin{center}
\includegraphics[width=0.8\textwidth]{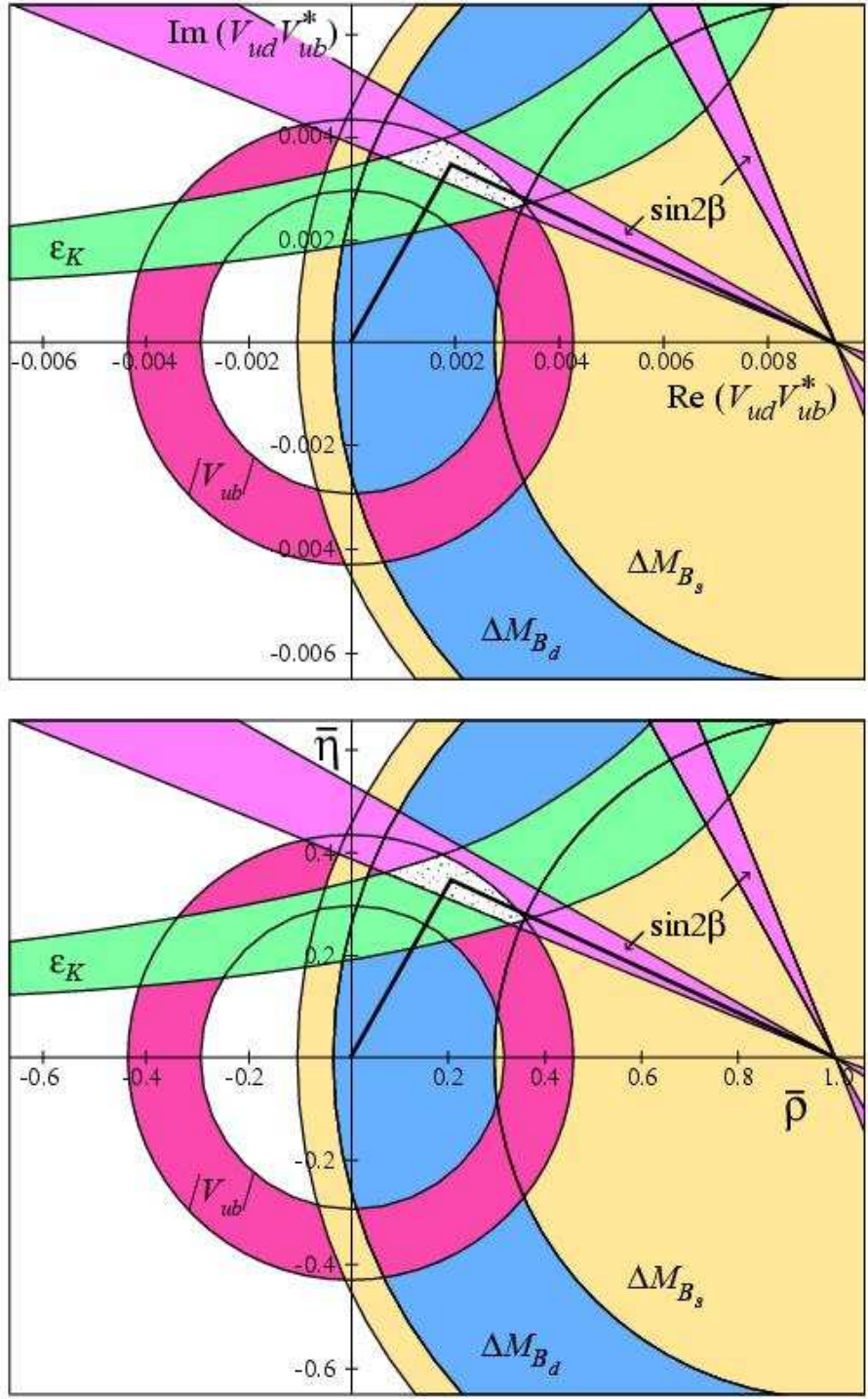}
\end{center}
\caption{Constraints from the text on the position of the apex, A,
of the unitarity triangle following from $|V_{ub} |$,$B$ mixing,
$\epsilon$, and $\sin 2\beta$. A possible unitarity triangle is
shown with A in the preferred region.}
\label{fig:UnitarityTriangleConstraints}
\end{figure}
The constraints on the apex of the unitarity triangle that follow
from Eqs.~(\ref{eq:Vub02}), (\ref{eq:BdMixingVtd}),
(\ref{eq:BsMixingVtd}), (\ref{eq:sine2beta}), and $\epsilon$ are
shown in Figure 2.  Both the limit on $\Delta M_s$ and the value
of $\Delta M_d$ indicate that the apex lies in the first rather
than the second quadrant. All constraints nicely overlap in one
small area in the first quadrant with the sign of $\epsilon$
measured in the K system agreeing with the sign of $\sin 2\beta$
measured in the B system.  Both the constraints from the lengths
of the sides (from $|V_{ub}|$, $|V_{cb}|$, and $|V_{td}|$) and
independently those from $CP$-violating processes ($\epsilon$ from
the $K$ system and $\sin2\beta$ from the $B$ system) indicate the
same region for the apex of the triangle.

From a combined fit using the direct measurements,
B mixing, $\epsilon$, and $\sin 2\beta$, we obtain:
\begin{eqnarray}
{\rm Re}~V_{td} &=& 0.0071 \pm 0.0008 \\
{\rm Im}~V_{td} &=& -0.0032 \pm 0.0004 \\
\bar{\rho} &=& 0.22 \pm 0.10  ~, \\
\bar{\eta} &=& 0.35 \pm 0.05  ~.
\end{eqnarray}

All processes can be
quantitatively understood by one value of the CKM phase
$\delta_{_{13}} = \gamma = 59^o \pm 13^o $.  The value of
$\beta = 24^o \pm 4^o$ from the overall fit is consistent
with the value from the $CP$ asymmetry measurements of
$26^o \pm 4^o$.  The invariant measure of $CP$ violation
is $J = (3.0 \pm 0.3) \times 10^{-5}$.

\def\F{{\cal F}}
\def\DRV{\Delta_{\mbox{\tiny R}}^{\mbox{\tiny V}}}

\title*{Superallowed $0^{+} \rightarrow 0^{+}$ Beta Decay: \protect\newline
Current Status and Future Prospects \protect\newline}
\toctitle{Superallowed $0^{+} \rightarrow 0^{+}$ Beta Decay: \protect\newline
Current Status and Future Prospects}
%
%
\titlerunning{Superallowed $0^{+} \rightarrow 0^{+}$ Beta Decay}
%
\author{J.C. Hardy, I.S. Towner}
\authorrunning{J.C. Hardy and I.S. Towner}
%
%
\institute{Cyclotron Institute, Texas A\&M University, College Station, TX 77843, USA}

\maketitle              

\begin{abstract}
The value of the $V_{ud}$ matrix element of the
Cabibbo-Kobayashi-Maskawa matrix can be derived from nuclear
superallowed beta decays, neutron decay and pion beta decay.
Today, the most precise value of $V_{ud}$ ($\pm 0.05\%$) comes
from the nuclear decays; and its precision is limited not by
experimental error but by the estimated uncertainty in theoretical
corrections, which themselves are of order 1\%.  When combined
with the best values of $V_{us}$ and $V_{ub}$, the results differ
at the 98\% confidence limit from the unitarity condition for the
CKM matrix.  This talk outlines the current status of both the
experimental data and the calculated correction terms, and
presents an overview of experiments currently underway to reduce
the uncertainty in those correction terms that depend on nuclear
structure.
\end{abstract}

\section{Introduction}
Superallowed $0^{+} \rightarrow 0^{+}$ nuclear $\beta$-decay
depends uniquely on the vector part of the weak interaction.  When
it occurs between $T=1$ analog states, a precise measurement of
the transition $ft$-value can be used to determine $G_V$, the
vector coupling constant. This result, in turn, yields $V_{ud}$,
the up-down element of the Cabibbo-Kobayashi-Maskawa (CKM) matrix.
At this time, it is the key ingredient in one of the most exacting
tests available of the unitarity of the CKM matrix, a fundamental
pillar of the minimal Standard Model.

\section{Current status}

Currently, there is a substantial body of precise $ft$-values determined for
such transitions and the experimental results are robust, most input data having
been obtained from several independent and consistent measurements \cite{TH98,Ha90}.
In all, $ft$-values have been determined for nine $0^{+} \rightarrow 0^{+}$
transitions to a precision of $\sim 0.1\%$ or better.  The decay parents --
$^{10}$C, $^{14}$O, $^{26m}$Al, $^{34}$Cl, $^{38m}$K, $^{42}$Sc, $^{46}$V, $^{50}$Mn and
$^{54}$Co -- span a wide range of nuclear masses; nevertheless, as anticipated
by the Conserved Vector Current hypothesis, CVC, all nine yield consistent
values for $G_V$, from which a value
of
\begin{equation}
V_{ud} = 0.9740 \pm 0.0005
\label{Vud}
\end{equation}
is derived.  The unitarity test of the CKM
matrix, made possible by this precise value of $V_{ud}$, fails by more than two
standard deviations \cite{TH98}: {\it viz.}
\begin{equation}
V_{ud}^2 + V_{us}^2 + V_{ub}^2 = 0.9968 \pm 0.0014.
\label{unitarity}
\end{equation}
In obtaining this result, we have used the Particle Data Group's \cite{PDG00}
recommended values for the much smaller matrix elements, $V_{us}$ and $V_{ub}$.
Although this deviation from unitarity is not completely definitive statistically, it is
also supported by recent, less precise results from neutron decay \cite{Ab02}.  If the
precision of this test can be improved and it continues to indicate non-unitarity,
then the consequences for the Standard Model would be far-reaching.

The potential impact of definitive non-unitarity has led to considerable recent activity, both
experimental and theoretical, in the study of superallowed $0^{+} \rightarrow 0^{+}$
transitions, with special attention being focused on the small correction terms that must
be applied to the experimental $ft$-values in order to extract $G_V$.  Specifically,
$G_V$ is obtained from each $ft$-value via the
relationship \cite{TH98}
\begin{equation}
\F t \equiv ft (1 + \delta_R^{\prime} + \delta_{NS})(1 - \delta_C ) =
\frac{K}{2 {G_V}^2 (1 + \DRV )} ,
\label{Ft}
\end{equation}
\noindent where $K$ is a known constant, $f$ is the statistical
rate function and $t$ is the partial half-life
for the transition. The correction terms -- all of order 1\% or less -- comprise
$\delta_C$, the isospin-symmetry-breaking
correction, $\delta_R^{\prime}$ and $\delta_{NS}$, the transition-dependent parts
of the radiative correction and $\DRV $, the transition-independent
part.  Here we have also defined $\F t$ as the ``corrected"
$ft$-value.  Note that, of the four calculated correction terms, two -- $\delta_C$ and $\delta_{NS}$ --
depend on nuclear structure and their influence in Eq.(\ref{Ft}) is effectively in the form
($\delta_C - \delta_{NS}$).

With Eq.(\ref{Ft}) in mind, it is now valuable to dissect the contributions to
the uncertainty obtained for $V_{ud}$ in Eq.(\ref{Vud}).  The contributions to
the overall $\pm 0.0005$ uncertainty are 0.0001 from
experiment, 0.0001 from $\delta_R^{\prime}$,
0.0003 from ($\delta_C  -  \delta_{NS}$), and
0.0004 from $\DRV$.  Thus, if the unitarity test is to be
sharpened, then the most pressing objective must be to
reduce the uncertainties on $\DRV$ and ($\delta_C  -  \delta_{NS}$).
It is important to recognize that the former also appears in the extraction of $G_V$ from
neutron decay, and thus it will also ultimately limit the precision achievable from neutron
decay to approximately the same level as the current nuclear result, regardless of
the precision achieved in the neutron experiments.  Improvements in $\DRV$ are
a purely theoretical challenge, the solution of which will not depend on further experiments.
However, experiments can play a role in improving the next most important contributor to
the uncertainty on $V_{ud}$, namely ($\delta_C  -  \delta_{NS}$).  Clearly this correction applies
only to the results from superallowed beta decay and, in the event that improvements are made in
$\DRV$, will then limit the precision with which $V_{ud}$ can be determined by this route.
Recently, a new set of consistent calculations for ($\delta_C  -  \delta_{NS}$) have appeared \cite{TH02}
not only for the nine well known superallowed transitions but for eleven other superallowed
transitions that are potentially accessible to precise measurements in the future.
Experimental activity is now focused on probing these nuclear-structure-dependent
corrections with a view to reducing the uncertainty that they introduce into the
unitarity test.

\section{Future prospects}
\renewcommand{\bottomfraction}{0.9}
\renewcommand{\textfraction}{0.1}
\begin{figure}[b]
\begin{center}
\includegraphics[width=90mm]{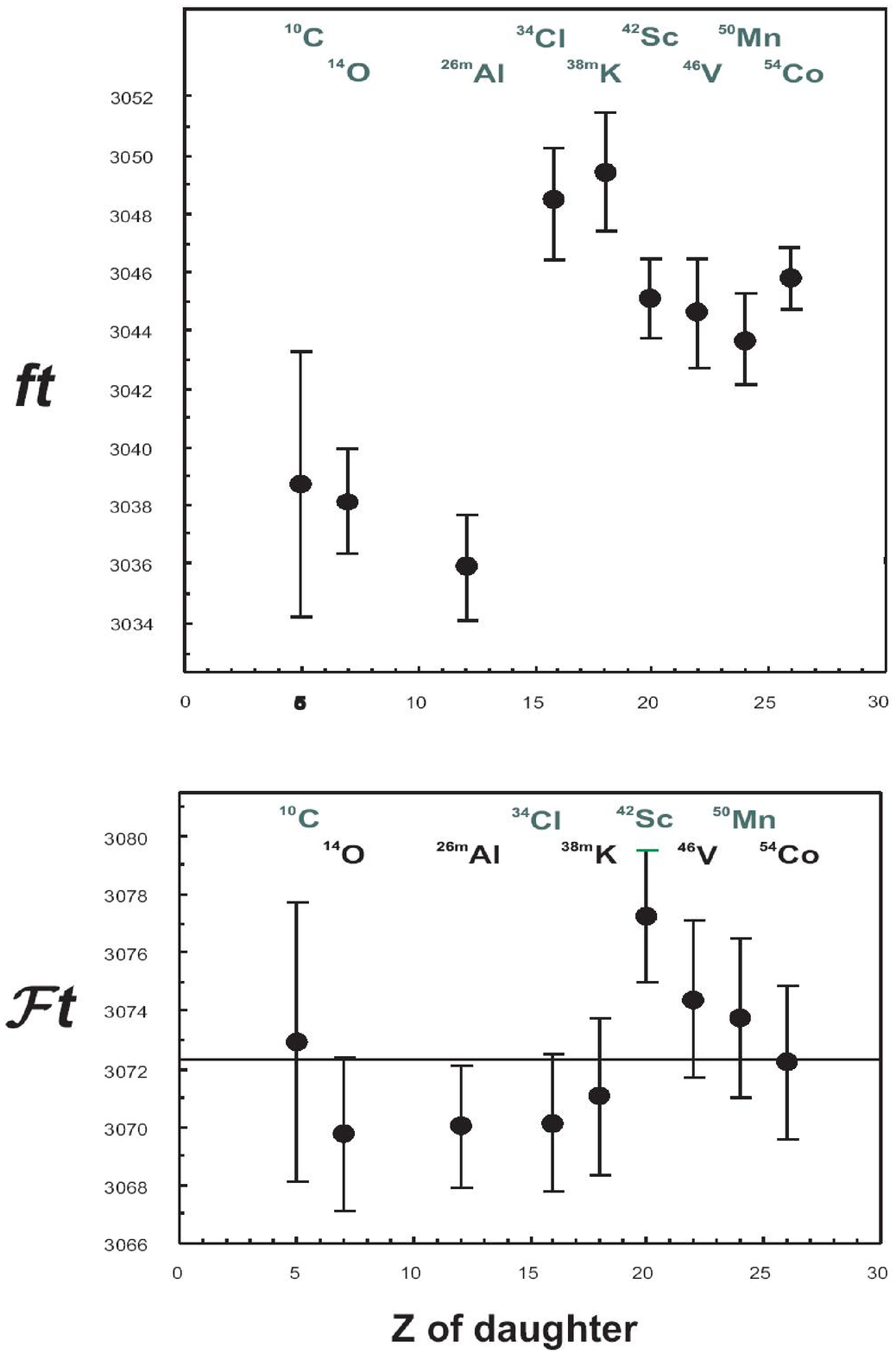}
\end{center}
\caption[]{Comparison of experimental {\it ft}-values and the
corrected $\F t$-values for the nine well-known superallowed
transitions.  This illustrates the effect of the calculated
nucleus-dependent corrections, which change from transition to
transition.  (The effect of $\delta_R^{\prime}$ is virtually the
same for all cases.)} \label{fig:1}
\end{figure}
The essential approach being taken by current experiments is best explained with reference
to Fig.~\ref{fig:1}.  The upper panel shows the uncorrected
experimental $ft$ values and the lower panel the corrected $\F t$ values with the
average indicated by a horizontal line.  If the experimental $ft$ values were left uncorrected,
their scatter would be quite inconsistent with a single value for the vector coupling
constant, $G_V$.  Once corrected, though, the resulting $\F t$ values are in excellent agreement with
this expectation ($\chi^2/\nu = 0.6$).  This, in itself, provides powerful validation of the
calculated corrections used in their derivation.  However, extending this concept to measurements of
other superallowed decays, we can continue to use CVC to test the validity of the
nuclear-structure-dependent corrections, ($\delta_C  -  \delta_{NS}$) at an even more demanding level.
By choosing transitions where it is predicted that the structure-dependent corrections are much
larger, we can achieve a more sensitive test of the accuracy of the calculations.

Of course it is only the {\em relative} values of ($\delta_C - \delta_{NS}$)
that are confirmed by the absence of transition-to-transition variations in the
corrected $\F t$-values.  However, $\delta_C$ itself represents
a difference -- the difference between the parent and daughter-state
wave functions caused by charge-dependent mixing.  Thus, the
experimentally determined variations in $\delta_C$ are actually
second differences.  It would be a pathological fault indeed that
could calculate in detail these variations ({\em i.e.} second differences)
in $\delta_C$ while failing to obtain their {\em absolute} values
({\em i.e.} first differences) to comparable precision.

Experimental attention is currently focused on two series of $0^{+}$
nuclei: the even-$Z$, $T_z = -1$ nuclei with $18
\leq A \leq 42$, and the odd-$Z$, $T_z = 0$ nuclei with $A \geq 62$.
The main attraction of these new regions is that the calculated
values of ($\delta_C - \delta_{NS}$) for the superallowed transitions \cite{TH02}
are larger, or show larger variations from nuclide to nuclide, than the
values applied to the nine currently well-known transitions.
These are just the properties required to test the
accuracy of the calculations.  It is argued that if the
calculations reproduce the experimentally observed variations where
they are large, then that must surely verify their reliability for the
original nine transitions whose $\delta_C - \delta_{NS}$ values are considerably
smaller.

Of the heavier $T_z = 0$ nuclei, $^{62}$Ga and $^{74}$Rb are receiving the
greatest attention at this time (see ref. \cite{HT02} and experimental references therein).  It
is likely, though, that the required experimental precision will take some time to achieve.
The decays of nuclei in this series are of higher energy than any previously studied and
each therefore involves numerous weak Gamow-Teller transitions in addition
to the superallowed transition\cite{HT02}.  Branching-ratio measurements are thus very
demanding, particularly with the limited intensities likely to be available initially for
most of these rather exotic nuclei.  In addition, their half-lives are considerably shorter
than those of the lighter superallowed emitters; high-precision mass measurements
($\pm 2$ keV) for such short-lived activities will also be very challenging.

More accessible in the short term are the $T_z = -1$ superallowed emitters with $18
\leq A \leq 42$.  There is good reason to explore them.  For example, the calculated
value of ($\delta_C - \delta_{NS}$) for $^{30}$S decay, though smaller than those expected
for the heavier nuclei, is actually 1.13\% -- larger than for any other case currently
known -- while $^{22}$Mg has a low value of 0.51\%.  Furthermore, the nuclear model space
used in the calculation of ($\delta_C - \delta_{NS}$) for these nuclei is exactly the same
as that used for some of the nine transitions already studied.  If the wide range of values
predicted for the corrections are confirmed by the measured $ft$-values, then it will do
much to increase our confidence (and reduce the uncertainties) in the corrections already
being used.   To be sure, these decays also provide an experimental challenge, particularly
in the measurement of their branching ratios, but sufficiently precise results have
just been obtained \cite{Ha02} for the half life and superallowed branching ratio for the
decay of $^{22}$Mg and work on $^{34}$Ar decay is well advanced. New precise $ft$-values should
not be long in appearing.  It would be virtually impossible for them to have any effect on the
central value already obtained for $V_{ud}$ but they may be expected ultimately to lead to
reduced uncertainties on that value.

The work of JCH was supported by the U.S. Department of Energy under Grant number
DE-FG03-93ER40773 and by the Robert A. Welch Foundation; he would also like to thank the
Institute for Nuclear Theory at the University of Washington for its hospitality and support
during part of this work.

\title*{New Analysis \\of Neutron $\beta$-Decay (Radiative Corrections) \\and Implications for CKM Unitarity}
\toctitle{New Analysis of Neutron $\beta$-Decay (Radiative Corrections) \protect\newline and Implications for CKM Unitarity}
%
%
\titlerunning{New analysis of neutron $\beta$-decay (radiative
corrections)}
%
\author{W.J. Marciano}
 \institute{Brookhaven National
Laboratory Upton, New York 11973, USA}
%
%
%

\maketitle              

\begin{abstract}
This article gives a brief summary of radiative corrections
with a new analysis of neutron $\beta$-decay.
\end{abstract}
\section{One and Two Loop Electroweak Corrections}
Modulo the Fermi function, electroweak radiative corrections to superallowed
$(O^{+}\rightarrow O^{+})$ nuclear beta decays are traditionally factored into
two
contributions called inner and outer corrections.  The outer (or long distance)
correction
is given by
\begin{equation}
1~+~\frac{\alpha}{2\pi}(g(E, E_{max})~+ 2C_{NS})~+ \delta_{2}(Z, E)
\label{al2pi}
\end{equation}
where $g(E, E_{max})$ is the universal Sirlin function \cite{aso} which
depends on the
nucleus
through $E_{max}$, the positron or electron end point energy.
$C_{NS}$ is a
nuclear
structure dependent contribution induced by axial-current nucleon-nucleon
interactions \cite{itjh} and $\delta_{2}$ is an $O(Z\alpha^{2})$ correction
partly
induced
by
factorization of the Fermi function and outer radiative corrections \cite{wjgr}.

The contribution from $g(E, E_{max})$ is quite large $(\sim1.3\%~for~O^{14})$
due to a $3\ln(m_{p}/E_{max})$ term which generally dominates.  Summation of
$(\alpha \ln m_{p}/E_{max})^{n}$, $n = 2,3\dots$ contributions from higher
orders gives an additional $0.028\%$ correction \cite{acwm} while additional
$O(\alpha^{2})$ effects are estimated to be $<0.01\%$.

$C_{NS}$ and $\delta_{2}(Z, E)$ are nucleus dependent.  The leading contribution
to
$\delta_{2}$ is of the form $Z\alpha^{2}\ln m_{p}/E$ where $Z$ is the charge of
the
daughter nucleus.  Just as in the case of the Fermi function, $\delta_{2}$ is
usually
given for
positron emitters (since that is appropriate for superallowed decays).  For
electron emitters
the sign of $Z$ should be changed in both the Fermi function and $\delta_{2}(Z,
E)$.
Unfortunately, as pointed out by Czarnecki, Marciano and Sirlin \cite{acwm},
that sign
change was
not made in the case of neutron decay.  As a result, the often quoted $0.0004$
contribution from $\delta_{2}$ to neutron $\beta$-decay should be changed to $-
0.00043$, an
overall shift of $-0.083\%$.  With those corrections, the overall uncertainty in
the outer
radiative corrections is now estimated to be about $\pm0.01\%$.

The inner radiative correction factor is given (at one loop level) by
\begin{equation}
1~+~\frac{\alpha}{2\pi}(4 \ln \frac{m_{Z}}{m_{p}} + \ln \frac{m_{p}}{m_{A}} +
A_{g} + 2C)
\label{lnmp}
\end{equation}
where the $\frac{2\alpha}{\pi}\ln m_{Z}/ m_{p} \simeq 0.0213$ universal short-
distance correction dominates \cite{rmp}.  The contributions induced by axial-
vector effects
are relatively small but carry the bulk of the theoretical uncertainty
\begin{equation}
\frac{\alpha}{2\pi}[\ln \frac{m_{p}}{m_{A}} + A_{g} + 2C] {\simeq} {-–}0.0015
\pm
0.0008
\label{magc}
\end{equation}
It stems from an uncertainty in the effective value of $m_{A}$ that should be
employed.
The quoted uncertainty in eq. (3) allows for a conservative factor of
2 uncertainty in that
quantity.  It would be difficult to significantly reduce the uncertainty for
nuclei or the
neutron.  In the case of pion beta decay, the uncertainty is likely to a factor
of 2 or more
smaller.

High order $(\alpha \ln m_{Z}/m_{p})^{n}, n = 2,3\dots$ leading log
contributions
are
expected to dominate the multi-loop effects.  They have been summed by
renormalization
group techniques [6], resulting in an increase in eq. (9) by $0.0012$.  Next to
leading logs of
$O(\alpha^{2}\ln m_{Z}/m_{p})$ have been estimated to give $-0.0002\pm0.0002$
while $O(\alpha^{2})$ effects are expected to be negligible.  In total, a recent update finds
\cite{acwm}
\begin{equation}
Inner R. C. Factor = 1.0240\pm0.0008 \label{inrc}
\end{equation}
which is essentially the same as the value given by Sirlin in 1994
\cite{lwsi}. It leads to
\begin{equation}
|V_{ud}|=0.9740\\\pm0.0001\pm0.0001\pm0.0003\pm0.0004 \label{vud7}
\end{equation}
extracted from super-allowed beta decays, where the errors stem
from the experimental uncertainty, the two transition dependent
parts of the radiative corrections $\delta_R^{\prime}$ and
$\delta_C - \delta_{NS}$, and the inner radiative correction
$\DRV$ respectively.

In the case of neutron decay, the radiative corrections carry a
similar structure and uncertainty.  Correcting for the sign error
in the $Z\alpha^{2}$ effect, one finds the master formula
\cite{acwm}
\begin{equation}
|V_{ud}|^{2}=\frac{4908 \pm 4 sec}{\tau_{n}(1 +3\lambda^2)}
\label{4908}
\end{equation}
Employing $\tau_{n} = 885.7(7)$s and $\lambda$ = 1.2739(19) then
implies
\begin{equation}
|V_{ud}| = 0.9717\pm0.0004\pm0.0012\pm0.00004\pm0.0004
\label{0971}
\end{equation}
where the errors stems from the experimental uncertainty in the
neutron lifetime, the $\beta$-asymmetry $A$ and the theoretical
outer and inner radiative correction $\delta_R^{\prime}$ and
$\DRV$ respectively.  In the case of pion beta decay, the theory
uncertainty in $|V_{ud}|$ is probably $\pm0.0002$ or smaller, but
the small $(\simeq 10^{-8})$ branching ratio makes a precision
measurement very difficult.


\title*{An Overview of Neutron Decay}
\author{J. Byrne}

\institute{Physics \& Astronomy Subject Group School of Chemistry,
Physics \& Environmental Science, University of Sussex, Brighton
BN1 9QJ, U.K.}
\maketitle

\section{Neutron Decay in the Context of Nuclear Physics}

\subsection{The Weak Interaction in Nuclei}

According to the Standard Model of particle physics the charged
weak current is purely left-handed, i.e. it is an equal admixture
of polar vector (V) and axial vector (A) currents of quarks and
leptons with appropriate relative sign. In nuclear physics vector
currents give rise to Fermi $\beta$ -transitions with coupling
constant $G_V$ and spin-parity selection rule for allowed
transitions:
\begin{equation}
\Delta I=0,\;\textrm{no parity change} \label{2.1a}
\end{equation}

\noindent Axial currents give Gamow-Teller $\beta$-transitions
with coupling constant $G_A$ and spin-parity selection rule for
allowed transitions:
\begin{equation}
\Delta I=0,\,\pm 1,\;\textrm{ no }0\Rightarrow 0,\;\textrm{no
parity change} \label{2.1b}
\end{equation}
\noindent The $\beta$-decay of free neutrons into protons
\begin{equation}
n\Rightarrow p+\;e^{-}+\stackrel{\_}{\nu }_e, \quad\frac
12^{+}\Rightarrow \frac 12^{+}  \label{2.1c}
\end{equation}

\noindent is allowed by both selection rules and is described as a
mixed transition. One can therefore observe parity-violating
effects in neutron decay associated with vector/axial vector
interference.
\subsection{ Neutron Decay Parameters}\label{secdecay}

The principal kinematic parameters which govern neutron decay are:
\begin{eqnarray}
\Sigma =(m_n+m_{p)}c^2=1877.83794\,MeV;\Delta
=(m_n-m_p)c^2=1.29332\,MeV \label{2.2a}
\\
\textrm{Kinetic energy of electrons}:\;0\leq T_e\leq 783\,keV
\label{2.2b}
\\
\textrm{Kinetic energy of protons}:\;0\leq T_p\leq 751\,eV
\label{2.2c}
\\
\textrm{Recoil parameter}:\;\delta =\Delta \,/\,\Sigma < 10^{-3}
\label{2.2d}
\end{eqnarray}

\noindent Because the recoil parameter $\delta$ is so small it
follows that the momentum transfer dependence of all form factors
may be neglected. This is also the reason why the neutron lifetime
is so long. The current best value of the neutron lifetime is
\cite{pdg2}:
\begin{equation}
\tau _n=885.7\pm 0.8\,sec.  \label{2.2e}
\end{equation}

\noindent This is greater by a factor of $\sim 4.10^8$ than the
lifetime of the muon which is the next longest lived elementary
particle.

\subsection{ Measurement of the Neutron Lifetime}
 Neutron lifetime experiments may be separated into two groups: the
classical 'beam' methods and the more modern 'bottle' methods. In
beam methods the number of decaying neutrons in a specified volume
of neutron beam is recorded. These methods rely on the
relationship:
\begin{equation}
\frac{dN(t)}{dt}=-\frac{N(t)}{\tau _n}  \label{2.3a}
\end{equation}

\noindent where N(t) is the number of neutrons in the source
volume V at time t. To proceed further we require two additional
relations:
\begin{equation}
\Bigl\langle \frac{dN(t)}{dt}\Bigr\rangle =n_d\frac{4\pi }{\Omega
\varepsilon } \label{2.3b}
\end{equation}

\noindent and
\begin{equation}
\langle N(t)\rangle =\rho _nV  \label{2.3c}
\end{equation}

\noindent where n$_d$ is the number of neutron decays recorded per
unit time in a detector of known solid angle $\Omega$ and
efficiency $\varepsilon$, and $\rho_n$ is the neutron density.
Assuming a 4$\pi $ collection solid angle, as in all recent
variants of the technique, and unit efficiency $\varepsilon$ for
recording the number N$_d$ of decays occurring per second in a
known length L of beam, the value of $\tau _n$ is given by
\begin{equation}
\tau _n=\frac{N_n\textrm{ L}}{N_d\,\sigma _0\,v_0\,\eta }
\label{2.3d}
\end{equation}

\noindent Here N$_n$ is the number of neutron-nucleus reactions
detected per unit time in a neutron counter, $\sigma _0$ is the
cross section at some standard neutron velocity v$_0$ (usually
2200m./sec.) and $\eta$ is the surface density of neutron detector
isotope. This result does not depend on the neutron velocity v,
provided $\sigma$(v) scales as v$^{-1}$. Suitable reactions are:
\begin{equation}
^{10}B(n,\alpha )^7Li\textrm{ }(\sigma _0=3836\pm 8b.),
\label{2.3e}
\end{equation}

\begin{equation}
^6Li(n,\alpha )^3H\textrm{ }(\sigma _0=941\pm 3b.)  \label{2.3f}
\end{equation}

\noindent and
\begin{equation}
^3He(n,p)^3H\textrm{ }(\sigma _0=5327\pm 10b.)  \label{2.3g}
\end{equation}

\noindent 'Bottle' methods for the determination of $\tau_n$ on
the other hand rely on the integrated form of (\ref{2.3a}), i.e.
\begin{equation}
N(t)=N(0)\,e^{-t/\tau _n}  \label{2.3h}
\end{equation}
\noindent  where N(t) is determined by recording the number of
neutrons surviving to time t as a function of the number N(0)
present in a fixed source volume at zero time. This is to be
contrasted with the beam methods where it is the number of
neutrons which fail to survive in a continually replenished source
of neutrons which is recorded. Ever since the identification of a
storable ultra-cold component of energy $\leq 2.10^{-7}$ eV in the
Maxwellian tail of the thermal flux from a reactor, the bottle
methods have been favored since they do not rely on the
performance of a number of subsidiary experiments,e.g.
determination of absolute
cross-sections or the precise isotopic composition of neutron counters.%

There are two principal neutron storage methods, magnetic
confinement or storage in a closed vessel made from a material
with suitable Fermi pseudo-potential. Magnetic confinement relies
on the force
\begin{equation}
\mathbf{F}=-\nabla \{\mathbf{\mu }_n.\mathbf{B}(\mathbf{r})\}
\label{2.3i}
\end{equation}
\noindent which is exerted on the neutron magnetic moment
$\mathbf{\mu }_n$ in an inhomogeneous magnetic field
$\mathbf{B}(\mathbf{r})$. Since the sense of the force depends on
the sign of the spin quantum number only one sign of the spin can
be confined which means that, in principle, neutrons can always be
lost from the source volume by spin-flipping which is a difficult
loss mechanism to control. Alternatively in the case of storage in
a material bottle the ideal relation (\ref{2.3h}) must be replaced
by
\begin{equation}
N(t)=N(0)e^{-t(1/\tau _n+1/\tau _w)}   \label{2.3j}
\end{equation}
\noindent where $\tau_w($v$)$ represents the lifetime for neutron
loss through absorption or inelastic collisions of ultra-cold
neutrons with the walls of the vessel. In general this is given by
a relation of the form
\begin{equation}
\tau _w(v)^{-1}=\langle \mu (v)\rangle v/\lambda  \label{2.3k}
\end{equation}
\noindent  where $\langle\mu($v$)\rangle$ is the loss rate per
bounce averaged over all angles of incidence and the mean free
path $\lambda $ is a function of the geometry of the containing
vessel. A number of techniques have been developed to estimate
$\tau_w($v$)$ by using variable geometry and/or counting the
number of up-scattered neutrons.

\subsection{Neutron Lifetime and the Big Bang}

The free neutron lifetime is also of significance in big bang
cosmology, where it directly influences the relative abundance of
primordial helium synthesized in the early universe. This is
determined by the ratio of the neutron lifetime to the expansion
time from that epoch at which neutrinos decouple from hadronic
matter to the onset of nucleosynthesis \cite{schramm}.

The argument goes briefly as follows. At times t $< 10^{-2}$sec.
and temperatures T $> 10^{11}$K the populations of neutrons and
protons are kept in a state of thermal equilibrium, i.e.
\begin{equation}
X_n/X_p=e^{-(m_n-m_p)c^2/kT}  \label{2.4a}
\end{equation}

\noindent   through the weak interactions
\begin{equation}
n+e^{+}\rightleftharpoons p+\overline{\nu
}_e\,;\;p+e^{-}\rightleftharpoons n+\nu _e  \label{2.4b}
\end{equation}

\noindent  At t $\simeq1$ sec. the freeze-out temperature T$\simeq
10 ^{10} $\thinspace K is reached where the leptons decouple from
the hadrons and neutrons begin to decay into protons according to
(\ref{2.1c}). This process continues until a time t $\simeq 180$
sec when the temperature has fallen to a value T$\simeq $10$^9\,$K
and deuterium formed by the capture of neutrons on protons remains
stable in the thermal radiation field. This is followed by a
sequence of strong interactions whose net effect is the conversion
of all free neutrons into helium. Using the current value of the
neutron lifetime, these considerations result in a relative helium
abundance
in the present day universe of about 25\% in good agreement with observation.%
\subsection{ Application to Solar Astrophysics}

The main source of solar energy derives from the proton-proton
cycle of thermonuclear reactions, the end-point of which is the
fusion of four protons into a helium nucleus with the release of
positrons, photons and neutrinos. In the first step two protons
interact weakly to form deuterium
\begin{equation}
p+p\Rightarrow ^2\!\!\!H+e^{+}+\nu _e  \label{2.5a}
\end{equation}

\noindent An alternative reaction is the weak $p-e-p$ process which occurs
with a branching ration of approximately 0.25\%
\begin{equation}
p+e^{-}+p\Rightarrow ^2\!\!\!H+\nu _e  \label{2.5b}
\end{equation}

\noindent That the timescale is determined by the neutron lifetime
stems from the fact that the governing reaction (\ref{2.5a}) is
just inverse neutron decay with the spectator proton providing the
energy,while the $p-e-p$ interaction (\ref{2.4b}) is the
corresponding electron capture process \cite{bahcall}. However
since the two protons can interact weakly only in the $^1S_{0}$
state because of the Pauli principle, and since the deuteron can
exist only in the triplet state, it follows that the vector
contribution to the underlying inverse neutron $\beta$-decay is
forbidden and the weak capture of protons on protons proceeds at a
rate proportional to ${\vert}G _A|^2$. To compute this rate it is
therefore necessary to determine individual values for the weak
coupling constants $G_V$ and $G_A$.
\subsection{Determination of the Weak Coupling Constants}

The neutron lifetime $\tau _n=t_n/\ln(2)$, where the half-life
$t_n$ is commonly employed in nuclear physics, is given by the
formula
\begin{equation}
ft_n=\frac{2\pi ^3\ln(2)\hbar ^3}{m_e^5c^4}\cdot[|G_{V\textrm{ }}|^2+3|G_A|^2]^{-1}=%
\frac K{|G_{V\textrm{ }}|^2}\cdot[1+3|\lambda |^2]^{-1}
\label{2.6a}
\end{equation}

\noindent where K=(8120.271$\pm
0.012)\cdot10^{-10}\,$GeV$^{-4}\,$sec., and
\begin{equation}
\lambda =G_A/G_{V\textrm{ }}  \label{2.6b}
\end{equation}

\noindent The factor f is the integral of the Fermi
Coulomb-corrected phase space function $F(E_e)$ which, including
the outer radiative corrections $\delta _R>0$, has the value
\cite{towner}
\begin{equation}
f(1+\delta _R)=1.71489\pm 0.00002  \label{2.6c}
\end{equation}

If isospin invariance of the strong interactions and conservation
of the weak vector current are assumed, then ${\vert}G_{V\;}|$ may
be determined from the ft-values of the sequence of pure Fermi superallowed 0%
$^{+}\Rightarrow 0_{}^{+}$nuclear positron emitters through the formula
\begin{equation}
\overline{ft(1-\delta _C)(1+\delta _R)(0^{+}\Rightarrow 0_{}^{+})}=K/|G_{V%
\textrm{ }}|^2  \label{2.6d}
\end{equation}

\noindent where each nuclear decay has been individually
corrected, incorporating factors (1-$\delta _C)\prec 1$ for
isospin symmetry-breaking and (1+$\delta _R)\succ 1$ for the
nucleus-dependent radiative correction. It follows that the values
of ${\vert}G_{V\;}|$ and ${\vert}G_A|$ can be determined from a
combination of equations (\ref{2.6a}) to (\ref{2.6d}). To
determine the relative sign of $G_V$ and $G_A$ it is necessary to
observe some phenomenon which relies on Fermi/Gamow-Teller
interference and this requires the availability of polarized
neutrons. Such phenomena allow the direct determination $\lambda$
and thus $G_V$ and $G_A$ can each be determined in both sign and
magnitude from neutron decay alone, in which case uncertainties
associated with nuclear structure effects do not arise.

\section{Neutron Decay in the Context of Particle Physics}

\subsection{The Cabibbo-Kobayashi-Maskawa Matrix}

The neutron and proton form the components of an isospin doublet
and are the lightest constituents of the lowest SU(3) flavor
octet, each of whose sub-multiplets is characterized by its
isospin (I) and its hypercharge (Y). A quantum number alternative
to hypercharge is the strangeness S=Y-B where the baryon number B
has the value unity. Flavor SU(3) symmetry is based on neglect of
the difference in mass between the u- and d-quarks on the one
hand, and the s-quark on the other, and is severely broken.
Because of the near equality of the u- and d-quark masses the
isospin SU(2) symmetry is much more closely realized, a result
which is derived from a dynamic global gauge symmetry of the QCD
Lagrangian which is expressed in the conservation of the weak
vector current.
\\ In increasing order of mass the octet contains
an isodoublet \{n,p; I=1/2, Y=1\}, an isosinglet \{$\Lambda ^0;$
I=0, $Y=0\}$, an isotriplet \{$\Sigma ^{-},\Sigma ^0,\Sigma ^{+};$
I=1, $Y=0\}$ and a heavy isodoublet, the so-called cascade
particles \{$\Xi ^{-},\Xi ^0,$ I=1/2, $Y=-1\}$. The $\Sigma ^0$
decays electromagnetically into the $\Lambda ^0$ which has the
same value of Y and differs only in the value of I which is not
conserved by the electromagnetic interaction. Semi-leptonic weak
decays within the octet are
characterized according to whether they are hypercharge conserving (e.g. $%
n\Rightarrow p,\Sigma ^{-}\Rightarrow \Lambda ^0$ and $\Sigma
^{+}\Rightarrow \Lambda ^0$), or hypercharge violating (e.g.
$\Sigma ^{-}\Rightarrow n,\Sigma ^{+}\Rightarrow n$ and $\Xi
^{-}\Rightarrow \Lambda ^0).$

It was Cabibbo's original insight to note and appreciate the
significance of the fact that the vector coupling constants
corresponding to hypercharge conserving weak decays G$_{V\textrm{
}}(\Delta Y=0),$and hypercharge non-conserving weak decays
G$_{V\textrm{ }}(\Delta Y=1)$ satisfied the empirical relations
\begin{equation}
G_{V}\,(\Delta Y=0)=G_F\cdot \cos(\theta _c);\;G_{V}\,(\Delta
Y=1)=G_F\cdot\sin(\theta _c)  \label{3.1a}
\end{equation}

\noindent where the Fermi coupling constant $G_F$ is determined
from the lifetime of the muon and the Cabibbo angle $\theta
_c\simeq 0.23$. The result (\ref{3.1a}) is interpreted to mean
that the charged vector bosons $W^{\pm }$ which mediate the weak
interaction couple to the mixtures of quark mass eigenstates
\begin{equation}
d^{\prime }=d\cdot\cos(\theta _c)+s\cdot\sin(\theta
_c)\;;\;s^{\prime }=-d\cdot\sin(\theta _c)+s\cdot\cos(\theta _c)
\label{3.1b}
\end{equation}

\noindent  rather than to the mass eigenstates of the down ($d$) and strange ($%
s$) quarks themselves.

In the Standard Model of Particle Physics these ideas are extended
to three quark generations where the couplings effective for the
weak semi-leptonic decays of quarks are described by the
Cabibbo-Kobayashi-Maskawa (CKM) matrix \cite{donoghue}, which
rotates the quark mass eigenstates $(d,s,b)$ to the weak
eigenstates$(d^{^{\prime }},s^{^{\prime }},b^{^{\prime }})$:
\begin{equation}
\left(
\begin{array}{c}
d^{^{\prime }} \\
s^{^{\prime }} \\
b^{\prime }
\end{array}
\right) =\left(
\begin{array}{ccc}
V_{ud} & V_{us} & V_{ub} \\
V_{cd} & V_{cs} & V_{cb} \\
V_{td} & V_{ts} & V_{tb}
\end{array}
\right) \left(
\begin{array}{c}
d \\
s \\
b
\end{array}
\right)  \label{3.1c}
\end{equation}

\noindent  where
\begin{equation}
V_{ud}\simeq V_{cs}\simeq \cos(\theta _c)\;;\;V_{us}\simeq
-V_{cd}\simeq \sin(\theta _c)  \label{3.1d}
\end{equation}

Since, assuming that no more than three quark generations exist,
the CKM matrix must be unitary, its nine elements can be expressed
in terms of only four real parameters, three of which can be
chosen as real angles and the fourth as a phase. If this phase is
not an integral multiple of $\pi $, then CP symmetry is violated.
For this to be possible the number of quark generations must be at
least three. In the present context the unitarity of the CKM
matrix requires that
\begin{equation}
|V_{ud}|^2+|V_{us}|^2+|V_{ub}|^2=1  \label{3.1e}
\end{equation}

\noindent  and the role of neutron $\beta $-decay centers on the
determination of the largest matrix element $V_{ud}$.

\subsection{ Neutron Decay in the Standard Model}

In the Standard Model the weak interaction responsible for neutron
decay is given as the contraction of a leptonic current $J_\mu
^l(x)$ and a hadronic current $J_\mu ^h(x)$, where, in the
convention that the operator (1-$\gamma _5)/2$ projects out the
left-handed field components,
\begin{equation}
J_\mu ^l(x)=\overline{e}\gamma _\mu (1-\gamma _5)\nu _{e\;};\;J_\mu ^h(x)=%
\overline{d}\cdot V_{ud}\gamma _\mu (1-\gamma _5)u_{\;}
\label{3.2a}
\end{equation}

\noindent  Since the leptons have no strong interactions the matrix element
of the weak leptonic current is relatively simple, i.e.
\begin{equation}
\langle e^{-}\overline{\nu }_e|J_\mu ^l(0)|0\rangle =\langle \overline{u_e}%
|\gamma _\mu (1-\gamma _5)|u_{\nu _e}\rangle  \label{3.2b}
\end{equation}

\noindent  where u$_e$ and u$_{\nu _e}$ are Dirac spinors describing
electron and neutrino respectively. However since the quarks are strongly
interacting particles confined in nucleons the hadronic matrix elements are
in principle limited only by the requirements of Lorentz invariance and
maximal parity violation. Thus we find for the matrix element of the vector
current

\begin{equation}
\langle p|J_\mu ^{h,V}(0)|n\rangle =\langle
\overline{v_p}|g_V(q)\gamma _\mu -i\frac \hbar
{2m_pc}g_{WM}(q)\sigma _{\mu \nu }q_\nu +\frac \hbar
{2m_pc}g_S(q)q_\mu |v_n\rangle  \label{3.2c}
\end{equation}

\noindent  where $v_n$ and $v_p$ are neutron and proton spinors
respectively, q$_\mu $ is the 4-momentum transfer and g$_i(i=V,$
$WM,$ $S)$ represent form factors corresponding to the bare
vector, induced weak magnetism and induced scalar interactions
respectively. As noted in section \ref{secdecay}, for neutron
decay all form factors may be evaluated at $q=0$. Conservation of
the vector current then requires that
\begin{equation}
g_V(0)=1,\quad g_{WM}(0)=\kappa _p-\kappa _n=3.70,\quad g_S(0)=0.
\label{3.2d}
\end{equation}

\noindent  where $\kappa _p=1.79,$and $\kappa _n=-1.91,$ are the \textit{%
anomalous} magnetic moments of neutron and proton respectively,
expressed in units of the nuclear magneton. Since weak magnetism
is a term of recoil order it makes only a very small correction to
the vector matrix element in neutron decay and is totally absent
in pure Fermi decays. An alternative test of the conserved weak
vector current theorem in action outside the regime of baryon
decays is the pure Fermi 0$^{-}\Rightarrow 0_{}^{-}$ $\beta
$-decay $\pi ^{+}\Rightarrow \pi ^0+e^{+}+\nu _e$. The induced
scalar interaction is also ruled out on the separate grounds that,
having the wrong transformation properties under the G-parity
transformation, it is second class and therefore does not
contribute to $\beta -$decays within an isospin multiplet
\cite{holstein}.

The corresponding axial matrix element is
\begin{equation}
\left\langle p\;|J_\mu ^{h,A}(0)|n\right\rangle =\left\langle
\overline{v_p}|g_A(q)\gamma _\mu \gamma _5-i\frac \hbar
{2m_pc}g_T(q)\sigma _{\mu \nu }q_\nu \gamma _5+\frac \hbar
{2m_pc}g_P(q)q_\mu \gamma _5|v_n\right\rangle  \label{3.2e}
\end{equation}

\noindent  The axial current is not conserved which means that the
form factor $g_A(0)$ is nucleon structure dependent and has to be
determined experimentally. Since the induced tensor form factor
g$_T(0)$ is also ruled out as second class, and the operator
$q_\mu\gamma_5$ does not contribute to allowed decay between
nuclear states of the same parity, it follows that the axial
matrix element depends only on the single constant $g_A(0)$ which,
given that $g_V(0)=1,$ becomes identical with the empirical
constant $\lambda$ introduced in (\ref{2.5b}).
\section{The Correlation
Coefficients in Neutron Decay}

\subsection{Polarized Neutron Decay}

In a pure Fermi transition nuclear polarization is not possible
and in a pure Gamow-Teller transition only the M=$\pm 1$ lepton
magnetic substates contribute to the correlation between the
nuclear spin and the lepton momenta. This is the origin of the
parity violation phenomenon first observed in the decay of
$^{60}Co$. However, in a mixed transition such as in neutron
decay, interference can arise between the singlet and triplet
magnetic substates with M=0. As a consequence, depending on the
sign of $\lambda$, either the electron or the antineutrino
asymmetry will be enhanced as compared with pure Gamow-Teller
decay, the other being reduced in proportion.

Experimental study of the angular and polarization coefficients
which characterizes the decay of unpolarized and polarized
neutrons offers an alternative route to the determination of
$\lambda$. These involve carrying out measurements of the neutron
spin polarization $\mathbf{\sigma }_n,$ and perhaps the electron
polarization $\mathbf{\sigma }_e,$ together with some combination
of the energies $E_e,E_{\bar{\nu}},E_p$ and momenta $\mathbf{p}
_e,\mathbf{p}_{\bar{\nu}}$, $\mathbf{p}_p$ of the three particles
in the final state. The transition rate for a polarized neutron
can then be written \cite{jackson}:

\[
dW(\mathbf{\sigma },\mathbf{p}_e,\mathbf{p}_{\bar{\nu}})\propto
F(E_e)d\Omega _ed\Omega _{\bar{\nu}}\{1+a\,\frac{\mathbf{p}_e\,\mathbf{.\,p}%
_{\bar{\nu}}}{E_eE_{\bar{\nu}}}+\frac{bm_e}{E_e}+
\]
\begin{equation}
+\langle\mathbf{\sigma }_n\mathbf{\rangle(}A\frac{\mathbf{p}_e}{E_e}+B\frac{\mathbf{p}_{%
\bar{\nu}}}{E_{\bar{\nu}}}+D\frac{\mathbf{p}_e\,\times \,\mathbf{p}_{\bar{\nu%
}}}{E_eE_{\bar{\nu}}}\mathbf{+}R\frac{\,\mathbf{\sigma }_e\,\times \,\mathbf{%
p}_e\,}{E_e}+\ldots)\}  \label{4.1a}
\end{equation}

\noindent where the neutron polarization $\mathbf{\sigma }_n$ has
been averaged over all wavelengths and positions within the
neutron beam and some less significant correlations have been
omitted. The three correlation coefficients $a,A$ and $B$, which
have finite values within the Standard Model, are given in lowest
order by the relations:
\begin{equation}
a=\frac{1-|\lambda |^2}{1+3|\lambda |^2},\quad A=-2\frac{|\lambda |^2+%
\mathit{Re}(\lambda )}{1+3|\lambda |^2},\quad B=2\frac{|\lambda |^2-\mathit{%
Re}(\lambda )}{1+3|\lambda |^2}  \label{4.1b}
\end{equation}

\noindent  where the possibility has been left open that the
coupling constant ratio $\lambda $ might be complex signalling a
break-down of time reversal invariance in the weak interaction.
Each of these coefficients has to be corrected by inclusion of
radiative corrections plus additional terms of recoil order
including weak magnetism. However certain linear combinations of
these coefficients exist which are independent of radiative
corrections to lowest order in the fine structure constant $\alpha
$ omitting cross terms of order $\alpha q$ or $\alpha
(E_e/m_p)\ln(m_p/E_e)$. These relations are \cite{garcia}:
\begin{equation}
f_1=1+A-B-a=0;\quad f_2=aB-A^2--A=0  \label{4.1c}
\end{equation}

The possibility of a breakdown in T-invariance is tested in a measurement of
the T-odd, P-even triple correlation coefficient D which is given by the
expression
\begin{equation}
D=\frac{2\mathit{Im}(\lambda )}{1+3|\lambda |^2}  \label{4.1d}
\end{equation}

\noindent In order to establish a violation of T-invariance it is
necessary to identify some feature of the decay which changes sign
under reversal of the time but not under inversion of the
coordinate system. The term\textbf{\ }$\mathbf{\sigma
}_n\,\cdot(\mathbf{p}_e\times \,\mathbf{p}_{\bar{\nu}})$ possesses
the desired property.
\subsection{Non-Standard Model Contributions to
the Correlation Coefficients}

The leading coefficients $a,A$ and $B$ are each sensitive to
right-handed contributions to the weak interaction irrespective of
any possible contribution from scalar or tensor couplings. For
example in left-right symmetric models the coefficient A takes the
form\cite{carnoy}
\begin{equation}
A=-2\frac{|\lambda |^2(1+y^2)+\mathit{Re}(\lambda )(1-xy)+T_1}{%
1+x^2+3|\lambda |^2(1+y^2)+T_2}  \label{4.2a}
\end{equation}

\noindent where T$_1$and T$_2$ are small terms of recoil order, x $\simeq
\delta -\zeta ,$ $y\simeq \delta +\zeta ,\delta $ is the square of the ratio
of the mass of the light W-boson which couples to left-handed currents to
the mass of the postulated heavy W-boson coupling to right handed currents
and $\zeta $ is the mixing angle. In these models D is linear in $\zeta $
and is particularly sensitive to a T-violating coupling of a left-handed
lepton to a right-handed quark.

When the possibility is allowed for contributions from scalar and tensor
couplings then both the Fierz interference coefficient $b$ and the
T-violating coefficient $R$ receive finite contributions. Specifically
\begin{equation}
b=b_F+b_{GT}\propto \mathit{Re}(G_VG_{S^{}}^{*}+G_V^{\prime
}G_{S^{}}^{\prime \,*})-3\mathit{Re}(G_AG_{T^{}}^{*}+G_A^{\prime
}G_T^{\prime \,*})  \label{4.2b}
\end{equation}

\noindent and

\[
R=R_F+R_{GT}\propto -\mathit{Im}(G_A^{\prime }G_S^{*}+G_A^{}G_S^{\,*})+
\]
\begin{equation}
+\mathit{Im}(3Re(G_AG_{T^{}}^{*}+G_A^{\prime }G_T^{\prime
\,*})+G_V^{\prime }G_T^{*}+G_V^{}G_T^{\prime \,*})  \label{4.2c}
\end{equation}

\noindent where in this case it is necessary quite generally to
distinguish between coupling constants which are P-conserving
(e.g. $G_V$) and P-non-conserving (e.g. $G_V^{\prime }$). The Fermi coefficients b$_F$ and R$%
_F$ are particularly sensitive to the scalar coupling of a
right-handed lepton to any quark while the Gamow-Teller
coefficients are sensitive to the tensor coupling of a
right-handed lepton to a left-handed quark \cite{deutsch}.

\section{Measurement of the Correlation Coefficients}

\subsection{ The Electron-Antineutrino Angular Correlation
Coefficient $a$}

Since the electron spectrum in allowed $\beta $-decay is determined by the
Fermi phase space factor F(E$_e)$ alone, it is insensitive to the details of
the weak interaction. Thus, up to the discovery of parity violation, the
correlation coefficient $a$ was the only parameter available to provide such
information. Also since the operator\textbf{\ p}$_{e\,}.\mathbf{p}_{\bar{\nu}%
} $ commutes with the total angular momentum of the leptons, and therefore
does not mix singlet and triplet operators, it follows that the correlation
coefficient
\begin{equation}
a=\frac{1-|\lambda |^2}{1+3|\lambda |^2};\qquad \frac{\delta |\lambda |}{%
|\lambda |}\simeq 0.27\frac{\delta a}a\simeq 1\%  \label{5.1a}
\end{equation}

\noindent contains no Fermi/Gamow-Teller interference terms apart
from small terms of recoil order.

It is, of course, impracticable to measure the correlation between
the electron and antineutrino momenta directly, since efficient
detectors of antineutrinos do not exist. In practice therefore
only two indirect methods have been have been employed. These are
(a) measuring the momentum spectrum of electrons emitted into a
given range of angles referred to the proton momentum and (b)
measuring the proton spectrum \cite{nachti}. The experimenter is
therefore presented with a choice between electron spectroscopy
and proton spectroscopy and both methods have been explored...

It turns out that, up to the present, the measurement of the
proton spectrum has proved the more fruitful and two studies of
this nature have been completed. These have used (a) proton
magnetic spectroscopy and (b) a Penning trap with adiabatic
focusing.Both experiments have required the addition of post
acceleration of the protons to energies of order 20-30 keV and
have each reached precisions on $a$ at the level of 5\%. Because
this correlation measures the anomaly in ${\vert}\lambda |$ rather
than ${\vert}\lambda |$ itself the resultant error in ${%
\vert}\lambda |$ is reduced to $\simeq $1.4\%.

Angular correlation measurements have the great advantage that it
is not necessary that the neutrons be polarized and this route to
the determination of ${\vert}\lambda |$ has yet to achieve its
true potential.

\subsection{ The Electron-Neutron Spin Asymmetry Coefficient $A$}

The correlation coefficient
\begin{equation}
A=-2\frac{|\lambda |^2+\mathit{Re}(\lambda )}{1+3|\lambda |^2};\qquad \frac{%
\delta \lambda |}\lambda \simeq 0.24\frac{\delta A}A\simeq 0.23\%
\label{5.2a}
\end{equation}

\noindent has been subjected to an enormous amount of experimental
study going back to the 1950's. It has provided the most precise
value for the parameter $\lambda $ both in magnitude and sign$,$
and therefore for the CKM matrix element V$_{ud}^{}$ based on
neutron decay alone \cite{hartmut}. This information has been
largely derived from studies over the past $\simeq $15 years at
the ILL, Grenoble using the electron spectrometer PERKEO in its
various forms. The current world average value for $\lambda $ is
\cite{pdg2}:
\begin{equation}
\lambda =-1.2670\pm 0.0030\   \label{5.2b}
\end{equation}

Like the $a$-coefficient, the $A$-coefficient has the great advantage the it
measures the anomaly in $\lambda$. However it relies critically on $\simeq $%
1 MeV electron spectroscopy, and, although this is in general
easier to perform than $\simeq$ 1 keV proton spectroscopy, it has
not proved possible to extend the electron spectrum down to the
lowest energies. However the measurement of $A$ suffers from the
great disadvantage that the neutrons must be polarized and the
neutron polarization must be measured to an accuracy $\geq 99\%$
and this is not easy. Fortunately discrepancies between the values
of the polarization derived using polarizer/analyser combinations
based on supermirrors and $^3He$ filters appear to have been
satisfactorily resolved.

\subsection{ The Antineutrino-Neutron Spin Asymmetry Coefficient
B}

The correlation
\begin{equation}
B=2\frac{|\lambda |^2-\mathit{Re}(\lambda )}{1+3|\lambda |^2};\qquad \frac{%
\delta \lambda }\lambda \simeq 2.0\frac{\delta B}B  \label{5.3a}
\end{equation}

\noindent is quite insensitive to the value of $\lambda$. Its
measurement has the disadvantages that it requires both that the
neutrons be polarized and that proton spectroscopy be performed.
For both these reasons it has tended to be neglected as a topic
for study. However, for the same reason that it is insensitive to
the precise value of $\lambda$, it is very sensitive to
contributions from right-handed bosons and recent measurements
have succeeded in setting a limit $m_{BR}> 284.3\,GeV/c^2$ for the
mass of the heavy W-boson which is postulated to couple to
right-handed currents \cite{serebrov}.

A recent encouraging development has been the simultaneous
measurement of $A$ and $B$ whose ratio is therefore independent of
neutron polarization \cite{14}.
\subsection{ The Triple Correlation Coefficient D}

This coefficient
\begin{equation}
D=\frac{2\mathit{Im}(\lambda )}{1+3|\lambda |^2}  \label{5.4a}
\end{equation}

\noindent is measured by counting coincidences between electrons
and protons detected in counters set at appropriately selected
angles for a given sign of the neutron spin. The spin is then
reversed and the relevant counting rate asymmetry is recorded.

The D-coefficient is of second order in the T-violating phase in
the CKM matrix and is expected to be vanishingly small. Currently
it is known to vanish at a level of about 0.1\% from neutron
decay, and to marginally better precision from the decay of
$^{19}Ne$. However, since the T-symmetry is non-unitary and is
generated by a non-linear operator, a violation can be mimicked by
final state electromagnetic interactions which in this instance
appear at a level of about 0.001\%.

\section{Additional Experimental Possibilities}

\subsection{The Proton-Neutron Spin Asymmetry Coefficient $\alpha$}

The individual coefficients A and B each have terms in ${\vert}
\lambda |^2$ deriving from the axial vector interaction, in
addition to terms in \textit{Re}($\lambda)$ generated through
polar vector/axial vector interference. Suppose, instead, one were
to measure the correlation $\alpha \mathbf{\sigma
}_n\cdot\mathbf{p}_p$ by detecting the complete range of proton
energies but without recording electron coincidences. Then, since
this is a parity-violating term and no lepton is detected, it
satisfies the conditions of Weinberg's interference theorem
\cite{15}, and is therefore
proportional to \textit{Re}($\lambda )$ with no term in ${\vert}%
\lambda |^2$. The corresponding expression for the coefficient
$\alpha $ is given by \cite{16}:
\begin{equation}
\alpha =C\frac{4\lambda }{1+3|\lambda |^2},\qquad C=0.27484,\qquad \frac{%
\delta \lambda }\lambda \simeq 1.5\frac{\delta \alpha }\alpha
\label{6.1a}
\end{equation}

\noindent where the kinematic constant C comes from the double
integral over electron and proton energies, and includes Coulomb,
recoil order and radiative corrections. Since in lowest order the
correlation $\alpha$ is
proportional to (A+B) it is also relatively insensitive to the value of $%
\lambda$.

The principle of an experiment is quite straightforward. Recoil
protons from the decay of longitudinally polarized neutrons are
collected in a magnetic field of order 5T, where the maximum
radius of the cyclotron orbit is $\prec 1mm$. If $N^{+}(N^{-})$
denote the numbers of protons with momenta parallel
(anti-parallel) to the neutron spin, then $N^{\pm }=N_0\{1\pm \alpha \langle%
\mathbf{\sigma }_n\rangle/2\}$ and the appropriate counting rate
asymmetry can be computed.\

To measure $N^{\pm }$, $s$et the orientation of the neutron spin parallel to
the magnetic field and reflect the protons from a $\simeq $1kV electrostatic
potential barrier so that protons of both senses of momentum enter the
detector which is maintained at about $-30\,kV$. Thus the counting rate is
given by
\begin{equation}
C_1=N^{+}+N^{-}+b  \label{6.1b}
\end{equation}

\noindent  where $b$ is the background. When the reflecting
potential barrier is removed the new counting rate is
\begin{equation}
C_2=N^{+}+\beta N^{-}+b  \label{6.1c}
\end{equation}

\noindent  where $\beta\ll1$ represents that fraction of protons
initially moving away from the detector which is reflected back
into the detector by magnetic mirror action. The procedure is now
repeated with the neutron spin direction reversed giving
corresponding counting rates $C_1^{^{\prime }},C_2^{^{\prime }}$
and background $b^{\prime}$. The counting rate asymmetry is then
given by
\begin{equation}
\frac{(C_1^{^{\prime }}-C_2^{^{\prime
}})-(C_1-C_2)}{(C_1^{^{\prime }}-C_2^{^{\prime
}})+(C_1-C_2)}=\alpha \langle\mathbf{\sigma }_n\rangle
\label{6.1d}
\end{equation}

The experiment only works on the assumption that the proton
counter background in the energy range $\leq 30\,keV$ is weak in
comparison to the signal strength which is certainly not true in
the case that the neutrons are polarized using a supermirror.

\subsection{ Two-Body Decay of the Neutron and Right-Handed
Currents}

When a neutron undergoes $\beta-$decay there is a small branching ratio $%
\simeq $4.10$^{-6}$ that the final state should contain an
antineutrino and a hydrogen atom i.e.
\begin{equation}
n\Rightarrow H+\stackrel{\_}{\nu }_e,  \label{6.2a}
\end{equation}

\noindent  where the hydrogen atom is created in an S-state. Since
this is a two-body decay, momentum conservation ensures that
antineutrino and hydrogen atom each carry off unique energies with
\begin{equation}
\mathbf{p}_{\bar{\nu}}+\mathbf{p}_H=0,\qquad
E_{\bar{\nu}}=783\,keV,\qquad T_H=352\,eV  \label{6.2b}
\end{equation}

Although the higher S-levels decay spontaneously, hydrogen atoms
created in the metastable 2S state can exist in one of four
decoupled hyperfine levels $ |M_e,M_p\rangle$ with populations
W$_i(i=1-4)$, where $\mathbf{n}_H=\mathbf{p}_H/p_H $ and
\begin{eqnarray}
\left|\frac {1}{2},\frac {1}{2}\right\rangle; \quad\;\;\;
W_1=2(1+|\lambda |^2)\{1+\mathbf{\sigma }_n\mathbf{.n}_H\}\simeq
0.57\%\quad \mathrm{when}\quad \mathbf{\sigma }_n\mathbf{.n}_H=0
\label{6.2c}
\\
\left|-\frac {1}{2},\frac {1}{2}\right\rangle;\quad W_2=8|\lambda
|^2\{1-\mathbf{\sigma }_n\mathbf{.n} _H\}\simeq 55.13\%\quad
\mathrm{when}\quad \mathbf{\sigma }_n\mathbf{.n}_H=0 \label{6.2d}
\\
\left|\frac {1}{2},-\frac {1}{2}\right\rangle;\quad
W_3=2(1-|\lambda |^2)\{1-\mathbf{\sigma }_n\mathbf{.n}_H\}\simeq
44.28\%\quad \mathrm{when}\quad \mathbf{\sigma }_n\mathbf{.n}_H=0
\label{6.2e}
\\
\left|-\frac {1}{2},-\frac {1}{2}\right\rangle;\;W_4=2(1+|\lambda
|^2)\{1+\mathbf{\sigma }_n\mathbf{.}
\stackrel{}{\mathbf{n}}_H\}\equiv 0  \label{6.2f}
\end{eqnarray}

The population $W_4$ vanishes identically in the case that the
weak interaction is purely left-handed, and this is a result which
depends on conservation of angular momentum only. Thus exploiting
the neutron polarization to suppress the populations $W_2$ and
$W_3$, observation of a finite population $W_4$ $\neq 0$ would
provide an unambiguous signature for the existence of right-handed
currents \cite{17}.

\subsection{Radiative Neutron Decay}

Radiative decay of the free neutron
\begin{equation}
n\Rightarrow p+\;e^{-}+\stackrel{\_}{\nu }_e+\gamma ,
\label{6.3a}
\end{equation}

\noindent also described as inner bremsstrahlung, has a branching
ratio at the level of 0.1\%. The matrix element for the process
consists of two terms; a term describing electron photon emission
and a term describing proton photon emission. Both terms contain
infra-red divergences which cancel. However because
${\vert}\lambda |\neq 1$, contrary to the situation in the case of
the muon which has no strong interactions, the total radiative
correction depends on the ultra-violet cut-off parameter $\Lambda
$. Thus the simplest experiments designed to detect the inner
bremsstrahlung provide a measure of the outer radiative correction
only.

Experiments designed to measure the branching ratio for radiative
neutron decay by detecting triple coincidences between electron,
proton and gamma are currently under way at the ILL Grenoble
\cite{18}.
\section*{Acknowledgements}

I should like to thank the organizers of the Quark Mixing-CKM
Unitarity Workshop at Heidelberg for the invitation to present
this overview of neutron $\beta$-decay. I am also very happy to
acknowledge the benefit of conversations on these topics with
Hartmut Abele, Philip Barker, Lev Bondarenko, Ferenc Gluck, John
Hardy, Paul Huffman, Nathal Severijns, Boris Yerolozimsky and
Oliver Zimmer.



\def\F{{\cal F}}
\def\DRV{\Delta_{\mbox{\tiny R}}^{\mbox{\tiny V}}}
\title*{Neutron Lifetime Value \\ Measured by Storing Ultra-Cold Neutrons with
Detection of Inelastically Scattered Neutrons}

\toctitle{Neutron Lifetime Value Measured by Storing Ultra-Cold Neutrons with
Detection of Inelastically Scattered Neutrons}
\titlerunning{Neutron Lifetime}

\authorrunning{L. Bondarenko et al.}
\tocauthor{S. Arzumanov, L. Bondarenko, W. Drexel, A. Fomin, P. Geltenbort, \\ V. Morozov, Yu. Panin, J. Pendlebury, K. Schreckenbach} 
\author{S. Arzumanov\inst{1}, L. Bondarenko\inst{1}, Chernyavsky\inst{1}, W. Drexel\inst{2}, A. Fomin\inst{1}, \\ P. Geltenbort\inst{2}, V. Morozov\inst{1}, Yu. Panin\inst{1}, J. Pendlebury\inst{3}, K. Schreckenbach\inst{4}}

\institute{ RRC Kurchatov Institute, 123182, Moscow, Russia\and
 Institute Laue Langevin, BP 156, F-38042
Grenoble Cedex 9, France \and University of Sussex, Brighton\and
BN1 9QH, Sussex, U.K.\and Technical University of Munich, D-85747
Garching, Germany}
\maketitle

\begin{abstract}
The neutron life time $\tau _{n}$ was measured by storage of ultracold
neutrons (UCN) in a material bottle covered with Fomblin oil. The
inelastically scattered neutrons were detected by surrounding neutron
counters monitoring the UCN losses due to upscattering at the bottle walls.
Comparing traps with different surface to volume ratios the free neutron
life time was deduced. Consistent results for different bottle temperatures
yielded $\tau _n\left[ \sec \right]=885.4\pm 0.9_{stat}\pm 0.4_{syst}.$
\end{abstract}

\section{Introduction}

A detailed description of this experiment was published in \cite{Arzu}.
In a simple quark picture of the free neutron beta decay a $d$ quark
transforms into an $u$ quark under emission of a virtual $W$-boson which in
turn decays into an electron and an electron antineutrino. A breakthrough in
precision of the neutron lifetime has been achieved by storage experiments
of ultra cold neutrons \cite{1,2,3}. Including results from the
correlation coefficients between the decay partners, in particular the beta
asymmetry coefficient A \cite{4,5,6} the vector and axial vector coupling
constants $g_{V}$ and $g_{A}$ were deduced from the neutron decay alone. The
obtained value for $g_{V}$ was compared with data from muon decay and
superallowed beta-decays yielding stringent limits on possible deviations
from the universality of the weak interaction coupling constants, on right
handed currents and on the unitarity of the CKM matrix \cite{4,7,8}.

In neutron lifetime measurements by UCN storage the UCN's are contained by
material walls due to the Fermi-pseudo potential, by gravity or by the
interaction on the neutron's magnetic moment with a magnetic field gradient.
Conceptually those experiments are quite simple. UCN's are filled in a
storage volume with suitable walls. After a storage period the surviving
neutrons are counted. Repeating this experiment with different storage times
yields the decay curve of the neutrons. The major problem encountered in
this method is caused by losses of UCN in collisions with the trap walls.
Extrapolation to infinite trap size yielded $\tau _{n}=1/\lambda _{n}$.

In the present experiment new method was used to separate wall losses from
beta decay. The main loss process was monitored during storage by measuring
the relative flux of inelastically scattered UCN by a set of neutron
detectors surrounding the vessel, see Fig.1. The survival probability of the
UCN was measured by the usual UCN storage and disappearance method of the
neutrons in the trap. The trap was arranged such that the UCN could be
stored in two different sections with different surface to volume ratios and
hence different total UCN survival times. Comparing the survival time and
upscattering rates for the two volumes yielded the value of $\tau _{n}$.

\section{Basic idea of the experimental method}

When monoenergetic UCN are in the trap, the number of neutrons $N(t)$ in the
trap changes exponentially during the storage time, i.e. $%
N(t)=N_{0}e^{-\lambda t}$. The value $\lambda $ is the total probability per
unit time for the disappearance of UCN due to both the beta-decay and losses
during UCN-wall collisions. In turn, losses are equal to the sum of the
inelastic scattering rate constant $\lambda _{ie},$ and that for the neutron
capture at the wall, $\lambda _{cap}$.
\begin{equation}
\lambda =\lambda _{n}+\lambda _{loss}=\lambda _{n}+\lambda _{ie}+\lambda
_{cap}
\end{equation}

The ratio $\lambda _{cap}/\lambda _{ie}$ is to a good approximation equal to
the ratio of the UCN capture and inelastic scattering cross sections for the
material of the wall surface since both values are proportional to the wall
reflection rate of UCN in the trap:
\begin{equation}
a=\lambda _{loss}/\lambda _{ie}=1+\lambda _{cap}/\lambda _{ie}=1+\sigma
_{cap}/\sigma _{ie}
\end{equation}
and is constant for the given conditions, i.e. same wall material and
temperature. During storage the upscattered neutrons are recorded with an
efficiency $\varepsilon _{th}$ in the thermal neutron detector surrounding
the storage trap. The total counts in the time interval $T$ is equal to
\begin{equation}
J=\varepsilon _{th}\lambda _{ie}\left( N_{0}-N_{T}\right) /\lambda
\end{equation}

Here $N_{0}$ and $N_{T}$ are the UCN populations in the trap at the
beginning and the end of the storage time $T$, respectively. The UCN
themselves are measured with an efficiency $\varepsilon $ such that the
detected UCN at the beginning (normalization measurement) and the end of the
storage time are equal to $N_{i}=\varepsilon N_{0}$ and $N_{f}=\varepsilon
N_{T}$ respectively. We have then :
\begin{equation}
\lambda _{ie}=\frac{J\lambda }{\left( N_{i}-N_{f}\right) }\frac{\varepsilon
}{\varepsilon _{th}}
\end{equation}
and
\begin{equation}
\lambda =\frac{1}{T}\ln (N_{i}/N_{f})
\end{equation}

The experiment is repeated with a different value for the wall loss rates
with constant value $a$. Thus $\lambda _{n}$ is given by
\begin{equation}
\lambda _{n}=\frac{\xi \lambda ^{(1)}-\lambda ^{(2)}}{\xi -1}
\end{equation}
where
\begin{equation}
\xi =\lambda _{ie}^{(2)}/\lambda _{ie}^{(1)}
\end{equation}

The indices refer to the two measurements with different $\lambda _{loss}$.
The result contains then only the ratios of the directly measured quantities
$J,N_{i},N_{f}$ since the efficiencies of the neutron detection cancel. {\bf %
It is very important that this method is thus relative and asks
only the time interval absolute measurement.}

\section{Method for a broad UCN energy spectrum}

In a real experiment it is necessary to take into account the energy
distribution of UCN since the scattering and capture cross section{\bf s}
are in general energy dependent and losses are different for different parts
of the UCN energy spectrum. It makes the measurement more complex.
Nevertheless, {\bf the main idea - the relativity of the experiment, remains
the same}. In this case the result evaluation includes $\lambda $-values,
averaged over the storage period $T$ ($\bar{\lambda}_{ie}$, $\bar{\lambda}$%
), as well as new parameters (ratios $\varepsilon _{i}/\varepsilon _{f}$,
coefficients $k$ in the time function for $\lambda _{ie}$etc) which were
measured during the storage period or at additional experiments. The ratio ${%
\varepsilon _{th}^{(1)}}/{\varepsilon _{th}^{(2)}}$ of the thermal
neutron detection was calculated by the Monte Carlo method that
was evaluated by the Monte-Carlo method using the neutron cross
sections measured at special experiments for neutrons that were
inelastically scattered during storing time.

Now the UCN populations decay at different rates in the trap \cite{9} and:
\begin{equation}
N(t)=N_{0}\exp \left( \int\limits_{0}^{t}-\lambda (t^{\prime })dt^{\prime
}\right)
\end{equation}

The rate $\lambda (t)=\lambda _{n}+\lambda _{loss}(t)\ $ and $\lambda
_{ie}(t)=\lambda _{ie0}(1-\gamma t)$. The quantity $\gamma $ is order of $%
10^{-{\bf 4}}s^{-1}$ when $T\ll 1/\lambda _{loss}$.

Using the above parameter definition we have $\lambda _{loss}=\lambda
_{ie}(t)\cdot a$ and the mean value of $\lambda (t)$ over the time interval $%
T$ is then given by
\begin{equation}
\bar{\lambda}=1/T\int \lambda (t)dt=\lambda _{n}+\lambda _{ie0}(1-\frac{%
\gamma T}{2})a\ \hspace{1in}
\end{equation}

and
\begin{equation}
\bar{\lambda}=\frac{1}{T}ln(\frac{N_{i}}{N_{f}}\cdot \frac{\varepsilon _{f}}{%
\varepsilon _{i}})
\end{equation}
where the ratio of the UCN detection efficiency $\varepsilon _{i}$, $%
\varepsilon _{f}$ varies slightly with $T$. The value of $a$ in
the case where trap walls are coated by a layer of hydrogenfree
oil (Fomblin type) is close to unity: $a-1<2\cdot 10^{-2}$ and
temperature dependent.

The full counts of the thermal neutron detector during storage is equal to
\begin{equation}
J=\varepsilon _{th}\lambda _{ie0}N_{0}\cdot \int_{0}^{T}(1-\gamma t)\exp
[-\left( \bar{\lambda}+\lambda _{ie0}a\gamma (T-t)/2\right) t]dt
\end{equation}

Expanding the second part of the exponent, neglecting the $\gamma^2$ terms
and solving for $\lambda_{ie0}$ gives
\begin{equation}
\lambda _{ie0}=\frac{\bar\lambda J}{(N_i-N_f)\Phi }\cdot \frac \varepsilon
{\varepsilon _{th}}
\end{equation}
where
\begin{equation}
\Phi =1-\frac \gamma {\bar\lambda} (1+\frac 12\lambda _{ie0}aT)I_2/I_1+\frac
12\lambda _{ie0}\frac{\gamma a}{\bar\lambda ^2}I_3/I_1
\end{equation}
and $I_1=\int_0^{\bar\lambda T}e^{-x}dx$; $I_2=\int_0^{\bar\lambda
T}e^{-x}xdx$; $I_3=\int_0^{\bar\lambda T}e^{-x}x^2dx$.

The measured value
\begin{equation}
\bar{\lambda}_{ie}=\frac{J\bar{\lambda}}{(N_{i}-N_{f})}\cdot \frac{%
\varepsilon }{\varepsilon _{th}}
\end{equation}
and performing the pair of measurements with two different loss rate values,
the $\lambda _{n}$ value is derived as

\begin{equation}
\lambda _n =\frac{\xi \bar\lambda ^{(1)} -\bar\lambda ^{(2)}}{\xi -1}
\end{equation}

where the $\xi $-value is determined by:

\begin{equation}
\xi =\frac{\bar\lambda _{ie}^{(2)}}{\bar\lambda _{ie}^{(1)}} \cdot \frac{%
\left( 1-\frac{\gamma^{(2)}T^{(2)}}2\right) \Phi ^{(1)}a^{(2)}}{\left( 1-%
\frac{\gamma^{(1)}T^{(1)}}2\right) \Phi ^{(2)}a^{(1)}}
\end{equation}

The correction terms relative to the monoenergetic case are quite small if
cleaning times $(t_{cl}^{(1),(2)},$ see next section$)${\bf \ }are chosen
properly to have almost the same stored UCN spectra. In addition the product
$\gamma T$ is constant to a good approximation, when storage times are
scaled such that the same average number of wall reflections occur in $T$.
The deviation of{\bf \ }$\Phi ${\bf \ }and $a$ from unity are in the percent
range{\bf , if the specification of the wall (temperature, type of wall,
etc.) is the same,} and depend only on the surface to volume ratio via the
development of the UCN spectrum during storage. The values for{\bf \ }$\Phi $%
{\bf \ }and $a$ can be determined in particular from the time dependence $%
j(t)$ of the upscattering rate during the storage time.

The UCN detection efficiency $\varepsilon $ includes also the UCN losses in
the vessel during the emptying time of the vessel into the UCN detector. To
take into account a slight change of $\varepsilon $, the counting rate{\bf \
}$n(t)$ of the UCN detector during the emptying phase has to be corrected
for the decay rate $\bar{\lambda}$.

\section{The experiment}

The experiment was carried out at the UCN source of the ILL High Flux
Reactor in Grenoble.

The storage vessel was composed of two coaxial horizontal
cylinders which walls were coated with a thin layer of Fomblin.
The storage vessel was placed inside the vacuum housing which had
the cooling system stabilized the bottle temperature in the range
$+20^{\circ}$C to $-26^{\circ}$C. Residual gas pressure in  the storage
vessel was about $(1\div 5)\cdot 10^{-6}$ torr. The set-up was
surrounded by the thermal neutron detectors and supplied with the
UCN detector, both $^{3}$He filled. The whole installation was
placed inside the shielding: Cd of 1mm thick and boron
polyethylene of 16cm thick.

The UCN were stored either in the inner cylinder or in the annular gap
between both cylinders thereby changing the UCN loss rate by a factor of
about 4 without breaking the vacuum. The construction allowed to refresh the
oil layers on the cylinder walls also without a vacuum break.

In the experiment there were measured counting rates of both detectors: $j(t)
$, $n(t)$ during storage time interval $T$ as well as integral counts$\
J,N_{i},N_{f}.$ The elementary run consisted of two consequent measurements
with the UCN storage in both vessels. These elementary runs were repeated as
many times as was necessary to get sufficient statistics.

As discussed above the storage time intervals $T^{(1)},T^{(2)}$ in the inner
vessel ($^{(1)}$) and the annular vessel ($^{(2)}$) as well as $%
t_{cl}^{(1)},t_{cl}^{(2)}$ were chosen to make almost the same evolution of
the UCN spectra.

Groups of experimental runs were performed at the different temperature $+20$%
, $-8$, $-9$ and $-26$ $^{\circ}$C, respectively.

\section{Evaluation of the data and result}

A data evaluation was performed on the base of the method developed for a
spread UCN spectrum. Some important details of calculations:
\begin{enumerate}

\item The parameter $a$ was constant for a run at the same wall temperature as
the temperature of the vessel was stabilized and constant over the wall
surface within 0.1$^{\circ}$C (for room temperature) and 1.5$^{\circ}$C (for the
lower temperatures).

\item An important correction for $\tau _{n}$ of $-3.10\pm 0.36$s
arose from the ratios of the UCN detection efficiencies
$\varepsilon ^{(1)}/\varepsilon ^{(2)}$ and $\varepsilon
_{f}/\varepsilon _{i}$.These ratios were experimentally determined
from the counting rates $n(t)$ combined with measured values for
$\bar{\lambda}$.

\item The ratio $\varepsilon _{th}^{(1)}/\varepsilon _{th}^{(2)}$ of the thermal
neutron detection was calculated by the Monte Carlo method using mean values
for the capture and scattering cross sections (also measured in special
experiments) for neutrons that were inelastically scattered during the
storing time. The correction in $\tau _{n}$  was $0.6\pm 0.3$s. The
systematic error in this calculation reflects the uncertainty of the
geometry, the upscattering cross sections and the spectrum of upscattered
neutrons.

\item Time distributions $j(t)$ were used to determine the $k$-parameter for $%
\bar{\lambda}_{ie}$ and the $\xi $-values. Compared to a monoenergetic UCN
spectrum the correction in $\tau _{n}$ was $-2.0\pm 0.3$s.

\item Since the different bottle temperatures lead to very different
loss rates, the consistency of the set of neutron life time values
with divergence of error bar, obtained for the different bottle
temperatures, gives confidence in the experimental method used.

\item The final result: $\tau _{n}$[sec]=885.4$\pm$ 0.9(stat)$\pm$
0.4(system).

\item The uncertainties in $\tau _{n}$ due to the ratios $\varepsilon
^{(1)}/\varepsilon ^{(2)}(0.36$s) and $\varepsilon
_{f}/\varepsilon _{i}(0.3$s) as well as due to $k(0.3$s) are
included in the statistical error since they are based on the
measured time spectra of detectors counting rates.

\item The possible systematic error for $\tau _{n}$ is composed of
\begin{enumerate}
\item the uncertainty in $\varepsilon _{th}^{(1)}/\varepsilon _{th}^{(2)}$ $%
(0.3$s),

\item the influence of the UCN scattering at the residual gas
$(0.2$s),

\item the epi-Fomblin neutron impurity in the UCN spectrum $(0.2$s)

\item the temperature difference over the walls $(0.15$s),
\end{enumerate}
and was estimated (added up in quadrature) as $0.4$s.

The present experimental result is in agreement with the recent evaluation
of earlier data on the neutron life time of 886.7(1.9)s by the Particle
Data Group. New one (PDG-2001) is equal to 885.7(0.8)s
\end{enumerate}
\section{Conclusion}
\begin{enumerate}
\item Presented method of the neutron lifetime measurement is relative and does
not demand the absolute calibrations of the apparatus except of the timer.

\item The most part of the necessary parameters (the UCN detection
efficiencies, the UCN spectrum change rates etc) were measured
during the storage experiment. The other parameters were specially
measured under the storage experiment conditions.

\item Monte Carlo evaluation of the ratio of the detection
efficiencies for thermal neutrons was based on these measured
parameters.

\item The final result changed the world mean value on one second and,
seems, concluded the precision possibility of our method.
\end{enumerate}
\section{Acknowledgments}

The authors like to express their deep gratitude to S.T.Belyaev for his
interest and fruitful discussions and support of this investigation. We are
pleased to thank Dr.P.Iaydjiev for his help and cooperation during the long
period of measurements at the UCN source of the ILL. Authors are grateful to
Prof.A.Steyerl and Dr.E.Korobkina for attentive discussions and fruitful
criticisms of the presentation. The experiment could not have been carried
out without the intensive and very skilful assistance of H.Just to whom we
would like to express our thanks. The authors like to thank very much the
ILL reactor personnel for much help and informal assistance.

This experiment was made possible due to supports of INTAS (grant 93-298)
and of {\bf Russian Foundation for Basic Research} (grant {\bf RFBR}
93-02-3927).

\title*{Is the Unitarity of the \\Quark-Mixing CKM Matrix \\Violated in Neutron $\beta$-Decay?  \protect\newline}
\toctitle{Is the Unitarity of the Quark-Mixing CKM Matrix Violated
in Neutron $\beta$-Decay?}
%
%
\titlerunning{Is the Unitarity of the CKM Matrix Violated?}
%
\author{H. Abele}
 \institute{Physikalisches Institut der
Universit\"at Heidelberg, \\ Philosophenweg 12, 69120 Heidelberg,
Germany}
\authorrunning{H. Abele}
%
%

\maketitle              

\begin{abstract}
 Measurements by various international groups of researchers
determine the strength of the weak interaction of the neutron,
which gives us unique information on the question of the quark
mixing. Neutron $\beta$-decay experiments now challenge the
Standard Model of elementary particle physics with a deviation,
2.7 times the stated error.
\end{abstract}

\section{The Standard Model, Quark-Mixing and the CKM Matrix}
This article is about the interplay between the Standard Model of
elementary particle physics and neutron $\beta$-decay. Since the
Fermi decay constant is known from muon decay, the Standard Model
describes neutron $\beta$-decay with only two additional
parameters. One parameter is the first entry $|V_{ud}|$ of the
CKM matrix. The other one is $\lambda$, the ratio of the vector
coupling constant and the axial vector constant. In principle, the
ratio $\lambda$ can be determined from QCD lattice gauge theory
calculation, but the results of the best calculations vary by up
to 30\%. In neutron decay, several observables are accessible to
experiment, which depend on these parameters, so the problem is
overdetermined and, together with other data from particle and
nuclear physics, many tests of the Standard Model become possible.
The chosen observables for determining $|V_{ud}|$ are the neutron
lifetime $\tau$ and a measurement of the $\beta$-asymmetry
parameter $A_0$.

As is well known, the quark eigenstates of the weak interaction do
not correspond to the quark mass eigenstates. The weak eigenstates
are related to the mass eigenstates in terms of a 3 x 3 unitary
matrix $V$, the so called Cabibbo-Kobayashi-Maskawa (CKM) matrix.
By convention, the u, c and t quarks are unmixed and all mixing is
expressed via the CKM matrix ${\it V}$ operating on d, s and b
quarks. The values of individual matrix elements are determined
from weak decays of the relevant quarks. Unitarity requires that
the sum of the squares of the matrix elements for each row and
column be unity. So far precision tests of unitarity have been
possible for the first row of ${\it V}$, namely
\begin{equation} |V_{ud}|^2 + |V_{us}|^2 + |V_{ub}|^2 = 1-\Delta
\end{equation} In the Standard Model, the CKM matrix is unitary
with $\Delta$ = 0.

A violation of unitarity in the first row of the CKM matrix is a
challenge to the three generation Standard Model. The data
available so far do not preclude there being more than three
generations; CKM matrix entries deduced from unitarity might be
altered when the CKM matrix is expanded to accommodate more
generations \cite{Groom,Marciano1}. A deviation $\Delta$ has been
related to concepts beyond the Standard Model, such as couplings
to exotic fermions \cite{Langacker1,Maalampi}, to the existence of
an additional Z boson \cite{Langacker2,Marciano2} or to the
existence of right-handed currents in the weak interaction
\cite{Deutsch}. A non-unitarity of the CKM matrix in models with
an extended quark sector give rise to an induced neutron electric
dipol moment that can be within reach of next generation of
experiments \cite{Liao}.

Due to its large size, a determination of $|V_{ud}|$ is most
important. It has been derived from a series of experiments on
superallowed nuclear $\beta$-decay through determination of phase
space and measurements of partial lifetimes. With the inclusion of
nuclear structure effect corretions a value of $|V_{ud}|$ =
0.9740(5)~\cite{Hardy} emerges in good agreement of different,
independent measurements in nine nuclei. Combined with $|V_{us}|$
= 0.2196(23) from kaon-decays and $|V_{ub}|$ = 0.0036(9) from
B-decays, this lead to $\Delta$ = 0.0032(14), signaling a
deviation from the Unitarity condition by 2.3 $\sigma$ standard
deviation. The quoted uncertainty in $|V_{ud}|$, however, is
dominated by theory due to amount, size and complexity of
theoretical uncertainties. Although the radiative corretions
include effects of order Z$\alpha^2$, part of the nuclear
corrections are difficult to calculate. Further, the change in
charge-symmetry-violation for quarks inside nuclei results in an
additional change in the predicted decay rate which might lead to
a systematic underestimate of $|V_{ud}|$. A limit has been reached
where new concepts are needed to progress. Such are offered by
studies with neutron and with limitations with pion $\beta$-decay.
The pion $\beta$-decay has been measured recently at the PSI. The
pion has a different hadron structure compared with neutron or
nucleons and it offers an other possibility in determining
$|V_{ud}|$. The preliminary result is
$|V_{ud}|$=0.9771(56)~\cite{Pocanic}. The somewhat large error is
due to the small branching ratio of 10$^{-8}$.

Further information on the CKM matrix and the unitarity triangle
are based on a workshop held at CERN \cite{Battaglia} and a
workshop held at Heidelberg \cite{HADM}.
\section{Neutron-$\beta$-Decay}
In this article, we derive $|V_{ud}|$, not from nuclear
$\beta$-decay, but from neutron decay data. In this way, the
unitarity check of (1) is based solely on particle data, i.e.
neutron $\beta$-decay, K-decays, and B-decays, where theoretical
uncertainties are significantly smaller. So much progress has been
made using highly polarized cold neutron beams with an improved
detector setup that we are now capable of competing with nuclear
$\beta$-decays in extracting a value for $V_{ud}$, whilst avoiding
the problems linked to nuclear structure.
\begin{figure}
\hbox to\hsize{\hss
\includegraphics[width=\hsize]{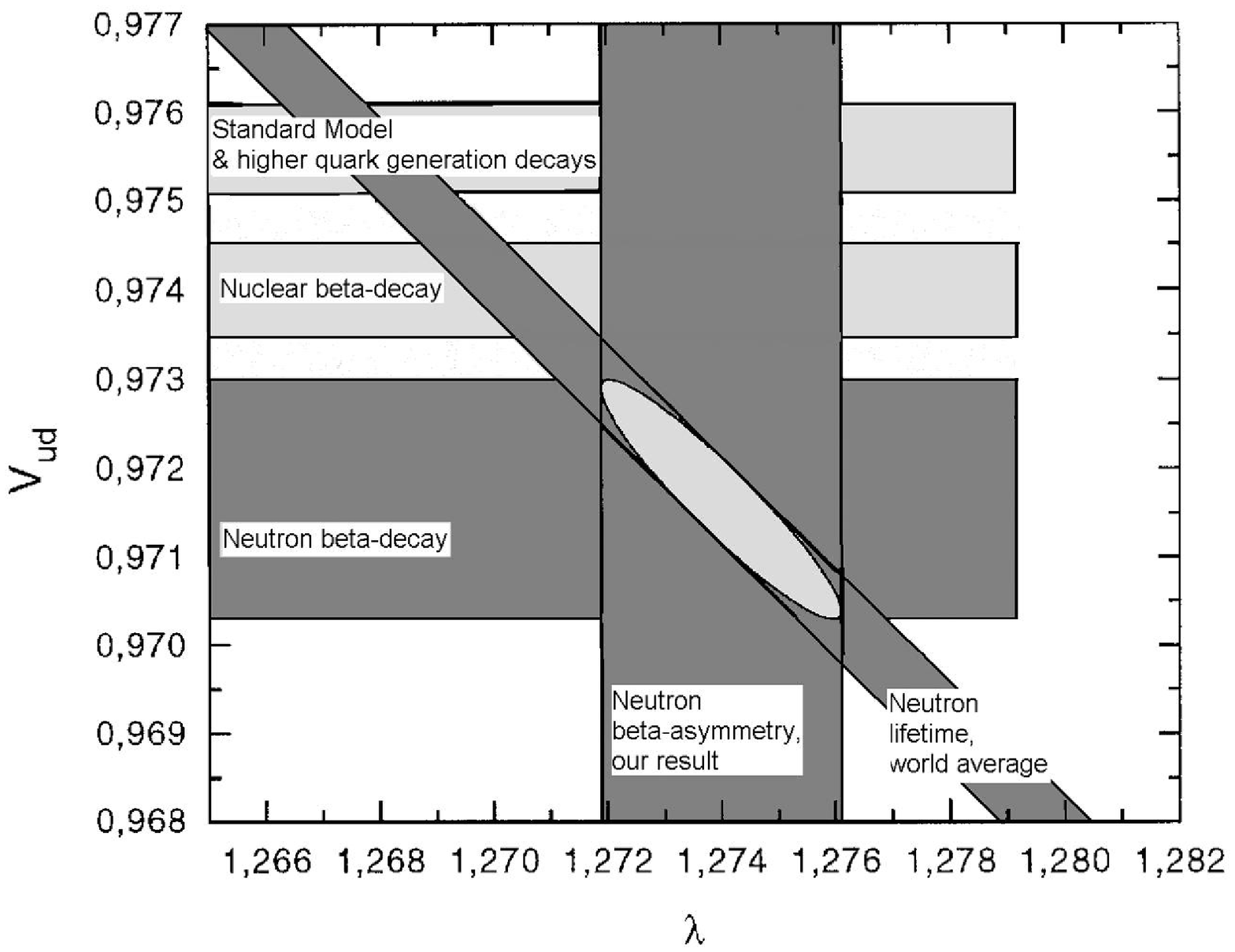}
\hss} \caption{$|V_{ud}|$  vs. $\lambda$. $|V_{ud}|$ was derived
from $Ft$ values of nuclear $\beta$-decays, higher quark
generation decays, assuming the unitarity of the CKM matrix, and
neutron $\beta$-decay.}
\end{figure}
A neutron decays into a proton, an electron and an electron
antineutrino. Observables are the neutron lifetime $\tau$ and
spins $\sigma_e$, $\sigma_\nu$, $\sigma_p$, and momenta $p_e$,
$p_\nu$, $p_p$ of the electron, antineutrino and proton
respectively. The electron spin, the proton spin and the
antineutrino are not usually observed. The lifetime is given by
\begin{equation}
\tau^{-1}=C |V_{ud}|^2(1+3\lambda^2) f^R(1+\Delta_R),
\end{equation}
where $C=G_F^2 m_e^5/(2 \pi^3)=1.1613\cdot10^{-4} s^{-1}$ in
$\hbar=c=1$ units, $f^R$ =1.71335(15) is the phase space factor
(including the model independent radiative correction) adjusted
for the current value of the neutron-proton transition energy and
corrected by Marciano \cite{MarcianoHA}. $\Delta_R$ = 0.0240(8) is
the model dependent radiative correction to the neutron decay rate
\cite{Towner1}. The $\beta$-asymmetry $A_0$ is linked to the
probability that an electron is emitted with angle $\vartheta$
with respect to the neutron spin polarization $P$ = $<\sigma_z>$:
\begin{equation} W(\vartheta) = 1 +\frac{v}{c}PA\cos(\vartheta),
\end{equation} where $v/c$ is the electron velocity expressed in
fractions of the speed of light. ${\it A}$ is the
$\beta$-asymmetry coefficient which depends on $\lambda$. On
account of order 1\% corrections for weak magnetism, $g_V-g_A$
interference,
and nucleon recoil, ${\it A}$ has the form $A$ = $A_0$(1+$A_{{\mu}m}$($A_1W_0+A_2W+A_3/W$)) 
with electron total energy $W = E_e /m_ec^2+1$ (endpoint $W_0$).
$A_0$ is a function of $\lambda$
\begin{equation}  A_0=-2\frac{\lambda(\lambda+1)}{1+3\lambda^2},
\end{equation} where we have assumed that $\lambda$ is real. The coefficients $A_{{\mu}m}$, $A_1$, $A_2$, $A_3$ are from \cite{Wilkinson1} taking a different $\lambda$ convention into consideration. In addition, a further small radiative
correction \cite{Gluck} of order 0.1\% must be applied. For
comparison, information about $|V_{ud}|$ and $\lambda$ are shown
in Fig. (1). The bands represent the one sigma error of the
measurements. The $\beta$-aymmetry $A_0$ in neutron decay depends
only on $\lambda$, while the neutron lifetime $\tau$ depends both
on $\lambda$ and $|V_{ud}|$. The intersection between the curve,
derived from $\tau$ and $A_0$, defines $|V_{ud}|$ within one
standard deviation, which is indicated by the error ellipse. Other
information on $|V_{ud}|$, derived from nuclear $\beta$-decay and
higher quark generation decays are shown, too. As can be seen from
Fig. (1), both the nuclear $\beta$-decay result from~\cite{Hardy}
and the neutron $\beta$-decay from~\cite{Abele2} do not agree with
this unitarity value.

\section{The Experiment PERKEO and the Result for $|V_{ud}|$}

The following section is about our measurement of the neutron
$\beta$-asymmetry coefficient $A$ with the instrument PERKEOII,
and on the consequences for $|V_{ud}|$. The strategy of PERKEOII
followed the instrument PERKEO~\cite{Bopp} in minimizing
background and maximizing signal with a $4\pi$ solid angle
acceptance over a large region of the beam. Major achievements of
the instrument PERKEO are:
\begin{itemize}
\item The signal to background ratio in the range of interest is
200. \item The overall correction of the raw data is 2.04\%. \item
The detector design allows an energy calibration with linearity
better than 1\%. \item New polarizers and developments in
polarization analysis led to smaller uncertainties related to
neutron beam polarization.
\end{itemize}
\begin{figure}
\hbox to\hsize{\hss
\includegraphics[width=\hsize]{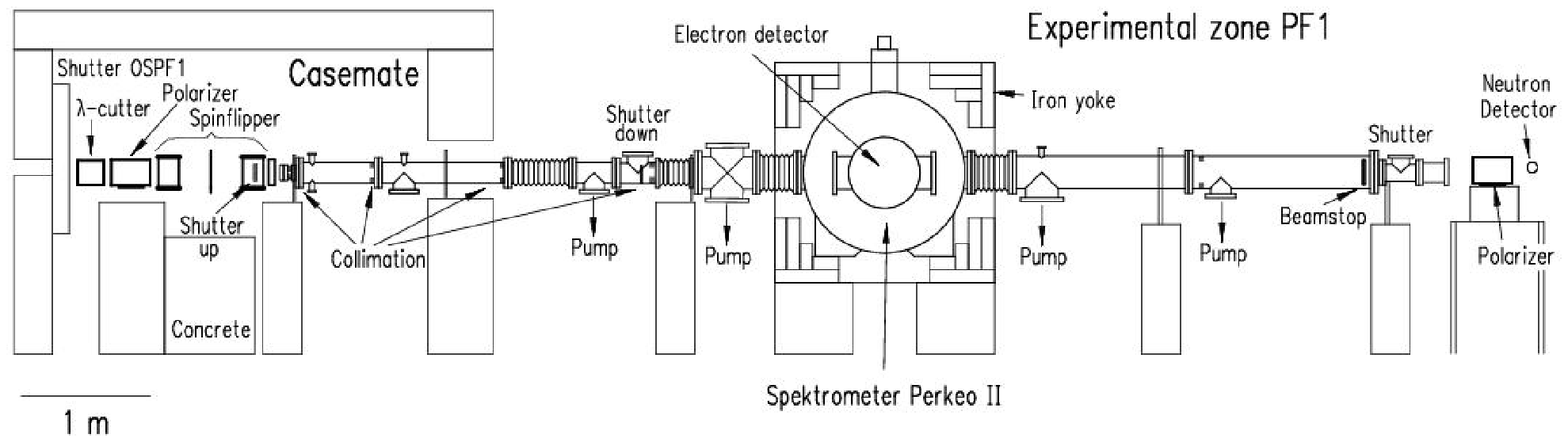}
\hss} \caption{A schematic view of the whole setup at the ILL.}
\label{setup}
\end{figure}

For a measurement of $\beta$-asymmetry $A_0$, the instrument
PERKEO was installed at the PF1 cold neutron beam position at the
High Flux Reactor at the Institut Laue-Langevin, Grenoble. Cold
neutrons are obtained from a 25 K deuterium cold moderator near
the core of the 57 MW uranium reactor. The neutrons are guided via
a 60 m long neutron guide of cross section 6 $\times$ 12 cm$^2$ to the
experiment and are polarized by a 3 $\times$ 4.5 cm$^2$ supermirror
polarizer. The de Broglie wavelength spectrum of the cold neutron
beam ranges from about 0.2 nm to 1.3 nm. The degree of neutron
polarization was measured to be P = 98.9(3)\% over the full cross
section of the beam. The polarization efficiency remained constant
during the whole experiment. The neutron polarization is reversed
periodically with a current sheet spin flipper. The main component
of the PERKEO II spectrometer is a superconducting 1.1 T magnet in
a split pair configuration, with a coil diameter of about one
meter. Neutrons pass through the spectrometer, whereas decay
electrons are guided by the magnetic field to either one of two
scintillation detectors with photomultiplier readout. The detector
solid angle of acceptance is truly 2$\times$2$\pi$ above a threshold of
60 keV. Electron backscattering effects, serious sources of
systematic error in $\beta$-spectroscopy, are effectively
suppressed. Technical details about the instrument can be found in
\cite{Abele,Reich}. The measured electron spectra
$N^\uparrow_i(E_e)$ and $N^\downarrow_i(E_e)$ in the two detectors
(i=1,2) for neutron spin up and down, respectively, define the
experimental asymmetry as a function of electron kinetic energy
$E_e$ and are shown in Fig. 3. \begin{equation}
A_{i_{exp}}(E_e)=\frac{N^\uparrow_i(E_e) -
N^\downarrow_i(E_e)}{N^\uparrow_i(E_e) + N^\downarrow_i(E_e)}.
\end{equation}
By using (3) and with $<\cos(\vartheta)>$ = 1/2, $A_i{_{exp}}(E)$
is directly related to the asymmetry parameter
\begin{equation}
A_{exp}(E_e)=A_{1_{exp}}(E_e)-A_{2_{exp}}(E_e)=\frac{v}{c}APf.
\end{equation}
\begin{figure}
\hbox to\hsize{\hss
\includegraphics[width=\hsize,clip]{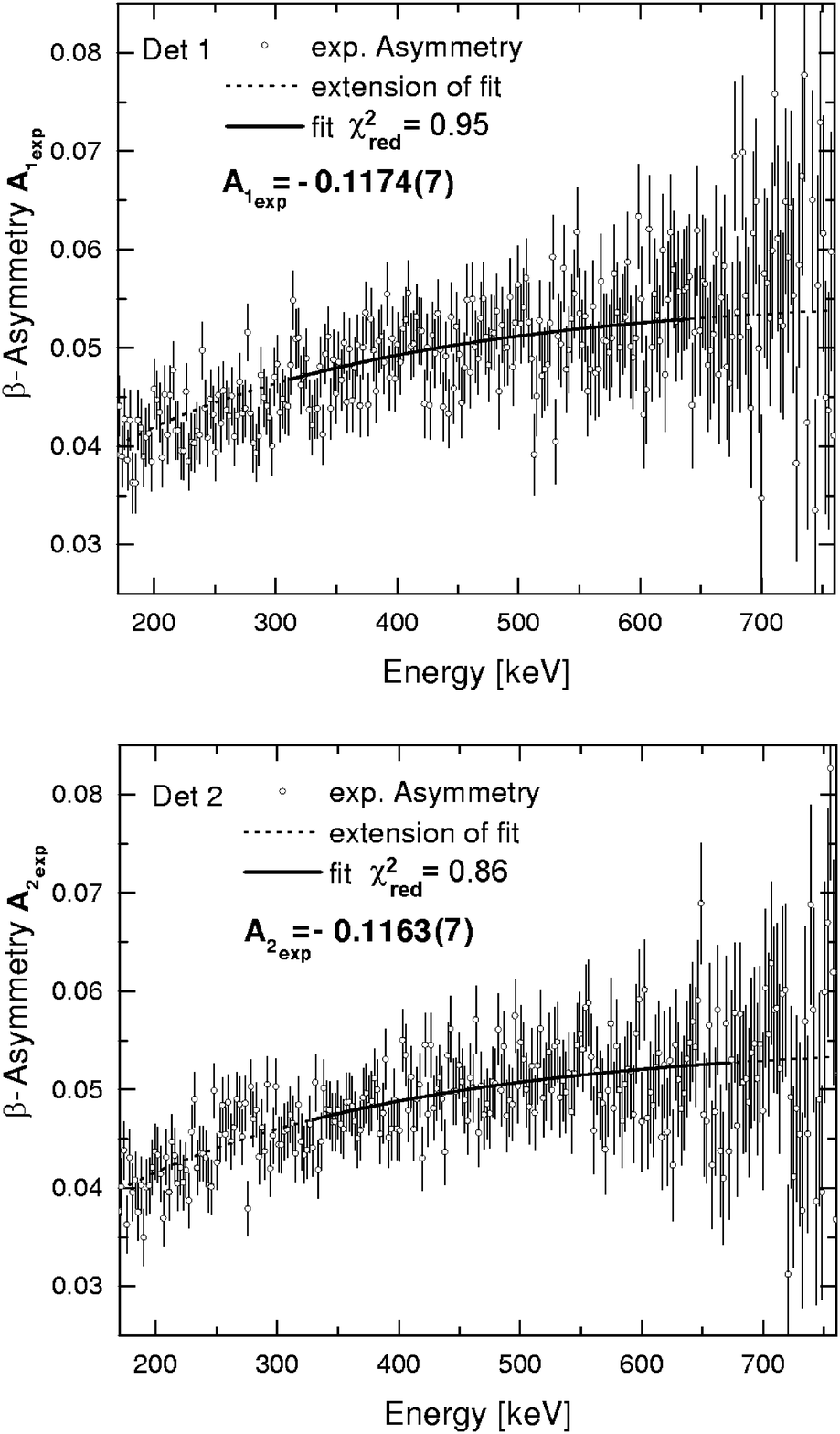}
\hss} \caption{Fit to the experimental asymmetry $A_{exp}$ for
detector 1 and detector 2. The solid line shows the fit interval,
whereas the dotted line shows an extrapolation to higher and lower
energies.}\label{fit}
\end{figure}
The experimental function $A_{i_{exp}}(E_e)$ and a fit with one
free parameter $A_{i_{exp}}$ (the absolute scale of $A_0$) is
shown in Fig. \ref{fit}. The total correction for the small
experimental systematic effects is 2.04\%.

With recent experiments from the University of Heidelberg
\cite{Abele2,Abele}, we obtain $A_0$ = -0.1189(7) and $\lambda$ =
- 1.2739(19). With this value, and the world average for $\tau$ =
885.7(7) s, we find that $|V_{ud}|$ = 0.9717(13). With $|V_{us}|$
= 0.2196(23) and the negligibly small $|V_{ub}|$ = 0.0036(9), one
gets
\begin{equation} |V_{ud}|^2 +
|V_{us}|^2 + |V_{ub}|^2 = 1 - \Delta = 0.9924(28).
\end{equation} This value differs from the Standard
Model prediction by $\Delta$ = 0.0076(28), or 2.7 times the stated
error. Earlier experiments \cite{Bopp,Yerozolimsky,Schreckenbach}
gave significant lower values for $\lambda$. Averaging over our
new result and previous results, the Particle Data Group
\cite{Groom} arrives at a new world average for $|V_{ud}|$ from
neutron $\beta$-decay which leads to a 2.2 s deviation from
unitarity.

An independent test of CKM unitarity comes from W physics at LEP
\cite{Sbarra} where W decay hadronic branching ratios can be used. 
Since decay into the top quark channel is forbidden by energy
conservation one would expect $\sum{|V_{ij}|^2}$ to be 2 with a
three generation unitary CKM matrix. The experimental result is
2.032(32), consistent with (7) but with considerably lower
accuracy.
\section{The future}
The main corrections in the experiment PERKEO are due to neutron
beam polarization (1.1\%), background (0.5\%) and flipper
efficiency (0.3\%). The total correction is 2.04\%. With such
small corrections to the data, we start to see a deviation from
the Standard Model already in the uncorrected raw data. For the
future, the plan is further to reduce all corrections. In the
meantime, major improvements both in neutron flux and degree of
neutron polarization has been made: First, the new ballistic
supermirror guide at the ILL from the University of Heidelberg
gives an increase of a factor of 4 in the cold neutron
flux~\cite{Haese1}. Second, a new arrangement of two supermirror
polarizers allows to achieve an unprecedented degree of neutron
polarization $P$ of between 99.5\% and 100\% over the full cross
section of the beam~\cite{Soldner1}. Third, systematic limitations
of polarization measurements have been investigated: The beam
polarization can now be measured with a completely new method
using an opaque $^3$He spin filter with an uncertainty of 0.1\%
\cite{Heil,Zimmerha}. As a consequence, we are now in the lucky
situation to improve on the main uncertainties in reducing the
main correction of 1.1\% to less than 0.5\% with an error of
0.1\%. Thus, a possible deviation from the Standard Model, if
confirmed, will be seen very pronounced in the uncorrected data.
 
\section{Summary}
$|V_{ud}|$, the first element of the CKM matrix, has been derived
from neutron decay experiments in such a way that a unitarity test
of the CKM matrix can be performed based solely on particle
physics data. With this value, we find a 2.7 $\sigma$ standard
deviation from unitarity, which conflicts the prediction of the
Standard Model of particle physics.

Future trends have been presented on this workshop "Quark-mixing,
CKM Unitarity" in Heidelberg. Regarding the
Unitarity problem, about half a dozen new instruments are planed
or under construction to allow for beta-neutrino correlation $a$
and $\beta$-asymmetry $A$ measurements at the sub-10$^{-3}$ level. With
next generation experiments measurements with a decay rate of over 10$^5$s$^{-1}$ become feasible \cite{Dubbers3}.

\newcommand{\PLB}[3]  {Phys.\ Lett.\ \textbf{B#1} (#2) #3}
\newcommand{\ZPC}[3]  {Z.\ Phys.\ \textbf{C#1} (#2) #3}
\newcommand{\EPC}[3]  {Eur.\ Phys.\ J.\ \textbf{C#1} (#2) #3}
\newcommand{\NIMA}[3] {Nucl.\ Instr.\ Meth.\ \textbf{A#1} (#2) #3}
\newcommand{\PPE}[1]  {CERN-PPE/{#1}}
\newcommand{\PRLD}[3] {Phys.\ Rev.\ Lett.\ \textbf{D#1} (#2) #3}
\newcommand{\PRD}[3]  {Phys.\ Rev.\ \textbf{D#1} (#2) #3}
\newcommand{\NPB}[3]  {Nucl.\ Phys.\ \textbf{B#1} (#2) #3}
\newcommand{\CiP}[3]  {Comp.\ in Phys.\ \textbf{#1} (#2) #3}
\newcommand{\NPhys}   {Nucl.~Phys}
\newcommand{\CPC}[3]  {Comput.\ Phys.\ Commun.\ \textbf{#1} (#2) #3}
\newcommand{\JPH}[3] {J.~Phys.\ \textbf{#1} (#2) #3}
\def\etal{\mbox{{\it et al.}}}
\newcommand{\Rc}{\ensuremath{R_{\rm{c}}^{\rm{W}}}}
\newcommand{\epem}{\ensuremath{\rm e^+ e^-}}
\newcommand{\mpmm}{\ensuremath{\mu^+\mu^-}}
\newcommand{\tptm}{\ensuremath{\tau^+\tau^-}}
\newcommand{\lplm}{\ensuremath{\ell^+\ell^-}}
\newcommand{\Zz}{\ensuremath{{\rm{Z}^0}}}
\newcommand{\Hz}{\ensuremath{{\rm{H}^0}}}
\newcommand{\WpWm}{\ensuremath{\rm{W}^+\rm{W}^-}}
\newcommand{\Wboson}{\ensuremath{\rm{W}}}
\newcommand{\Zboson}{\ensuremath{\rm{Z}}}
\newcommand{\Wp}{\ensuremath{\rm{W}^+}}
\newcommand{\Wpm}{\ensuremath{\rm{W}^\pm}}
\newcommand{\Wm}{\ensuremath{\rm{W}^-}}

\newcommand{\eeWW}{\ensuremath{\epem\rightarrow\WpWm}}
\newcommand{\qq}{\ensuremath{\rm{q\overline{q}}}}
\newcommand{\bb}{\ensuremath{\rm{b\overline{b}}}}
\newcommand{\glgl}{\ensuremath{\rm{gg}}}
\newcommand{\pp}{\ensuremath{\rm{\overline{p}p}}}
\newcommand{\ff}{\ensuremath{\rm{f\overline{f}}}}
\newcommand{\lnu}{\ensuremath{\ell\overline{\nu}_{\ell}}}
\newcommand{\lv}{\lnu}
\newcommand{\lpnu}{\ensuremath{\ell^+ \nu_{\ell}}}
\newcommand{\lmnu}{\ensuremath{{\ell^{\prime}}^-\overline{\nu}_{\ell^{\prime}}}}
\newcommand{\enu}{\ensuremath{\rm{e\overline{\nu}_{e}}}}
\newcommand{\mnu}{\ensuremath{\mu\overline{\nu}_{\mu}}}
\newcommand{\tnu}{\ensuremath{\tau\overline{\nu}_{\tau}}}
\newcommand{\nue}{\ensuremath{\rm{\overline{e}\nu_{e}}}}
\newcommand{\num}{\ensuremath{\overline{\mu}{\nu}_{\mu}}}
\newcommand{\nut}{\ensuremath{\overline{\tau}{\nu}_{\tau}}}
\newcommand{\taunu}{\ensuremath{\overline{\nu}_{\tau}}}
\newcommand{\ataunu}{\ensuremath{{\nu}_{\tau}}}
\newcommand{\mumu}{\ensuremath{\mu^+\nu_{\mu}\mu^-\overline{\nu}_{\mu}}}
\newcommand{\muel}
  {\ensuremath{\mu^+\nu_{\mu}\rm{e}^-\overline{\nu}_{\rm e}}}
\newcommand{\qqqq}{\ensuremath{\qq\qq}}
\newcommand{\qqgg}{\ensuremath{\qq{\rm{gg}}}}
\newcommand{\qqln}{\ensuremath{\qq\lnu}}
\newcommand{\qqll}{\ensuremath{\qq\lplm}}
\newcommand{\qqvv}{\ensuremath{\qq\nu\overline{\nu}}}
\newcommand{\eell}{\ensuremath{\epem\lplm}}

\newcommand{\qqmn}{\ensuremath{\qq\mnu}}
\newcommand{\qqmv}{\qqmn}
\newcommand{\qqtn}{\mbox{\qq\tnu}}
\newcommand{\qqtv}{\qqtn}
\newcommand{\Vij} {\ensuremath{|V_{ij}|}}
\newcommand{\Vud} {\ensuremath{|V_{\rm{ud}}|}}
\newcommand{\Vus} {\ensuremath{|V_{\rm{us}}|}}
\newcommand{\Vcd} {\ensuremath{|V_{\rm{cd}}|}}
\newcommand{\Vcb} {\ensuremath{|V_{\rm{cb}}|}}
\newcommand{\Vub} {\ensuremath{|V_{\rm{ub}}|}}
\newcommand{\Vcs} {\ensuremath{|V_{\rm{cs}}|}}
\newcommand{\phzs}       {\phantom{^{0*}}}
\newcommand{\phz}        {\phantom{0}}
\newcommand{\phzz}       {\phantom{00}}
\newcommand{\phm}        {\phantom{-}}
\newcommand{\CoM}        {centre--of--mass}
\newcommand{\Aleph}      {\mbox{A{\sc leph}}}
\newcommand{\Delphi}     {\mbox{D{\sc elphi}}}
\newcommand{\Ltre}       {\mbox{L{\sc 3}}}
\newcommand{\Opal}       {\mbox{O{\sc pal}}}
\newcommand{\sww}        {\sigma_\rm{WW}}
\newcommand{\wwbr}       {{\cal{B}}}
\newcommand{\oa}         {{\cal{O}}(\alpha)}
\newcommand{\Wtolnu}     {\mbox{$\rm{W}\rightarrow
                                 \ell\overline{\nu}_{\ell}$}}
\newcommand{\Wtoenu}     {\mbox{$\rm{W}\rightarrow
                                 \rm{e\overline{\nu}_{e}}$}}
\newcommand{\Wtomnu}     {\mbox{$\rm{W}\rightarrow
                                 \mu\overline{\nu}_{\mu}$}}
\newcommand{\Wtotnu}     {\mbox{$\rm{W}\rightarrow
                                 \tau\overline{\nu}_{\tau}$}}
\newcommand{\BWtoenu}    {\mbox{$\mathcal{B}(\rm{W}\rightarrow
                                 \rm{e\overline{\nu}_{e}})$}}
\newcommand{\BWtomnu}    {\mbox{$\mathcal{B}(\rm{W}\rightarrow
                                 \mu\overline{\nu}_{\mu})$}}
\newcommand{\BWtotnu}    {\mbox{$\mathcal{B}(\rm{W}\rightarrow
                                 \tau\overline{\nu}_{\tau})$}}
\newcommand{\BWtolnu}    {\mbox{$\mathcal{B}(\rm{W}\rightarrow
                                 \ell\overline{\nu}_{\ell})$}}
%
\title*{CKM Unitarity and $|\rm V_{cs}|$ from W Decays}
\toctitle{CKM Unitarity and $|\rm V_{cs}|$ from W Decays}
%
%
\titlerunning{CKM Unitarity and $|\rm V_{cs}|$ from W Decays}
%
\author{E. Barberio}
\authorrunning{E. Barberio}
%
%
\institute{Physics Department,
Southern Methodist University, Dallas, TX, USA}

\maketitle              

\begin{abstract}
Decays of W$^\pm$ bosons produced at LEP2 have been used to
measure the Cabibbo-Kobayashi-Maskawa matrix element  $|\rm
V_{cs}|$ with a precision of 1.3\% without the need of a form
factor. The same data set has been used to test the unitarity of
the first two rows of the matrix at the 2\% level. At a future
$\rm e^+ e^-$ linear collider, with a data sample of few million
of W decays a precision of 0.1\% can be reached.
\end{abstract}

\section{Introduction}
Within the framework of the Standard Model of electroweak
interactions, the elements of the Cabibbo-Kobayashi-Maskawa (CKM)
\cite{ckm} mixing matrix are free parameters, constrained only by
the requirement that the matrix be unitary. The values of the
matrix elements can only be determined by experiment.

In general, CKM elements are derived from the measurement of
weak hadronic decays. Since quarks are bound in hadrons
 QCD symmetries need to be employed to parameterize
the non-perturbative aspects of QCD. Hence, for most of these elements
the principal error is no longer experimental but rather theoretical:
it reflects theoretical models and the assumption invoked.

In the Standard Model the branching fraction of the W boson decays
depend on the six CKM matrix elements which do not involve the top
quark. Measuring the W production rates for different flavors
gives access to the individual CKM matrix elements without
  parameterization of non-perturbative QCD:
$$
  \Gamma(\rm W \to q' \bar{q}) = \frac{C(\alpha_s)G_F M^3_W}{6\sqrt{2\pi}}
          |V_{ij}|^2 = (707 \pm 1) |V_{ij}|^2 \rm MeV $$
where
$$
          C(\alpha_s) = 3 \bigg[ 1 +
          \sum_{i=1,3}
          \frac{a_i \alpha_s(\rm{M}^2_{\rm{W}})}{\pi}
          \bigg]
$$
is the QCD color factor, up to the third order in
 $\alpha_s(\rm{M}^2_{\rm{W}})$, the strong coupling
constant.

Furthermore,  'on shell' W bosons decay before the hadronization
process starts,
and  the quark transition occurs in a perturbative QCD regime.
Hence, W boson decays offer a complementary way to determine the
CKM matrix elements.

From 1997 to 2000 the LEP \epem collider has been operated at
energies above the threshold for W-pair production. This offered a
unique opportunity to study the hadronic decays of W boson in a
clean environment and to investigate the coupling strength of W
boson bosons to different quark flavors.

\section{\Vcs ~from hadron decays}
Values of \Vcs ~can be obtained from neutrino scattering
production of charm and form semileptonic D decays with
theoretical uncertainties larger than 10\%. The value of \Vcs
~extracted from neutrino scattering production of charm depends on
assumption about the s-quark sea density in partons. This method
gives a lower bound of 0.59.

Using semileptonic D decays one measures:
$$ { \rm \Gamma  ( D \to K e \nu) =}
|f^D_+ (0)|^2 \rm ~\Vcs^2 ~(1.54\cdot10^{11})~ s^{-1},$$ where $
|f^D_+ (0)|$ is the D$_{e3}$ form factor. The most recent
evaluation of the magnitude of \Vcs ~from this method yields $\Vcs
= 1.04 \pm 0.16$~\cite{pdgbarber}, dominated by theoretical
uncertainties which exceed 10\%.

\section{\Vcs ~from W decays}
Each one of the four LEP experiment collected a data sample of
about 15000 W-pair events. This sample is not large enough to measure a
   the six CKM elements not involving the top quark,
but large enough to improve the knowledge of the matrix element
\Vcs. This was the least well known CKM element. It is of order
one, while the production of bottom quarks is in fact highly
suppressed due to the small magnitude of \Vub ~and \Vcb ~and the
large mass of the top quark, the weak partner of the bottom quark.
The same data set can be  used to test unitarity of the first two
rows of the CKM matrix at the few percent level.

Using W decays, the magnitude of the CKM matrix element \Vcs ~can be
derived {\it indirectly} from either the W boson leptonic
branching fraction or the a direct measurement of the production fraction
of charm in W decays. \Vcs ~can
then be derived from the charm production rate or
the W boson leptonic
branching fraction, using the knowledge of the
other CKM matrix elements.
A {\it direct} measurement can be done from the direct measurement
of the $\rm W \to cs $ transition.

The  {\it indirect} measurement provides a high statistical sample
and use the fact that so far direct measurements of \Vcs ~have
limited precision compared to the other CKM matrix elements
important in W boson decays, i.e. \Vud, \Vus, and \Vcd.

The {\it direct} observation of the $\rm W \to cs $ transition requires
either very good particle identification or a large WW sample (much larger
than the LEP one).

Experimentally quarks from W decays form jets and to measure all
six CKM elements the original flavor of the quark which originated
the jet needs to be identified. This is possible, due to the
strong correlation between the primary quark flavor and the jet
properties.

The identification of charm or the beauty quarks is relatively easy.
Their tagging is based on well understood and unambiguous special
properties: long lifetime, higher jet multiplicity, etc.
The identification of light quark (u-,d-, s-quark) to a precision
needed for a meaningful measurement of the CKM elements is more
difficult and requires large statistics and/or good particle identification.

The  observation of $\rm W \to cx$ transition
requires the identification of a charm jet.
The  observation of $\rm W \to cs$ transition
requires the identification of a charm jet and a strange jet.

\subsection{\Vcs ~from charm production rate}
At LEP
the production rate of charm in W bosons
can be measured without a separation
of charm and bottom quarks decays.
Charm hadron identification is
based upon jet properties in particular on lifetime information
and semileptonic decay products in the events.
The measured value of \Rc is then used to determine \Vcs.
$$
 \Rc = \frac{\Gamma \rm ( W \to c\,X)}{
\Gamma\,( W  \to hadrons)} =
\frac{\Vcd^2 +\Vcs^2 + \Vcb^2 }{\Vud^2 +\Vus^2 + \Vub^2
+\Vcd^2 +\Vcs^2 + \Vcb^2 }.
$$
In the  Standard Model  $\Rc=0.5$, due to the CKM unitarity.

Charm jets are identified using a multi-dimensional estimator
based on charm jet characteristics: the most powerful
discriminator are the lifetime
    information and leptons produced in charm decays.
Weakly decaying charm hadrons have lifetimes between 0.2\,ps and
1\,ps~\cite{pdgbarber}, leading to typical decay lengths of a few
hundred microns to a few millimeters at LEP2 energies. These
relatively long--lived particles produce secondary decay vertices
which are significantly displaced from the primary event vertex.
About $20\,\%$ of all charm hadrons decay semi-leptonically and
produce an electron or a muon in the final state. Because of the
relatively large mass and hard fragmentation of the charm quark
compared to light quarks, this lepton is expected to have a larger
momentum than leptons from other sources (except the small
contribution from semileptonic bottom decays in background
events). Therefore, identified electrons and muons can be used as
tags for charm hadrons in W boson boson decays.

Three LEP collaborations \cite{rcwaleph}, \cite{rcwdelphi}
and \cite{rcwopal}  measured \Rc using only a  part of their data set.
Their average gives:
$$ \Rc = 0.49 \pm 0.03_{stat} \pm 0.03_{sys}, $$
where the systematics error is dominated by the Monte Carlo
calibration of the analysis. This result is consistent with the
Standard Model prediction that the \Wboson\ boson couples to up
and charm quarks with equal strength. There is no much gain in
using the full LEP W-pairs data set for this method as the total
error is limited by systematics errors correlated  between the LEP
experiments.

Using direct measurements of the other CKM matrix
elements and $\Gamma^W_{tot} = 2.135 \pm 0.069$ GeV \cite{mass},
\Vcs ~can be calculated from \Rc:
$$\Vcs =  \rm \frac{\Rc  Br(W \to  had) \Gamma^W_{tot}}
{( 701 \pm 1)\rm MeV} $$
 yielding $\Vcs = 0.976 \pm 0.88_{\Rc}\pm 0.003_{CKM}$.

\subsection{\Vcs ~from W leptonic branching fraction}
This method provide the most precise determination
of \Vcs ~and is the one which provided the smallest error.


The leptonic branching fraction of the W boson
$\mathcal{B}(\Wtolnu)$ is related to the six CKM elements not
involving the top quark by:
$$
  \frac{1}{\mathcal{B}(\Wtolnu)}\quad = \quad
C(\alpha_s(\rm{M}^2_{\rm{W}}))
          \sum_{\tiny\begin{array}{c}i=(u,c),\\j=(d,s,b)\\\end{array}}
          |\rm{V}_{ij}|^2 .
$$
Using  $\alpha_s(\rm{M}^2_{\rm{W}})$=$0.121\pm0.002$,
the measured leptonic branching fraction of the W yields
\begin{equation}
  \sum_{\tiny\begin{array}{c}i=(u,c),\\j=(d,s,b)\\\end{array}}
  |\rm{V}_{ij}|^2
  \quad = \quad
  2.039\,\pm\,0.025_{\wwbr (Wtolnu)}\pm\,0.001_{\alpha_s},
\label{res}
\end{equation}
where the first error is due to the uncertainty on the branching
fraction measurement and the second to the uncertainty on
$\alpha_s$ \cite{lep}. This result is consistent with the
unitarity of the first two rows of the CKM matrix at the 1.5\%
level, as in the Standard Model:
\begin{equation}
          \sum_{\tiny\begin{array}{c}i=(u,c),\\j=(d,s,b)\\\end{array}}
          |\rm{V}_{ij}|^2 = 2.
\label{th}
\end{equation}
 No assumption on the values of the single CKM elements are made.

Using the experimental knowledge~\cite{pdgbarber} of the sum
$|\rm{V}_{ud}|^2+|\rm{V}_{us}|^2+|\rm{V}_{ub}|^2+
 |\rm{V}_{cd}|^2+|\rm{V}_{cb}|^2=1.0477\pm0.0074$,
the above result can also be interpreted as a measurement of
$|\rm{V}_{cs}|$:
$$
  |\rm{V}_{cs}|\quad=\quad0.996\,\pm\,0.013.
$$
The error includes
a $\pm0.0006$ contribution from the uncertainty on $\alpha_s$
and a $\pm0.004$ contribution from the uncertainties
on the other CKM matrix elements,
the largest of which is that on $|\rm{V}_{cd}|$.
These contributions are negligible compared to
the experimental error
from the measurement of the W branching fractions, $\pm0.013$.


The  W decay branching fractions,
\mbox{$\mathcal{B}(\rm{W}\rightarrow\textrm{f}\overline{\textrm{f}}')$},
are determined
from the cross sections for the individual
WW$\rightarrow\rm{4f}$ decay channels
measured by the four experiments at all energies above 161 GeV.
These branching fractions can be derived
with and without the assumption of lepton universality.
In the fit with lepton universality,
the branching fraction to hadrons is determined from that to leptons
by constraining the sum to unity.

\begin{figure}[tbhp]
\begin{center}
\hspace*{-0.2truecm}
   \epsfig{figure=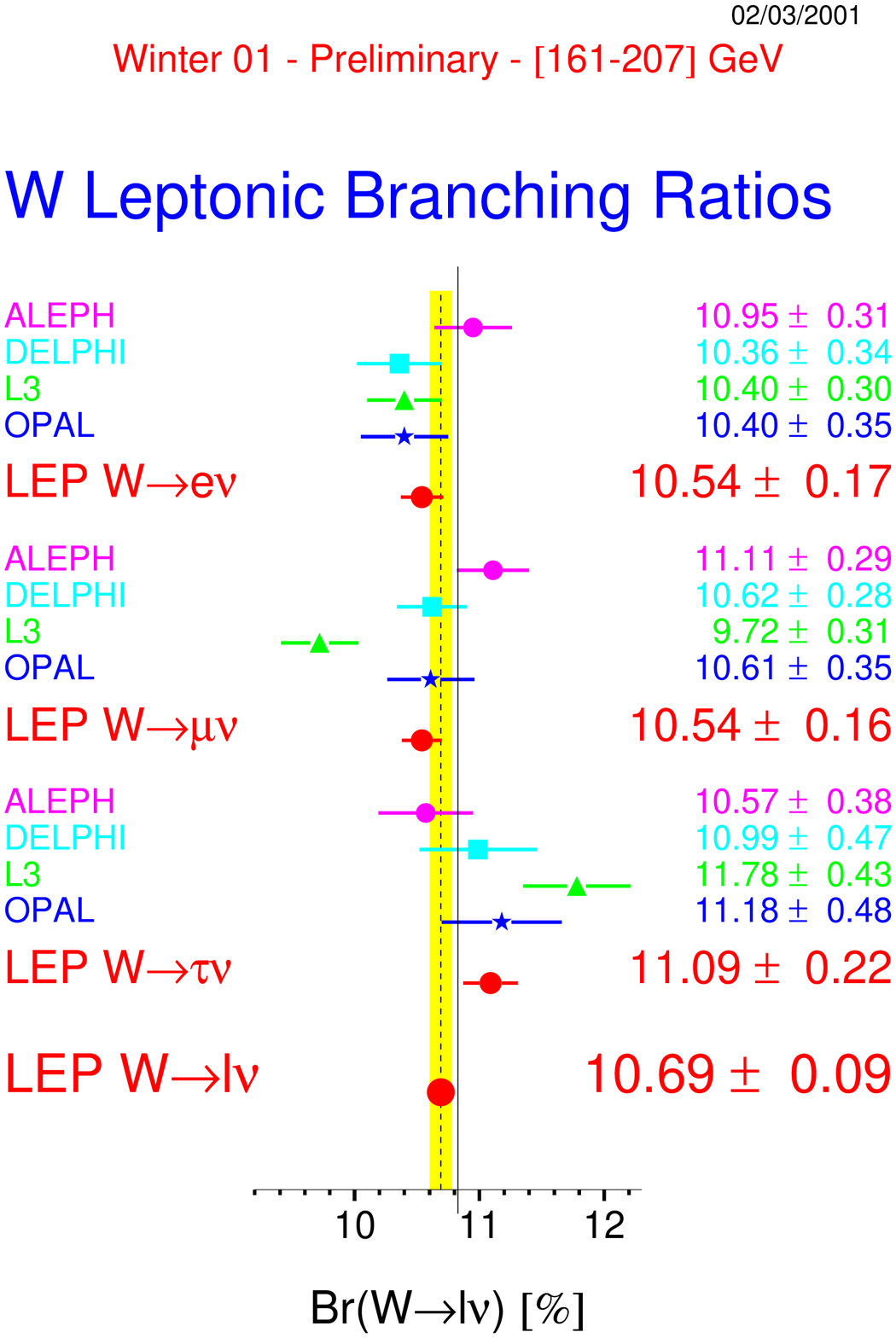,width=0.62\textwidth}
    \vspace*{-1.0truecm}
  \hspace*{0.02\textwidth}
  \epsfig{figure=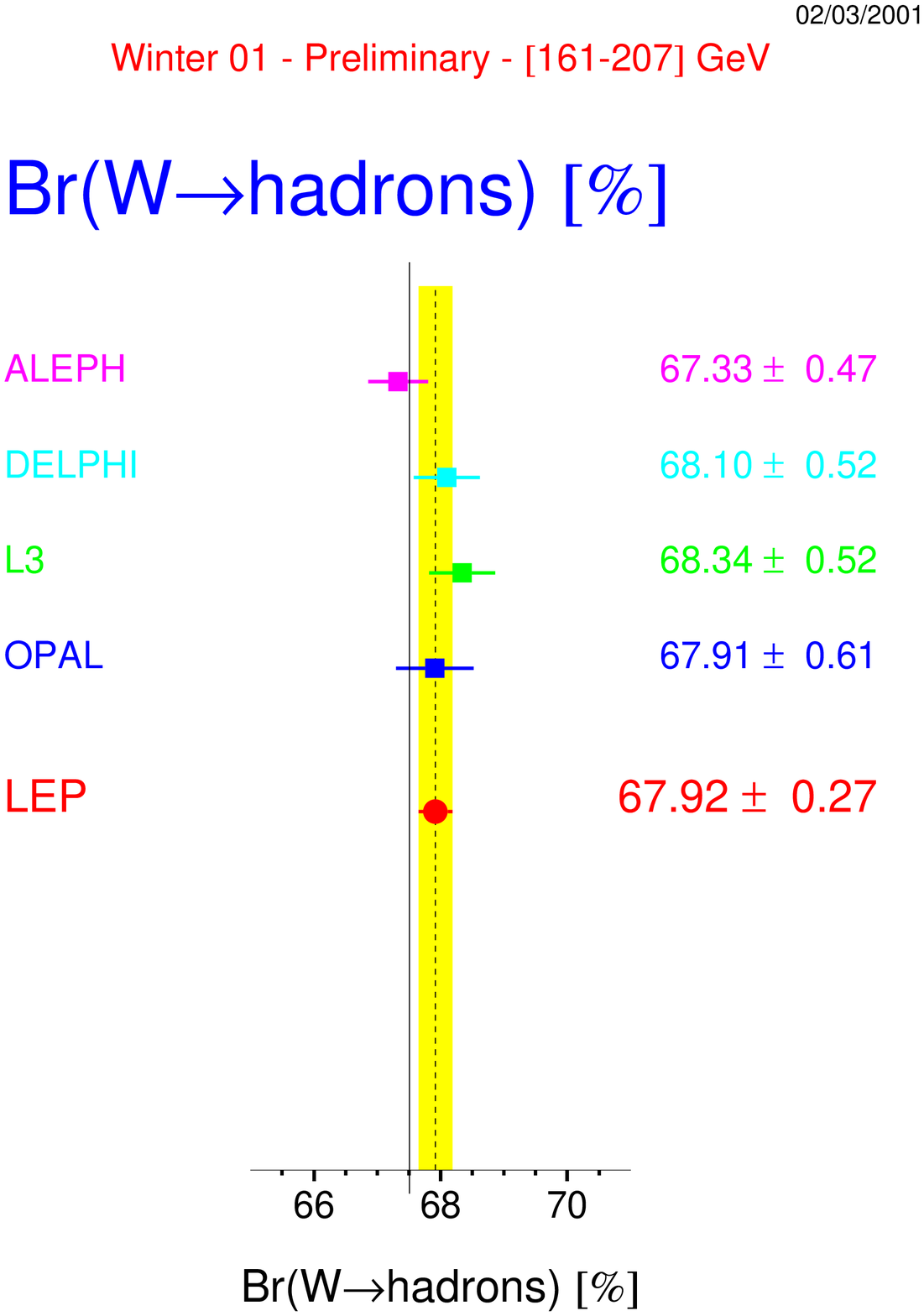,width=0.62\textwidth}
    \vspace*{-1.0truecm}
  \vspace{1.5cm}
\caption{
  Summary of leptonic and hadronic W branching fractions
  derived from preliminary W-pair production
  cross sections measurements up to 207 GeV \CoM\ energy.}
\label{4f_fig:wwbra}
\end{center}
\end{figure}

Results from the experiments \cite{lep} and the LEP combined values
are given in Table~\ref{wwbra}.
Figure~\ref{4f_fig:wwbra} shows the results of the LEP combination.
\renewcommand{\arraystretch}{1.2}
\begin{table}
\caption{\small
  Summary of leptonic and hadronic
  W branching fractions derived from preliminary W-pair production
  cross sections measurements up to 207 GeV \CoM\ energy.
}

\begin{center}
\begin{tabular}{|c|c|c|c|c|}
\cline{2-5}
\multicolumn{1}{c|}{$\quad$} & \multicolumn{3}{|c|}{Lepton} & Lepton \\
\multicolumn{1}{c|}{$\quad$} & \multicolumn{3}{|c|}{non--universality} & universality \\
\hline
Experiment
         & \wwbr(\Wtoenu) & \wwbr(\Wtomnu) & \wwbr(\Wtotnu)
         & \wwbr({\mbox{$\rm{W}\rightarrow\rm{hadrons}$}}) \\
         & [\%] & [\%] & [\%] & [\%]  \\
\hline
\Aleph\  & $10.95\pm0.31$ & $11.11\pm0.29$ & $10.57\pm0.38$ & $67.33\pm0.47$ \\
\Delphi\ & $10.36\pm0.34$ & $10.62\pm0.28$ & $10.99\pm0.47$ & $68.10\pm0.52$ \\
\Ltre\   & $10.40\pm0.30$
                       & $\phz9.72\pm0.31$ & $11.78\pm0.43$ & $68.34\pm0.52$ \\
\Opal\   & $10.40\pm0.35$ & $10.61\pm0.35$ & $11.18\pm0.48$ & $67.91\pm0.61$ \\
\hline
LEP      & $10.54\pm0.17$ & $10.54\pm0.16$ & $11.09\pm0.22$ & $67.92\pm0.27$ \\
\hline
$\chi^2/\textrm{d.o.f.}$ & \multicolumn{3}{|c|}{14.9/9} & 18.8/11 \\
\hline\end{tabular}
\label{wwbra}
\end{center}
\vspace*{-0.5truecm}
\end{table}
\renewcommand{\arraystretch}{1.}

Assuming lepton universality, the measured hadronic branching
fraction is $67.92\pm0.17\rm{(stat.)}\pm0.21\rm{(syst.)}\%$ and
the measured leptonic branching fraction is
$10.69\pm0.06\rm{(stat.)}\pm0.07\rm{(syst.)}\%$. These results are
consistent with the Standard Model expectations, 67.51\% and
10.83\% respectively~\cite{yellowreport}. In this case, the high
$\chi^2$ of 18.8 for 11 degrees of freedom is mainly due to the
\Ltre\ results for W decays to muons and taus.

\subsection{\Vcs from $\rm W \to cs $}

The element \Vcs ~can be extracted directly from the $\rm W \to cs $
transition
without knowledge on the other CKM matrix elements.

The measurement of the $\rm W \to cs$ branching fraction,
requires the identification of a charm jet and a strange jet.
While charm or beauty jets can be relatively easily distinguished
from  light quark jets (u, d, s), the separation of
u, d and s jets is much more difficult.

Jets from strange quarks can be tagged using high momentum charged
kaons. Fast kaons are very likely to be originated from the
primary s-quark, while high momentum pions may come from u or d
jets. This naive picture is spoiled by the jet development where
many light quark are produced and by particle miss-identification.
At the energies which the analysis is performed, many protons can
be miss-identified as fast kaons if dE/dx is used for particle
identification. In this case a meaningful measurement with this
method requires a large data sample. If a RICH counter is used for
particle identification, \Vcs ~can be measured from  $\rm W \to
cs$ at the 10\% level from a data sample of 15000 W-pairs.

At LEP, only DELPHI identify particles using a RICH counter.
Exploiting the RICH detector and the fact that Cabibbo-suppressed
matrix elements can be neglected for the statistics collected at
LEP, DELPHI attempted to measure directly \Vcs ~\cite{ab}.

DELPHI uses events in which a charm and a strange jet are tagged and
to improve the signal purity exploit the V-A structure of the W decays.
Using only 120 hadronic W decays DELPHI gets:
$$  \Vcs  = 0.97 \pm 0.37, $$
where the uncertainty is dominated by the statistical error. This
uncertainty could go down to 5-6\% if all DELPHI W-pairs data set
would be used.
  The analysis is in progress and  results may be ready next summer.


\section{Conclusion and future prospective}

Using the full data set of W decays collected at LEP, \Vcs can be
extracted with an error of less than 2\%. The unitarity of the CKM
matrix for the part not involving the top quark can be measured at
the same level.

The LEP measurement are limited by statistics and future a $\rm e^+e^-$
machine can exploit W decays to measure all the six CKM elements
not involving the top quark with a good precision \cite{pm}.


\title*{PIBETA: \\A Precise Measurement
          \\of the Pion Beta Decay Rate \protect\newline}
\toctitle{PIBETA: A Precise Measurement
          of the Pion Beta Decay Rate}
%
%
\titlerunning{PIBETA: A Precise Measurement of the Pion Beta Decay
Rate} 
%
\author{D. Po\v{c}ani\'c, for the PIBETA Collaboration}
\authorrunning{D. Po\v{c}ani\'c}
%
%
\institute{Department of Physics, University of Virginia,
Charlottesville, VA 22904-4714, USA}

\maketitle              

\begin{abstract}
\index{abstract}We report preliminary working results of the PIBETA
experiment analysis for pion beta decay ($\pi\beta$),
$\pi^+\to\pi^0e^+\nu$, and for radiative pion decay (RPD) $\pi^+\to
e^+\nu\gamma$.  The former is in excellent agreement with the SM
predictions at the 1\,\% accuracy level.  The latter, an important
background for the $\pi\beta$ channel, shows an intriguing departure
from the basic V$-$A description.

\end{abstract}

\section{Experiment Goals and Motivation}

The PIBETA experiment \cite{pibeta} at the Paul Scherrer Institute
(PSI) is a comprehensive set of precision measurements of the rare
decays of the pion and muon.  The goals of the experiment's first
phase are:

\begin{itemize}

\item[(a)] To improve the experimental precision of the pion beta
decay rate, $\pi^+ \to \pi^0 e^+ \nu$ (known as $\pi_{e3}$, or
$\pi\beta$), from the present $\sim 4\,\%$ to $\sim 0.5\,\%$.  The
improved experimental precision will begin to approach the theoretical
accuracy in this process, and thus for the first time enable a
meaningful extraction of the CKM parameter $V_{ud}$ from a
non-baryonic process.

\item[(b)] To measure the branching ratio (BR) of the radiative decay
$\pi\to e\nu\gamma$ ($\pi_{e2}R$, or RPD), enabling a precise
determination of the pion form factor ratio $F_A/F_V$, and, hence, of
the pion polarizability.  Due to expanded phase space coverage of the
new measurement, we also aim to resolve the longstanding open question
of a nonzero tensor pion form factor.

\item[(c)] A necessary part of the above program is an extensive
measurement of the radiative muon decay rate, $\mu\to e \nu \bar{\nu}
\gamma$, with broad phase space coverage.  This new high-statistics
data sample is conducive to a precision search for non-\,(V$-$A)
admixtures in the weak Lagrangian.

\item[(d)] Both the $\pi\beta$ and the $\pi_{e2}R$ decays are
normalized to the $\pi\to e \nu$ (known as $\pi_{e2}$) decay rate.
The first phase of the experiment has, thus, produced a large sample
of $\pi_{e2}$ decay events.  The second phase of the PIBETA program
will seek to improve the $\pi_{e2}$ decay branching ratio precision
from the current $\sim 0.35\,\%$ to under 0.2\,\%, in order to provide
a precise test of lepton universality, and thus of certain possible
extensions to the Standard Model (SM).

\end{itemize}

Recent theoretical work \cite{jaus01,ciri02} has demonstrated low
theoretical uncertainties in extracting $V_{ud}$ from the pion beta
decay rate, i.e., a relative uncertainty of $5\times 10^{-4}$ or less,
providing further impetus for continued efforts in improving the
experimental accuracy of this process.

\section{Experimental Method}\label{sec:exp_met}

\begin{figure}[htb]
\hbox to\hsize{\hss
\includegraphics[width=0.8\hsize]{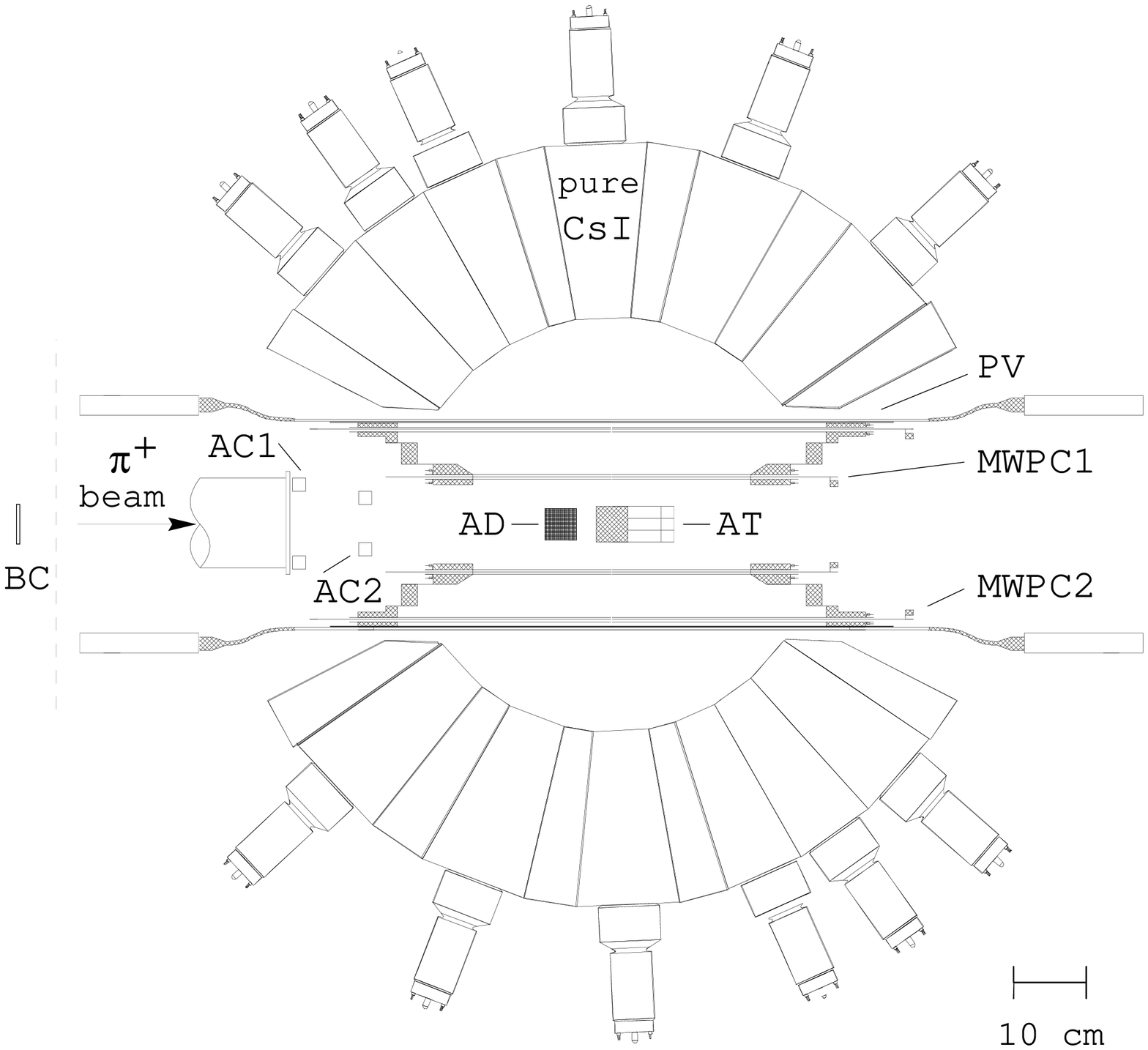}
\hss}
\caption{A schematic cross section of the PIBETA detector system.
Symbols denote: BC--thin upstream beam counter, AC1,2--active beam
collimators, AD--active degrader, AT--active target, MWPC1,2--thin
cylindrical wire chambers, PV--thin 20-segment plastic scintillator
barrel.  BC, AC1, AC2, AD and AT detectors are also made of plastic
scintillator.}
\label{fig:xsect}
\end{figure}

The $\pi$E1 beam line at PSI was tuned to deliver $\sim 10^6$
$\pi^+$/s with $p_\pi \simeq 113\,$MeV/c, that stop in a segmented
plastic scintillator target (AT).  The major detector systems are shown
in a schematic drawing in Fig.~\ref{fig:xsect}.  Energetic charged
decay products are tracked in a pair of thin concentric MWPC's and a
thin 20-segment plastic scintillator barrel detector (PV).  Both
neutral and charged particles deposit most (or all) of their energy in
a spherical electromagnetic shower calorimeter consisting of 240
elements made of pure CsI.  The CsI radial thickness, 22\,cm,
corresponds to 12$\,X_0$, and the calorimeter subtends a solid angle
of about 80\,\% of $4\pi\,$sr.

The basic principle of the measurement is to record all non-prompt
large-energy (above the $\mu \to e\nu\bar{\nu}$ endpoint)
electromagnetic shower pairs occurring in opposite detector
hemispheres (non-prompt two-arm events).  In addition, we record a
large prescaled sample of non-prompt single shower (one-arm) events.
Using these minimum-bias sets, we extract $\pi\beta$ and $\pi_{e2}$
event sets, using the latter for branching ratio normalization.  In a
stopped pion experiment these two channels have nearly the same
detector acceptance, and have much of the systematics in common.

A full complement of twelve fast analog triggers comprising all
relevant logic combinations of one- or two-arm, low- or high
calorimeter threshold, prompt and delayed (with respect to $\pi^+$
stop time), as well as a random and a three-arm trigger, were
implemented in order to obtain maximally comprehensive and unbiased
data samples.

\begin{figure}[b]
\hbox to\hsize{\includegraphics[width=0.47\hsize]{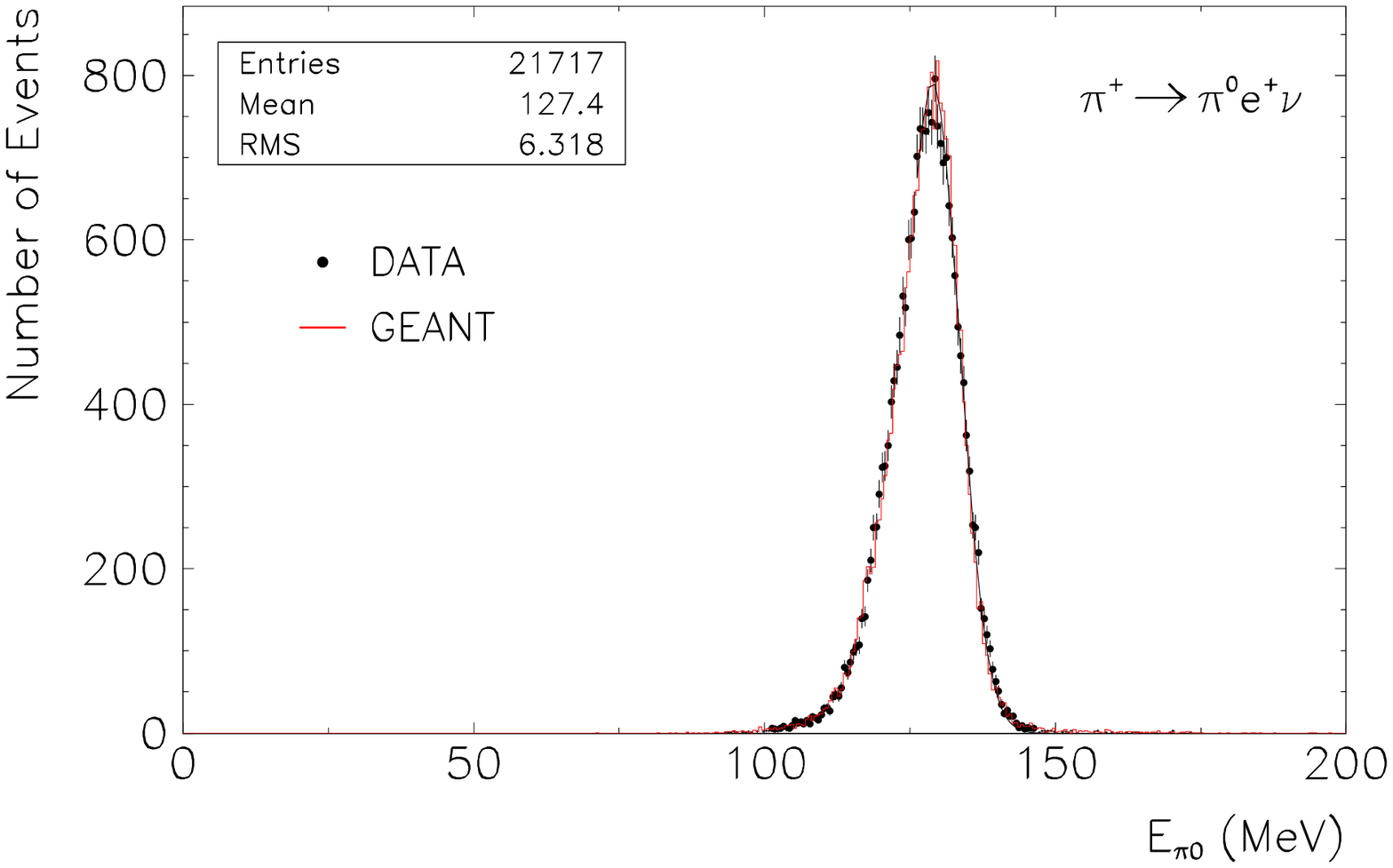}\hss
                \includegraphics[width=0.47\hsize]{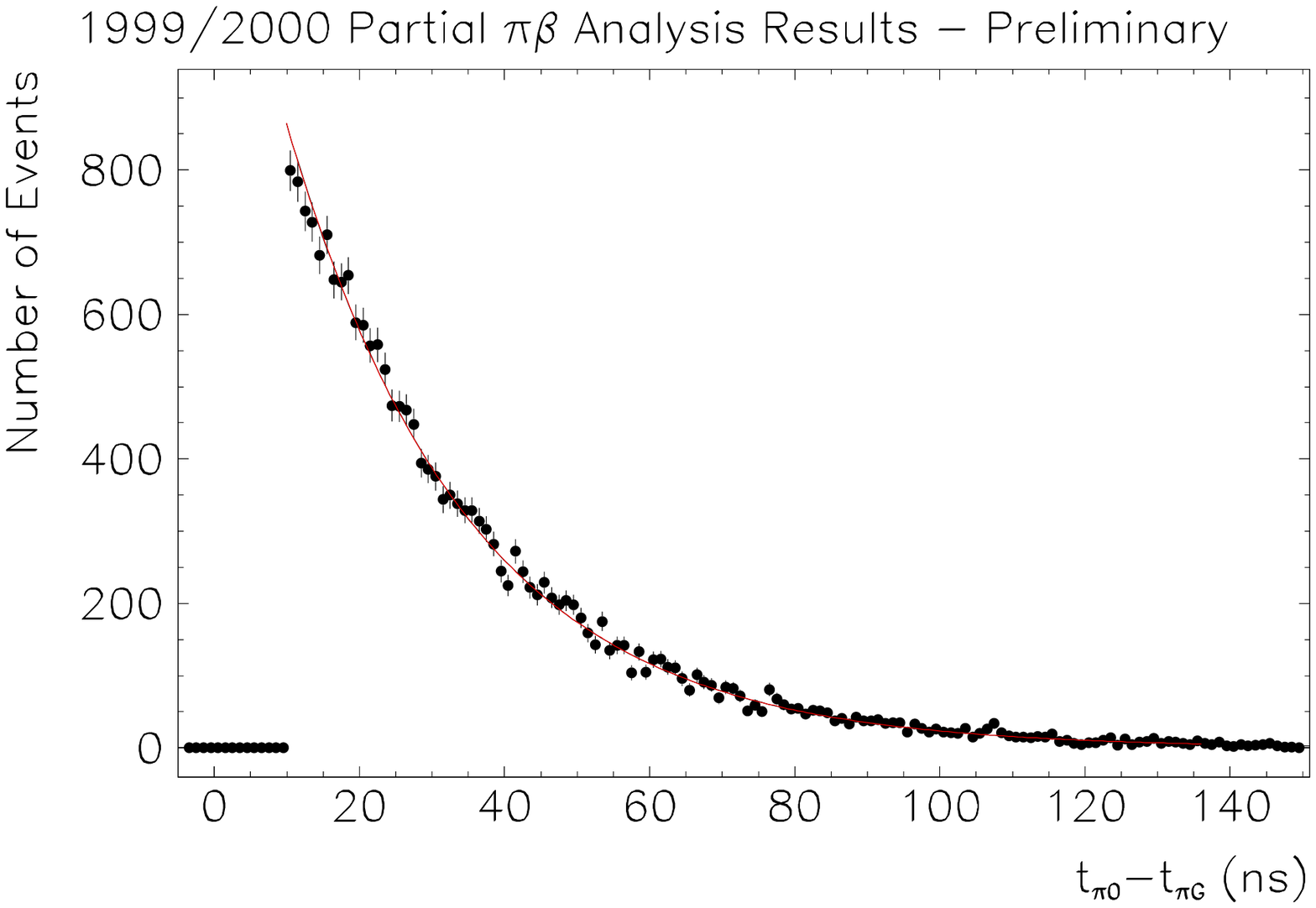}}
\hbox to\hsize{\includegraphics[width=0.47\hsize]{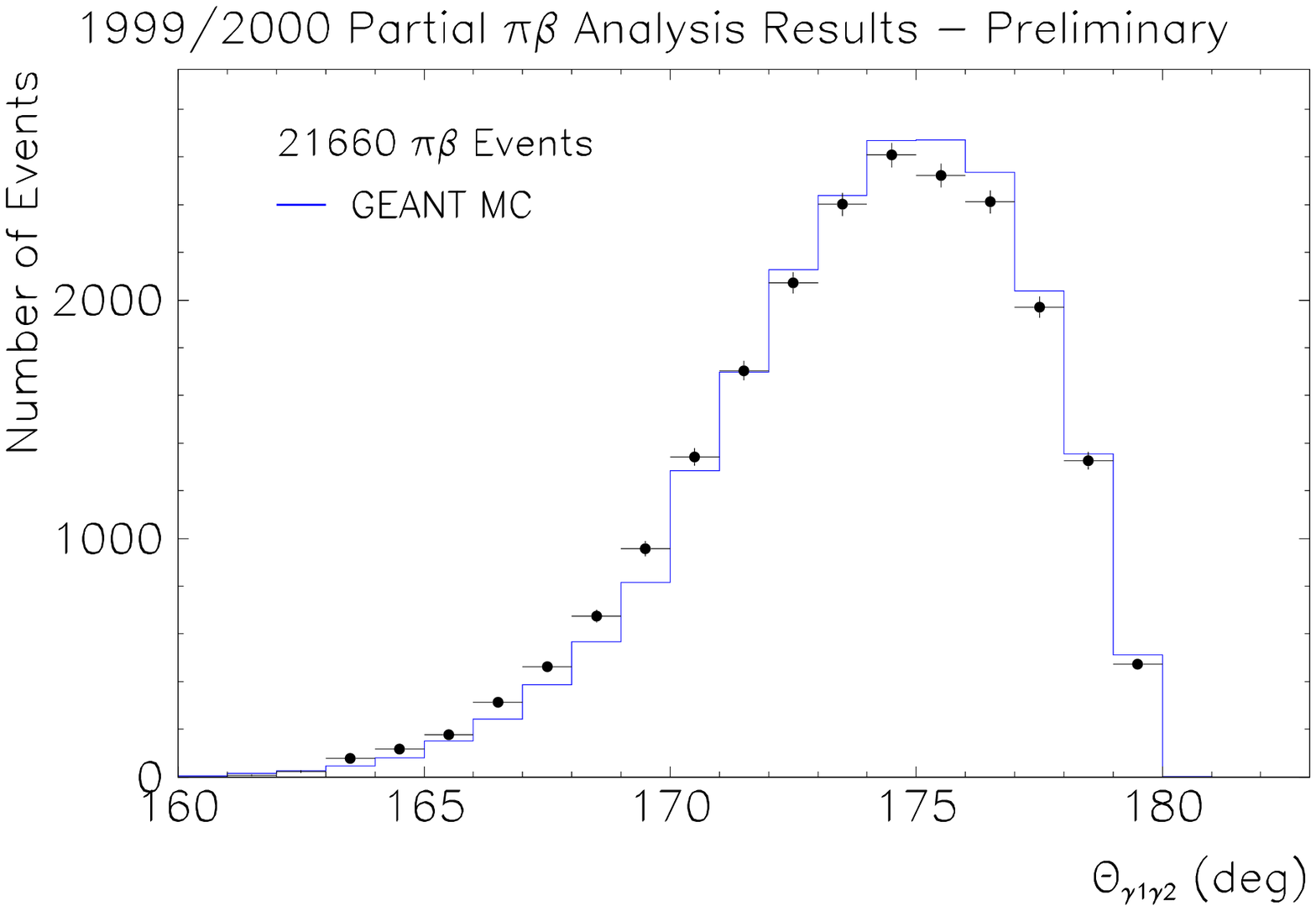}\hss
                \includegraphics[width=0.47\hsize]{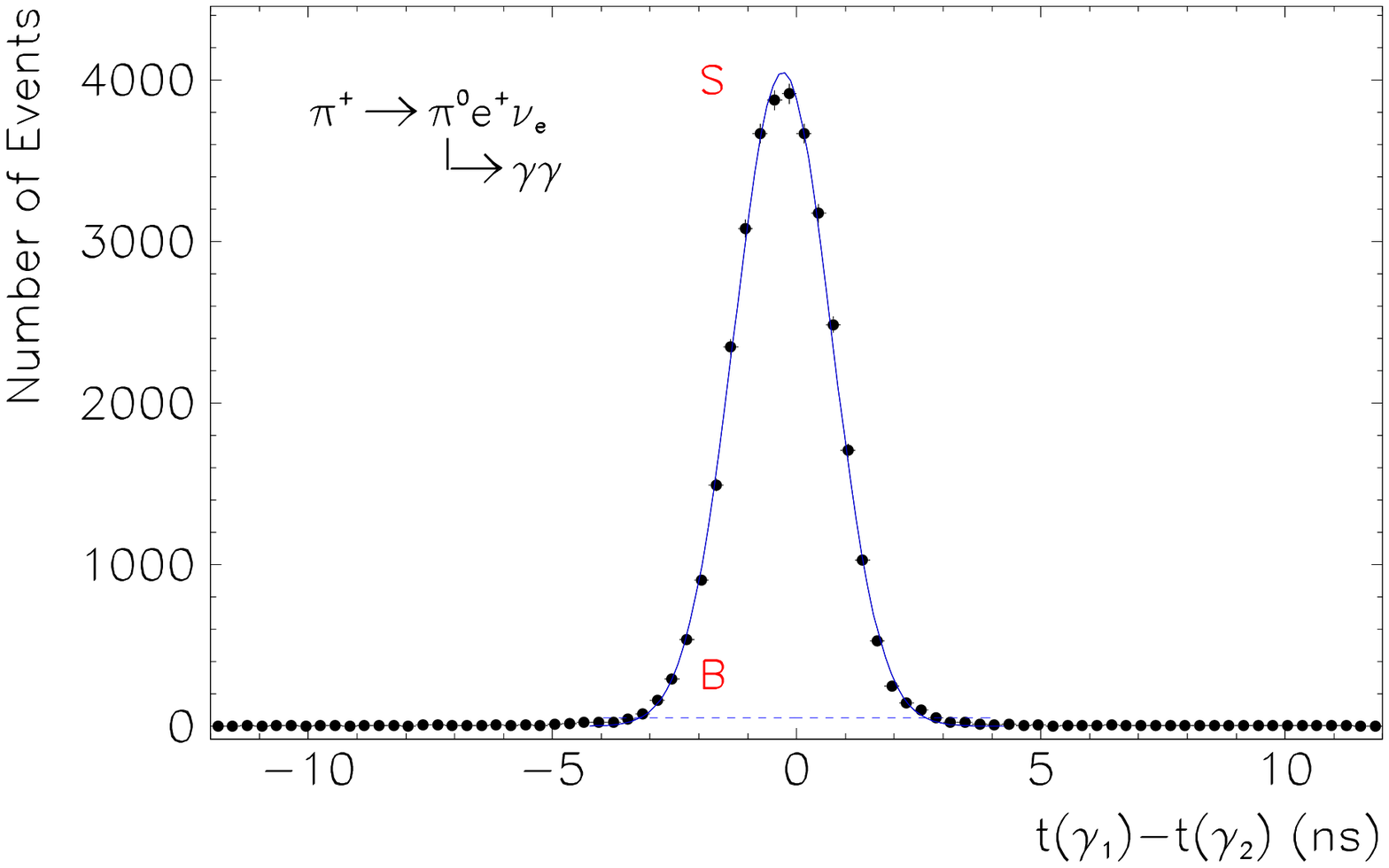}}
\caption{
Upper left: $\pi^0$ energy spectrum in $\pi\beta$ decay;
solid curve: GEANT simulation.  
Upper right: Histogram of time differences between the beam pion stop
and the $\pi\beta$ decay events (dots); curve: pion lifetime
exponential curve.  A software cut at 10\,ns was applied.  
Lower left: Histogram of the measured $\gamma$-$\gamma$ opening angle
in $\pi\beta$ decay events; solid curve: GEANT simulation.  
Lower right: Histogram of $\gamma$-$\gamma$ time differences for
$\pi\beta$ decay events (dots); curve: fit.  Signal to background
ratio exceeds 250.
All plots reflect a subset of the $\pi^+\to\pi^0e^+\nu$ decay
data measured in 1999/2000.}
\label{fig:pb_raw}
\end{figure}

The high quality of the PIBETA data is demonstrated in the histograms
of the calorimeter energy and event timing (following the $\pi^+$ stop
time), as well as of the $\gamma$-$\gamma$ opening angle and time
difference for a subset of the recorded pion beta decay events, shown
in Fig.~\ref{fig:pb_raw}.  In particular, the low level of accidental
background is evident in the $\gamma$-$\gamma$ relative timing
histogram; the peak to background ratio exceeds 250.  The histogram of
recorded $\gamma$-$\gamma$ opening angles for pion beta events
provides possibly the most sensitive test of the Monte Carlo
simulation of the apparatus, and of the systematics related to the
geometry of the beam pion stopping distribution.  The latter is the
single largest contributor to the overall uncertainty in the
acceptance, and, hence, in the branching ratio.

\section{First Results: Pion Beta Decay}

The first phase of measurements took place in 1999, 2000 and 2001,
resulting in some 60,000 recorded pion beta events.  The plots of
Fig.~\ref{fig:pb_raw} are based on a data subset acquired in 1999 
and 2000.  Our current {\sl\bfseries preliminary working} result for
the pion beta decay branching ratio, extracted from the above
analysis, is
\begin{equation}
   BR \simeq 1.044 \pm 0.007{\rm (stat.)} \pm  0.009{\rm (syst.)} 
             \times 10^{-8}\ .       \label{eq:pb_br_exp}
\end{equation}
Our result is to be compared with the previous most accurate
measurement of McFarlane et al. \cite{McF85}: 
$$
   BR = 1.026 \pm 0.039 \times 10^{-8}\ ,
$$
as well as with the SM Prediction (Particle Data Group,
2002 \cite{PDG02proca}): 
\begin{center}
\begin{tabular}{r@{\extracolsep{0.2em}}l}
   $BR =$ & $1.038 - 1.041 \times 10^{-8} \rm \quad (90\% C.L.)$ \\
        & $(1.005 - 1.008 \times 10^{-8} \rm \quad excl.\ rad.\ corr.)$\\
\end{tabular}
\end{center}
We see that our working result strongly confirms the validity of the
CVC hypothesis and SM radiative corrections \cite{Mar86,jaus01,ciri02}.
Another interesting comparison is with the prediction based on the
most accurate evaluation of the CKM matrix element $V_{ud}$ using the
CVC hypothesis and the results of measurements of superallowed Fermi
nuclear decays (Particle Data Group 2002 \cite{PDG02proca}):
$$
   BR = 1.037 \pm 0.002 \times 10^{-8}\ .
$$
Thus, our current preliminary working result is in very good agreement
with the predictions of the Standard Model and the CVC hypothesis.
The quoted systematic uncertainties are being reduced as our analysis
progresses.  To put this result into broader perspective, we can
compare the central value of $V_{ud}$ extracted from our data with
that listed in PDG 2002 \cite{PDG02proca}:
\begin{center}
\begin{tabular}{rl}
        {\rm PDG\ 2002:}  & $V_{ud} = 0.9734 (8)$,  \\
   {\rm PIBETA\ prelim:}  & $V_{ud} = 0.9771 (56)$. 
\end{tabular}
\end{center}

Table \ref{tab:pb_unc} summarizes the main sources of uncertainties
and gives their values both in the current analysis, and those that
are expected to be reached in a full analysis of the entire dataset
acquired to date.  We have temporarily enlarged the systematic
uncertainty quoted in Eq.~\ref{eq:pb_br_exp} pending a resolution of
the discrepancy found in the RPD channel and discussed in the
following section.

\begin{table}[ht]
\caption{Summary of the main sources of uncertainty in the extraction
of the pion beta decay branching ratio.  The column labeled
``Partial'' corresponds to the present analysis based on a portion of
the data taken in the years 1999 and 2000.  }

\begin{center}

\begin{tabular}{llcc}

\hline \\[-2ex]
  &  \multicolumn{3}{c}{Uncertainties in \%}           \\   
  \multicolumn{2}{c}{Dataset analyzed:}
           &  Partial  &    Full    \\[0.5ex]
\hline \\[-1ex]
 external:
    &  pion lifetime           &  0.019   & 0.019       \\
    &  $BR(\pi \to e\nu)$      &  0.33    
                                & $\sim 0.1^{\mathrm a}$  \\
    &  $BR(\pi^0\to\gamma\gamma)$ &  0.032   & 0.032       \\[1ex]
 internal:
     & $A(\pi\beta)/A(e\nu)$   &  0.5     & $<0.3$      \\
     & $\Delta t(\gamma - e)$  &  0.03    & 0.03        \\
     & E threshold             &  $<0.1$  & $<0.1$      \\[1ex]
 statistical: &  &  0.7 & $\sim 0.4$ \\[1ex]
 total:       &   &  $\sim 0.9$  
                                       & $\lesssim 0.5$ \\[1ex]
\hline\\[-1ex]

\multicolumn{4}{l}{$^{\mathrm a}$ Requires a new measurement.}    \\[-2ex]

\end{tabular}

\label{tab:pb_unc}

\end{center}

\end{table}

\section{First Results: Radiative Pion Decay}

As was already pointed out, we have recorded a large data set of
radiative decays: $\pi^+\to e^+\nu\gamma$ and $\mu^+\to
e^+\nu\bar{\nu}\gamma$.  To date we have analyzed both pion and muon
radiative decays, though with more attention devoted to the former, as
it is an important physics background to the $\pi\beta$ decay.  The
radiative pion decay analysis has given us the most surprising result
to date, and has commanded significant effort on our part to resolve
the issue.

Unlike previous experiments, the different one- and two-arm event
triggers used in our experiment are sensitive to three distinct
regions in the RPD phase space, resulting in broad coverage.  Without
going into details, we can loosely label the three phase-space regions
according to the positron and gamma energy thresholds ($E^t_e$,
$E^t_\gamma$) in each region: A (high, high), B (low, high), and C
(high, low).  Here the low threshold corresponds typically to 20\,MeV
or less, while the high threshold lies above the Michel decay
endpoint, typically 55\,MeV or more.

Together, the three regions overconstrain the Standard Model
parameters describing the decay, and thus allow us to examine possible
new information about the pion's hadronic structure, or non-(V$-$A)
interactions.  Appropriate analysis of these data is involved and
nuanced, requiring a lengthy presentation; we therefore present here
only a brief summary of our results.

Our analysis indicates a measurable departure from SM predictions.
Standard Model with the V$-$A electroweak sector requires only two
pion form factors, $F_A$ and $F_V$, to describe the so-called
structure-dependent amplitude in RPD.  The remainder of the decay
amplitude is accounted for by QED in the inner-bremsstrahlung (IB)
term.  The pion vector form factor is strongly constrained by the CVC
hypothesis, while existing data on the radiative pion decay (PDG
2002 \cite{PDG02proca}) suggest that $F_A \simeq 0.5\,F_V$, yielding
$$
    F_V = 0.0259 \pm 0.0005 \ , \qquad {\rm and} \qquad
    F_A \simeq 0.012\ .
$$
Simultaneous as well as separate fits of our data in the three phase
space regions confirm the above ratio of $F_A/F_V \simeq 0.5$.
However, they show a statistically significant deficit in RPD yield in
one region of phase space, corresponding to high $E_\gamma$ and lower
$E_e$ (mostly in region B), compared to predictions based on the above
values of the pion form factors.

A larger deficit in RPD yield, though less statistically significant
than our result due to far fewer events, was first reported by the
ISTRA collaboration \cite{Bol90a,Bol90b}.  This first observation was
interpreted by Poblaguev \cite{Pob90,Pob92} as indicative of the
presence of a tensor weak interaction in the pion, giving rise to a
nonzero tensor pion form factor $F_T \sim -6 \times 10^{-3}$.
Subsequently, Peter Herczeg \cite{Her94} found that the existing
experimental evidence on beta decays could not rule out a small
nonzero value of $F_T$ of this order of magnitude.  Tensor interaction
of this magnitude would be consistent with the existence of
leptoquarks \cite{Her94}.

\begin{figure}[b]
\hbox to\hsize{\hss\includegraphics[width=0.7\hsize]{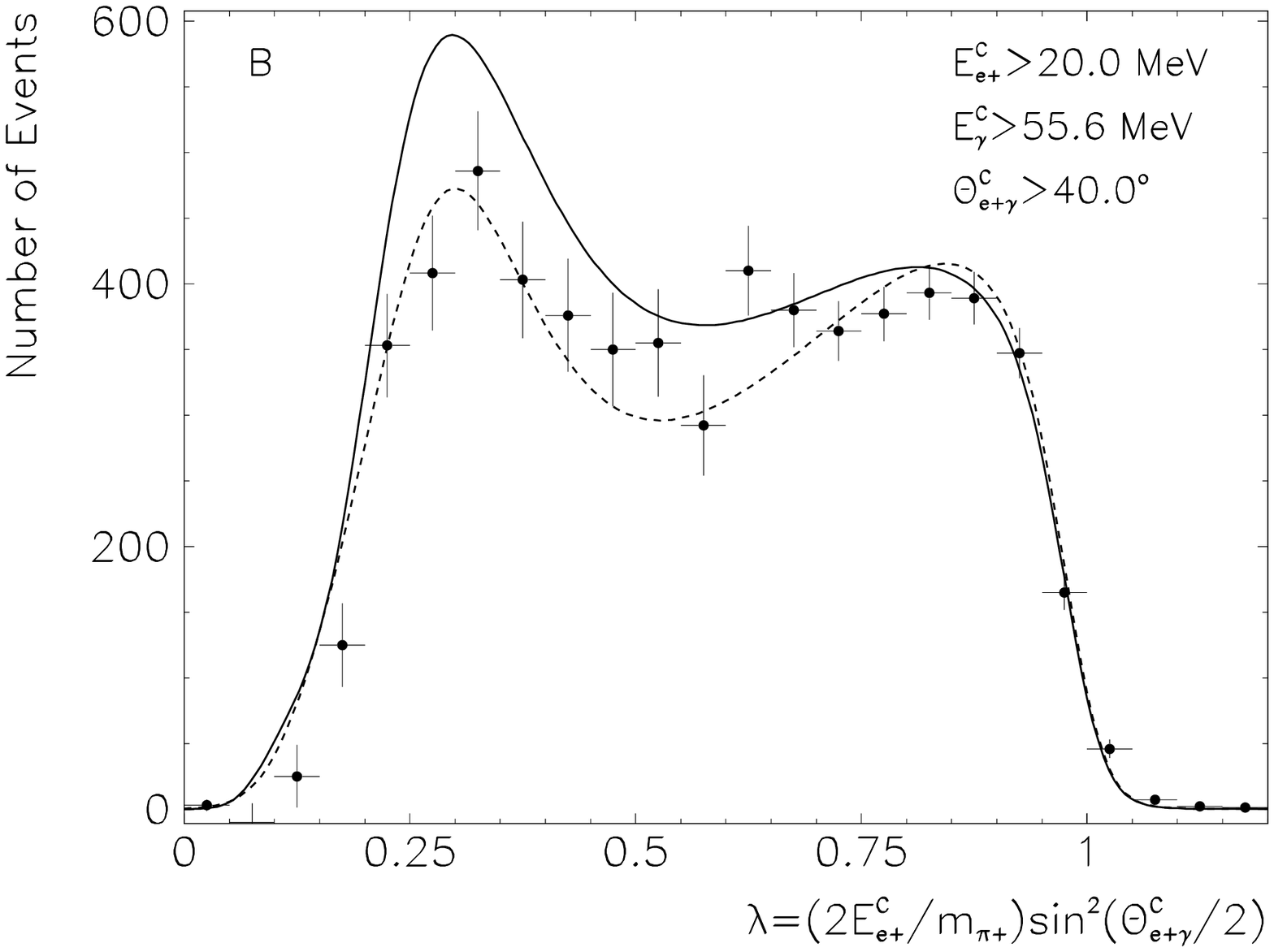}\hss}
\caption{\small Measured spectrum of the kinematic variable $\lambda =
(2E_{e+}/m_{\pi+})\sin^2(\theta_{e\gamma}/2)$ in $\pi^+ \to
e^+\nu\gamma$ decay for the kinematic region B, with limits noted in
the figure.  Solid curve: fit with the pion form factor $F_V$ fixed by
the CVC hypothesis, $F_T=0$, and $F_A$ free.  Dashed curve: as above,
but with $F_T$ also released to vary freely, resulting in $F_T =
-0.0016\,(3)$.  This work is in progress.}
\label{fig:pienug:2a_l1}
\end{figure}

We illustrate our working results in Fig.~\ref{fig:pienug:2a_l1} which
shows a projected one-dimensional distribution of $\lambda$, a
convenient kinematic variable based on $E_e$ that ranges from 0 to 1
regardless of $E_\gamma$.  It is clear that for lower values of
$\lambda$ (and therefore of $E_e$), an SM fit with only $F_V, F_A \neq
0$ overestimates the experimental yield.  Adding a nonzero tensor form
factor of $F_T \sim -0.0016$ produces statistically significantly
better agreement with the data.  The fits are two-dimensional and
encompass all three kinematic regions.  This work is in progress, and
the results are subject to change---we are currently refining the
analysis as well as the fit strategies.

This working result may be indicating either the existence of a tensor
weak interaction, or, alternatively, that the standard treatment
of the RPD may not at this time correctly incorporate all known SM
physics.  Radiative corrections seem to be a particularly good
candidate for reexamination.

\section{Conclusions}

We have extracted an experimental branching ratio for the pion beta
decay at the 1\,\% uncertainty level, and expect to reduce the
uncertainty by an additional factor of two in the near future.  Our
result agrees with the CVC hypothesis and radiative corrections for
this process, and it opens the way for the first meaningful extraction
of the CKM parameter $V_{ud}$ from a non-baryonic process.

Our analysis of the $\pi\to e\nu\gamma$ decay confirms that $F_A/F_V
\simeq 0.5$, in agreement with the world average.  However, events
with a hard $\gamma$ and soft e$^+$ are not well described by standard
theory, requiring ``$F_T \neq 0$''.  We can, though, rule out a large
``$F_T$'', as reported in analyses of the ISTRA data.

The high statistics and broad coverage of our RPD data in principle
guarantee extraction of pion weak form factor values with
exceptionally low uncertainties.  However, it appears that theoretical
treatment of RPD may have to be revisited before the full potential of
the PIBETA data is realized.


\title*{Semileptonic Kaon and Hyperon Decays:\\
What Do They Tell Us about V$_{us}$?}
\toctitle{Semileptonic Kaon and Hyperon Decays: \protect\newline
What Do They Tell Us about $V_{us}$?}
%
%
\titlerunning{Semileptonic Kaon and Hyperon Decays}
%
\author{H.-W. Siebert}
\authorrunning{H.-W. Siebert}
%
%
\institute{Physikalisches Institut, Univ. Heidelberg \protect\newline
Philosophenweg 12, D-69120 Heidelberg, Germany}

\maketitle              

\begin{abstract}
From $K^+_{e3}$ and $K_{L,e3}$ decays, $V_{us}$ can be determined to be
 $0.220\pm 0.003$. If $SU_3$ symmetry is invoked, semileptonic hyperon
decays offer an independent determination $V_{us}\, = \, 0.223 \pm 0.004$.
The unknown effects of $SU_3$ symmetry breaking make this result less
safe than the $K_{e3}$ result.

\end{abstract}

\section{Introduction}
Our best sources of information on $V_{us}$ are semileptonic decays
(s.l.d.)
of kaons and hyperons. The leptonic part of the matrix element of
s.l.d. is unambiguous.
In contrast, the hadronic part is modified by $SU_3$ symmetry
breaking.
For kaon s.l.d.,
which are pure V transitions, this is a minor problem, since for
V transitions $SU_3$ breaking  effects
are of second order only (Ademollo-Gatto theorem).
On the other hand, s.l.d. of hyperons are mixed vector and axialvector
transitions, and there is at present no agreement on the theoretical
side how to handle $SU_3$ breaking here.

\section{Semileptonic kaon decays}

Both $K^+_{e3}$ and $K_{L,e3}$ decay rates are well-known
experimentally since more than 20 years. The experimental
situation \cite{pdg3} is summarized in table \ref{table:kl3}. It
is seen, that in $K^+_{e3}$ decay, the experimental error of the
decay rate is dominated by the error of the branching ratio, while
in $K_{L,e3}$ decay, the errors of the $K_L$ lifetime and the
$K_{L,e3}$ branching ratio contribute about equally.

Already in 1984, a value $V_{us}\,  =\, 0.2196 \, (\pm 1.1 \% )$
was extracted from the combined $K^+$ and $K_L$ data
\cite{leutwyler}. Since then, more investigations of the radiative
corrections and the $q^2$-dependence of the formfactors have not
changed the situation: The experimental error is $\approx 0.8$\%,
the theoretical error is $<1\%$, and we find a safe value {\bf
$V_{us}\, =\, 0.220\pm 0.03$} (see also the discussion in ref.
\cite{pdg3} and the talk by J. Marciano in these Proceedings).

\section{Semileptonic hyperon decays}

The experimental information on
strangeness-changing hyperon s.l.d. is summarized in table
\ref{table:Y}. Most of the results are more than 10 years old
\cite{bour1,bour2,bour3,wise80,hsueh88}
with the exception of the new results from the KTEV experiment
on  $\Xi^0 \to \Sigma^+ e\overline{\nu}$ decay \cite{alavi01}.

The decay rate $\Gamma$ of each decay is proportional to $|V_{us}|^2$,
but depends also on the formfactor ratio
 $g_1 /f_1$.
Experimentally, the decay rates are calculated from the measured
decay branching ratios, using the much better known lifetimes of the
mother hyperons.
 $g_1 /f_1$ has been determined from the
Dalitz plot distributions and in some cases from polarization
asymmetries.
Three cases can be distinguished here: Using a beam of hyperons
polarized at production \cite{hsueh88},
using $\Lambda $s from $\Xi^-$ decays \cite{bour2},
which have
a longitudinal polarization $P\, =\, \alpha_{\Xi^-} \, =\, 0.64$
or analyzing the polarization of the daughter hyperon
\cite{bour1,alavi01}.

There are four branching ratio measurements with errors
between 2\% and 6\% and three measurements of the
formfactor ratio $g_1 /f_1$ with errors between 0.015 and 0.05.
(The experimental errors of the corresponding hyperon lifetimes are
always
smaller by at least a factor of 2).
In good approximation, the decay rates are
 $\Gamma_{if} \, =\, const.\cdot V_{us}^2 \cdot (1 \,  +\,  3 g_1 /f_1 )$.
From CVC and $SU_3$ symmetry, one obtains for the decays
 $B_i \to B_j e^- \overline{\nu}$
the relation $g_1 /f_1 \, =\, f_{ij}\cdot F \, + \, d_{ij}\cdot D$,
where $f_{ij}$ and $d_{ij}$ are $SU_3$
Clebsh-Gordan coefficients.
Therefore, all measured decay rates and $g_1 /f_1$ ratios
within the baryon ground state
octet can be described by the  parameters
 $V_{us}$, $F$ and $D$.
The dependence of $g_1 /f_1$ on $F$ and $D$
is given in the last column of table \ref{table:Y}.

\begin{table}
\caption{Experimental data on kaon s.l. decays.}
\begin{center}
\renewcommand{\arraystretch}{1.4}
\setlength\tabcolsep{5pt}
\begin{tabular}{|l|c|c|c|}   \hline\noalign{}
decay & $\tau$ [10$^{-8}$ s] & BR & $\Gamma$ [10$^{-15}$MeV]  \\ \hline
 $K^+_{e3}$  & 1.2384 ($\pm$ 0.2\% ) & 0.0487  ($\pm$ 1.2\% ) &
 2.590 ($\pm$ 1.25\% ) \\
\hline\noalign{}
 $K_{L,e3}$  & 5.17 ($\pm$ 0.8\% ) & 0.388  ($\pm$ 0.7\% ) &
 4.938 ($\pm$ 1.05\% ) \\
\hline\noalign{}
\end{tabular}
\end{center}
\label{table:kl3}
\end{table}

\begin{table}[h]
\caption{ Experimental data on strangeness-changing hyperon
s.l. decays.}
\begin{center}
\renewcommand{\arraystretch}{1.4}
\setlength\tabcolsep{5pt}
\begin{tabular}{|l|c|c|c|}   \hline\noalign{}
decay & 10$^4 \cdot$BR & $g_1 /f_1 (exp.)$ & $g_1 /f_1$ ($SU_3$)
\\ \hline\noalign{}
 $\Lambda \to pe\overline{\nu} $ & 8.32 $\pm$ 0.14 & 0.718 $\pm$ 0.015
&     F + D/3
\\ \hline\noalign{}
 $\Sigma^- \to ne\overline{\nu}$ & 10.17 $\pm$ 0.34 & -0.340 $\pm$
0.017
&  F - D
\\ \hline\noalign{}
 $\Xi^- \to \Lambda e\overline{\nu}$ & 5.63 $\pm$ 0.31 & 0.25 $\pm$
0.05
&  F - D/3
\\ \hline\noalign{}
 $\Xi^- \to \Sigma^0 e\overline{\nu}$ & 0.87 $\pm$ 0.17 &
&  F + D
\\ \hline\noalign{}
 $\Xi^0 \to \Sigma^+ e\overline{\nu}$ & 2.62 $\pm$ 0.17 & 1.3$\pm$0.2
&  F + D
\\ \hline\noalign{}

\end{tabular}
\end{center}
\label{table:Y}
\end{table}

Are those results consistent?

\begin{figure}[h]
\begin{center}
\mbox{\epsfxsize=9cm\epsffile{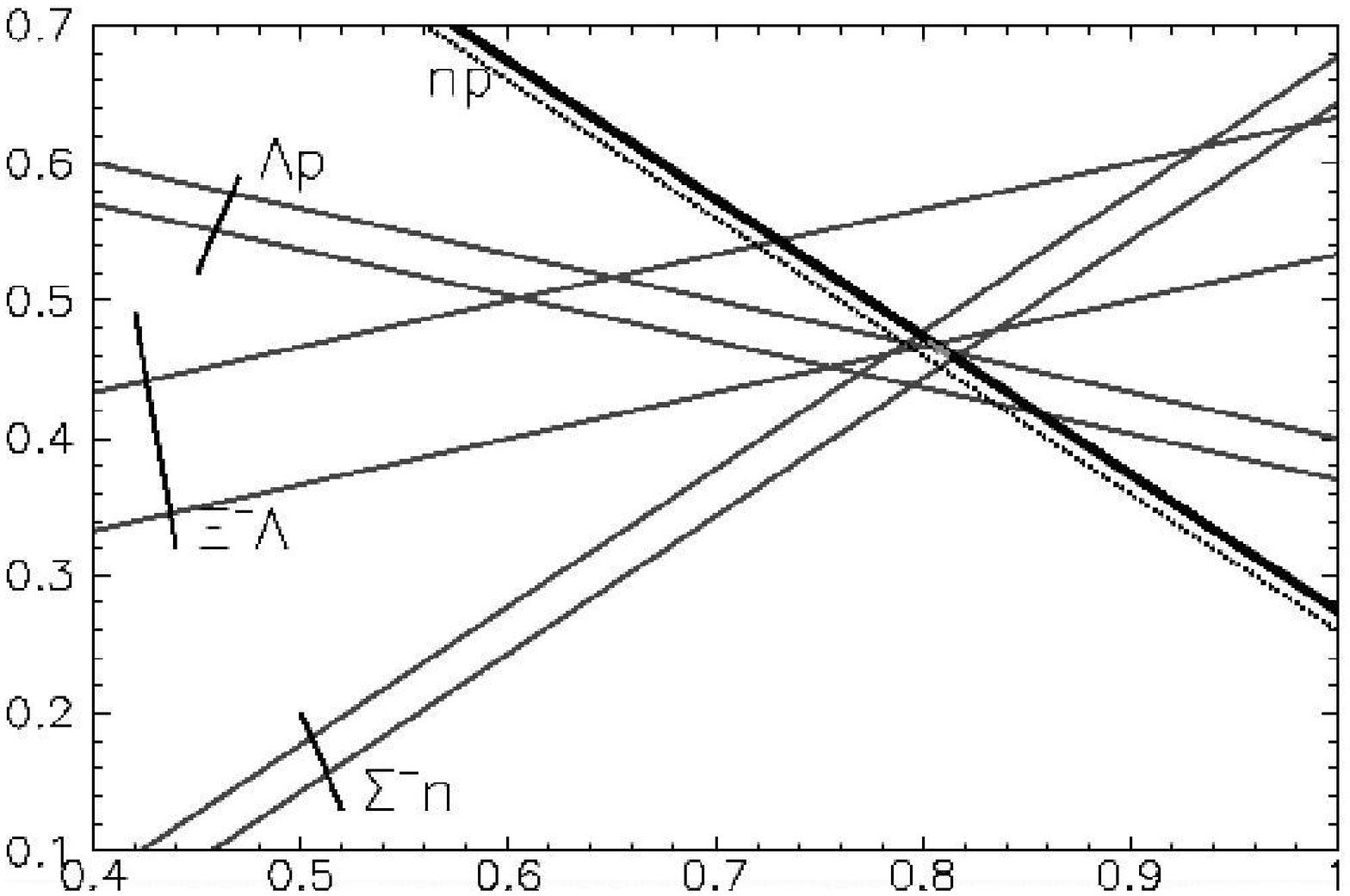}} \caption{ F and D as
determined from $g_1 /f_1$ measurements. The bands from the
hyperon s.l. decays correspond to the $\pm \, 1\sigma$ contours.
The dashed line is for  F+D=1.26.} \label{fig:g1f1}
\end{center}
\end{figure}

First we  fit four branching ratios and three formfactor ratios from
table \ref{table:Y} to the three parameters
F+D, D/(F+D) and $V_{us}$,
excluding the much less precise results for the
 $\Xi^- \to \Sigma^0 e\overline{\nu}$ branching ratio
and the  $\Xi^0 \to \Sigma^+ e\overline{\nu}$ formfactor.
The result is listed as
``fit 1'' in table \ref{tab:fits}. If we include the much more precise value of
 $g_1 /f_1$ in $n\to p e\overline{\nu}$ decay (fit 2),
F+D is pulled towards the neutron decay value and
 $V_{us}$ moves toward the value found in kaon s.l.d., but the $\chi^2$
 of the fit becomes worse.
Let us look just at the $g_1 /f_1$ values in fig.~\ref{fig:g1f1}:
Here for each decay the $\pm 1\sigma$ contour in the F,D plane
from the measured value of  $g_1 /f_1$ is shown.
There is good consistency between the hyperon decay
and the neutron decay values.
We also see, that the neutron decay result produces a
very strong correlation between F and D,
therefore we prefer the parameters F+D and  D/(F+D),
which are almost uncorrelated in the fit.
To demonstrate the precision of the neutron decay result,
a dashed line for F+D=1.260 is also drawn.
The results are given as ``fit 3'' in table   \ref{tab:fits}.

\begin{table}[h]
\caption{ Fit results.}
\begin{center}
\renewcommand{\arraystretch}{1.4}
\setlength\tabcolsep{5pt}
\begin{tabular}{|c|c|c|c|c|}   \hline\noalign{}
 & F+D & D/(F+D) & $V_{us}$ & $\chi^2$/NDF
\\ \hline\noalign{}
fit 1 & 1.225$\pm$0.022 & 0.638$\pm$0.005 & 0.227$\pm$0.003 & 11.5/4
\\ \hline\noalign{}
fit 2 &1.2736$\pm$0.0018 & 0.639$\pm$0.005 & 0.223$\pm$0.002 & 16.1/5
\\ \hline\noalign{}
fit 3 & 1.2737$\pm$0.0013 &  0.635$\pm$0.006 & & 2.5/2
\\ \hline\noalign{}
\end{tabular}
\end{center}
\label{tab:fits}
\end{table}

\begin{figure}[t]
\begin{center}
\mbox{\epsfxsize=8cm\epsffile{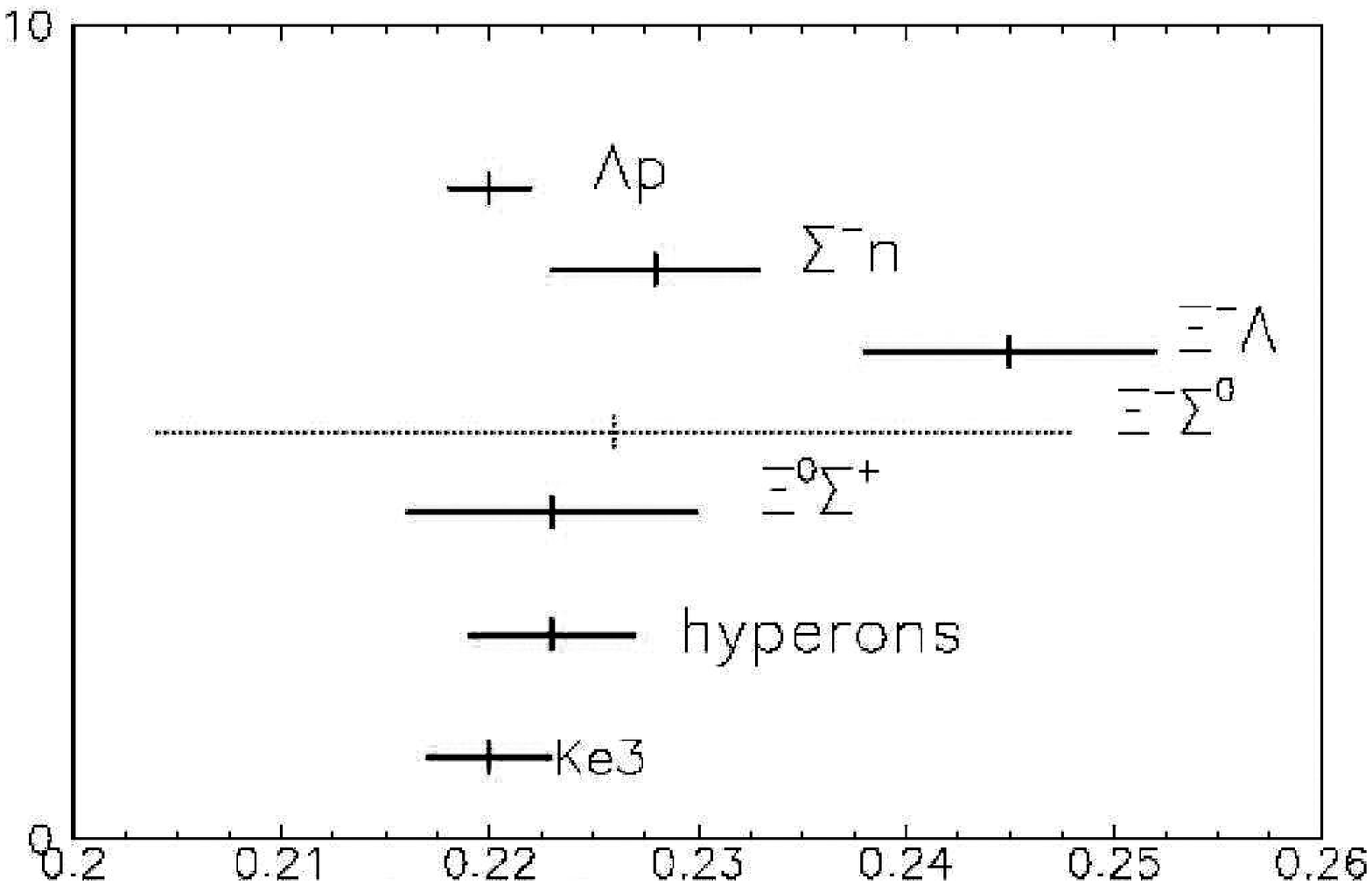}} \vspace{5pt} \caption{ Determination of
$V_{us}$ from Kaon and hyperon s.l.d. rates.
          The dashed result from  $\Xi^- \to \Sigma^0 e\overline{\nu}$
    decay is not included in the average labelled ``hyperons''.}
\label{fig:vus}
\end{center}
\end{figure}

The ``fit 3'' result  then is used to calculate $V_{us}$
for each hyperon s.l.d. from the measured branching ratio.
The results are shown in fig. \ref{fig:vus}.
The mean of the four  $V_{us}$ values drawn as solid lines is
0.223$\pm$0.004, marked as ``hyperons''
in fig. \ref{fig:vus}, with  $\chi^2$/NDF = 12.7/3.
where the error of the mean was increased by a factor
 $\sqrt{\chi^2/NDF}$.
This value is in good agreement with the $K_{e3}$ result.

However, this analysis was done without taking
 $SU_3$ symmetry breaking into account.
There is at present no agreement between the different
attempts to predict the effect of  $SU_3$ symmetry breaking
on $f_1$ and $g_1$, except that these changes could
be up to 10\% in some cases. There is not even
agreement on whether a given formfactor should become larger or
smaller.
Table \ref{tab:symbreak}, taken from ref. \cite{flores98},
lists some calculations by different authors
on the change in $f_1$, which would directly
affect the determination of $V_{us}$ from
a given s.l. decay rate (apologies to all authors neglected here).
Also, in the calculation of the decay the radiative corrections
from ref.  \cite{Toth} have been used. Are they really
safe (see ref. \cite{glueck}, for instance) ?

\begin{table}[h]
\caption{ Theoretical estimates of symmetry breaking effects
on $f_1$. The ratio $f_1 /f_1^{SU(3)}$ is listed. }

\begin{center}
\renewcommand{\arraystretch}{1.4}
\setlength\tabcolsep{5pt}
\begin{tabular}{|l|c|c|c|c|c|}   \hline\noalign{}
Transition & ref.~\cite{flores98} &ref.~\cite{anderson93} &
ref.~\cite{donoghue87} & ref.~\cite{krause90} & ref.~\cite{schlumpf95}
\\ \hline\noalign{}
 $\Lambda \to pe\overline{\nu} $ & 1.02$\pm$0.02 & 1.024 & 0.987 &
0.943 & 0.976 \\
$\Sigma^- \to ne\overline{\nu}$ & 1.04$\pm$0.02 & 1.100 & 0.987 &
 0.987 & 0.975 \\
 $\Xi^- \to \Lambda e\overline{\nu}$ & 1.10$\pm$0.04 & 1.059 &
0.987 & 0.957 & 0.976 \\
 $\Xi^- \to \Sigma^0 e\overline{\nu}$ & 1.12$\pm$0.05 & 1.011 0.987 &
 0.943 & 0.976 &
\\ \hline\noalign{}
\end{tabular}
\end{center}
\label{tab:symbreak}
\end{table}

To conclude, we find agreement on $V_{us}$ from   $K_{e3}$ and
from hyperon s.l.d. The special feature of  $K_{e3}$ decays being
pure vector transitions makes the determination  of  $V_{us}$ from
the  $K_{e3}$ data much safer than from the hyperon s.l. decay
data. Unless new experimental data differ radically from the
existing results, or a much more precise understanding of  $SU_3$
symmetry breaking and radiative corrections in hyperon s.l.d.
emerges, the  $K_{e3}$ result will be our best bet on  $V_{us}$.
So it seems unlikely that $V_{us}$ will change sufficiently to
erase the intriguing unitarity deficit...

\title*{Project of a New Measurement of the Electron-Antineutrino-Correlation $a$ Coefficient in Neutron Beta Decay}
\toctitle{Project of a New Measurement of the
Electron-Antineutrino-Correlation $a$ Coefficient in Neutron
Beta Decay}
%
%
\titlerunning{New Measurement of Electron-Antineutrino-Correlation $a$}
%
\author{B.G. Yerozolimsky}
\authorrunning{B.G. Yerozolimsky}
%
%
\institute{Harvard University High Energy Physics Laboratory \\ 42 Oxford Street, Cambridge MA 02138, USA}

\maketitle

\begin{abstract}
The project which is the subject of this talk is to be carried out
by a collaboration of several groups of scientists working in the
Institutes of USA and Russia\footnote{List of participants:\\
F. Wietfeldt, principal investigator of this project, and C. Trull
 \textit{ (Tulane University, USA)}\\
Yu. Mostovoy, S. Balashov and V. Fedunin \textit{ (Kurchatov
Institute, Russia)}\\ B. Yerozolimsky, L. Goldin and R. Wilson
\textit{(Harvard University, USA)}\\
M. S. Dewey, F. Bateman, D. Gilliam, J. Nico and A. Thompson
\textit{(NIST, USA)}
\\
A. Comives \textit{(De Pauw University, USA)}
\\
B. Collett and G. Johns \textit{ (Hamilton College, USA)}
\\
M. Leuschner \textit{(Indiana University, USA)}}.\end{abstract}

Up to date, there were only three attempts to measure this angular
correlation coefficient: the first was made in 1967 in Moscow
\cite{Grigoryev}, the second was finished in 1978 in Zeibersdorf
\cite{Stratova} and the last has been done recently at ILL and
published in 2002 \cite{byrne}.

The data derived in these experiments are consistent with each
other, but their accuracy does not exceed 5{\%}:
\begin{eqnarray*}
\mathbf{a} = - 0.091\;\; \pm 0.039 \qquad\;\: \cite{Grigoryev}\\
\mathbf{a} = - 0.1017 \pm 0.0051 \qquad\cite{Stratova}\\
\mathbf{a} = - 0. 1054 \pm 0.0055  \qquad\cite{byrne}
\end{eqnarray*}
This situation looks particularly poor if one takes into account
that the values of all other parameters which characterize this
fundamental beta decay process (lifetime $\mathbf{\tau}$, angular
correlation coefficients \textbf{A} and \textbf{B}) are known with
much higher precision. This stops possible investigations of
validity of the Standard Model Theory of Week Interactions. Thus,
measurements of the value of \textbf{a} made with improved
accuracy are now urgently needed. Discussions which took place
during this Workshop confirm this conclusion.

From our point of view, the main reason of poor accuracy of
\textbf{a}--measurements is the methods used in all previous
experiments, which required too precise spectrometry of the decay
products (electrons and recoil protons). For instance, in the
experiment of Prof. J. Byrne, the proton spectrum was measured,
and it is too insensitive to the \textbf{a}--value. In order to
reach 1{\%} accuracy in \textbf{a}--value the proton spectrum had
to be measured with $\sim 3\cdot 10^{-4}$ precision.

The goal of our new approach is to arrange the experimental device
in such a manner that events with opposite directions of
antineutrino would be reliably separated. Then the value of
\textbf{a} can be simply derived by comparing numbers of events in
each group, and no precise spectrometry will be needed.

The method of measurement we are going to use is based on two ideas:

First of them was proposed many years ago in our laboratory at
Kurchatov Institute in Moscow. It is to arrange two detectors (one
for beta-electrons and the second for recoil protons) in 180
degree geometry relative to decay region of the neutron beam.
These detectors are switched in coincidences with one another. It
can be easily understood that if solid angles of these detectors
are sufficiently small and the energy of the detected electrons is
not too low ($T_{e} >$ 250 keV) the proton energy spectrum will
consist of two groups belonging to decay events with opposite
antineutrino directions. These two groups can be easily separated
with the help of usual TOF technique without any serious
precision.

Unfortunately, in spite of the fact that using this approach one
can detect decay events corresponding to definite cones of
antineutrino escape angles, the antineutrino--electron angular
correlation can not be evaluated from these data because solid
angles of antineutrino emission in these two groups are
principally different. This method of selecting events with
definite antineutrino escape solid angles has been only used in
experiments with polarized neutrons (measurements of \textbf{B}
correlation coefficient) where the asymmetries measured were
connected with the spin-flip, and antineutrino solid angles did
not change.

\begin{figure}[ht]
\begin{center}
\includegraphics[scale=0.5]{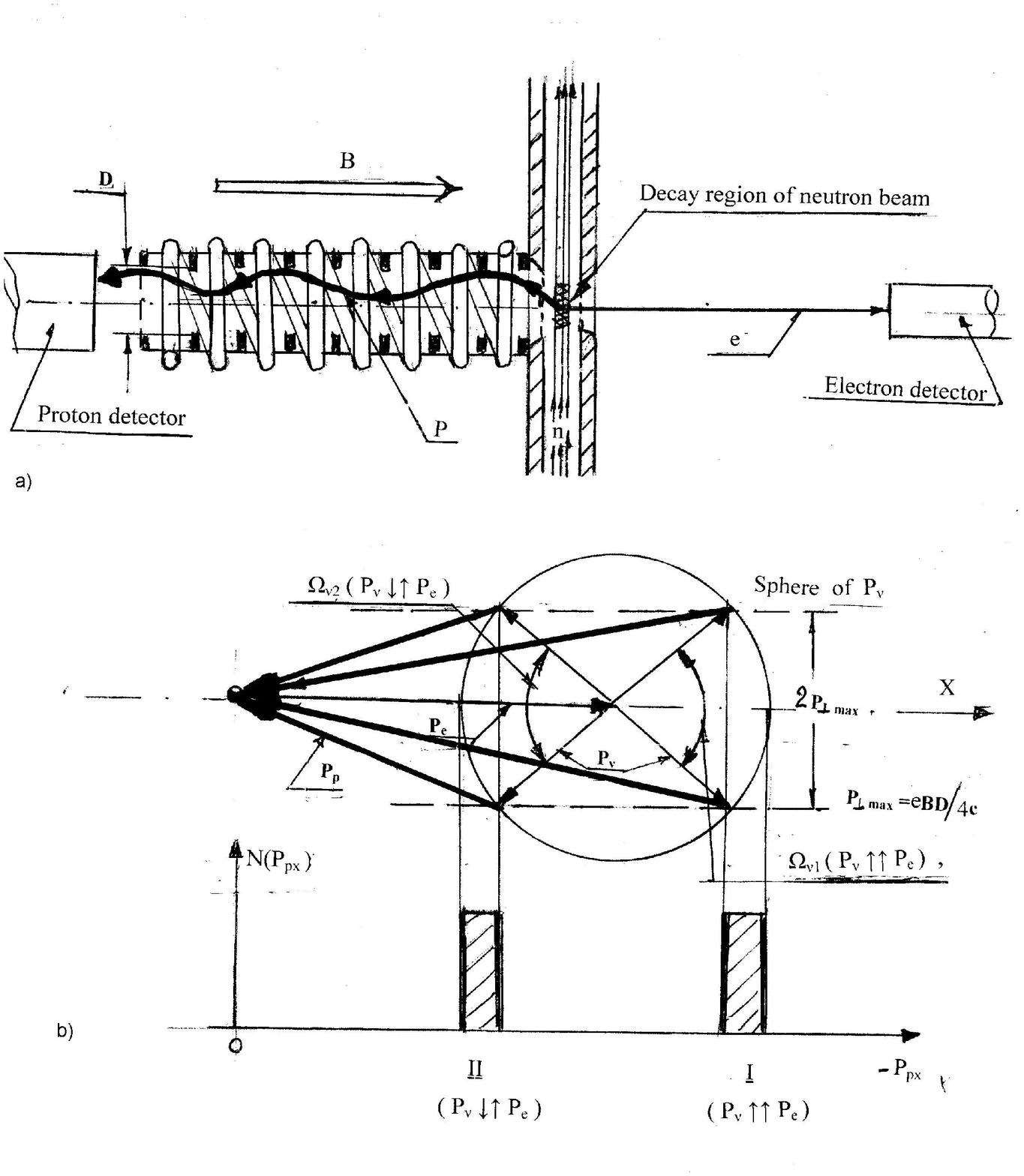}
\end{center}
\vskip -10mm
\caption[]{a) The arrangement with the longitudinal magnetic field b) Momentum diagram.}
\label{figur1}
\end{figure}

The second idea was proposed by Yu. Mostovoy in 1994
\cite{balashov}. He has understood how to make these two
antineutrino solid angles equal. He proposed to use a distributed
set of diaphragms which form a ``proton guide'' (about 1 m long)
leading protons from the neutron decay region to the detector and
a strong axial magnetic field inside. In such a case to come to
the detector protons must have not a limited flight angle but a
limited transverse component of momentum P$_{p\perp}$, and as a
result, both solid angles of antineutrino become identically
equal. This situation is illustrated with the help of a momentum
diagram in Fig \ref{figur1}, where a schematic sketch of the
measurement arrangement including both ideas is presented too. The
value of the transverse momentum limit depends upon the diameter
of the diaphragms and the strength of the longitudinal magnetic
field. If, for example, a small proton source is disposed on the
central axis

\begin{equation}
P_{p\perp}(max)= (eBD)/4c
\end{equation}

Where B is the magnetic field in Gs, D -- diameter of the
diaphragms in cm, e -- elementary electric charge and c --
velocity of light.

The equality of solid angles of both groups of antineutrino which
is evidently seen on the momentum diagram would lead to equality
of integrals N$_{1}$ and N$_{2}$ of the P$_{px}$ spectrum groups
if the angular correlation between the electron and antineutrino
momentums would be absent. Thus, in order to measure the
\textbf{a}--coefficient one has simply to compare the numbers of
events in these groups of the proton TOF spectrum. The measured
asymmetry $x = ( N_{1 }- N_{2 }) / (N_{1 }+ N_{2} )$ is connected
with \textbf{a}--value by a simple formula
\begin{equation}
x = \mathbf{a}\cdot v/c \langle \cos\theta
_{e\nu}\rangle\end{equation}

In this equation v/c must be averaged over the spectrum of
electrons, and the Cosine of the angle between electron and
antineutrino flight directions must be averaged too. To make these
calculations one must know the spectrum of electrons detected.
Therefore the electron detector must be able to measure these
energies. But the requirement for precision of measurements in
this case is very moderate: estimations show that uncertainty in
the knowledge of electron energy on the level of several keV will
cause a methodical error in \textbf{a} essentially lower than the
accuracy planned.

All these features look very promising, but there is one problem
which can spoil this elegant approach. As it is well seen from the
momentum diagram in Fig \ref{figur1}, in order to get the two
proton groups of the TOF spectrum well separated one imperative
condition must be fulfilled: the maximum transverse proton
momentum has to be less than the antineutrino momentums in all
decay events recorded. This means that the spectrum of electrons
detected must have an upper limit. This requirement which looks
from the first sight not very difficult to fulfill appears really
to be rather serious due to the presence of low amplitude ``tail''
in the response curve of all beta-detectors.
\begin{figure}[h!]
\begin{center}
\includegraphics[scale=0.5]{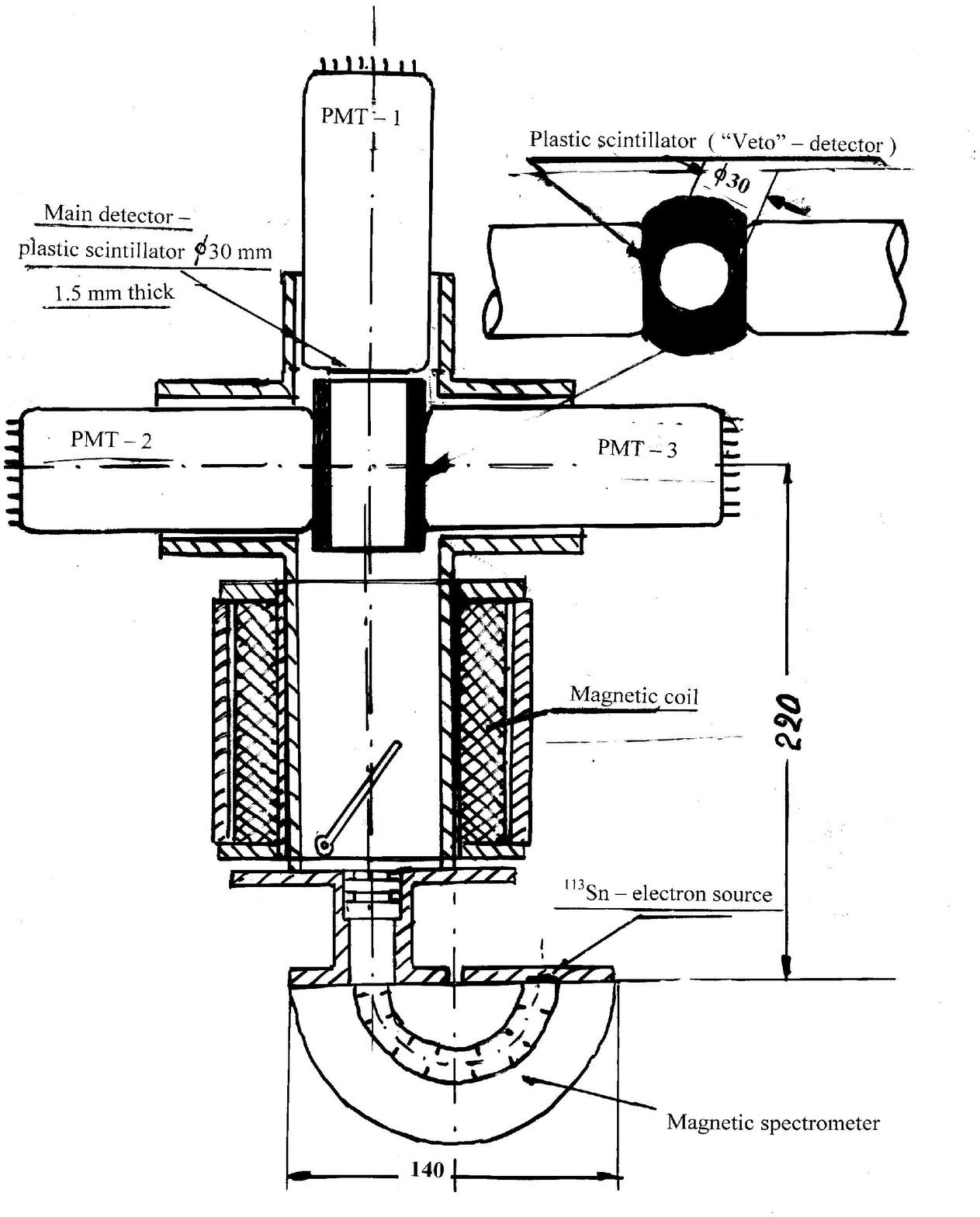}
\end{center}
\caption[]{A schematic sketch of the experimental setup at Harvard}
\label{figur2}
\end{figure}

The main reason of such property of detectors is backscattering of
electrons after hitting the detector - the process which causes
the loss of some part of electron energy, which had to be
transmitted to the detector. As a result, the two proton groups
corresponding to opposite directions of antineutrino are never
separated completely, what causes methodical uncertainty in
measured values of \textbf{a}. Monte Carlo computer simulations
carried out at the Kurchatov Institute in Moscow and at NIST in
USA confirmed these conclusions. This was the reason of special
investigations carried out at Harvard. Their goal was to
investigate this effect and to reduce it as much as possible. In
order to exclude the events when electrons are backscattered from
the detector we decided to install an additional detector disposed
in such a manner that scattered electrons will hit it. Such events
can be rejected by an anticoincidence module.

\begin{figure}[h!]
\begin{center}
\includegraphics[angle=-90,scale=.45]{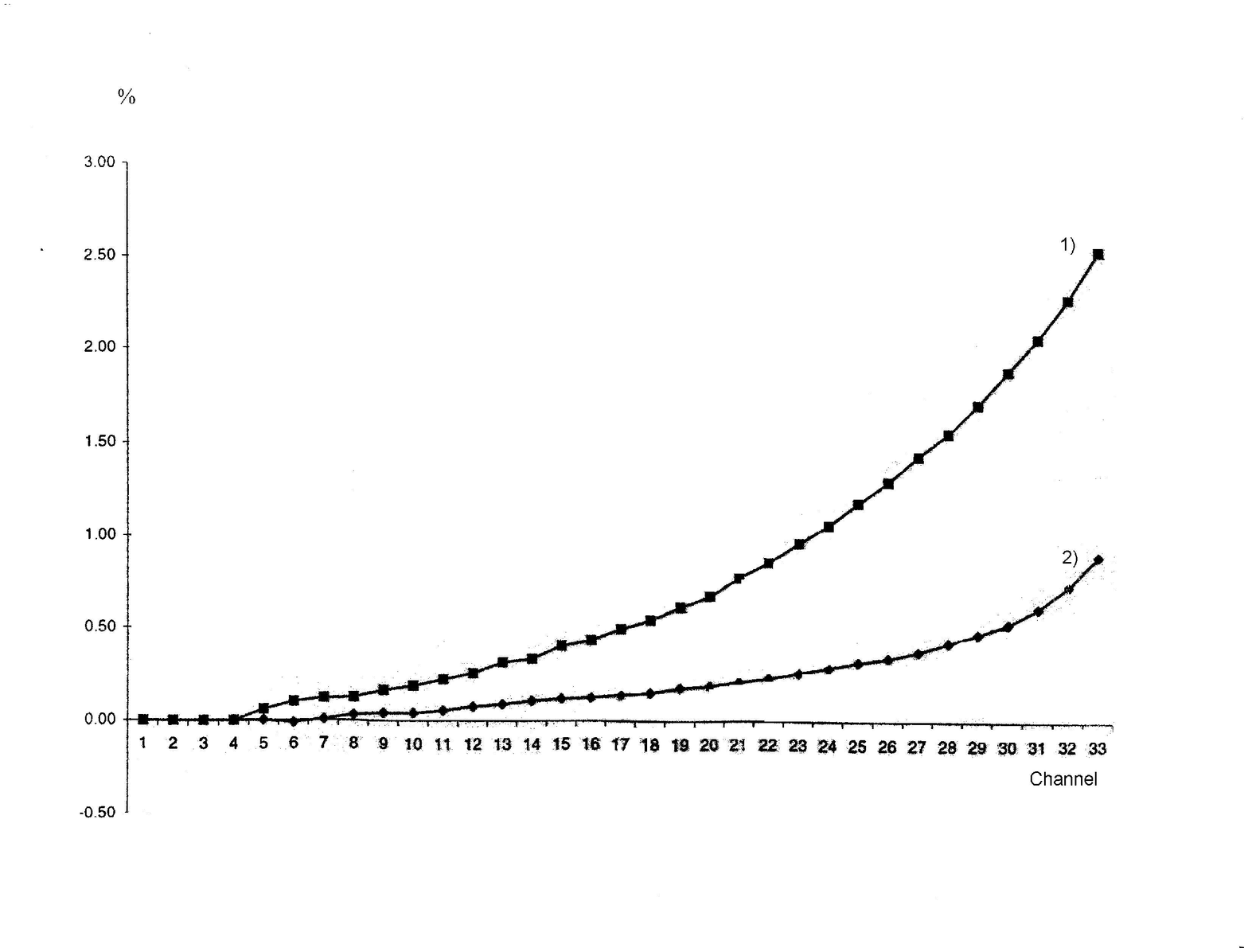}
\end{center}
\vskip -8mm
\caption[]{Experimental integral spectrum of low amplitue "tail" measured with plastic scintillators}
\label{figur3}
\end{figure}

This idea was checked during several years with a special
experimental set-up in our laboratory at Harvard University. A
sketch of one of the last versions of this set-up is shown in Fig
\ref{figur2}. A sample of Sn-113 activated in the MIT reactor was
used as a conversion electron source. The electron beam emitted by
the source was cleaned by a magnetic separator. After cleaning the
spectrum of the electrons reaching the detector consisted of a
single monochromatic line (360 keV). A solenoid with $\sim $ 250
Gs magnetic field was used to focus the electrons on the detector.
The latter was disposed at $\sim $ 30 cm from the exit of the
separator. We took measures to prevent electrons to be scattered
on the way to the main detector.

An additional detector intended to exclude cases of backscattering
(we will call it ``veto-detector'') was installed before the main
detector. It was a plastic scintillation detector viewed by two
photomultipliers. The electrons from the beam came to the main
detector through a cylindrical hole made along the axis. The solid
angle for catching backscattered electrons was $\sim $ 97{\%} of
2$\pi $. As it was mentioned, the signals from this
``veto''--detector activated an anticoincidence circuit, and so,
almost all cases when electrons were backscattered had to be
excluded.

We investigated several types of main detectors: plastic
scintillators, liquid scintillators, Stilben, Si(Li) semiconductor
detectors. Best results were obtained with simple plastic
detectors. The diameter of scintillator was 30 mm, thickness  1
mm, type of the plastic NE102.
\begin{figure}[b]
\begin{center}
\includegraphics[scale=0.5]{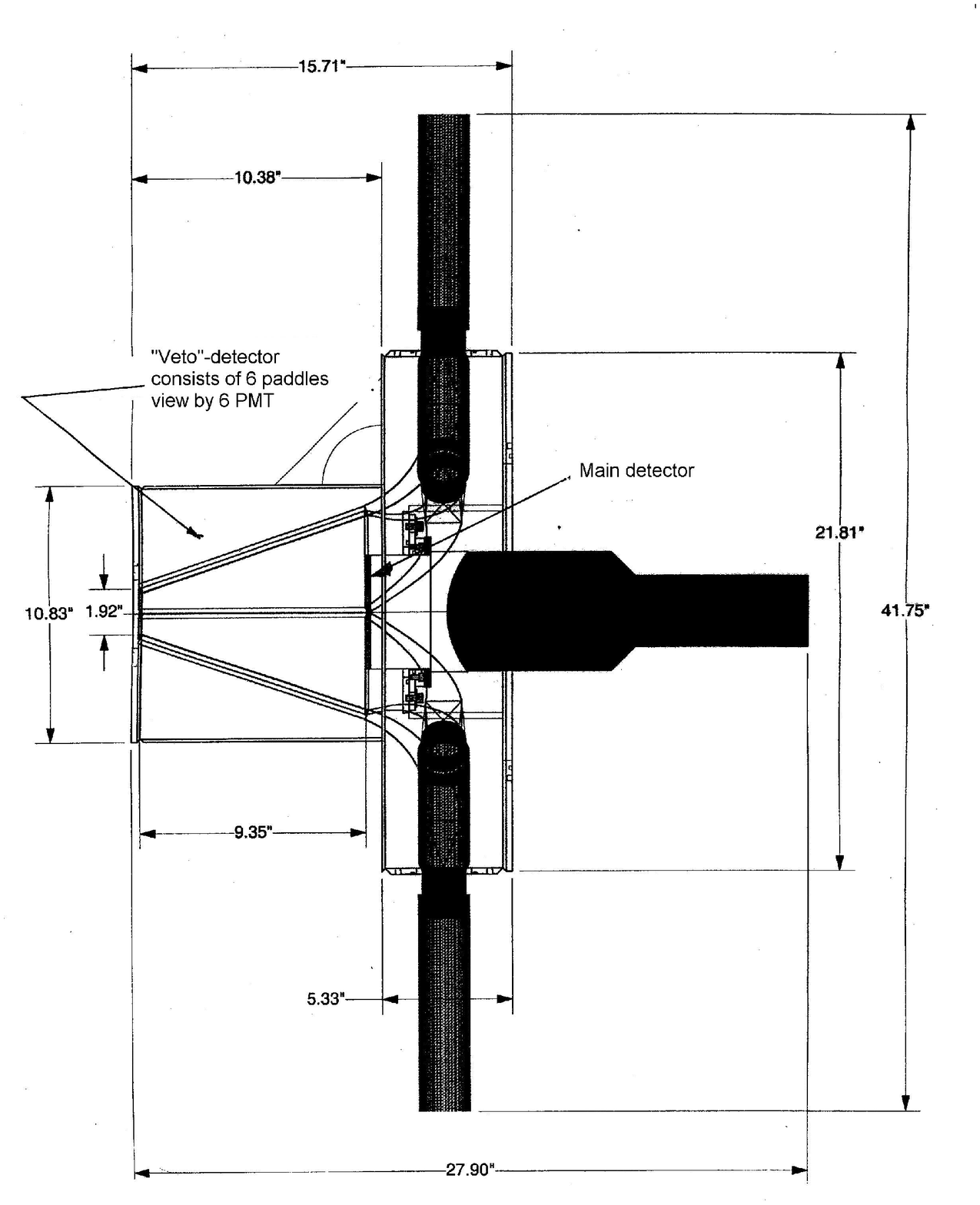}
\end{center}
\caption[]{Full-scale model of beta-detector with"Veto"-system build at NIST}
\label{figur4}
\end{figure}

The amplitude spectra in Fig \ref{figur3} demonstrate the effect
of the ``veto'' system.

Curves (1) and (2) present integral spectrums of ``tails''
$_{o}\smallint ^{n}y(n)dn$, where $y(n)$ are the numbers of counts
in the analyzer channel $n$, in {\%} of ``whole spectra'' without
and with anticoincidences with the ``veto''--detector. A 3 -- 4
fold reduction of the low-amplitude ``tail'' in the spectrum is
evidently seen.

The residual low--amplitude part of spectra is, most probably, not connected
with electron backscattering. Its origin is to be investigated. Preliminary
results of our recent experiments show that some part of this effect is
connected with the bremsstrahlung of electrons hitting the detector.
Calculations carried out at Harvard show that the value of the ``tail'' now
achieved is good enough to go on with the realization of the project.

I will speak now about very important results obtained at NIST. A
detailed computer Monte Carlo calculations were carried out there
in order to optimize all parameters of the experimental apparatus
which is to be built and to estimate all possible sources of
methodical errors in \textbf{a} on the level less than 0.5{\%}.
Estimations of the ``luminosity'' of the optimized measuring
device showed that the statistical possibilities available on the
NIST reactor and moreover at ILL secure the possibility of
obtaining the \textbf{a}--value with 1{\%} accuracy during a
reasonable period of time. A full-scale model of electron detector
with ``veto-detector'' system based on the same idea as in the
experiments at Harvard has been built at NIST too. Fig
\ref{figur4} shows the construction of this model. It was tested
on the Van de Graaf accelerator which was a source of
monochromatic electrons with the energy about 800 keV. The
spectrums obtained confirmed the reduction of the ``low-amplitude
tail'' with the help of ``veto'' system.

A big team of scientists (listed in the preface to this talk) is busy now
with the construction of main parts of experimental set-up needed for
realization of this project, and we hope to have grant application ready to
the end of this year.



\def \Kla#1{\left( #1 \right)}
\def \Klb#1{\left[ #1 \right]}
\def \Klc#1{\left\{ #1 \right\}}
\def \KlB#1{{\left| #1 \right|}}
\def \Kly#1#2#3{{\left<{\rm #1}\KlB{\vphantom{#1}#2\vphantom{#3}}{\rm #3}\right>}}
\def \Klz#1{{\left< #1 \right>}}
\def\x{\hbox{$\,$}}
\def \Unit#1{\x\hbox{\rm #1}}
\def \etal{et al$.$}
\def \aSPECT{{\it a}SPECT}

\title*{The Neutron Decay Spectrometer \aSPECT \protect\newline}
\toctitle{The Neutron Decay Spectrometer \aSPECT}
%
%
\titlerunning{The Neutron Decay Spectrometer \aSPECT}
%
\tocauthor{S. Bae\ss{}ler, S. Bago, J. Byrne, F. Gl\"uck, J. Hartmann, W. Heil, \\ I. Konorov, G. Petzoldt, Y. Sobolev, M. van der Grinten, O. Zimmer\inst{2}}
\author{S. Bae\ss{}ler\inst{1}\and S. Bago\inst{2}\and J. Byrne\inst{3}\and
F. Gl\"uck\inst{1}\and J. Hartmann\inst{2}\and W. Heil\inst{1}\and \\ I. Konorov\inst{2}
\and G. Petzoldt\inst{2}\and Y. Sobolev\inst{1}\and M. van der Grinten\inst{3}\and O. Zimmer\inst{2}}
\authorrunning{S. Bae\ss{}ler \etal}
%
%
\institute{ Institute of Physics, U. Mainz, Germany \and Physics
Department E18, TU M\"unchen, Germany\and University of Sussex,
Falmer, Brighton, UK}

\maketitle              

\begin{abstract}
In this paper we briefly describe the motivation and construction
of our new neutron decay spectrometer \aSPECT. The goal is to
enable us to measure the neutrino-electron-correlation coefficient
$a$ in the decay of the free neutron with unprecedented accuracy.
We summarize the systematic uncertainties of our spectrometer.
\end{abstract}

\section{Introduction}
Measurements of the lifetime $\tau_{\rm n}$ and the beta asymmetry
$A$ of the free neutron, combined with the muon lifetime, enable
us to determine the coupling constants of the weak interaction and
the upper left element of the Cabbibo-Kobayashi-Maskawa-Matrix
(CKM), $V_{\rm ud}$. The other entries of the upper line of the
CKM matrix, $V_{\rm us}$ and $V_{\rm ub}$, are known from high
energy physics. Therefore a test of the unitarity of the CKM
matrix is possible,
\begin{displaymath}
\KlB{V_{\rm ud}}^2 + \KlB{V_{\rm us}}^2 + \KlB{V_{\rm ub}}^2 = 1
\end{displaymath}

Traditionally, this test uses $V_{\rm ud}$ as determined by
nuclear $0^+\to 0^+$-decays. Since recently, the above mentioned
neutron decay data can be used to calculate an independent value
for $V_{\rm ud}$ to a precision that is comparable to that of the
traditional method. In both cases, the sum of the squares of the
matrix elements of the equation above is too small by $2-3\sigma$
\cite{PDG2002baessler,Hard02,Abele02}.

A violation of the unitarity of the CKM matrix would call for the
Standard Model to be extended. Although the theoretical
corrections to the experimental results could be wrong, the
experimental situation also needs to be scrutinised. The most
accurate experiments which measured the beta asymmetry $A$
disagree with each other. Several groups are preparing new
measurements to improve the accuracy on the beta asymmetry
\cite{Abele02,Dubbers02,Young02}.

A different approach is offered by measuring the electron neutrino
correlation coefficient $a$ in neutron decay together with the
neutron lifetime $\tau_{\rm n}$ to determine $V_{\rm ud}$. The
experimental and part of the theo\-re\-ti\-cal systematics are
different from those of a beta asymmetry measurement, so that such
a measurement of $a$ can give independent information.
Unfortunately the present knowledge of $a$ (see
\cite{Strat78,Byrne02}) is too poor to add meaningful information
to the above mentioned problem.

\section{Description of the spectrometer \aSPECT}
The aim of our collaboration is to use the neutron decay
spectrometer \aSPECT{} to improve the precision of the knowledge
of $a$ by more than an order of magnitude. In the standard model a
mesurement of $A$ is equivalent to a measurement of $a$. The
desired precision of \aSPECT{} corresponds to an improvement in
the beta asymmetry $A$ by a factor of about 5. A measurement of
$a$ to that precision can remove the current experimental
ambiguity surrounding this problem.

\begin{figure}[ht]
\begin{center}
\includegraphics{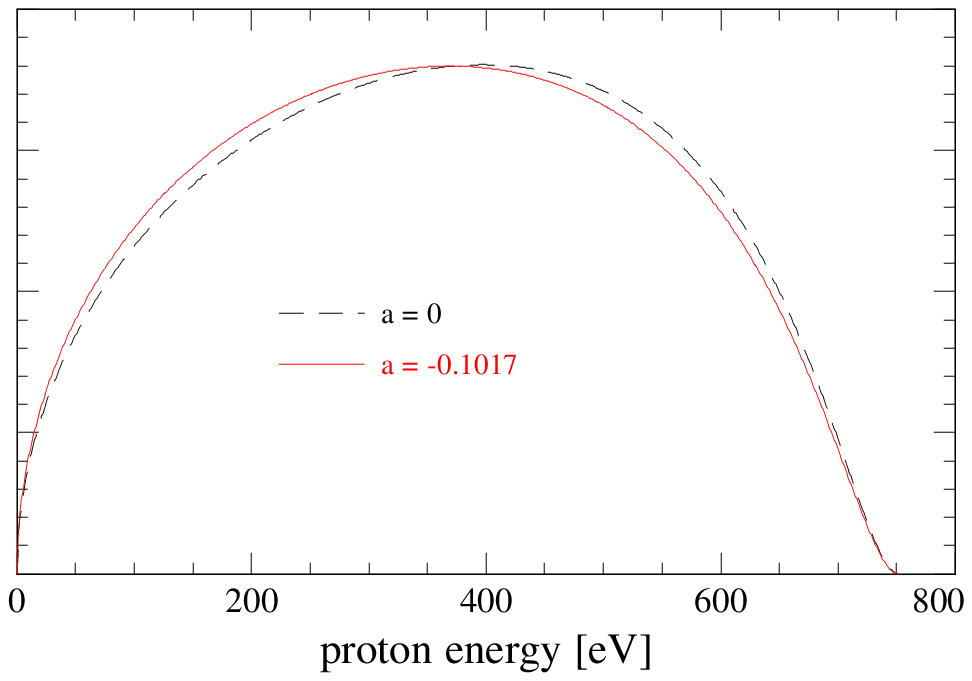}
\caption{Dependence of the Proton Spectrum on the value of the
neutrino-electron-correlation-coefficient $a$. As a comparison a
fictious proton spectrum assuming $a=0$ is given. With \aSPECT{}
we are able to measure the difference between these spectra with
high precision.} \label{ProtonSpek}
\end{center}
\end{figure}

The main idea of \aSPECT{} is that the recoil spectrum of the
proton in neutron decay is sensitive to the
neutrino-electron-correlation coefficient $a$. If the electron and
neutrino emerge with parallel momenta, the proton needs to have a
high momentum in the opposite direction, whereas otherwise the
proton momentum can be small. As a result the value of $a$ shows
up as a distortion of the proton recoil spectrum.

So far spectrometers to study nuclear recoils were magnetic
spectrometers which could be optimized either for high energy and
momentum resolution or for a high acceptance solid angle.
\aSPECT{} is a retardation spectrometer which integrates the
proton spectrum above an electrostatic barrier. Retardation
spectrometers combine a $4\pi$ solid angle with a high energy
resolution and are suitable for precision experiments even if the
source strength is low.

We define the ratio $r_h$ as the count rate of decay protons in
the proton detector while a barrier voltage $U$ of about
$400\Unit{V}$ is applied divided by the total count rate (no
barrier voltage). If $w(E)$ is the proton spectrum as shown in the
picture and $T(E)$ the transmission function of the electrostatic
barrier at the barrier voltage $U$, then $r_h$ is given by
\begin{displaymath}
r_{\rm h}= \frac{\int T(E)w(E)dE}{\int w(E)}\quad.
\end{displaymath}
If all decay protons would be emitted with momenta parallel to the
magnetic field, then $T(E)$ would be a simple step function. The
extracted value of $a$ depends on $r_{\rm h}$ in a way which is
analytically known.

\begin{figure}[ht]
\begin{center}
\includegraphics[width=12cm]{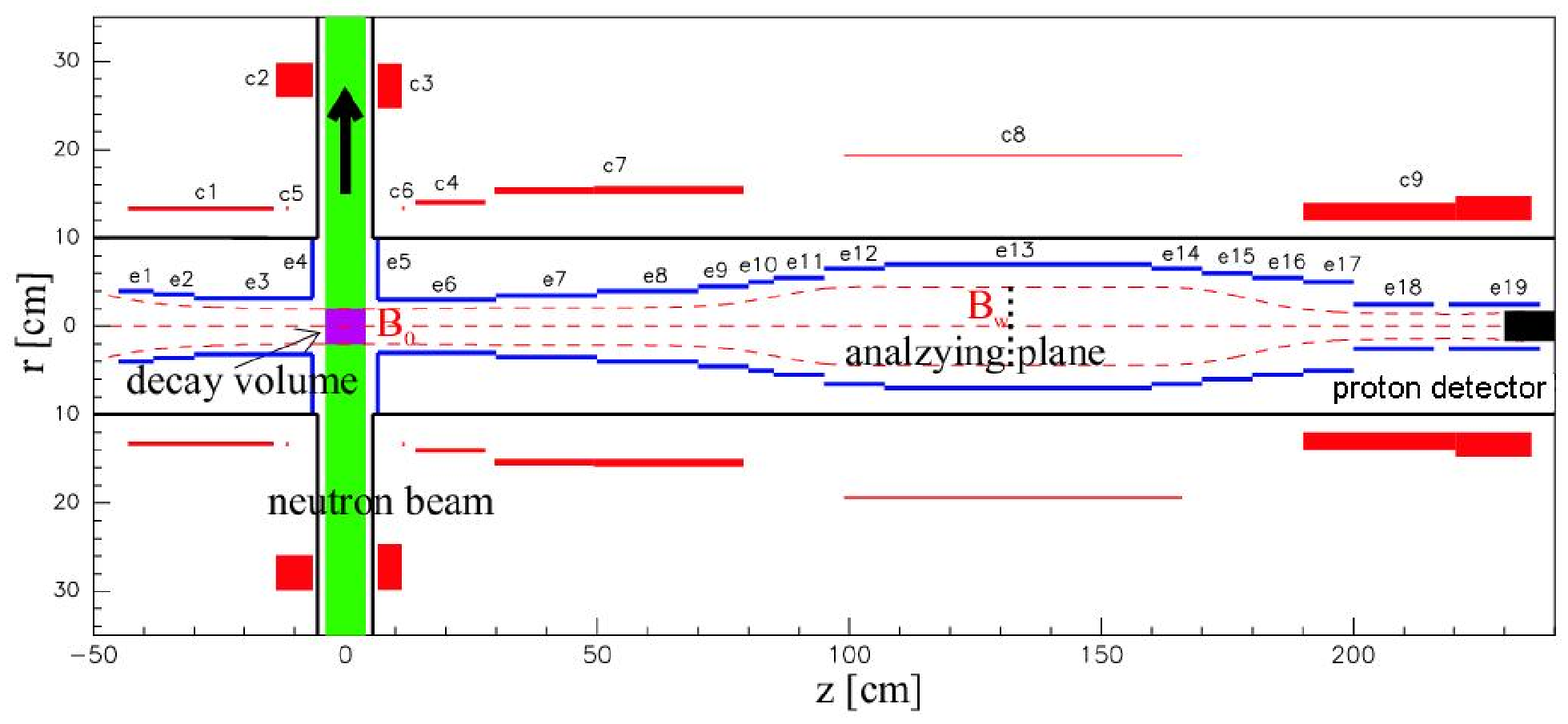}
\end{center}
\caption{Sketch of the {\it a}SPECT spectrometer. The coils are
denoted with c1 to c9, the electrodes are e1 to e19. Coils,
electrodes (except for e7 and e18) and cryostat are cylindrically
symmetric.} \label{Sketch}
\end{figure}

A sketch of the setup is shown in figure \ref{Sketch}. Unpolarized
neutrons are guided through the decay volume of the spectrometer.
About one out of $10^8$ neutrons decays in the decay volume, the
recoil proton is guided by the magnetic field lines. Protons
emitted to the left direction are reflected by an electrostatic
mirror (electrode e1) so that all protons finally are directed
towards the analyzing plane and detector. At the analyzing plane
there will be the above mentioned barrier voltage. Protons
traveling from the high field $B_0 \sim 3\Unit{T}$ to the weak
field $B_{\rm w} \sim 0.5\Unit{T}$ will turn their momenta due to
the inverse magnetic mirror effect to be nearly parallel to the
magnetic field lines. Therefore our transmission function $T(E)$
is close to the ideal step function. The proton detector which
will be a silicon drift detector counts the protons which pass the
barrier. Further details can be found in \cite{Zimmer00}.

\section{Expected systematic uncertainties}
Detailed studies were performed to keep the systematic
uncertainties at a level which allows for a determination of $a$
with an uncertainty of $\Delta a/a \sim 10^{-3}$.

In our analysis we rely on the transmission function to be
precisely known. This requires the following conditions to be met:
\begin{itemize}
\item All protons from neutron decay events in the decay volume which are able to pass the analyzing plane can reach it. If so, we can compute the transmission function under the assumption that the proton emission angular distribution in the decay volume is isotropic.

The protons which are emitted in the wrong hemisphere are
reflected back by the mirror electrode e1.

The magnetic mirror effect is the reflection of protons in an
increasing magnetic field. Since the reflection depends on the
emission angle of the proton, the magnetic mirror effect
introduces a cut in the emission angles. In our setup the magnetic
field is decreasing slightly in the decay volume in the direction
of the analyzing plane so that the magnetic mirror effect is
completely avoided.

A shallow magnetic field maximum at the analyzing plane helps to
ensure that the protons which should pass the analyzing plane are
not reflected too early.
\item The proton movement can be described in the so-called adiabatic approximation. This is true if the electric and magnetic fields are changing sufficiently slow on a proton trajectory. We checked this assumption for our setup in precise trajectory calculations.
\item The electric potential in the analyzing plane is uniform and known accurately, as well as the magnetic fields in analyzing plane and decay volume.
\item The detection efficiency of the proton detector does not vary strongly with impact energy, angle, or position.
\item Protons which cannot pass the analyzing plane are trapped between it and the electric mirror e1. They have to be removed from the trap before they interact with the rest gas. Our solution is that electrode e7 is a dipole electrode producing en electric field across the magnetic field direction. Due to the $E\times B$ drift the protons are removed from the trap after a few tens of oscillations.
\end{itemize}
We can thus determine the transmission function by measuring the
electric potential in the analyzing plane and the magnetic fields
in decay volume and analyzing plane only. The above-mentioned
level of systematic uncertainties can be reached if we know the
electric potential in the analyzing plane $\Delta U$ better than
$10\Unit{mV}$ and the magnetic fields in decay volume and
analyzing plane to a relative accuracy of $\Delta B/B \sim
10^{-4}$. The vacuum has to be as good as $p\sim
10^{-9}\Unit{mbar}$ to avoid that interactions with the rest gas
have a significant impact on the transmission function.

The background count rates can be efficiently determined in
auxilliary measurements with an electric potential in the
analyzing plane which is higher than the maximum proton energy.
The remaining problem is the electron background, caused by
electrons emitted in the same decay events as the protons. The
electron signal might overlap with the proton signal and this
introduces errors in the proton counting. With a second, much
stronger $E\times B$ drift electrode (e18) we can move the protons
far enough sideways that the positions sensitivity of our detector
ensures that electrons and protons coming from the same neutron
decay are counted as two separate events. Then electron events can
be removed from the count rates in the same way as any other
background events.

\title*{Neutron Decay Correlation Measurements \\ in Pulsed Beams \protect\newline}
\toctitle{Neutron Decay Correlation Measurements in Pulsed
Beams}
%
%
\titlerunning{Neutron Decay Correlation Measurements in Pulsed Beams}
%
\author{D. Dubbers}
 \institute{Physikalisches Institut der
Universit\"at Heidelberg, \\ Philosophenweg 12, 69120 Heidelberg,
Germany}
\authorrunning{D. Dubbers}
%
%

\maketitle              

\begin{abstract}
 We describe an instrument "The New Perkeo" which is under development in Heidelberg and which will serve to measure neutron decay correlation coefficients using a pulsed cold neutron beam. The new scheme allows to eliminate the four leading error sources typical for such experiments, while vastly increasing statistical accuracy.
\end{abstract}

\section{The New PERKEO}
In a recent publication \cite{Abele1} we discussed the CKM matrix
element $V_{ud}$ as derived from neutron decay measurements. We
came to the conclusion that unitarity of quark mixing is violated
in the first row of the CKM matrix at the level of about 3
$\sigma$. The main error in this analysis stems from the neutron
beta decay asymmetry $A$. Other quantities which enter the
analysis, like the neutron lifetime $\tau$, the strange and bottom
matrix elements $V_{us}$ and $V_{ub}$, and the radiative
corrections would have to be wrong by about 8 $\sigma$ in order to
explain the observed deviation from the Standard Model prediction
on unitarity.

As this finding is the result of many years of very careful work,
what is needed now is a very significant improvement in the
quality of neutron correlation experiments in order to resolve
this problem in a satisfactory way. In the following we present an
instrument scheme which will strongly improve both the statistical
as well as the systematic errors in neutron decay correlation
experiments.

Our strategy is the following: Today, in neutron decay,
count-rates of up to 300 s$^{-1}$ can be achieved \cite{Abele1}.
However, with the new "ballistic" supermirror cold-neutron guide
H113 which was designed and installed at ILL by our group
\cite{Haese}, the neutron decay rate within the  beam
(cross-section 20$\times$6 cm$^2$) is incredibly high, namely
2$\times$10$^6$ (unpolarized) neutron decays per second and per
meter of beam length. However, only a rate of about 10$^4$
s$^{-1}$ is manageable in the detectors, a rate which would
decrease the now dominant statistical error by one order of
magnitude. Therefore, if all electrons or protons from neutron
decay were collected at H113 over several meters of beam length,
then we would have an intensity which is several hundred times
higher than needed. As is well known, in this type of experiments
one usually can trade the parameter "count-rate" against
improvements in other parameters like background, neutron
polarization, or various "edge" effects. In the following we shall
discuss how we can realize such a trade-off with a new instrument
now under construction, called "The New Perkeo".

The principle of this instrument is rather simple. We apply a
magnetic field of B$_0$ = 1 Tesla along the neutron beam over a
length of 3m. The magnetic field serves as guide field for the
electrons and protons from neutron decay. They leave the solenoid
at both ends, upstream and downstream. There, by additional
magnetic fields applied at right angles to B$_0$, i.e. to the
neutron beam direction, the decay products can be spatially
separated from the neutron beam and guided to appropriate particle
detectors. With a continuous neutron beam passing through this
instrument (with a cross section of 6$\times$6 cm$^2$), the count-rate on
the detectors will be above 10$^6$ s$^{-1}$. Of course, this
count-rate is far too high to be useful, and systematic errors
would be similar as in earlier experiments.

I shall now discuss how to trade most of this count-rate for
better systematics. The most obvious trade-off is to sacrifice
neutron intensity for higher neutron polarization. Recent
investigations at ILL \cite{Soldner} have shown that with
supermirror polarizers one can obtain an average polarization of
99.5\% when sacrificing another factor of two in intensity (i.e.
10\% instead of the usual 20\% transmission).

More such trade-offs can be realized when a pulsed neutron beam is
used. Pulsing can be achieved with two beam choppers at 10m
distance to each other, with a duty cycle of 65\% and 25\%,
respectively, and a repetition period of 6ms. This set-up produces
short neutron pulses of 1.0$\times$10$^8$ neutrons in each pulse,
and leads to an (unpolarized) peak neutron decay rate of
1.1$\times$10$^5$ s$^{-1}$. This number includes losses due to
further beam-tailoring and corrections for chopper opening time

The detectors are gated "on" only while the neutron pulse is fully
contained within the fiducial volume of the spectrometer. For a 2m
long such volume, the duty cycle of the detectors is 13\%. The
time average neutron decay rate then is 1.4$\times$10$^4$
s$^{-1}$. Working with the highly polarized neutron beam mentioned
above would reduce these numbers by ten.

This beam-chopping scheme will eliminate several further error
sources. The other leading error in neutron correlation
measurements (besides neutron polarization) is undetected
background. In the previous Perkeo instrument, neutrons fly
freely, in a high vacuum and without touching any material
devices, from the last beam-tailoring orifice, situated 1m
upstream of the entrance of the instrument, all the way down to
the beam stop, situated 4m downstream of the exit of the
instrument. Background (of order 2\%) is determined off-line by
shutting the beam with a $^6$LiF plate after the last orifice. The
error induced by this method is twofold: firstly, this neutron
beam shutter itself produces a small background (of order
10$^{-4}$ of the incoming beam), of which a tiny amount can reach
the detectors and leads to an over- correction of background;
secondly, when the neutron beam is on, the beam stop produces a
similar amount of uncorrected beam-related background. In previous
experiments, these backgrounds could be estimated and required a
correction of (0.50 $\pm$ 0.25)\% on $A$.

In the new scheme, these background errors are completely
eliminated: with a pulsed beam, neutron decay is being measured
while the beam is closed by the chopper, that is under exactly the
same condition as when background is measured off-line. Secondly,
the beam stop can produce no additional background because at the
time when the pulsed beam reaches the beam stop the neutron
detectors will be gated off.

The next-to-leading error source in our previous experiments is
due to magnetic mirror effects in the inhomogeneous B0-field.
Electrons or protons spiralling under near 90° in a magnetic field
of increasing magnitude will be repelled by the field, and will be
counted in the wrong detector. In the new scheme, this effect can
be suppressed at will, as in the evaluation of the experiment the
length of the active volume, i.e. the region of high uniformity of
the B$_0$-field, can be chosen even after the end of the
experiment.

Another unavoidable error source in continuous beam experiments is
due to edge effects on the detector. Electrons or protons guided
to the detector by the magnetic field may hit the detector near
its edge, and be detected or not, depending on their precise
trajectory. Also this effect is completely eliminated in the new
scheme, as the neutron cloud is, by the magnetic field, projected
onto the active area of the detector without coming close to its
edge.

So, in "The New Perkeo", the four leading error sources of
previous experiments will be strongly suppressed or completely
eliminated. The remaining experimental errors in $A$ then are the
errors due to detector response, on which at present our group is
concentrating its efforts. Our aim then is to reduce the error in
the CKM-matrix element $V_{ud}$ which is due to the beta decay
asymmetry $A$ below the errors of all the other quantities which
enter the unitarity check, as mentioned at the beginning of this
article.

Due to the high count-rate of the new set-up, only single-rate
experiments can be done, but no coincidence measurements. Still,
with single-rate experiments one can measure all three allowed
correlation coefficients in neutron decay with high precision. In
Perkeo, the beta asymmetry $A$ was measured with single electrons
and gives information not only on the unitarity of the quark
mixing matrix, but also on possible right-handed currents (with
main sensitivity to their relative phase), and possibly also on
weak magnetism. The single-proton asymmetry from polarized neutron
decay gives the antineutrino asymmetry $B$, which is sensitive to
the mass of a right-handed W-boson. The electron-neutrino
correlation coefficient a is measured from the single-proton
intensity spectrum (with unpolarized neutrons), and is as
sensitive to the unitarity of quark mixing as is the beta decay
asymmetry. In the two latter cases the proton energy spectrum will
be measured by time-of-flight, by using an electrostatic chopper.

With the high count-rates available with this instrument also
other observables could become measurable which are not accessible
with present day's low count-rate instruments. One could do
magnetic spectrometry of the decay electrons to gain additional
information on radiative corrections or on the Fierz interference
term b, do Mott-scattering of the electrons to measure their
helicity spectrum, in order to gain information on the neutrino
helicity and on right-handed currents, or even measure the very
small helicity of the decay protons.

\title*{TRI$\mu$P -- a New Facility for Trapping Radioactive Isotopes
}

\toctitle{TRI$\mu$P -- a New Facility for Trapping Radioactive
Isotopes }
%
%
\titlerunning{TRI$\mu$P -- a New Facility for Trapping Radioactive Isotopes}
%
\author{K. Jungmann
\thanks{representing work of the TRI$\mu$P group \cite{trimp_group}
at KVI}}
\authorrunning{K. Jungmann}
%
%
\institute{Kernfysisch Versneller Instituut, Rijksuniversiteit
Groningen,
           The Netherlands}

\maketitle              

\begin{abstract}
At the Kernfysisch Versneller Instituut (KVI) in Groningen, NL,
a new facility (TRI$\mu$P) is under development. It aims for
producing, slowing down and trapping of
radioactive isotopes in order to  perform accurate measurements
on fundamental symmetries and interactions.
A spectrum of radioactive nuclids
will be produced in direct, inverse kinematics of fragmentation
reactions using heavy ion beams from the superconducting
AGOR cyclotron.
The research programme pursued by the local KVI group includes
precision studies of nuclear
$\beta$-decays through $\beta$--neutrino (recoil nucleus) momentum
correlations in weak decays and searches for permanent electric dipole
moments in heavy atomic systems.
The facility in Groningen will be open for use by
the worldwide community of scientists.
\end{abstract}

The new facility TRI$\mu$P ({\bf T}rapped {\bf R}adioactive {\bf I}sotopes:
{\large \bf $\mu$}icro-laboratories for fundamental {\bf P}hysics) at the
Kernfysisch Versneller Instituut (KVI) in Groningen,  The Netherlands,
has been designed with the central goal to provide cold radioactive
atoms and ions for precise measurements of fundamental symmetries
and fundamental interactions in physics
\cite{Jungmann_02}.
There is a common believe that the present Standard Model
is embedded in a larger theoretical framework.
The TRI$\mu$P facility will allow to contribute to the searches for such
a new theory by precision tests of the standard theory \cite{Langacker_95}.
The concept of TRI$\mu$P \cite{Wilschut_99}
exploits several major recent developments in different fields
of physics, in particular the production of intense radioactive beams and
very advanced techniques for cooling and trapping
of charged and neutral atomic systems in Penning, Paul or laser traps.

The TRI$\mu$P facility design is based on three main radioactive isotope production
mechanisms, i.e. direct, inverse kinematics fusion
and evaporation and fragmentation reactions. The best method, beam and target material,
will be chosen for each experiment depending on the anticipated production yields.
Heavy ion beams from
the superconducting AGOR cyclotron, which delivers beams up to the
100 MeV/nucleon region, are directed onto fixed targets. The foreseen
reactions favor in general proton rich nuclei.

The created isotopes of interest are separated from
the primary beam and other reaction products
in a novel designed combined fragment and (gas filled) recoil separator.
The ion optical  system consists of two pairs of dipole magnets
for the primary particle selection and quadrupoles
for accurate imaging.
It can be tuned for  mass and momentum selection.
Two possible target positions are provided within the arrangement of magnets:
one at its very entrance
for fragmentation reactions and another one between the two dipole pairs
for inverse kinematics reactions.
Gas filling of the separator is essential
for good imaging if the reaction products appear in a distribution
of electric charge states.
With appropriate gas densities concurrent electron capture and stripping
processes result in an effective charge for the ions,
which determines their trajectory in the ion optical system \cite{Leino_97}.

At the exit of the separator the product particles have typically
1 MeV/c momentum. For later trapping it is important that they be
slowed down. Moderation to eV energies is foreseen in a high
pressure gas cell. Presently detailed measurements of relevant
atomic physics processes during slow down in gases are under way
at KVI to determine the design parameters of this essential
device, where an as high as possible throughput needs to be
achieved. Despite high activity in the field of radioactive beam
facilities with similar goals there is at present little
information available on the behavior of the particles in a very
interesting energy range where their velocities are of order
$\alpha \cdot c$. This is the order of valence electron velocities
of the slower gas and where charge exchange and, in particular,
neutralization cross sections are high. Alternate slowing down
scenarios which are less generic are also explored.

\begin{figure*}[t]
 \unitlength 1.0cm
   \centering{
\includegraphics[width=12.2cm,angle=0,clip=false]{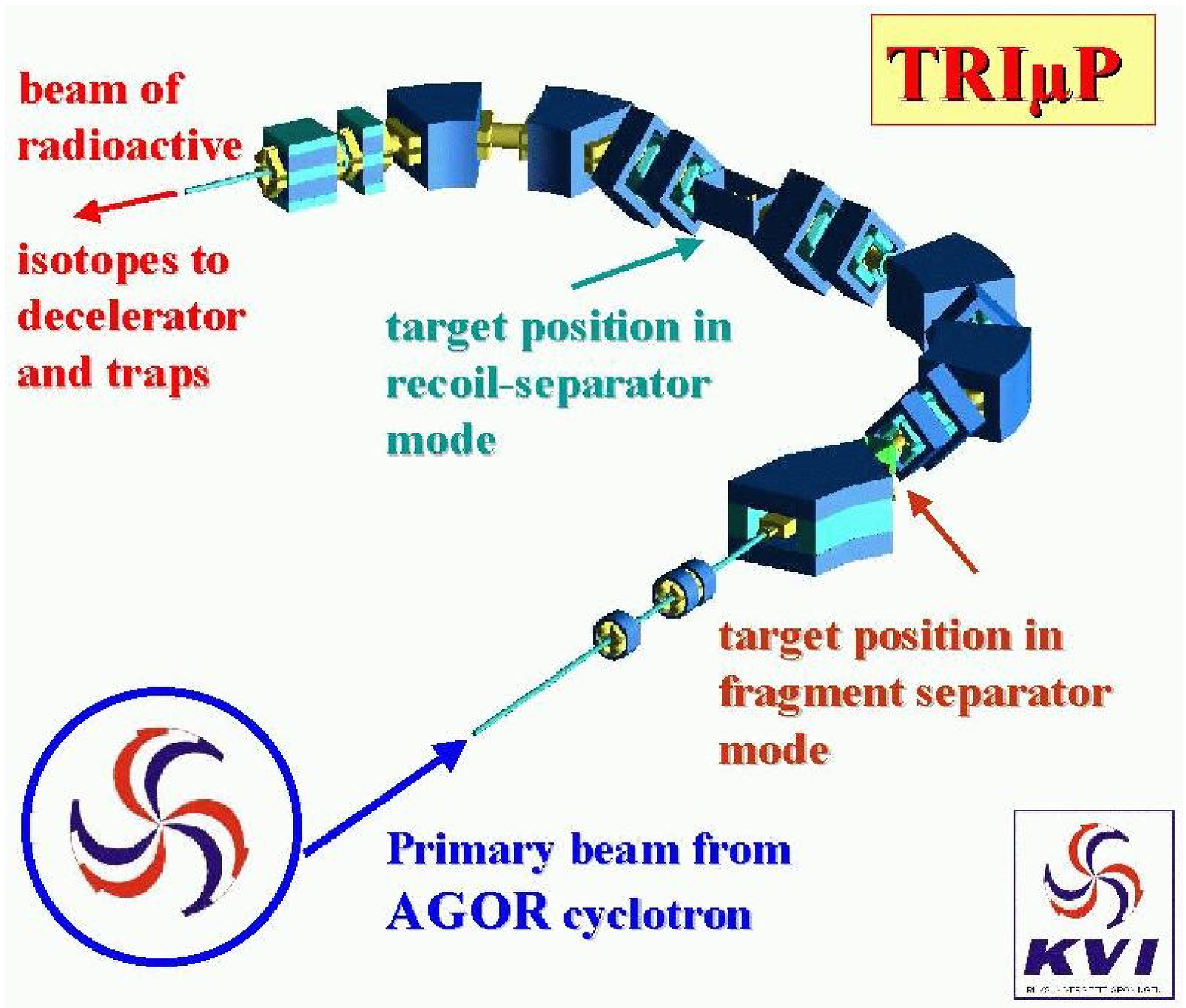}             }
\vskip -5mm
  \caption[]{The TRI$\mu$P combined fragment and recoil separator.
It is designed to access a large variety of proton-rich isotopes.}
\end{figure*}

Further cooling of the ions, which are expected to be singly charged upon exiting
the gas cell slower, occurs in a low density buffer gas.
Here the ions are longitudinally guided by an electrostatic potential gradient while
at the same time radially confined by a radio frequency quadrupole field.
They are collected in  a Paul Trap the end voltage of which can be rapidly
switched in order to allow
this device to act as a beam buncher.
After neutralization the atoms can be stored in atoms traps, e.g. a
magneto-optical trap.

Research using trapped radioactive atoms and ions is pursued
in many laboratories worldwide.
It covers a wide range of physics topics in atomic, nuclear and particle physics
\cite{Schweikhard_00}.
The local researchers at KVI concentrate therefore,
on two groups of experiments, which are optimal suited for the facility:
\begin{itemize}
\item
precision measurements of nuclear $\beta$-decays, and
\item
searches for permanent
electric dipole moments in atoms.
\end{itemize}
In standard theory the structure of weak interactions is of the
V-A type, where V and A are  vector and axial-vector currents with
opposite relative sign causing a left handed structure of the
interaction and parity violation \cite{Herczeg_01}. Other
possibilities which could explain four fermion interactions like
$\beta$-decays which are of scalar, pseudo-scalar and tensor type
would be clear  signatures of new physics. In particular,
right-handed currents would also give rise to deviations from
standard theory predictions. An observation of a particle behavior
in $\beta$-decay which is not V-A type can be expected to shine
some light onto the mysteries behind parity violation.

The double differential decay probability
$ d^2W/d\Omega_e d\Omega_{\nu}$for a $\beta$-radioactive nucleus is
related to the electron and neutrino momenta $\vec{p}$ and $\vec{q}$ through
{\small
\begin{eqnarray}
\label{diffprob}
&\frac{d^2W}{d\Omega_e d\Omega_{\nu}}  \sim  1 +  a ~\frac{\vec{p}\cdot\vec{q}}{E}
+  b ~~\sqrt{1-(Z \alpha)^2}~~\frac{m_e}{E}       \nonumber     \\
& + <\vec{J}>     \cdot \left[ A~~ \frac{\vec{p}}{E} + B~~\vec{q} + D~~\frac{\vec{p} \times ~\vec{q}}{E} \right]\\
& + <\vec{\sigma}> \cdot \left[ G~~ \frac{\vec{p}}{E} + Q~~\vec{J} + R~~ <\vec{J}> \times ~\frac{\vec{q}}{E} \right] \nonumber
\end{eqnarray}
}
where   $m_e$ is the $\beta$-particle mass,
        $E$ its energy,
        $\vec{\sigma}$ its spin,  and
        $\vec{J}$ is the spin of the decaying nucleus.
Among the coefficients in equation (\ref{diffprob}) D describes the correlation between
the neutrino and $\beta$-particle momentum vectors for spin polarized nuclei
and is time reversal violating in nature.
It renders a particularly high potential for further restricting parameters
in speculative models.
The coefficient R relates is highly sensitive within a smaller set of
models, since in this region there exist some already well established
constraints, e.g., from searches for a permanent electric dipole moment (edm)
of fundamental particles.
In such experiments,
the neutrino momentum cannot be determined directly in a meaningful way. Therefore
the recoiling nucleus will be detected instead. The neutrino
momentum can then be reconstructed from the  kinematics of the process.
Since the recoil nuclei have energies typically a few 10 eV,
precise measurements can only be performed, if the decaying isotopes are
suspended using sufficiently shallow potential wells. Such exist
in magneto-optical traps, where neutral atoms are stored at
temperatures below 1 mK. Since the atoms must have suitable spectral lines for
optical trapping, since also the nuclear properties must be such that
rather clean transitions can be observed, and since a precision determination
of parameters requires high statistics, i.e. high decay rates,
the isotopes of primary interest for
the KVI experimenters are $^{20,21}$Na and $^{18,19}$Ne.

An edm of any fundamental particle
violates both parity and time reversal symmetries \cite{Khrip_97}.
With the assumption of a generally valid
CPT invariance a permanent dipole moment also violates CP.
CP violation, such as it is known from the K and B systems, induces
edm's for all particles through higher order loops.
Their values are for any known system at least 4 orders of magnitude below the
present experimentally established limits. It should be noted
that the known sources of CP violation are not sufficient in
Sakharov's model for the baryon asymmetry, i.e. the dominance of matter
over anti-matter in the universe \cite{Trodden_99}. New sources of CP violation
need to be discovered. Indeed, a large number of speculative models
foresees edm's which could be as large as
the present experimental bounds.
Historically the non-observation of any edm has
ruled out more speculative models than any other experimental
approach in all of particle physics \cite{Ramsey_99}.

Permanent electric dipole moments have been
searched for in various systems with different sensitivities
\cite{Commins_94,Bailey_78,Aciari_98,Cho_89,Harris_99,Romalis_01}.
Since the searched for, yet unknown, new fundamental forces may act completely different
on various particles and systems, a comparison of experiments
solely on the basis of the smallest limit in units of $e\cdot cm$
is not possible. Careful and detailed analysis is indispensable.
In composed systems such as molecules
or atoms fundamental particle dipole moments can be enhanced
significantly \cite{Sandars_01}. Particularly in polarizable
systems where large internal fields can exist.
Ra atoms in excited states are of particular interest
for edm searches,
because of the rather close lying $7s7p^3P_1$ and $7s6d^3_2$ states
the a significant enhancement has been predicted  \cite{Dzuba_01}.
This gives a several orders of magnitude
advantage over the ${^{199}}$Hg atom, the system
which has given the best limits so far \cite{Romalis_01}.
Further enhancements have been predicted in Ra isotopes with
strongly octupole deformed nuclei.
Speculative models would allow to observe an atomic edm in Ra already
at an absolute value some three orders of magnitude below the
present limit set by Hg!
{
\begin{table}[bht]
\caption{Limits on Permanent Electric Dipole Moments $d$ for
         electrons ($e$) \cite{Commins_94},
         muons ($\mu$)   \cite{Bailey_78},
         tauons ($\tau$) \cite{Aciari_98},
         protons ($p$)   \cite{Cho_89},
         neutrons ($n$)  \cite{Harris_99}, and the
         mercury atom ($^199$Hg)    \cite{Romalis_01}.
         The new Physics Limits correspond to the predicted highest values
         among various models beyond standard theory.
}

\centering
\begin{tabular}{|c|rl|c|c|} \hline
& Present Limit & on              & Standard Model   & New Physics           \\
&  $|d|$        &                 & Prediction       & Limits               \\
& [10$^{-27}$&e\,cm]&  [10$^{-27}$e\,cm] & [10$^{-27}$e\,cm]                \\ \hline\hline
$e$        & $<1.6$          &{(90\% C.L.)}           & $\stackrel{ <}{\sim}10^{-11}$
                                                  & $\stackrel{ <}{\sim}1$   \\ \hline
$\mu$    & $<1.05\cdot10^9$  &{(95\% C.L.)}   & $\stackrel{ <}{\sim}10^{-8}$
                                                  & $\stackrel{ <}{\sim}200$ \\ \hline
$\tau$   & $<3.1\cdot10^{11}$&{(95\% C.L.)} & $\stackrel{ <}{\sim}10^{-7}$
& $\stackrel{ <}{\sim}1700$                                                  \\ \hline
$p$        & $-3.7\,(6.3)\cdot10^4$&              & $\sim10^{-4}$ &
$\stackrel{ <}{\sim}60$                                                      \\ \hline
$n$        &$<63$             &{(90\% C.L.)}              & $\sim10^{-4}$  &
$\stackrel{<}{\sim}60$                                                       \\ \hline
$^{199}$Hg&
$<0.21$                       &{(90\% C.L.)}              & $\sim10^{-6}$  &
$\stackrel{ <}{\sim}0.2$                                                     \\ \hline
\end{tabular}
\label{edm_limits}
\end{table}
}
From a technological point of view Ra atoms well accessible
spectroscopically and a variety of isotopes can be produced in
fusion and evaporation or in fission reactions. The advantage of
an accelerator based Ra experiment is apparent, because a nuclear
edm requires an isotope with spin. and all Ra isotopes with finite
nuclear spin are relatively short-lived.

TRI$\mu$P at KVI is expected to offer
new possibilities to study with high precision
fundamental interactions in physics and fundamental symmetries in
nature. The approach combines nuclear physics, atomic physics and
particle physics in experimental techniques as well as in
the conceptual approaches. The scientific approach chosen in TRI$\mu$P
can be regarded as complementary to such high energy physics.

For TRI$\mu$P there exists also a large variety of
possibilities for research with cold radioactive isotopes in  connection with applied
sciences. For example,
cold polarized $\beta$-emitters could be the basis for extending the method of
$\beta$-NMR, which is very successful in bulk material \cite{Fick_00},
to condensed matter surfaces, on which such atoms could be  softly deposited.

At KVI a user facility is created which is open to the worldwide
scientific community. TRI$\mu$P is jointly funded by FOM
(Stichting voor Fundamenteel Onderzoek der Materie, Dutch funding
agency) and the Rijksuniversiteit Groningen in the framework of a
managed programme. TRI$\mu$P is expected to receive a 50\% share
of the AGOR beam time. The time planning foresees that the
facility is set up by 2004 followed by an exploitation phase until
2013. First physics experiments are expected in 2004. The facility
is open for outside users worldwide and proposals are highly
welcome.

\title*{Progress Towards Measurement of the Neutron Lifetime Using
    Magnetically Trapped Ultracold Neutrons}

\toctitle{Progress Towards Measurement of the Neutron Lifetime Using
    Magnetically Trapped Ultracold Neutrons}

\titlerunning{Neutron Lifetime Using Magnetically Trapped Neutrons}

\author{P. R. Huffman\inst{{1}\, {2}}\and  K. J. Coakley\inst{3}\and
    S. N. Dzhosyuk\inst{1}\and  R. Golub\inst{4}\and  E. Korobkina\inst{4}\and
    S. K. Lamoreaux\inst{5}\and  C. E. H. Mattoni\inst{1}\and
    D. N. McKinsey\inst{1}\and  A. K. Thompson\inst{2}\and \\ G. L. Yang\inst{2}\and
    L. Yang\inst{1}\and J. M. Doyle\inst{1}}

\authorrunning{P. R. Huffman et al.}

\institute{Harvard University, Cambridge, MA 02138, USA \and
National Institute of Standards and Technology, Gaithersburg, MD
20899, USA \and National Institute of Standards and Technology,
Boulder, CO 80303, USA \and Hahn-Meitner-Institut, Berlin, Germany
\and Los Alamos National Laboratory, Los Alamos, NM 87545, USA}

\maketitle

\begin{abstract}
We report progress towards a measurement of the neutron lifetime using
magnetically trapped ultracold neutrons (UCN).  UCN are
produced by inelastic scattering of cold (0.89~nm) neutrons in a
reservoir of superfluid $^{4}$He and confined in a three-dimensional
magnetic trap.  As the trapped neutrons decay, recoil electrons
generate scintillations in the liquid He, which are detectable
with greater than 90~\% efficiency.  The number of
UCN decays \emph{vs.} time will be used to determine the neutron
beta-decay lifetime.
\end{abstract}

\section{Introduction}

We present a progress report on a measurement of the neutron
lifetime using three-dimen-sional magnetic confinement of neutrons
\cite{Doy94}. For detailed information on the experiment, the
reader is directed to the graduate theses of Carlo Mattoni
\cite{Mat02} and Daniel McKinsey \cite{McK02} and references
\cite{Doy94} and \cite{Huf00}.  This paper summarizes the
improvements made to the experiment since the demonstration of
three-dimensional magnetic confinement in 1999 \cite{Huf00} and
shows some preliminary diagnostic data taken with the new setup.

In the 1999 proof-of-principle experiment, approximately 500 UCN were
trapped per loading cycle and their decay was observed with a 31~\%
detection efficiency.  The neutron lifetime estimated from two months
running time was $660^{+290}_{-170}$~s \cite{Bro00}, which is
consistent with the accepted value of the neutron beta-decay lifetime
of 886~s.  The error in estimating the neutron lifetime was primarily
due to the fact that the signal to background ratio was approximately
1 to 20.

Various improvements in the apparatus and experimental techniques were
made in order to increase the signal to background ratio, increase the
signal, and improve the statistical sensitivity to the neutron
lifetime.  A large number of changes were made to the apparatus and a
careful assessment of the neutron activation and neutron-induced
luminescence properties of all materials exposed to the neutron beam
was performed.  In our work thus far, the entire cold neutron beam,
only a small fraction of which is in the wavelength region that
contributes to single phonon UCN production, was introduced into the
apparatus.  In an effort to increase signal to background ratio, a
0.89~nm monochromator was developed.  Changes made to the cryostat and
its contents include a larger magnetic trap and trapping region and a
new superfluid helium-filled heat link connecting the cell to the
dilution refrigerator.  The detection system was also substantially
overhauled.  Changes include larger light collection optics and
detectors, utilizing an optically clear neutron beamstop, and the
switch from embedding the wavelength shifter in a clear plastic matrix
optically coupled to a plastic tube to evaporating the wavelength
shifter onto a diffuse reflector.  A new data acquisition system was
installed.  The results from running the trapping experiment for about
one week are presented.

\section{Materials Activation and Luminescence}

We have studied the low temperature ($\sim 4.2$~K) activation and
luminescence properties of a variety of materials present in our
trapping apparatus \cite{Mat02}.  None of the weak neutron absorbers
studied (acrylic, graphite, wavelength shifter, or diffuse reflector)
displayed any detectable neutron-induced luminescence.

Strong luminescence, of the order of $10^{5}$~s$^{-1}$ initially after
capture of $\sim10^{11}$ cold neutrons, was observed for the neutron
absorbers boron nitride and lithium fluoride.  By comparing the rates
of luminescence on two photomultiplier tubes to the rate of their
coincidence, a limit on fraction of luminescence events producing more
than one correlated photon was set as $<10^{-3}$.  Although BN is an
appealing shielding material due to its availability in high purity
form and its easy machinability, its strong luminescence is a concern.
An alternative shielding material, boron carbide, did not display any
measurable luminescence.  Since B$_{4}$C is extremely difficult to
machine, we continued to use BN as a neutron absorber but shielded the
detectors from BN's luminescence light using a thin ($\sim0.5$~mm)
layer of graphite as a light absorber.

Of the optically transparent neutron absorbers tested -- LiF,
B$_{2}$O$_{3}$, and a boron/lithium glass -- strong luminescence was
observed only in LiF. The B$_{2}$O$_{3}$ displayed a time varying
signal consistent either with luminescence with $\sim 10^{-3}$
intensity relative to that of LiF or with activation.  The
boron/lithium glass did not display any measurable luminescence.  All
three of these materials are subject to some activation concerns.  The
fluorine in LiF becomes activated by the $^{19}$F(n,$\gamma$)$^{20}$F
reaction.  Since $^{20}$F decays with a lifetime of 16~s, which is
much shorter than the neutron beta-decay lifetime of 886~s, it may be
sufficient to simply wait for any activated fluorine to decay.  The
production of $^{18}$F from the $^{16}$O(T,n)$^{18}$F reaction (the
tritons are produced in neutron capture on lithium) in the
boron/lithium glass and $^{13}$N from the $^{10}$B($\alpha$,n)$^{13}$N
reaction (the alphas are produced in neutron capture on boron) in
B$_{2}$O$_{3}$ may be much more problematic due to their longer
lifetimes.  Due to the strong $^{18}$F decay signal observed in the
boron/lithium glass, and the intense luminescence observed in LiF,
B$_{2}$O$_{3}$ was chosen as the beamstop material for the neutron
trapping measurements.  Initial estimates indicated that the signal
from the B$_{2}$O$_{3}$ was consistent with the activation of
$^{13}$N. Separate high-flux measurements of this reaction indicate
that $^{13}$N production is roughly three orders of magnitude weaker.

\section{Monochromator}
The production of UCN by single phonon downscattering of cold neutrons
from superfluid helium (the ``superthermal process'' \cite{Gol91})
requires input neutrons only in a narrow wavelength band around
0.89~nm.  Only a small fraction ($< 1$~\%) of the spectrum of the cold
neutron beam at the NG-6 guide at the NIST Center for Neutron
Research, where the experiment is located, lies within the relevant
wavelength band for production.  Thus, for the neutron lifetime
measurement, all other cold neutrons contribute mainly to backgrounds.
The signal to background ratio of the experiment can be substantially
improved by filtering the cold neutron spectrum such that the only
neutrons entering the trapping apparatus are those that can produce
UCN by single phonon downscattering.

\begin{figure}[t]
    \centering
    \includegraphics[scale=0.8]{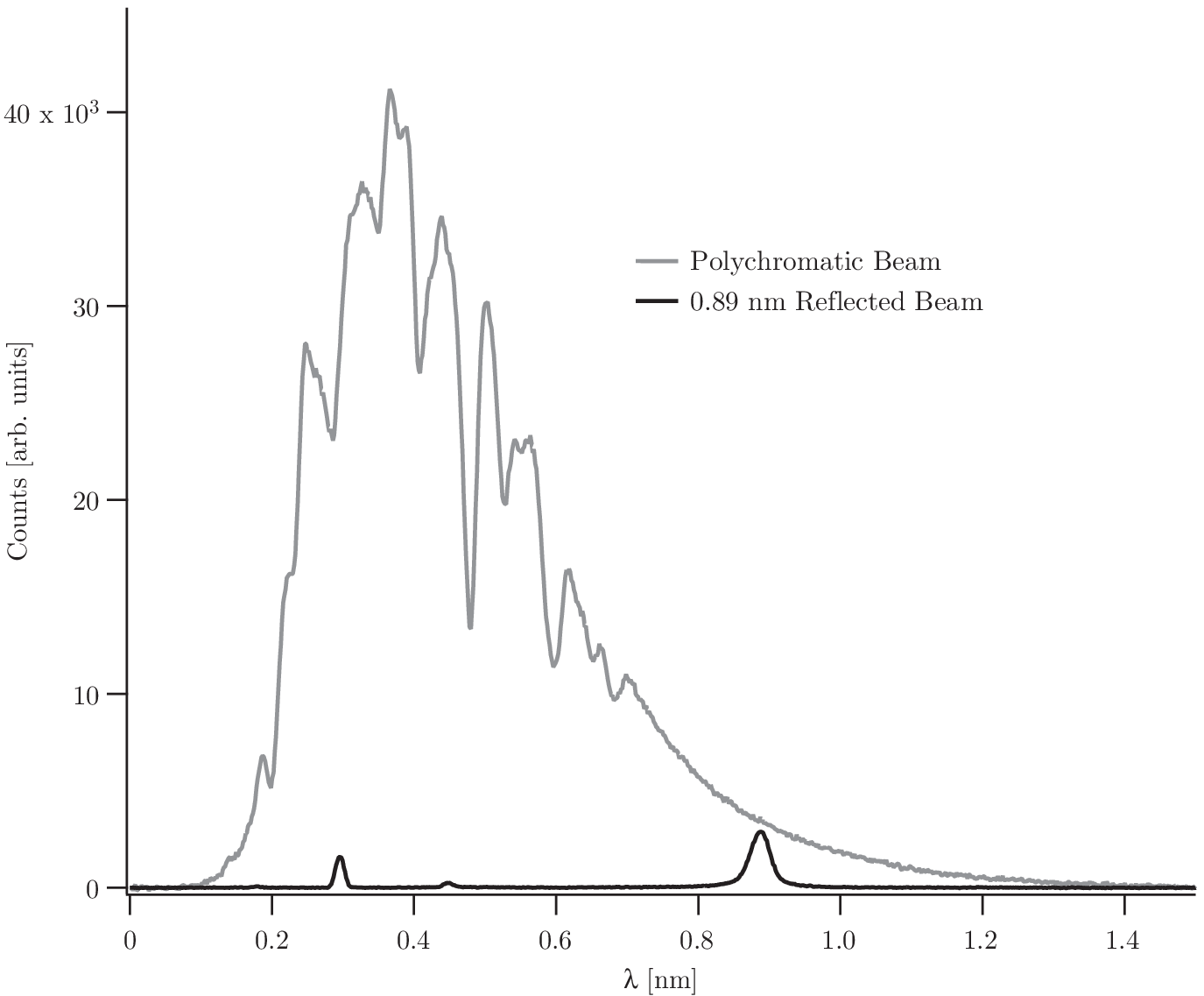}
    \caption{The time-of-flight spectrum of the incident and reflected
    neutron beams.  The reflected monochromatic beam has the $\lambda
    /2$ and $\lambda /3$ peaks filtered from the beam using pyrolytic
    graphite and polycrystalline bismuth.}
    \label{fig:monobeam}
\end{figure}

An 0.89~nm monochromator has been constructed using stage 2
potassium-intercalated graphite.  The monochromator, tiled from nine
pieces, has a total size of 6~cm by 15~cm.  The individual
monochromator pieces have high stage purity and mosaics varying from
1$^{\circ}$ to 2$^{\circ}$.  The monochromator reflects more than 80~\%
of the incident 0.89~nm neutrons, while reflecting less than 2~\% of
the total cold neutron beam.  The incident and reflected beams are
shown in Fig.~\ref{fig:monobeam}.  The signal to
neutron-induced-background in the magnetic trapping experiment is thus
improved by a factor of 40 through use of the monochromator.  A
detailed description of the monochromator can be found in
Ref.~\cite{Mat02}.

\section{Apparatus}

The cryogenic apparatus used for the proof-of-principle magnetic
trapping measurements is described in detail in \cite{Bro00}.  The
changes made to the apparatus since 1999 will be outlined below and
are described in detail in \cite{Mat02} and \cite{McK02}.  The
existing cryostat was modified to make room for a larger bore magnetic
trap and the accompanying light detection system.  The new magnet was
designed to be the largest Ioffe-type magnet assembly that would fit
in the existing cryostat.  Inside of the bore of this new magnet was
inserted a larger neutron trapping and detection cell.  The cell is
coupled to the dilution refrigerator using a superfluid heatlink that
provides considerably higher thermal conductance and lower eddy
current heating than the copper heatlink used previously.

\subsection{Heat Link}

The fill lines and heat link between the mixing chamber of the
dilution refrigerator and the horizontal trapping region were
replaced.  In the previous design, the heat link between the mixing
chamber and the cell was provided by copper rods and braids, while the
cell was filled with ultrapure helium through two 3.2~mm stainless
steel lines.  This design was susceptible to eddy-current heating of
the large copper mass when the magnet current was changed.
Furthermore, the heat link had poor thermal conductivity, limited by
the braid.  Since the new apparatus was designed for higher magnetic
fields and with a smaller separation between the magnet coils and the
heat link, estimates of the large size of the eddy current heating
necessitated a new design.

A continuous cylindrical volume of helium (approximately 5~cm$^{2}$ in
area) now extends from the mixing chamber to the horizontal cell.  The
use of superfluid helium rather than copper results in a much greater
thermal conductance for this heat link compared with the previous one.
The previous heat link had a measured conductance of $2.3 \times
10^{-4}$~W~K$^{-1}$ at 200~mK \cite{Bro00}.  A thermal conductance of
$0.81~T^{3}$~W~K$^{-1}$, where $T$ is the temperature of the mixing
chamber in Kelvin was measured for the new design \cite{McK02}.

A small buffer volume provides the thermal link between the superfluid
helium heat link and the mixing chamber of the dilution refrigerator.
This volume contains copper fins coated with silver sinter to maintain
high thermal conductivity between the mixing chamber and superfluid,
even at temperatures $< 100$~mK, where the Kapitza boundary resistance
due to phonon mismatch is significant.

\subsection{Magnet}

The magnetic trap used in this work is in the Ioffe configuration.
Four racetrack-shaped coils are arranged to produce the
radially-confining magnetic quadrupole.  Axial confinement is provided
by two solenoid assemblies with identical current senses.

The magnetic trap used to demonstrate magnetic trapping of UCN in 1999
had an inner bore of 50~mm, and ran at a maximum current of 180 A,
corresponding to a trap depth of 1.0~T. The newly constructed magnetic
trap was designed to have a considerably larger volume and maximize
the number of trapped neutrons.  The design was constrained by the
size of the helium bath inside the current dewar.  The new design was
constructed using coils of $2.5~\mbox{cm} \times 2.5$~cm
cross-section, with a bore diameter of 105~mm.  With a larger bore
size, a higher detection efficiency could be obtained, as light could
be more easily extracted from the trapping region.  Also, the larger
trap volume allowed the confinement of approximately ten times more
neutrons than in with the previous trap.

The trap depth is defined as the difference between the maximum and
minimum values of the magnetic field within the confinement region.
In our case the maximum magnetic field value is found at the radial
edge of the trap, which is defined by the location of the cell walls
at a radius of 4.2~cm.  The trap depth is 1.1~T for an operating
current of 170~A.

\subsection{The Detection Insert}

The detection insert is based on tetraphenyl butadiene (TPB)
evaporated on a 1~mm thick GoreTex\footnote{Certain trade names and
company products are mentioned in the text or identified in
illustrations in order to adequately specify the experimental
procedure and equipment used.  In no case does such identification
imply recommendation of endorsement by the National Institute of
Standards and Technology, nor does it imply that the products are
necessarily the best available for the purpose.} sheet, rolled up into
a tube and inner diameter of 8.4~cm.  The density of the evaporated
layer is between 200~$\mu$g~cm$^{-2}$ and 400~$\mu$g~cm$^{-2}$.
Extreme ultraviolet (EUV) scintillation light is produced by the
recoil of neutron decay electrons through the superfluid helium
filling the trapping region.  TPB absorbs photons over a broad
wavelength region from the soft UV to X-rays and emits blue light with
a spectrum peaked at 440~nm and a width of approximately 50~nm
\cite{Bur73}.  The fluorescence efficiency of an evaporated TPB film
is approximately 1.4.

The great majority ($>$ 80~\%) of the scintillation light from the
prompt singlet decay of the excited helium is emitted within 20~ns of
the ionization event.  Each neutron decay event creates a bright flash
of EUV light, which is converted to a pulse of blue light by the
wavelength shifter.  This visible light is transported to room
temperature through windows and light guides, and is detected by
photomultiplier tubes (PMTs) at room temperature.  In this way,
neutron decay events result in bursts of photoelectrons, which are
detected and recorded.

In order for the TPB fluorescence to be collected by light guides and
transported to room temperature, it must pass through an optically
transparent, neutron-absorbing disc of boron oxide located at the end
of the trapping region.  The difficulty with boron oxide is that it
clouds up when exposed to atmosphere (it is hygroscopic).  However, we
found that by minimizing its exposure to the atmosphere, its
transparency could be maintained within acceptable limits.  It was
polished, then stored under vacuum.  We found that it could be kept in
the atmosphere for up to 15~min.  before clouding significantly.

Following the beam dump is a ultraviolet transmitting grade acrylic
light guide of diameter 8.7 cm and length 40.7 cm that transports
light from the Gore-Tex tube to the end of the experimental cell.  The
light guide is wrapped in a 175~$\mu$m thick layer of Tyvek to
increase its light transport efficiency.  Scintillation light that
passes out of the light guide then passes through acrylic and quartz
windows at 4~K and into a 35~cm long, 11.4~cm diameter light guide
that transports the scintillation light from 77~K to 300~K. At room
temperature, the scintillation light is split into two
photomultipliers and detected in coincidence.

The detector was calibrated by placing a $^{113}$Sn beta source in the
center of the insert.  Using a single photomultiplier at the end of
the 77~K light guide, it was found that the 364~keV beta from the
$^{113}$Sn source caused an average signal of approximately 30
photoelectrons.

The efficiency of the light guide splitter was determined by comparing
the peak channel position with one photomultiplier to the peak channel
position with the splitter installed and two photomultipliers attached.
It was found that the light guide splitter is 94~\% efficient.

\section{Data Acquisition System}

The goal of the new DAQ is to record multi-photon scintillation events
originating within the cell, and reject events produced by cosmic ray
muons.  Hence, the data acquisition system triggers when coincident
pulses are observed on the two main detection photomultipliers which
are not coincident with an event on any of the muon paddles
surrounding the trapping apparatus.  For each trigger, the data
acquisition records the time of occurrence and the digitized waveforms
from the two main PMTs.  The waveforms are recorded using two 500~MHz
PCI digitizer cards.  The timing signals and diagnostic information
are recorded in a separate timer/counter card.  Each event trace is
recorded to disk for offline analysis.

\section{Progress to Date}

The neutron trapping apparatus was operated on several occasions in
the spring of 2001 and the fall of 2002.  This section contains data
during the 2002 runs.  It compromises approximately one week of
actual trapping data.  Note that although the data looks promising,
considerably more data must be taken to fully investigate systematic
effects.  This data collection is now in progress.
\begin{figure}[t]
    \centering
    \includegraphics[width=\hsize]{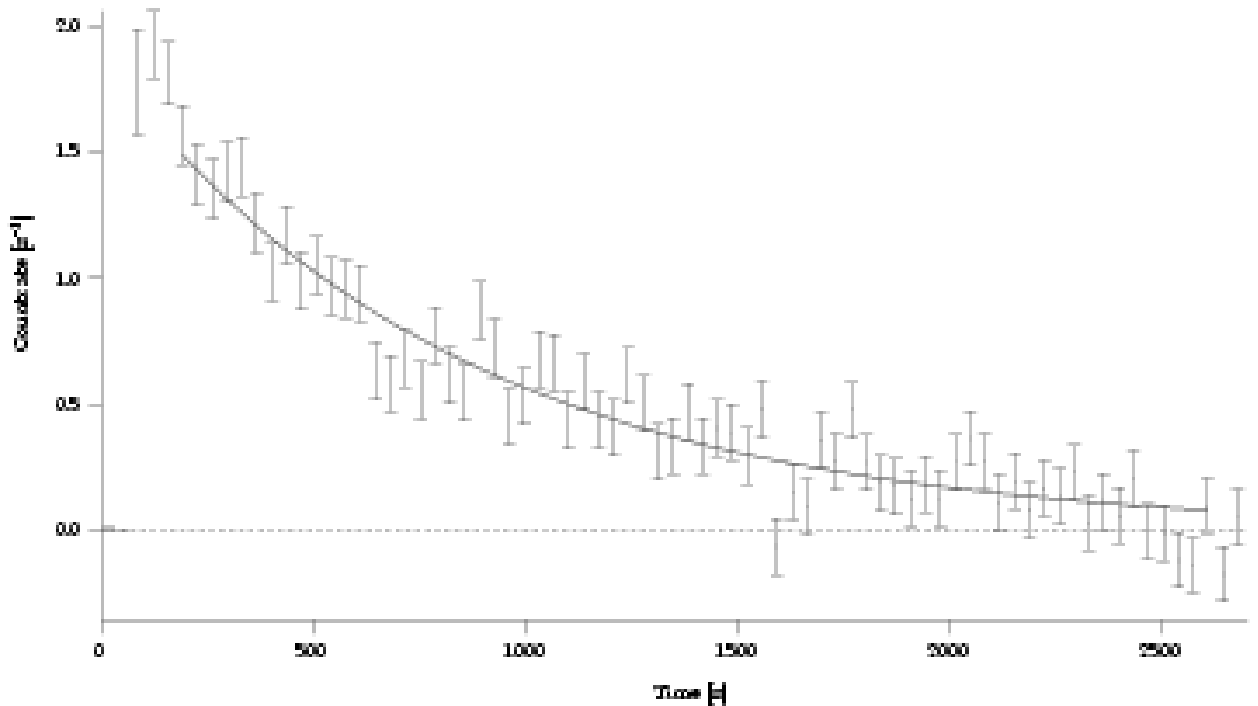}
    \caption{The difference of the positive and negative data collection
    runs for approximately one week of data.  The curve is a fit to the
    function $y = A exp(-t/\tau)$ with fit parameters $A = (1.88 \pm
    0.07)$~s$^{-1}$ and $\tau = (834 ^{+39}_{-36})$~s and a $\chi^{2}=
    1.2$.  If the data is fit to $y = y_{0} + A exp(-t/\tau)$, where the
    presumably zero constant offset is allowed to vary, one obtains values
    of $A = (1.87 \pm 0.07)$~s$^{-1}$, $\tau = (848 ^{+104}_{-84})$~s, and
    $y_{0} = (-0.008 \pm 0.052)$~s$^{-1}$.  Note that the one standard
    deviation errors quoted are purely statistical and the data is very
    preliminary.}
    \label{fig:diff}
\end{figure}

The presence of time-dependent backgrounds necessitates taking what we
refer to as ``positive'' and ``negative'' runs.  In positive runs, the
magnet is energized while neutrons are loaded into the trap.  After
the beam has been turned off, the neutron decay events are recorded.
In negative (or background) runs, the magnet is deenergized while the
beam is on, then raised to the full value as the neutron beam is
turned off.  In this negative case, the backgrounds from neutron
activation, etc.  should be similar.  A difference in the countrate
versus time between positive (trapped UCN + backgrounds) and negative
(backgrounds only) runs should be magnetically trapped UCN. If for
some reason the backgrounds are not identical in the positive and
negative runs, then the subtraction process will leave a residual
difference which could mimic a trapping signal.  Measurements made
with natural abundance helium can then be used to determine if a
putative trapping signal is caused by imperfect background
subtraction.

Analysis of the data is performed by integrating the pulse area of the
PMT signals and applying appropriate lower level cuts on the area of
the pulses.  Since luminescence is known to be present in the
coincidence data with thresholds at single photoelectron levels,
thresholds are typically set to require an area in each pulse to be
equivalent to at least three photoelectrons.  Data from all positive
runs and all negative runs are pooled and subtracted to yield a
difference curve as shown in Fig.~\ref{fig:diff}.

The absolute detection efficiency has not been determined for the
detector configuration used above, so the total number of neutrons
contained in the trap is not known.  We estimate, however, that the
number is consistent within a factor of two of the expected number
trapped.

\section{Future Directions}
Our present plan is to take a similar set of data using a natural
isotopic abundance helium sample in order to study backgrounds.  With
this sample, the lifetime of UCN within the trap is less than 1~s due
to absorption of the UCN by $^{3}$He.  Similar positive and negative
data will be taken to verify that the decaying signal seen above
arises from trapped neutrons.

Once complete, we hope to take a considerably longer set of trapping
data to lower the errors on the lifetime to roughly $\pm 10$~s.  This
should allow us to more carefully study systematic effects.  Studies
of the temperature dependence of the storage time due to phonon
upscattering is also planned.

Ultimately, any technique for measuring the neutron lifetime will be
limited by systematic as well as statistical errors.  The limiting
systematic errors of bottle and beam lifetime measurements, wall
interactions and flux measurement, respectively, are not relevant when
measuring the neutron lifetime using magnetically trapped neutrons.
The use of a magnetic trap prevents interactions with material walls,
and the continuous measurement of the neutron decay rate makes an
independent measurement of the neutron flux unnecessary.  The limiting
systematic errors for the magnetic trapping technique are expected to
be non-beta-decay loss mechanisms for trapped UCN. All known loss
mechanisms, including capture on $^{3}$He, single- and multi-phonon
upscattering, neutron depolarization (Majorana flips), and marginal
trapping, are expected to result in loss rates less than $10^{-5}$
times the beta-decay rate \cite{Doy94,Bro00}.  Another proposed
systematic error, the modification of the free neutron lifetime due to
nuclear interactions with the surrounding helium, is calculated to
have even less of an effect \cite{Bro00}.  In principle, the
measurement of the neutron lifetime using magnetically trapped UCN
should remain statistics limited until the fractional error reaches
$10^{-5}$.  The imperfect subtraction of time-dependent backgrounds
can also introduce a systematic error into the measurement.  This type
of effect will be studied in the coming months.

\section{Acknowledgments}

Neutron facilities used in this work were provided by the NIST Center
for Neutron Research.  This work was supported in part by the National
Science Foundation under grant No.\ PHY-0099400.

\title*{Measurement of the Neutron Lifetime \\ by Counting Trapped Protons}
\toctitle{Measurement of the Neutron Lifetime \protect\newline by Counting Trapped Protons}
\authorrunning{F.~E. Wietfeldt et al.}
\tocauthor{F.~E.~Wietfeldt, M.~S.~Dewey, D.~M.~Gilliam, J.~S.~Nico, X.~Fei, W.~M.~Snow, G.~L.~Greene, J. Pauwels, Eykens, A. Lamberty, \\ J. Van Gestel}
\author{F.~E.~Wietfeldt\inst{1}, M.~S.~Dewey\inst{2}, D.~M.~Gilliam\inst{2}, J.~S.~Nico\inst{2}, X.~Fei\inst{3}, W.~M.~Snow\inst{3}, 
G.~L.~Greene\inst{4}, J. Pauwels\inst{5}, Eykens\inst{5}, A. Lamberty\inst{5}, \\ and J. Van Gestel\inst{5}}
\institute{Tulane University, New Orleans, LA 70118, USA \and National Institute of Standards and Technology,
Gaithersburg, MD 20899, USA \and Indiana University, Bloomington, IN 47408, USA \and University of Tennessee/Oak Ridge
National Laboratory, \\ Knoxville, TN 37996, USA \and European Commission, Joint Research Centre, Institute
for Reference Materials and Measurements, 2440 Geel, Belgium}

\maketitle

\begin{abstract}
We have measured the neutron decay lifetime by the
absolute counting neutron decay recoil protons that were confined in a quasi-Penning trap.
The neutron beam fluence was measured by capture in
a thin $^{6}$LiF foil detector with known absolute efficiency. The combination of these
measurements gives the neutron lifetime:
$\tau_{n} = (886.8\pm1.2{\rm [stat]}\pm 3.2{\rm [sys]})$~s, which is the most precise neutron
lifetime determination to data using an in-beam method.
\end{abstract}


\par
\begin{figure}[t]
\includegraphics[width=\hsize,clip]{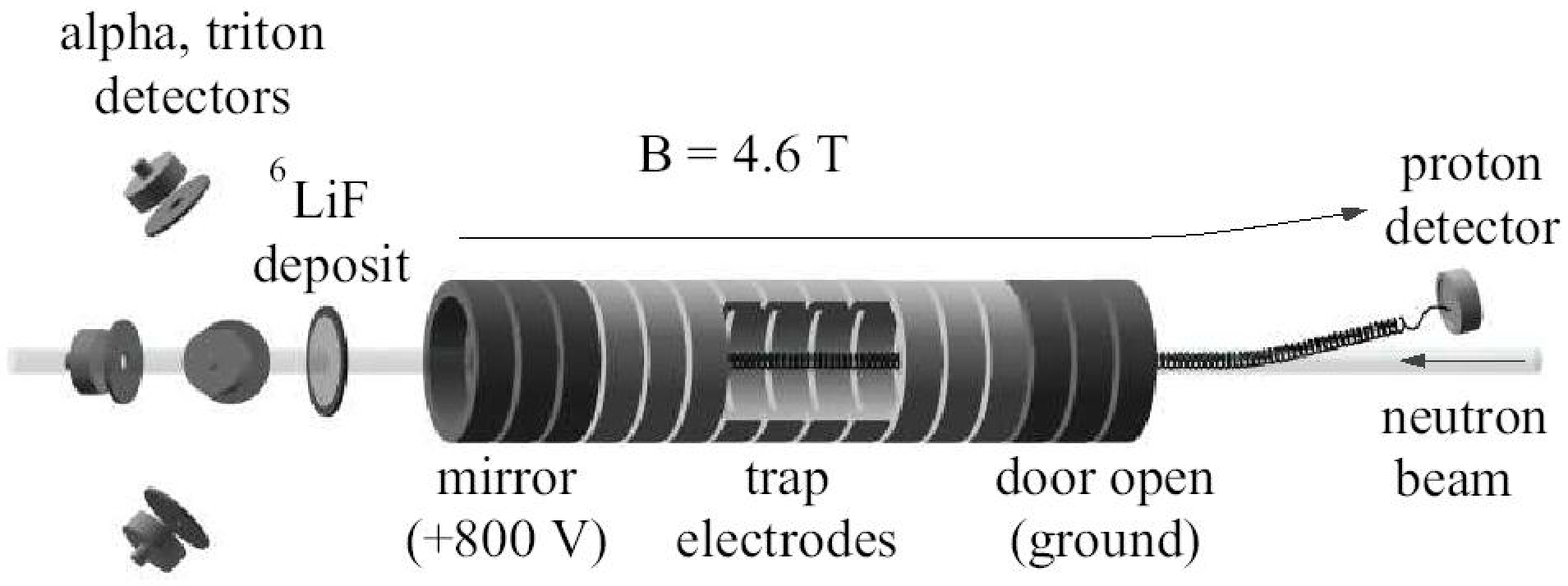}
\caption{\label{tNScheme}Schematic representation of the experimental method (not to scale).}
\end{figure}

We measured the neutron lifetime at the cold neutron beam NG6 at the National Institute of
Standards and Technology (NIST) Center for Neutron Research, using the quasi-Penning trap method
first proposed by Byrne,  {\em et al.}. This method is described in detail in previous publications \cite{PTmeth}.
Figure \ref{tNScheme} shows a  sketch of the experimental configuration.
A proton trap of length $L$ intercepts the entire width of the neutron beam.  Neutron
decay is observed by trapping and counting decay protons within the trap with an efficiency
$\epsilon_{p}$.  The neutron beam is characterized by a velocity
dependent fluence rate $I(v)$.  The rate $\dot{N_{p}}$ at which decay protons are detected is proportional
to the mean number of neutrons inside the trap volume:
\begin{equation}
    \dot{N_{p}}=\frac{\epsilon_{p} L}{\tau_n}\int_{A}da\, I(v)\frac{1}{v}
\label{eq:mean_p}.
\end{equation}
where $A$ is the beam cross sectional area,
After leaving the trap, the neutron beam passes through a
thin foil of $^6$LiF. The probability for absorbing a neutron in the foil
through the $^6{\rm Li}(n,t)^4$He reaction is inversely proportional to
the neutron velocity $v$. The reaction products, alphas or tritons, are counted by
a set of four silicon surface barrier detectors in a well-characterized geometry.
We define the efficiency for the neutron
detector, $\epsilon_{o}$, as the ratio of the reaction product rate to
the neutron rate incident on the deposit for neutrons with thermal velocity
$v_{o} = 2200$~m/s. The corresponding efficiency for neutrons of
other velocities is $\epsilon_{o}v_{o}/v$.  Therefore, the net reaction product
count rate $\dot{N_{\alpha}}$ is
\begin{equation}
    \dot{N_{\alpha}}=\epsilon_{o}v_{o}\int_{A}da\,I(v)\frac{1}{v}
\label{eq:mean_alpha}.
\end{equation}
The integrals in Eq.~(\ref{eq:mean_p}) and Eq.~(\ref{eq:mean_alpha})
are identical;  the velocity dependence of the neutron detector efficiency compensates for the fact that the
faster neutrons in the beam spend less time in the decay volume.  This cancellation is exact except for
a correction due to the finite thickness of the $^6$LiF foil (+5.4 s), and we obtain the neutron lifetime $\tau_n$
from the experimental quantities $\dot{N_{\alpha}} / \dot{N_{p}}$, $\epsilon_{o}$, $\epsilon_{p}$, and $L$.

\par
The proton trap was a quasi-Penning trap, composed
of sixteen annular electrodes, each 18.6 mm long with an inner diameter of 26.0~mm, cut from fused quartz
and coated with a thin layer of gold.
Adjacent segments were separated by 3 mm-thick insulating spacers of uncoated fused quartz.
The dimensions of each electrode and spacer
were measured to a precision of $\pm5$~$\mu$m using a coordinate measuring machine at NIST.
Changes in the dimension due to thermal contraction are below the 10$^{-4}$ level for
fused quartz.
The trap resided in a 4.6 T magnetic field and the vacuum in the trap was maintained below $10^{-9}$~mbar.
\par
\begin{figure}
\begin{center}
\includegraphics[width=4.5in]{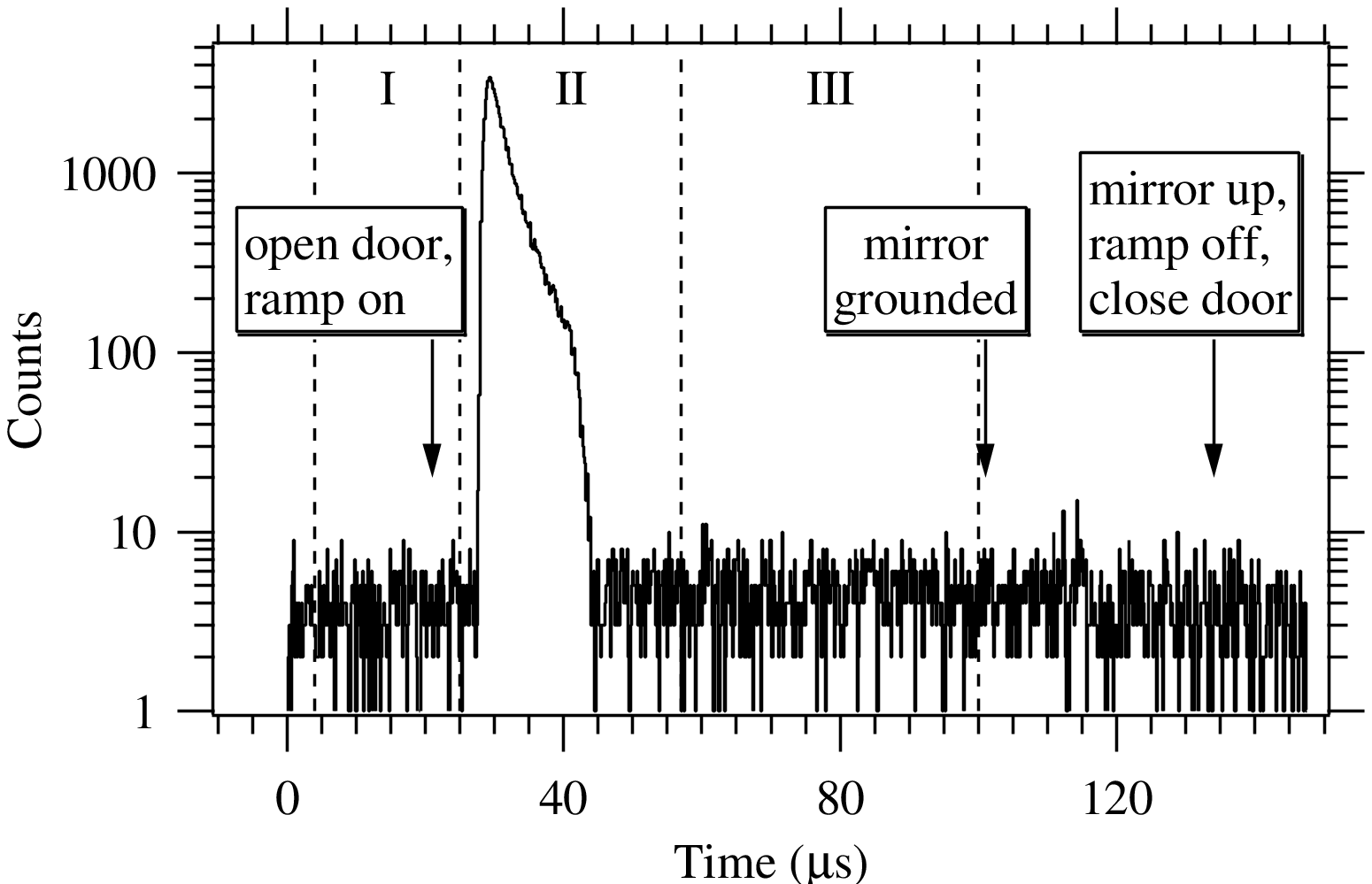}
\caption{\label{tPlot}A typical plot of proton detection time after gating on the detector ($t=0$).
Regions I and III are used to subtract background from the peak region II.
After correcting for deadtime in the time-to-digital converter, the resulting peak area gives the proton rate $\dot{N_p}$.}
\end{center}
\end{figure}

In trapping mode, the three upstream electrodes (the ``door'') were held
at +800 V, and a variable number of adjacent electrodes (the ``trap'') were held
at ground potential. The subsequent three adjacent electrodes (the ``mirror'')
were held at +800 V. We varied the trap length from 3 to 10 grounded electrodes. When a neutron
decayed inside the trap, the decay proton was trapped radially by the magnetic field
and axially by the electrostatic potential in the
door and mirror.
After some trapping period, typically 10 ms, the trapped protons were counted.
The door was ``opened'', i.e. the door electrodes were lowered
to ground potential, and a small ramped potential was applied to the trap electrodes
to assist the slower protons out the door.
The protons were then guided by a 9.5$^{\circ}$ bend in the magnetic
field to the proton detector held at a high negative potential (-27.5~kV to -32.5~kV).
After the door was open for 76 $\mu$s, a time
sufficient to allow all protons to exit the trap, the mirror was lowered to
ground potential. This prevented negatively charged particles, which may contribute to instability, from
accumulating in any portion of the trap. That state was maintained for 33 $\mu$s, after which the door and
mirror electrodes were raised again to +800 V and another trapping cycle began.
Since the detector needed to be enabled only during extraction, the counting background was reduced by the
ratio of the trapping time to the extraction time (typically a factor
of 125). Figure \ref{tPlot} shows a plot of proton detection time for a typical run.

\par
\begin{figure}
\begin{center}
\includegraphics[width=4.5in]{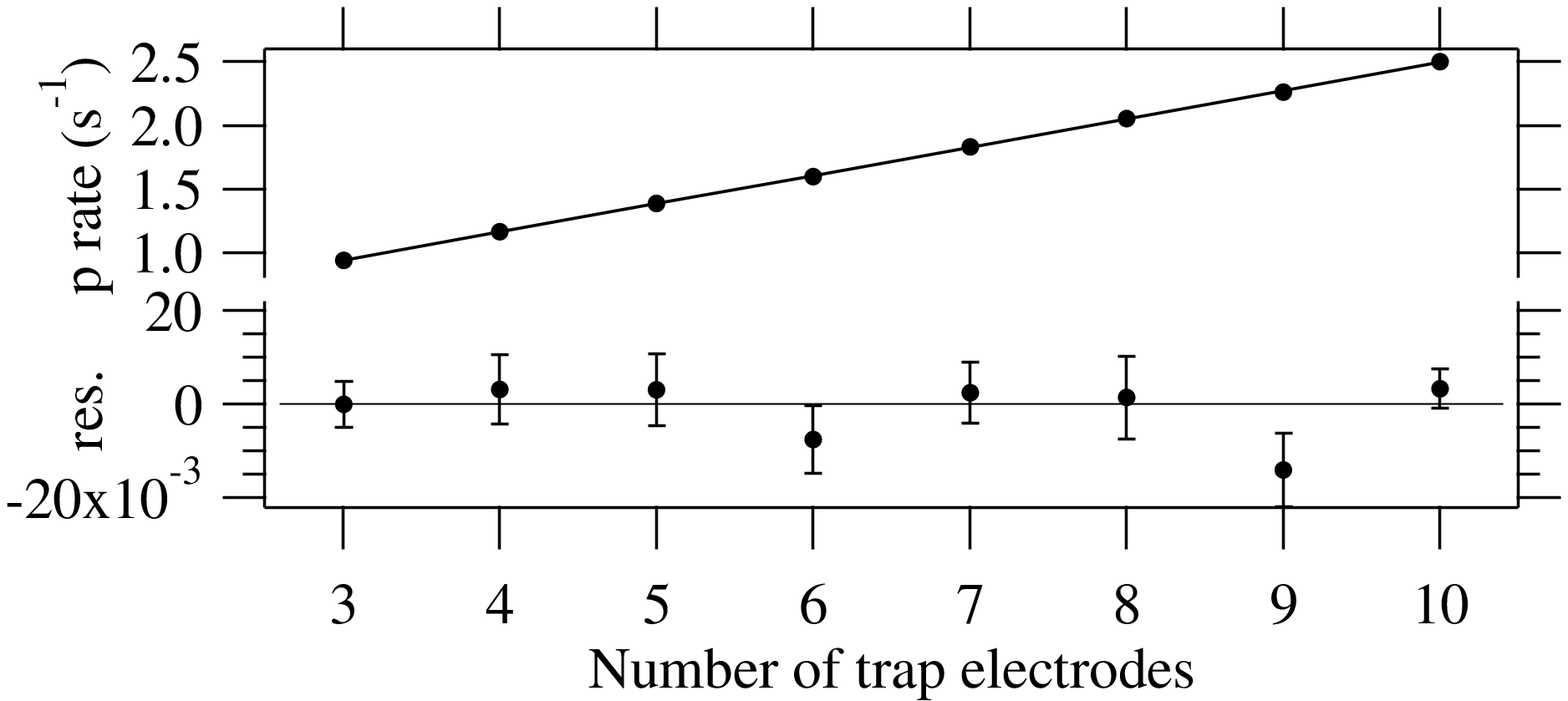}
\caption{\label{pRate}Typical raw proton count rate $\dot{N_p}$ {\em vs.} trap length data, fit to a
straight line (top), and residuals (bottom). These data have not yet been corrected for nonlinearities.}
\end{center}
\end{figure}

Protons that were born in the trap (grounded electrode region) were trapped with 100\% efficiency. However protons
that were born near the door and mirror (the ``end regions''), where the electrostatic potential is elevated,
were not all trapped. A proton born in the end region was trapped if its initial (at birth) sum of
electrostatic potential energy and axial kinetic energy was less than the maximum end potential.
This complication caused the effective length $L$ of the trap to be
difficult to determine precisely. It is for this reason that we varied the trap length.
The shape of the electrostatic potential near the door and mirror
was the same for all traps with 3--10 grounded electrodes, so
the effective length of the end regions, while unknown, was in principle constant.
The length of the trap can then be written $L = nl + L_{\rm end}$
where $n$ is the number of grounded electrodes and $l$ is the physical length of one electrode plus
an adjacent spacer. $L_{\rm end}$ is an {\em effective} length of the
two end regions; it is proportional to the physical length of the end regions {\em and} the
probability that protons born there will be trapped. From Eq. (\ref{eq:mean_p}) and (\ref{eq:mean_alpha})
we see that the ratio of proton counting rate to alpha counting rate is then
\begin{equation}
\frac{\dot{N_p}}{\dot{N}_{\alpha}} = \tau_n^{-1}\left( \frac{\epsilon_p}{\epsilon_0 v_o}\right)
(nl + L_{\rm end}).
\end{equation}
We fit $\dot{N_p} / \dot{N_{\alpha}}$ as a function of $n$ to a straight line and determine $\tau_n$ from
the slope, so there is no need to know the value of $L_{\rm end}$, provided that
it was the same for all trap lengths. Figure \ref{pRate} shows raw data from a typical run, proton count
rate vs. trap length $n$.

\par
Because of the symmetry in the Penning trap's design, $L_{\rm end}$ was approximately
equal for all trap lengths that we used.
However there were two trap-length-dependent effects that broke the
symmetry: the gradient in the axial magnetic field, and the divergence of the neutron beam. Each of these effects caused $L_{\rm end}$
to vary slightly with trap length.
A detailed Monte Carlo simulation of the experiment, based on the measured and calculated magnetic and electric field inside the trap, was developed in order to correct for these trap nonlinearities.
It gave a trap-length dependent correction that lowered the lifetime by 5.3s.
\par
Surface barrier (SB) and passivated
ion-implanted planar silicon (PIPS) detectors were used at various times for counting the protons.
The proton detectors were large enough so that all protons produced by neutron decay in the collimated beam,
defined by C1 and C2, would strike the 19.7-mm diameter active region after the trap was opened.
The detector was optically aligned to the magnetic field axis,
and the alignment was verified by scanning with a low energy electron source at the trap's center and with actual neutron decay protons.
When a proton hit the active region of the detector, the efficiency for proton detection was less than unity because:
1) a proton can lose so much energy in the inactive surface dead layer that it deposits insufficient
energy in the active region to be detected;  2) a proton can be stopped within
the dead layer and never reach the active region; 3) a proton can backscatter from the dead layer or the active region and deposit
insufficient energy. In the first two cases, the proton will definitely not be detected. We denote the fraction of protons lost
in this manner by $f_{\rm Lost}$.  In the last case, there is some probability that the backscattered proton will be reflected back
to the detector by the electric field and subsequently be detected.
We denote the fraction of protons that backscatter but still have some chance for detection by $f_{\rm Bsc}$.
\par
To determine the proton detection efficiency, we ran
the experiment with a variety of detectors with different dead layer thicknesses and different acceleration potentials.
The fraction of protons lost $f_{\rm Lost}$ and the fraction that backscatter $f_{\rm Bsc}$ were calculated using the SRIM 2000 Monte
Carlo program \cite{ZIE00}. We found that
$f_{\rm Lost}$ varied from $4.0(3)\times 10^{-5}$ to $8.0(6)\times 10^{-3}$, and $f_{\rm Bsc}$ from $1.83(13)\times 10^{-3}$ to $2.37(17)\times 10^{-2}$, depending on the detector and acceleration potential.
The proton counting rate $\dot{N_p}$ for each run was multiplied by $1 + f_{\rm Lost}$ to correct for lost protons. The correction for backscattered protons was not as simple because of the unknown probability for a backscattered proton to return and be detected, so we made an extrapolation of the measured neutron lifetime to zero backscatter fraction (see Figure \ref{TauVsBksc}).
\par
The neutron detector target was a thin (0.34~mm), 50-mm-diameter single crystal
wafer of silicon coated with a 38~mm diameter deposit of $^6$LiF, fabricated at the Institute
for Reference Materials and Measurements (IRMM) in Geel, Belgium.
The manufacture of deposits and characterization of the $^{6}$LiF areal density were exhaustively detailed in
measurements performed over several years~\cite{TAG91}.  The average areal density was $\rho = (39.30 \pm
0.10)~\mu\rm{g/cm^{2}}$. The $\alpha$ particles and tritons produced by the
the $^6{\rm Li}(n,t)^4$He reaction were detected
by four surface barrier detectors, each with a well-defined and carefully measured
solid angle. The geometry was chosen to make the solid angle subtended by the detectors insensitive to first order in the source position.
The parameter $\epsilon_{0}$ gives the ratio of detected alphas/tritons to incident thermal neutrons.
It was calculated using
\begin{equation}
\epsilon_{0} =\frac{\sigma_{0}}{4\pi}
\int\int\Omega(x,y)\rho(x,y)\theta(x,y) dx dy \label{eq:neutroneff},
\end{equation}
\noindent where $\sigma_{0}$ is the cross section at thermal ($v_{0} =
2200$~m/s) velocity, $\Omega(x,y)$ is the detector solid angle,
$\rho(x,y)$ is the areal mass density of the deposit, and
$\theta(x,y)$ is the areal distribution of the neutron intensity on
the target. The $^{6}$Li thermal cross section is
($941.0 \pm 1.3$)~b ~\cite{CAR93}. It is important to note that we take the ENDF/B-6 $1\sigma$ uncertainty from the evaluation, {\em not} the expanded uncertainty, to be the most appropriate for use with this precision experiment.
The neutron detector solid angle was measured in two independent ways: mechanical contact
metrology and calibration with $^{239}$Pu alpha source of known
absolute activity. These two methods agreed to within 0.1~\%.
\par
Proton and neutron counting data were collected for 13 run series, each with a different proton detector and acceleration potential. The corrected value of the neutron lifetime for each series was calculated and plotted vs. backscattering fraction, as shown in Figure \ref{TauVsBksc}. A linear
extrapolation to zero backscattering gave a result of $\tau_{n} = (886.8\pm1.2{\rm [stat]}\pm 3.2{\rm [sys]})$~s.   Our result would be improved by an independent absolute calibration of the neutron counter, which would significantly reduce the two largest systematic uncertainties, in the $^{6}$LiF foil density and $^{6}$Li cross section.  A cryogenic neutron radiometer that promises to be capable of such a calibration at the 0.1\% level has recently been demonstrated {{Zema}, and we are pursuing this method further. With this calibration, we expect that our experiment will ultimately achieve an uncertainty of less than 2 seconds in the neutron lifetime.
\par
We thank P.~Robouch of the IRMM for assistance in the
characterization of the $^{6}$LiF and $^{10}$B targets. We also thank J.~Adams, J.~Byrne, A.~Carlson, Z.~Chowdhuri,
P.~Dawber, C.~Evans, G.~Lamaze, and R.~D.~Scott
for their very helpful contributions, discussions, and
continued interest in this experiment. We gratefully acknowledge the support of NIST
(U.S. Department of Commerce), the U.S. Department of Energy (interagency agreement No.\ DE-AI02-93ER40784) and the
National Science Foundation (PHY-9602872).

\begin{figure}
\begin{center}
\includegraphics[width=4.5in]{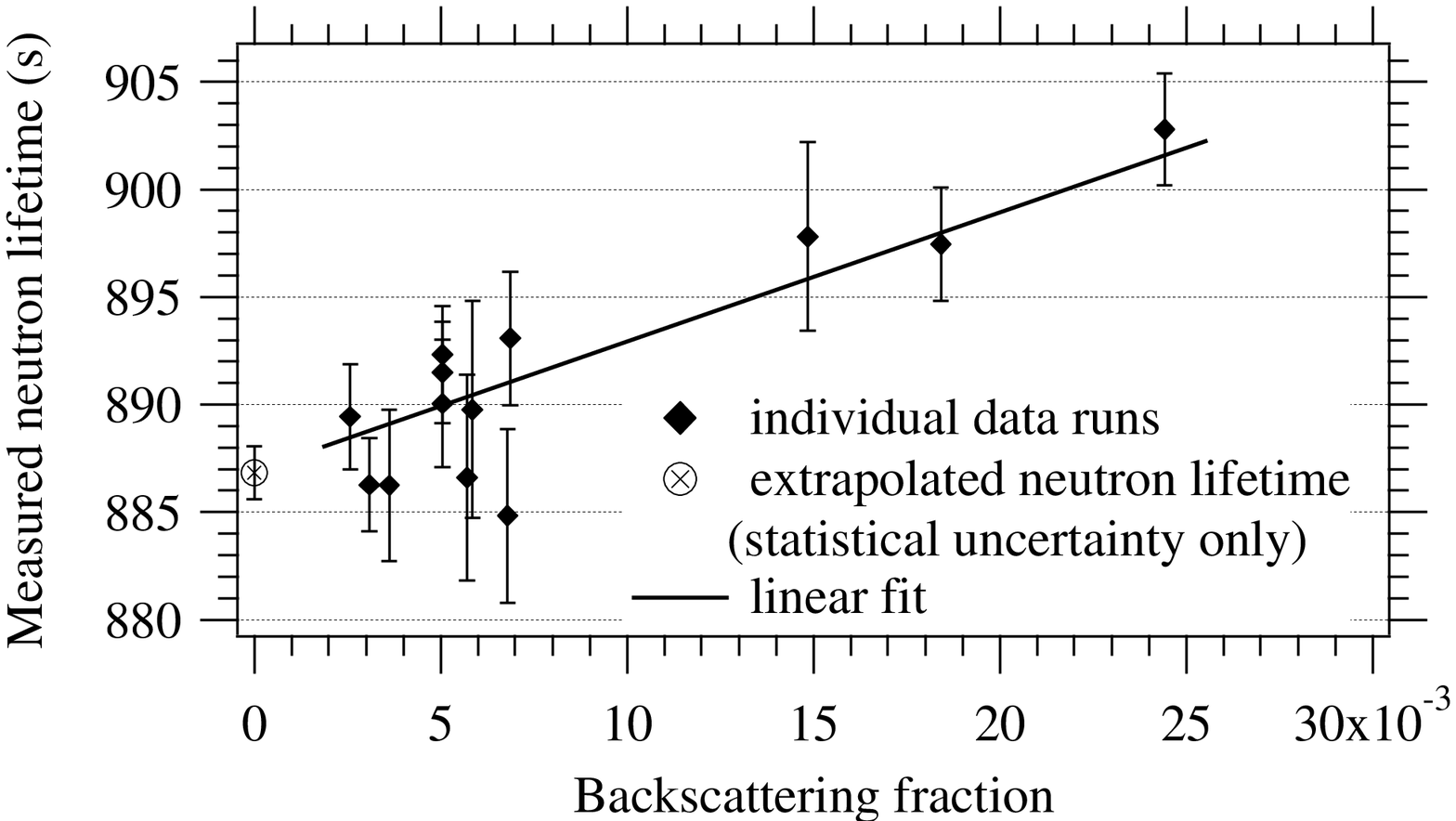}
\caption{\label{TauVsBksc}A linear fit of the measured neutron lifetime at varying values of the detector backscattering fraction.  The extrapolation to zero backscattering gives the free neutron lifetime.}
\end{center}
\end{figure}


\title*{A Magnetic Trap for Neutron-Lifetime Measurements\protect\newline}
\toctitle{A Magnetic Trap for Neutron-Lifetime Measurements}
%
%
\titlerunning{A magnetic storage device}
%
\author{F.J. Hartmann for the UCN-group: I. Altarev\and A. Frei\and F.J. Hartmann\and
S.~Paul\and G.~Petzoldt\and R.~Picker\and W.~Schott\and
D.~Tortorella\and U. Trinks\and O.~Zimmer}
\authorrunning{F.J. Hartmann {\em et al.}}
%
%
\institute{Physik-Department E18, Technische Universit\"at M\"unchen}

\maketitle              

\begin{abstract}
A  trap for neutrons with superconducting magnets - planned by the
UCN group at the Physics Department of Technische Universit\"at
M\"unchen - shall serve to measure the neutron lifetime. Magnetic
trapping is a method complementary to that usually employed in
recent years, the trapping of neutrons in bottles with material
walls. It avoids the problems with neutron losses by wall
collisions. With a volume of about 900\,dm$^3$ the arrangement
allows to store around $10^6$ neutrons at the high-flux reactor of
ILL, Grenoble, and orders of magnitude more with the new UCN
source of the reactor FRM-II at Arching.
\end{abstract}

\section{Introduction}
Two important weak-interaction parameters may be
determined from the $\beta$-decay of the free neutron:
the absolute value of the matrix element $V_{\mathrm{ud}}$
and the ratio $\lambda = g_{\mathrm{A}}/g_{\mathrm{V}}$
of the axial-vector and vector coupling constants
\cite{Abe00}.
Here the knowledge of the neutron lifetime $\tau_{\mathrm{n}}$
and one parameter of the correlation between
neutron direction or spin and direction and/or spin of the
decay products is sufficient.

\begin{table}[t]
\caption[-]{Experimental results for the neutron lifetime
$\tau_{\mathrm{n}}$.}
\renewcommand{\arraystretch}{1.4}
\setlength\tabcolsep{5pt}
\begin{center}
\begin{tabular}{|l|lcl|l|}
\hline
Method             &\multicolumn{3}{c}{$\tau_{\mathrm{n}}$ (s)}&Authors\\
\hline
Storage ring \& counters     &877  &$\pm$&10          &Paul {\em et al.} \cite{Pau89}\\
Gravitational trap \& counter&887.6 &$\pm$&3.0        &Mampe {\em et al.} \cite{Mam89}\\
Gravitational trap \& counter&888.4 &$\pm$&3.1 (stat.)&Nezvishevski {\em et al.} \cite{Nes92}\\
                             &      &     &1.1 (syst.)&\\
Gravitational trap \& counter&882.6 &$\pm$&2.7        &Mampe {\em et al.} \cite{Mam93}\\
Penning trap for $p$\&counter&889.2 &$\pm$&3.0 (stat.)&Byrne {\em et al.} \cite{Byr96}\\
                             &      &     &3.8 (syst.)&\\
Gravitational trap \& counter&885.4 &$\pm$&0.9 (stat.)&Arzumanov {\em et al.} \cite{Arz00}\\
                             &      &     &0.4 (syst.)&\\
Gravitational trap \& counter&881   &$\pm$&3          &Pichlmaier {\em et al.} \cite{Pic00}\\
{\bf Mean value}             &{\bf 885.7}&{\bf $\pm$}&{\bf 0.8}&Particle Data Group\cite{PDG02hart}\\
\hline
\end{tabular}
\end{center}
\label{LdMess}
\end{table}

The most recent measurements of $\tau_{\mathrm{n}}$
were performed by storing ultra-cold neutrons (UCN) in
material and gravitational traps and measuring the number of
surviving neutrons as a function of time.
Table \ref{LdMess} shows the results of the lifetime
experiments with $\sigma(\tau_{\mathrm{n}}) \le 10$\,s.

\begin{figure}[t]
\centering
\includegraphics[height=10cm]{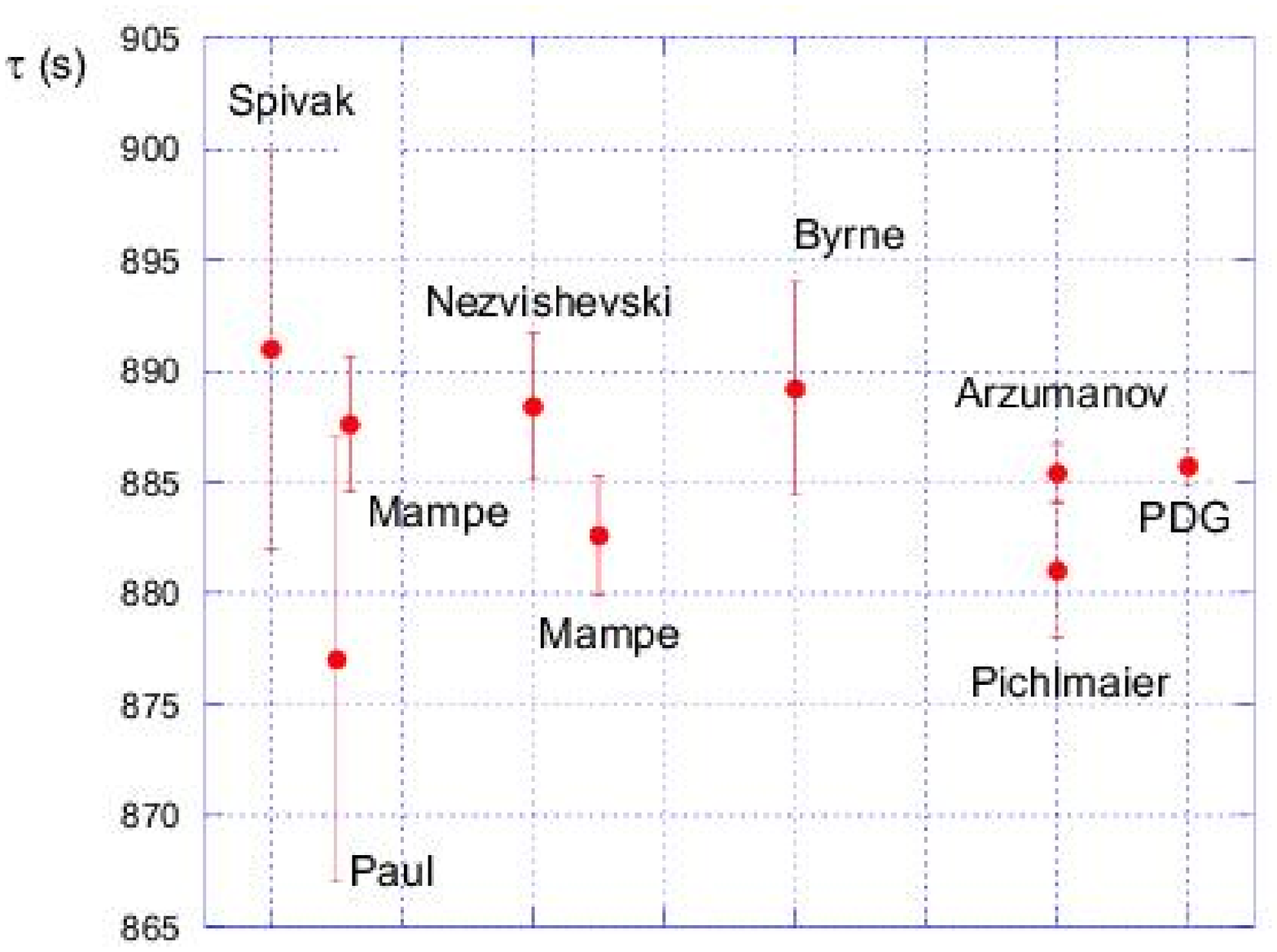}
\label{Fig:MeasLT}
\caption{Neutron-lifetime measurements with $\sigma(\tau_{\mathrm{n}}) \le 10$\,s}.
\end{figure}

The value for $\tau_{\mathrm{n}}$ adopted by the Particle Data Group \cite{PDG02hart}
is mainly determined by one measurement \cite{Arz00}. As may be seen from
Fig. \ref{Fig:MeasLT}, the results from different
groups scatter appreciably. This may be due to the fact that
the UCN stored in the trap hit the walls many times during the measuring
period and even a very low loss probability per wall collision leads to an
appreciable decrease in the {\em measured} neutron lifetime. The nature of these losses
at the trap walls is not completely clarified up to now and systematic errors
may show up during the necessary extrapolation to infinite trap volume.
Arzumanov et al. themselves state in their latest publication:
"The major problem ... is caused by losses of UCN in collisions with
the trap walls". Magnetic trapping avoids neutron contact with the trap walls and stays
free of the systematic uncertainties connected with this source of losses.
Hence experiments with {\em magnetic} traps are complementary to the
common method and a welcome addition.

As early as in 1951 W. Paul \cite{Pau51} proposed to store neutrons in magnetic
traps. The method is based on the interaction of the neutron magnetic moment
$\mu_n$ with an inhomogeneous magnetic field. Neutrons with the right orientation
of their magnetic moment, the {\em low-field seekers}, are driven away from
the regions of large magnetic fields at the trap walls.
The neutron-spin projection on the magnetic field is an adiabatic constant:
for sufficiently slow change of the direction of the magnetic field the spin
turns with the field and no spin flip occurs. The angular velocity of the rotation of the field lines
seen by the neutron has to be small compared with the frequency of
the neutron Larmor precession in the field,
$\omega_{\mathrm{Larmor}} = |\gamma| \cdot B$
(with the gyromagnetic ratio $\gamma = 183$\,MHz/Tesla).

\section{Experimental arrangement}
\subsection{Layout of the experimental set-up}

\begin{figure}[hbt]
\centering
\includegraphics[height=12cm]{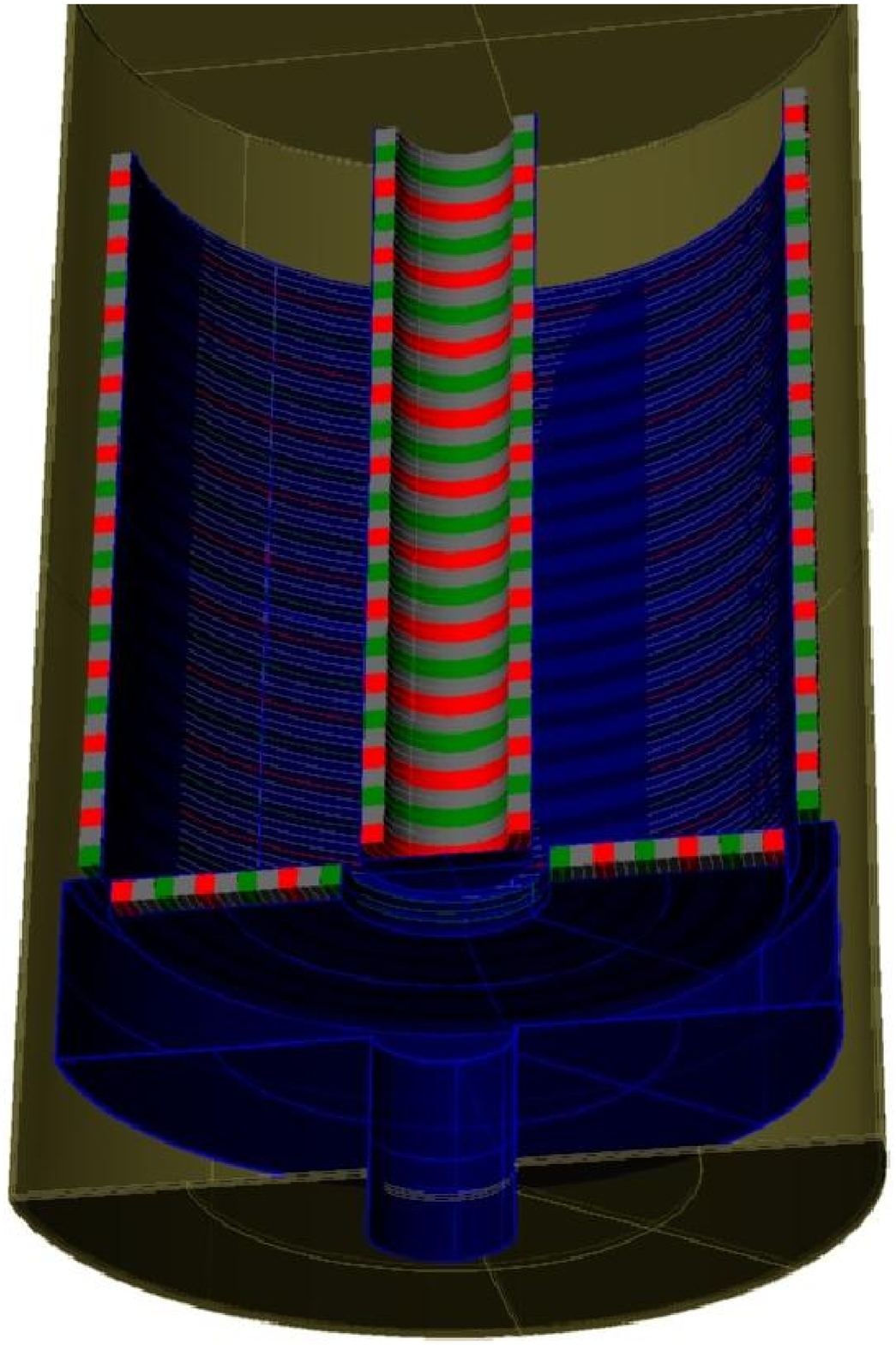}
\label{Fig:MagTra}
\caption{Schematic view of the magnetic storage device for UCN.
The superconducting coils are shown in red and green, the soft
iron layers in grey and the neutron reflectors and
guides in blue.}
\end{figure}

A schematic view of the planned trap with superconducting magnets
is shown in Fig. 2. A torus with 10\,cm
inner and about 50\,cm outer radius and with a height of around
1.2\,m is composed of sandwiches of superconducting ring coils
with rings of permendur (a NbFeV alloy) or soft iron in between.
The bottom of the trap is put together from concentric rings,
again from soft ferromagnetic material between superconducting coils.
The axis of the torus stands vertical and the top remains
open. A field at the walls of about 2\,Tesla is sufficient
to confine UCN with energies below 120\,neV. Gravitation prevents
these UCN from leaving through the top.
Two ring-shaped slits in the bottom allow filling the trap and
detecting depolarized neutrons. In order to fill the bottle
one (or several) coils around the slits are switched off.

The probability $w$ for non-adiabatic reorientation during
passage through regions of low magnetic field is given by
\cite{Vla61}
$$w = \exp \left (-\frac{\mu \cdot B_{\mathrm{min}}^2}
{\hbar \cdot |\mathrm{d}B/\mathrm{d}t|}\right )
=  \exp \left ( -\alpha \right )$$
Spin-flip in the low-field region inside the torus is
avoided by installing a straight bar in the torus axis
and letting a current of about 100\,A flow. This keeps the magnetic
field everywhere inside the trap larger than 0.2\,mTesla
and - at velocities below 5\,m/s - makes $\alpha$ stay well
above one thousand.

At a volume of about 900\,dm$^3$ and a UCN density of about
10/cm$^3$ at the high-flux reactor of Institut Laue-Langevin
in Grenoble we may expect to store more than $10^6$ UCN with the
right orientation of the magnetic moment in one filling.

\subsection{Experimental procedure}

In order to fill the UCN storage unit the trap is operated with
a weak magnetic field in the outer slit. To this end the coils
at the lower, outer edge (cf. Fig. 2 are switched off.
After filling with neutrons of one spin direction the entrance
slit is closed by a shutter covered with Be or $^{58}$Ni.
Now the current in the coils is increased again, slow enough to
leave the current in the other coils nearly unaffected. Afterwards
the shutter may be opened again.Under these circumstances
the time constant for filling through a 3-5\,cm slit is as low
as about 10\,s.

After trapping the neutrons we can determine their number as a function
of time. Three methods seem feasible
\begin{enumerate}
\item
At different times after filling a UCN detector is lowered into
the trap and counts the number of neutrons left at that moment.
This is possible as there are no coils at the top of the trap.
Semiconductor detectors with very low reflectivity (and
correspondingly high efficiency) for UCN have been developed
at our institute. The detector itself is covered by a thin
multilayer of $^6$LiF and $^{62}$Ni \cite{Pet01}. This guarantees
a negative Fermi potential down to lowest energies and
efficient detection of the products from the reaction
$$n + \mathrm{^6Li} \Rightarrow \mathrm{^3H} + \mathrm{^4He}.$$
Another possibility would be the use of a mixture of $^6$LiF
and $^7$LiF. The isotopic composition has
then to be adjusted to reach negative Fermi potential.
The preparation of the UCN detector would become much simpler
if one could - as has also been proposed - use a thin backing
for the neutron-sensitive layer and thus separate it from the
detector.
\item
Again at different times after filling the current in dedicated
superconducting coils is ramped down and the neutrons flowing
through the entrance/exit slit are registered by a UCN detector below
the trap. In this case a $^3$He counter may be used as the UCN
are accelerated by gravity.
\item
The protons from neutron decay are accelerated towards the top
of the setup by an electric field generated via field wires
at the torus walls. A voltage of about 5\,kV between bottom
and top seems sufficient. According to calculations more than
90\% of the protons reach the top of the torus; this because
part of them are reflected in the magnetic field near the
walls. After further acceleration to about 30 keV
these protons hit a thin foil of large area from a suitable
material (e.g. kapton with a thin Al layer). The electrons
from ion-induced electron emission are accelerated towards
an (again large) scintillator that is held at zero potential.
The (weak) scintillation light is detected with sensitive
photomultipliers or hybrid photomultiplier tubes. Several other
schemes of proton detection are considered at the moment, but
it is not yet clear, how feasible they are.
\end{enumerate}

\subsection{Possible systematic errors and their evaluation}

Important sources of systematic errors are i) the presence of
neutrons of wrong polarization (high-field seekers)
at measurement start and
ii) neutron spin flip during the measuring cycle. In both cases
neutrons may escape from the storage volume and are lost. In order to detect them,
the walls of the trap will be covered by beryllium or DLC (diamond-like carbon). Thus these
neutrons are reflected at the walls and finally reach the bottom
of the trap, where they leave it through the slits and are detected.
Monte-Carlo calculations revealed that the time $t_{\mathrm{detect}}$
between spin flip
and detection is of the order of 10 s. Figure 3
shows a typical delay-time distribution. 
\begin{figure}[t]
\centering
\hspace{12mm}
\includegraphics[width=9.5cm]{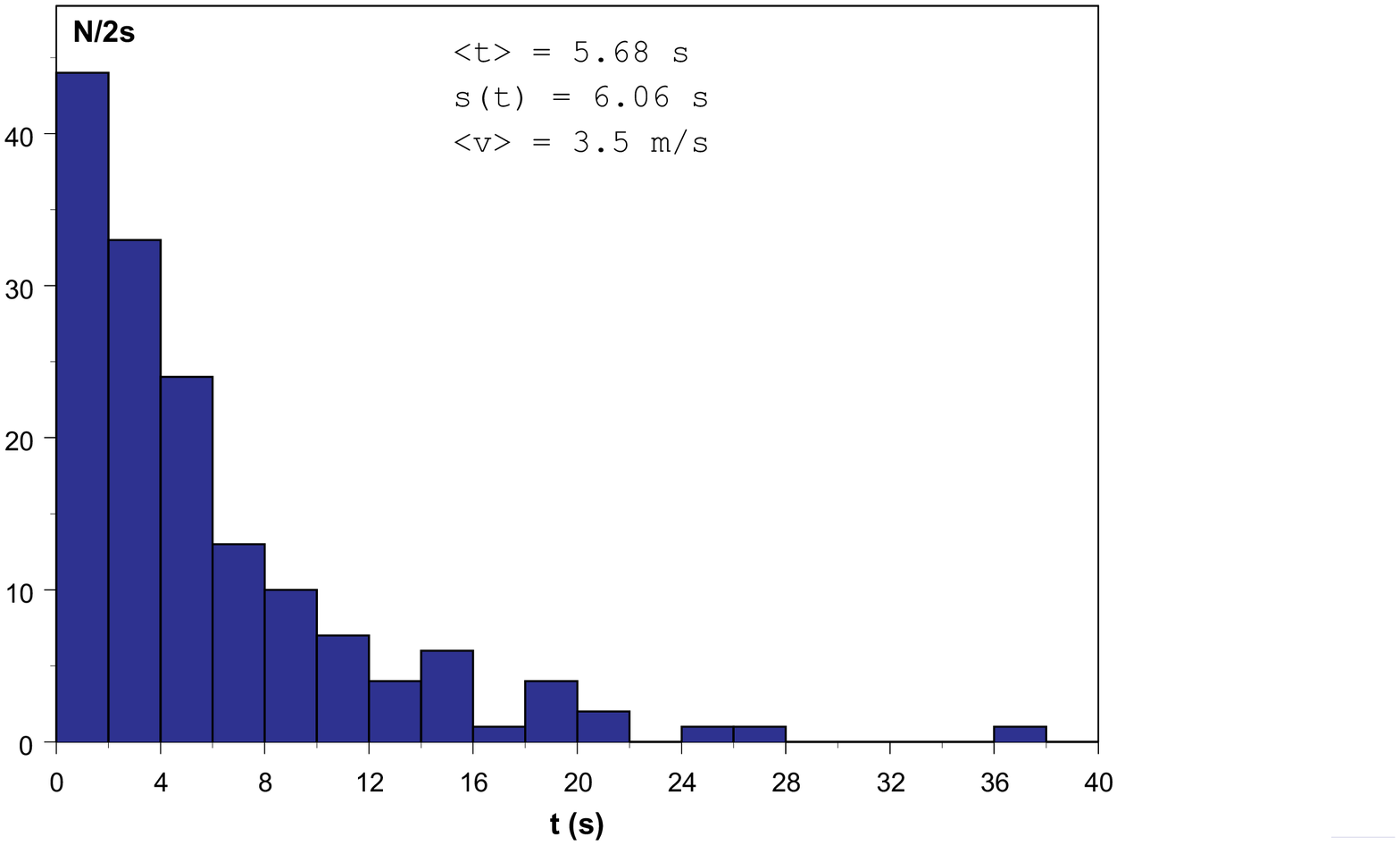}
\label{Fig:SpinFlipDet}
\caption{Distribution of the delay times between UCN spin flip and
detection of the flipped particles.}
\end{figure}
The mean value
$<t_{\mathrm{detect}}>$ is in good agreement with the value
found from the simple formula $t=V/4A \cdot 1/<v>$ that may be
derived from kinetic gas theory ($V$ and $A$ are the trap volume
and the total area of the exit slit, respectively and $<v>$ is the
mean UCN velocity. Figure 4 presents the
typical track of a depolarized neutron starting at $r = 0.26$\,m and
$z = 0.4$\,m; in this case the trap walls are assumed to be at
$r = 13$\,cm and $r = 49$\,cm.
\renewcommand{\bottomfraction}{0.9}
\renewcommand{\textfraction}{0.1}
\begin{figure}[b]
\centering
\hspace{6.5mm}
\includegraphics[width=9.5cm]{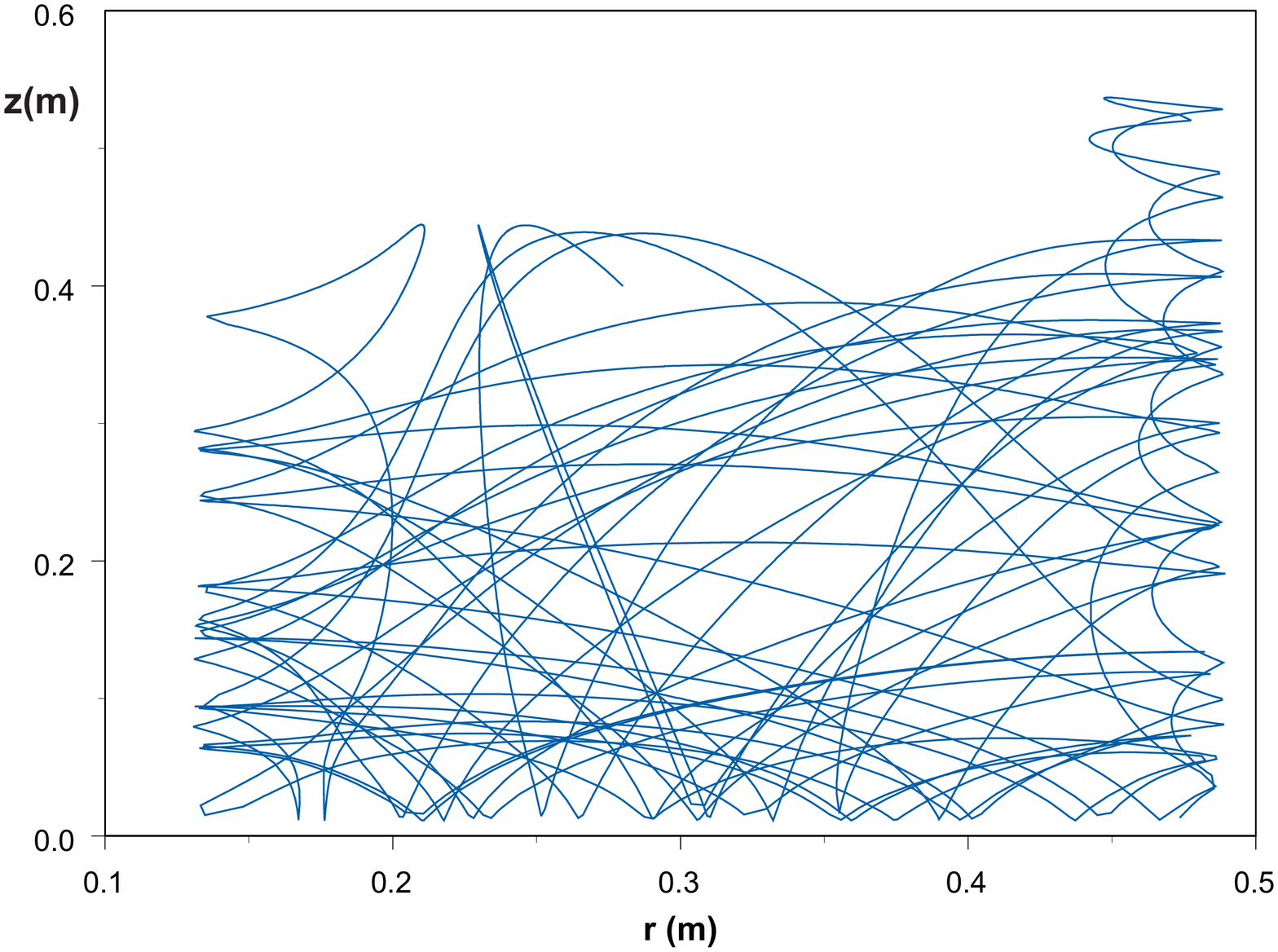}
\label{Fig:TrackDepol}
\caption{Typical track of a depolarized neutron from the middle of the trap to
the exit slit at the lower right corner.}
\end{figure}

Covering the trap walls with Be has a second advantage. If low-field
seeking UCN penetrate through the magnetic wall, they are
reflected by Be with very small losses and stay in the storage
volume. As the number of such events is expected to be low,
losses may be kept negligible.

Another source of systematic errors may be the variation
with time of the detection efficiency,
e.g. by a change of the neutron density distribution in
the storage volume with time. In the case of
proton detection it was already pointed out before, that more
than 90\% of the protons from the inner
part of the trap are detected. Hence the effect of efficiency variations
should be suppressed. In addition, the other methods for the lifetime
evaluation described above may be used for thorough cross-checks.

\section{Conclusion and outlook}
A trap with superconducting magnets will be built at the
Physik-Department of Technische Universit\"at M\"unchen.
The method chosen is complementary to those mainly employed
up to now. As the storage volume is large, a large number of
neutrons will be available at a time. This makes extensive
checks of possible systematic effects possible. Special care will be
taken to pin down systematic errors possibly caused by the
existence and generation of neutrons with wrong polarization.

The conceptual layout of the magnetic trap is almost ready and
construction work will start in 2003.


\title*{Towards a Perfectly Polarized Neutron Beam}
\toctitle{Towards a Perfectly Polarized Neutron Beam}
%
%
\titlerunning{Towards a Perfectly Polarized Neutron Beam}
%
\tocauthor{A.~Petoukhov, T.~Soldner, V.~Nesvizhevsky, M.~Kreuz, M.~Dehn, \protect\newline M.~Brehm}
\author{A.~Petoukhov\inst{1}, T.~Soldner\inst{1}, V.~Nesvizhevsky\inst{1}, M.~Kreuz\inst{{1}\,{2}}, M.~Dehn\inst{3}, \\ \and M.~Brehm\inst{2}}
\authorrunning{A.~Petoukhov et al.}
%
%
\institute{Institut Laue Langevin, BP156, F-38042 Grenoble Cedex
9, France\and Universit{\"a}t Heidelberg, Heidelberg, Germany\and
Universit{\"a}t Mainz, Mainz, Germany}

\maketitle              

\begin{abstract}
We propose a new method of double super mirror polarizers in
crossed geometry for neutron beam polarization. With such a
geometry a beam polarization of 99.60(5)\% without any significant
spatial and wavelength dependence between 3 and 10 {\AA} could be
demonstrated. \index{abstract}
\end{abstract}

\section{Motivation}

Many experiments in particle physics with cold neutrons require
both, a high degree of neutron polarization and its precise
measurement. The neutron polarization enters linearly in the
measurements of beta and antineutrino asymmetry in neutron beta
decay. Improvements of these experiments are particularly
important to test the unitarity of the CKM matrix and to search
for right handed currents. Their next generation aims to push the
precision below 0.1\%. This requires corresponding precision of
polarization. Other types of experiments require a neutron beam
with a very homogeneous and wavelength independent polarization
with variation of less than 10$^{-3}$ (whereas the absolute value
is less important here). Examples are spin rotation experiments
that search for parity or time reversal violation in the passage
of polarized neutrons through matter.
\begin{figure}[ht]
\begin{center}
\includegraphics[scale=0.44]{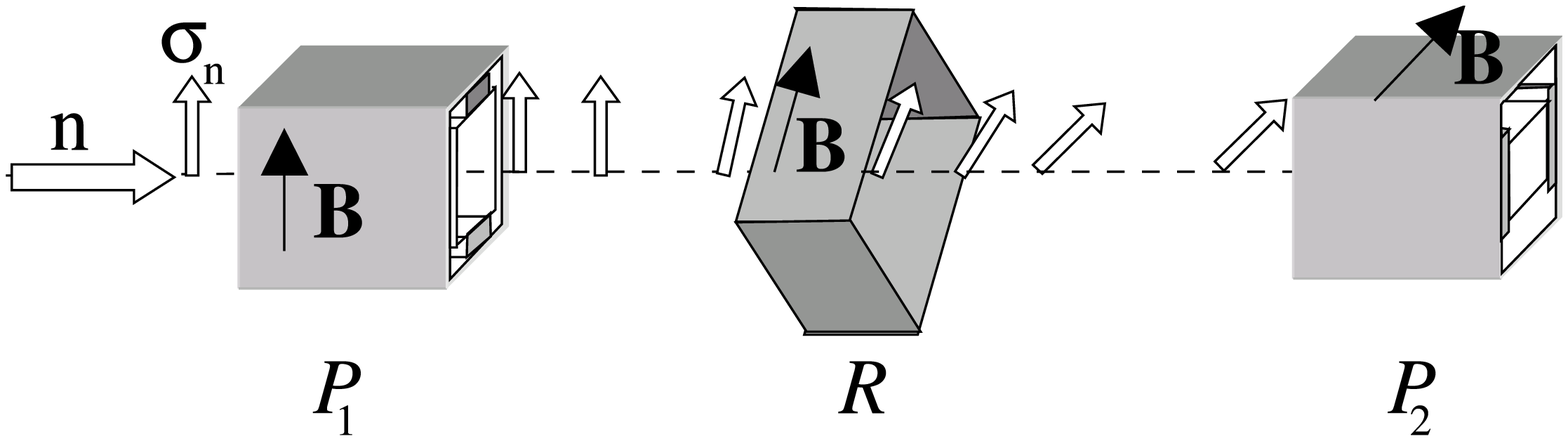}
\end{center}
\caption[]{Scheme of double SM polarizers in crossed geometry.}
\label{SchemeCrossed}
\end{figure}

\begin{figure}[t]
\begin{center}
\includegraphics[scale=0.35]{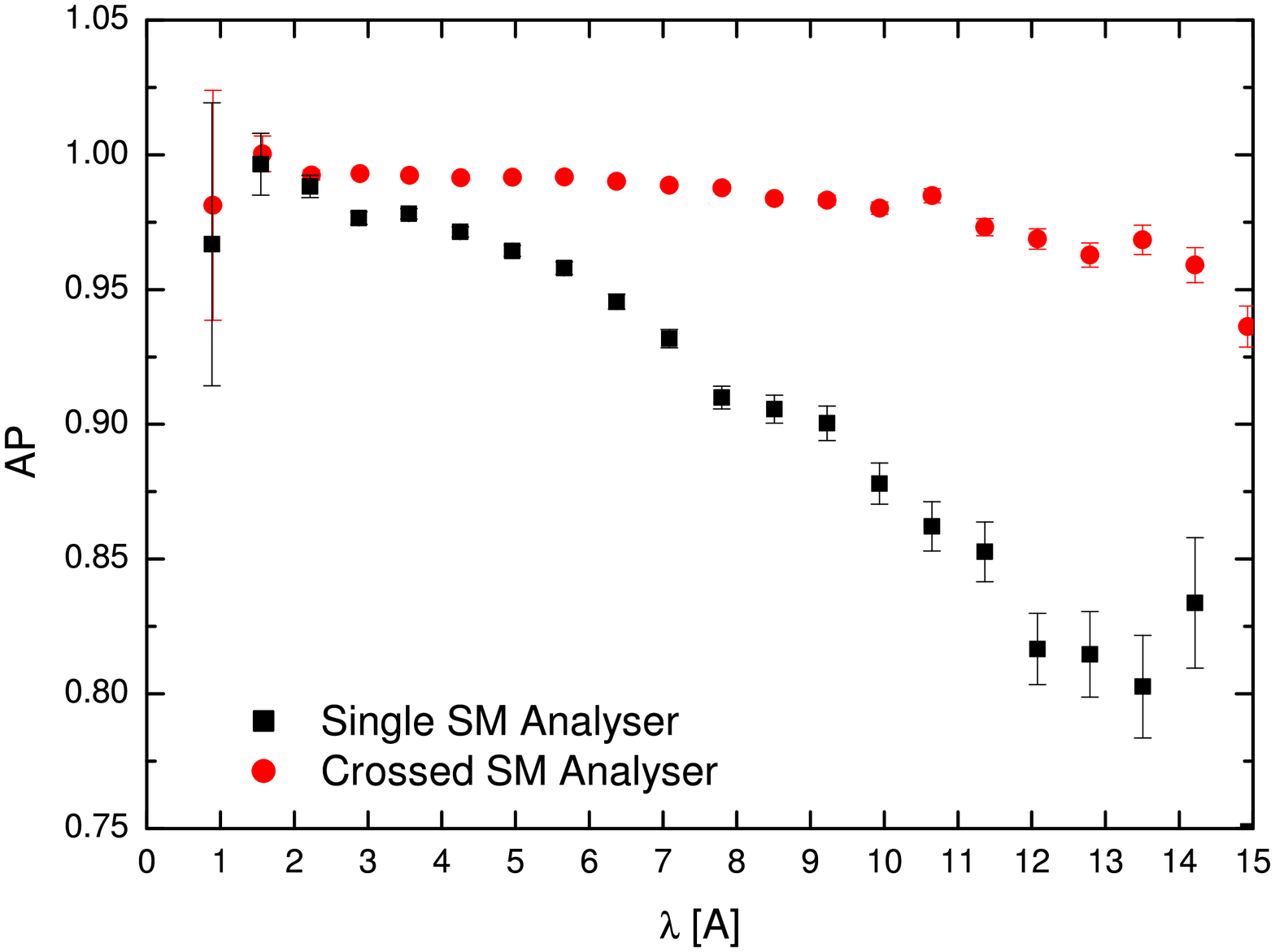}
\hfill\includegraphics[scale=0.35]{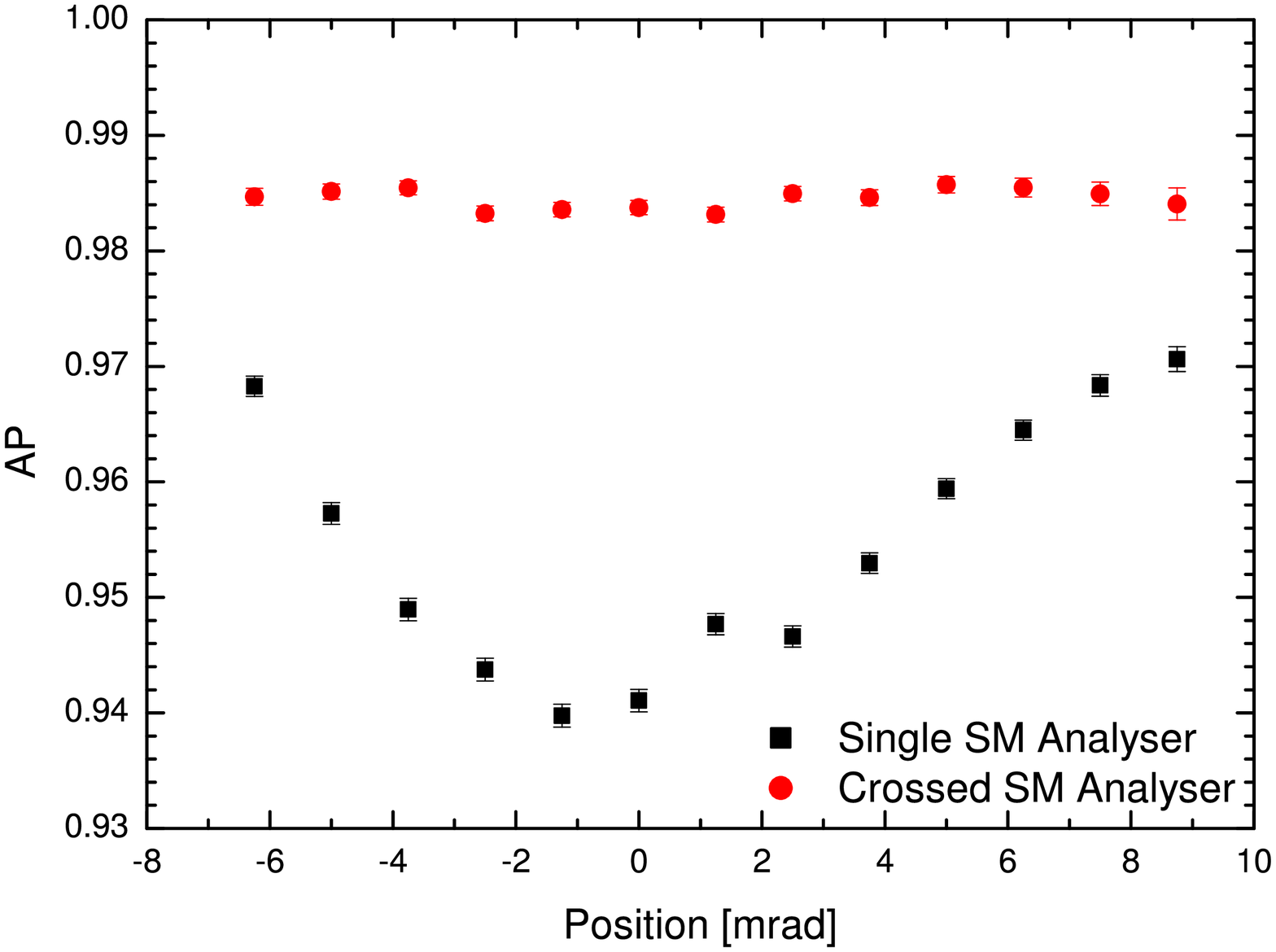}
\end{center}
\caption[]{Comparison of single analysed and double SM analysed
  in crossed geometry: (Top) wavelength, (Bottom) angular dependence.}
\label{WavelengthAngular} \label{Angular}
\end{figure}
\begin{figure}[b]
\begin{center}
\includegraphics[scale=0.34]{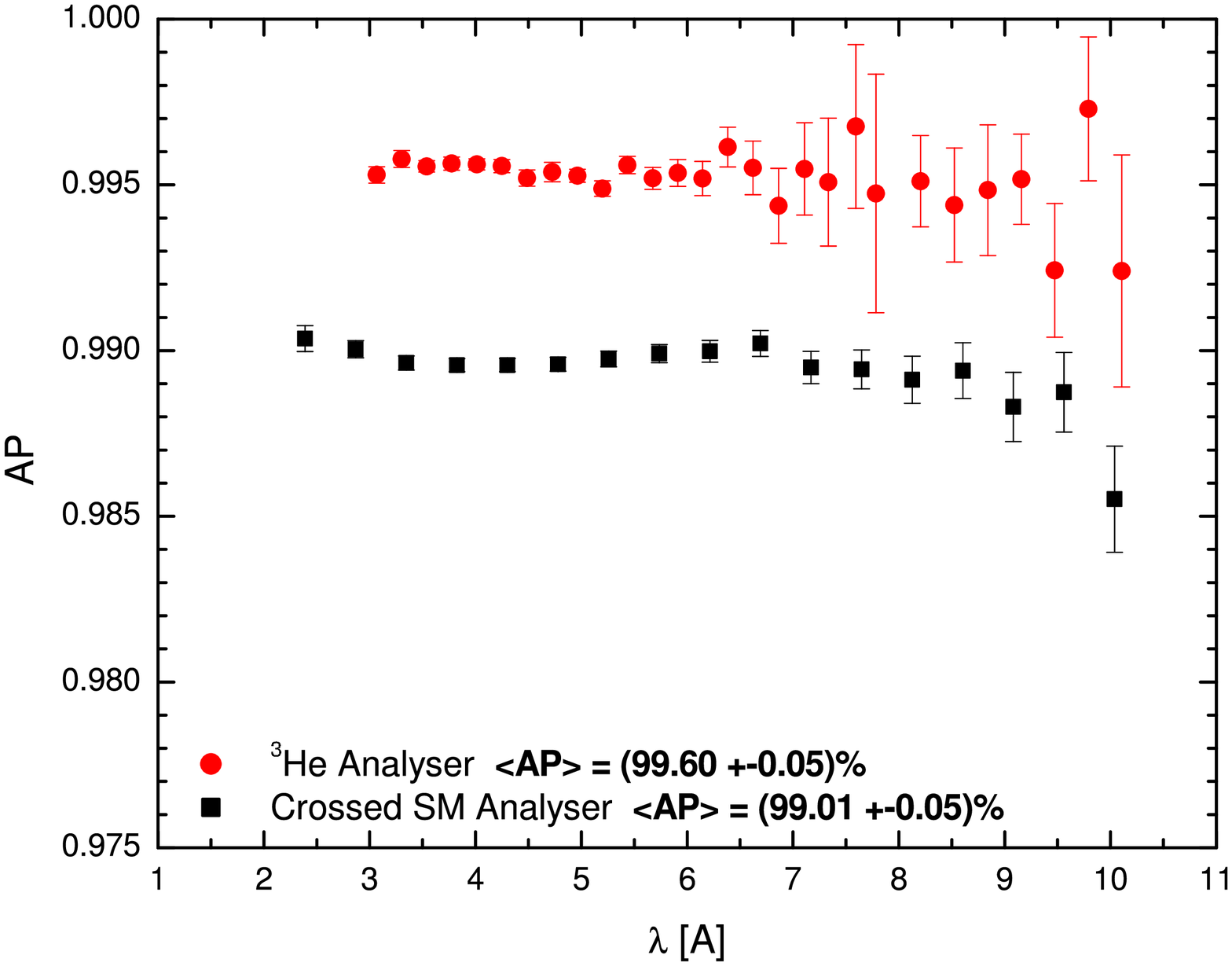}
\hfill\includegraphics[scale=0.34]{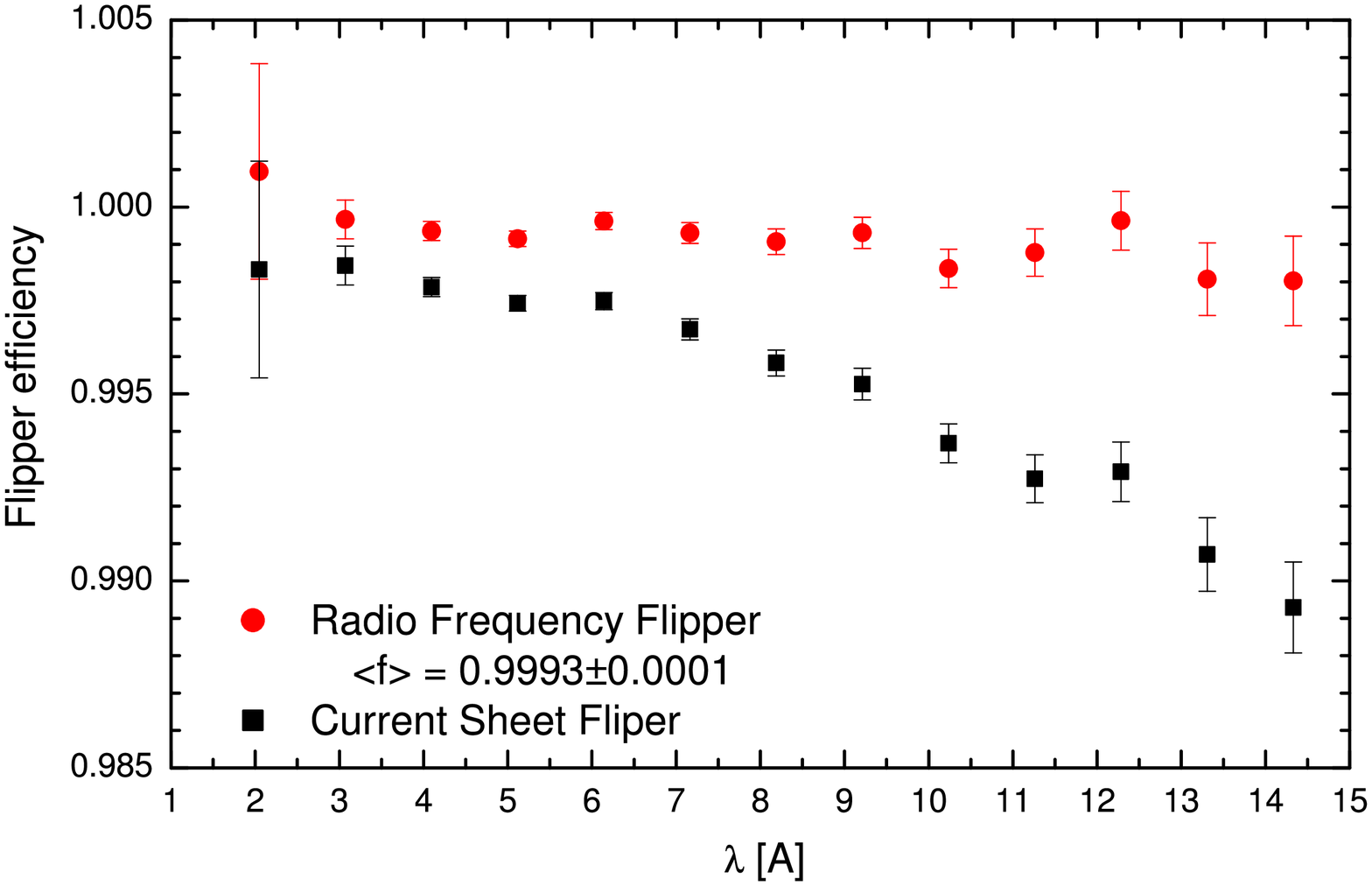}
\end{center}
\caption[]{Comparison of (Top) double SM analysed in crossed
geometry and $^3$He spin filter, (Bottom) radio frequency and
current sheet spin flipper.} \label{SM3HeFlippers}
\end{figure}
\section{Status of cold neutron beam polarization techniques}

Presently, super mirror (SM) polarizers
\cite{schaerpf1989,schaerpf1989a} are the most powerful tool to
polarize a white neutron beam. The polarization resulting from
spin dependent reflections on magnetized SMs depends on the
wavelength and the incident angle. Using this technique, a neutron
beam with a polarized intensity of $2\cdot10^9$ n/cm$^2$s and an
averaged polarization of 98.5\% is available at PF1b (Institut
Laue Langevin). However, the variation of the polarization in the
beam due to wavelength and position can be as high as 20\% and
3\%, respectively (see Fig.~\ref{WavelengthAngular}). This strong
variation limits the application of SM polarizers in spin rotation
experiments since a change of position or orientation of the
sample changes the effective neutron polarization.

In polarization analysis, only the convolution of polarizing and
analyzing power can be determined.  Therefore, for an absolute
measurement of the beam polarization - as required by neutron beta
decay experiments - the analyzing power has to be known
independently. This is not possible with SM analysers but can be
solved using opaque $^3$He spin filters
\cite{zimmer1999,zimmer1999a}. This technique provides 100\%
analyzing power and is not angle and spatial dependent. It can
therefore be used to average over the neutron beam (with several
practical problems like corrections for windows, the time
dependence of $^3$He polarization, etc.). However, it is limited
to a narrow wavelength range by a very low and wavelength
dependent transmission of less than $10^{-3}$. Moreover, for an
inhomogeneous beam, even the correct average polarization may
differ from the average seen from the experimental set-up.

A perfectly polarized neutron beam should therefore be homogeneous
on the level of the attempted experimental precision. Today, this
can be obtained neither with a single SM nor with a $^3$He
polarizer.

\section{The crossed geometry of two SM polarizers}

{\setlength{\parindent}{0pt}We propose to use double SM polarizers
in crossed geometry (Fig.~\ref{SchemeCrossed}). This geometry is
particularly important since the angular acceptance of one device
is independent of the angular acceptance of the other. Therefore,
both devices are independent. This allows to predict the
properties of the combined polarizer (polarization $P_{12}$ and
transmission $T_{12}$) from the properties of the single devices
($P_i$ and $T_i$) using the standard matrix formalism:
\begin{equation}
  P_{12}=\frac{P_1+P_2}{1+P_1P_2}\approx1-\frac{1}{2}(1-P_1)(1-P_2)
\end{equation}
and
\begin{equation}
  T_{12}=T_1T_2.
\end{equation}
Using typical SM polarizer properties ($P_i\approx98$\%), we
predict a very high polarization ($P_{12}>99.9$\%) for a wide
wavelength range and an angular variation of less then 10$^{-3}$.
The intensity should drop only to about 50\% of the single device
($T_1, T_2\approx50$\% for the ``good'' spin component).}

\section{Experimental tests}
For the experimental tests at PF1b the following set-up was used:
The polarized beam was created by double SM polarizers in crossed
geometry, followed by a radio frequency flipper ($f=50$ kHz) and a
current sheet flipper. The beam was analysed by either a single SM
analyzer, double SM analysers in crossed geometry, or a polarized
$^3$He spin filter. A time of flight set-up allowed a
wavelength-resolved analysis. All SMs were adjusted to maximum
transmission. To cover the wavelength range between 3 and 10
{\AA}, different $^3$He cells and pressures were used.

Fig.~\ref{WavelengthAngular}a (\ref{WavelengthAngular}b) compares
the wavelength (angular) dependence of the product AP of
polarization and analyzing power for the single and the crossed
analysed. Both dependencies are suppressed strongly for the
crossed geometry.

In Fig.~\ref{SM3HeFlippers}b the performances of the two flippers are
compared. The efficiency of the radio
frequency flipper was 99.93(1)\%, averaged between 2 and 15 {\AA} (only
statistical error given).

Fig.~\ref{SM3HeFlippers}a compares the AP
products obtained with the crossed analysed and with the $^3$He
spin filters. The values 99.01(5)\% and 99.60(5)\% differ
significantly. We explain this by a depolarization in the SM
analysed itself and attribute the difference between 99.60(5)\%
and 100\% to a depolarization in the SM polarizer. A
depolarization during the polarization transport can be excluded
since we use the same configuration and change only the analysed.
Moreover, any depolarization due to a non-adiabatic transport in
the magnetic field should be wavelength dependent. This was not
observed within the obtained precision of 10$^{-3}$. To check the
depolarization, we reduced the SM housing magnetic field from 35
mT to 20 mT in both, polarizer and analysed. The AP product
dropped by 1\%.

\section{Conclusions}

For absolute measurements of correlations in neutron beta decay
and for spin rotation experiments, double SM polarizers in crossed
geometry should be used to polarize the beam. For this set-up, an
average polarization of more than 99.5\% can be obtained for a
cold neutron beam. Spatial and wavelength dependencies can be
neglected on the 0.1\% level. The neutron flux is reduced by about
50\% only compared to a single SM polarizer. For such a perfect
beam the polarization can be measured with a precision of better
than 0.1\% using $^3$He spin filters. A spin flip efficiency of
more than 99.9\% can be reached with a radio frequency spin
flipper. We hope that the absolute value of polarization can be
increased in the near future using higher SM housing magnetic
fields.

\section*{Acknowledgements}
We thank H.~Humblot, D.~Jullien, and F.~Tasset for providing us
with polarized $^3$He cells, M.~Thomas for optimizing the magnetic
housing for the $^3$He cells, R.~G{\"a}hler for providing us his
nice new SM polarizers, and D.~Berruyer (all Institut Laue
Langevin) for preparing many components of the installation.


%
%

\part{Friday}

\title*{Order-$\alpha$ Radiative Corrections to Neutron, Pion and
Allowed Nuclear
$\beta$-Decays}
\toctitle{Order-$\alpha$ Radiative Corrections to Neutron, Pion and
Allowed Nuclear
$\beta$-Decays}
%
%
\titlerunning{Radiative Corrections }
%

%
\author{F. Gl\"uck}
\institute{Johannes Gutenberg University Mainz, Inst. Physics,
WA EXAKT, D-55099 Mainz, Germany,
e-mail: fglueck@uni-mainz.de;
On leave of absence from: Research Institute for Nuclear and Particle
Physics, H-1525 Budapest, POB 49, Hungary,
e-mail: gluck@rmki.kfki.hu
}
\maketitle
\begin{abstract}
  It is important to take into account the radiative corrections
in precision beta decay analysis. The photon bremsstrahlung
calculation in beta decays with small decay energy is free from
strong interaction uncertainties, it is mainly QED result. These
photons change the beta decay kinematics, it is important to take
into consideration this effect, in order to obtain meaningful
radiative corrections. The model independent part of the radiative
correction is reliable, sensitive to the experimental details, and
it is this part which changes the spectrum shapes and asymmetries.
The model dependent correction can be absorbed, to a good
approximation, into effective form factors. The dominant
asymptotic part is reliable and universal, but the smaller
non-asymptotic part contains non-perturbative strong interaction
uncertainties, and it can depend on the decay type.

\end{abstract}

\section{Introduction}

The theoretical analysis of neutron decay is more reliable
than the analysis of high energy beta decays. There are several
simplifications here:
\begin{enumerate}
  \item the $q^2$ dependence of the form factors is negligible;

  \item the effect of the vector form factor $f_3$ and the $g_2$ and
$g_3$ axialvector form factors is negligible;

  \item the value of the weak magnetism form factor $f_2$ is predicted
  by the conserved vector current hypothesis (CVC); this is tested by nuclear
  beta decays;

  \item we have $f_1(0)=1$ to high precision: this follows from the
  Behrends-Sirlin-Ademollo-Gatto theorem, and also from quark model
  computations.

\end{enumerate}

After these simplifications we have only 2 free parameters: the $up-down$
element $V_{ud}$ of the CKM-matrix, and the $\lambda=g_1(0)$
axialvector form factor.

In the 3-body pion beta decay $\pi^{\pm}\to \pi^0 e^{\pm} \nu_e$
the $q^2$ dependence is larger, but here an additional
simplification is the absence of the axialvector form factors.

In the case of the nuclear beta decays the theoretical analysis is not
so simple and reliable, due to the isospin-symmetry breaking
(charge dependent) corrections
to the matrix elements, and the larger recoil corrections in
beta decays with higher decay
energies. Nevertheless,   good nuclear structure
models are available nowadays which can be employed to compute
these corrections to the required
precision.

Low energy beta decays can be used to test some important aspects
of the Standard Model (SM): the unitarity of the CKM-matrix, the
absence of the right-handed, scalar, tensor and time-reversal
violating weak couplings. It is obvious now that any deviation
from the SM should be quite small. Therefore, very accurate
measurements are needed in order to test precisely these
deviations, and they require precise theoretical analysis.

 One of the
most important small corrections to the theoretical distributions
and quantities are the radiative corrections. These corrections
are the consequences of the presence of photons which are created
by the charged particles of the beta decay processes. These
photons can change slightly the various decay probabilities, but
some of them (the bremsstrahlung photons, which are in principle
detectable) can also change the kinematics of the decay processes.
These small changes are  important in the analysis of the
precision experiments. For example: neglecting these corrections
one could see some deviations from the Standard Model which in
reality is not present. An incomplete theoretical analysis could
result in a false interpretation of an experiment which has
otherwise no serious systematic errors.
\section{Bremsstrahlung and virtual photons}
The radiative corrections are the consequences of the interaction
of photons with charged particles. We know from classical
electrodynamics that accelerating particles create electromagnetic
field, and we know from quantum physics that this field is
quantized, with the photons representing its quantums. During beta
decay charged particles come into existence with large velocities,
and this mechanism creates photons.

It is important
to distinguish 2 different types of photons: bremsstrahlung and virtual.
The  bremsstrahlung photons
are on-shell, that is for them $E=|{\bf p}|$ (with the $c=1$ particle physics
convention), where $E$ is their energy, and ${\bf p}$ denotes their
three-momentum. In some sense, these bremsstrahlung photons are real:
they can go very far from the beta decay region, so they are in
principle detectable; although, most of these photons have very small energy,
so it is not easy to detect them.

On the other hand, the virtual photons are off-shell: for them
$E \not=|{\bf p}|$, they behave so as if they had a non-zero mass. These
virtual photons can live only for an extremely short time and within a
very small
region of space: they are constrained to the space-time vicinity of the
beta decay process.

The contribution of the bremsstrahlung photon events is represented by
Feynman diagrams where the photon is created in a vertex on a charged particle
line, and it goes out into infinity. On the other hand, the virtual
photon contributions are visualized by diagrams where the photon is created
in a vertex point on a charged particle line, and it is absorbed in another
vertex point on the same charged particle line (self-energy diagrams) or
another line (box or vertex diagrams).

The various measurable quantities of the beta decays can be calculated
from the decay amplitude
by using the diagrams of the decay processes and the Feynman rules of the
Standard Model.
Let ${\cal M}_0$ denote the decay amplitude without radiative corrections;
we call this zeroth-order amplitude. The zeroth-order part
of the measurable quantities are calculated from
$|{\cal M}_0|^2$ by summation and averaging over the spins and by
phase-space integration over the momenta of the particles participating in the
beta decay.
The ${\cal M}_{VIRT}$ virtual photon amplitude has to be added to the
zeroth-order one, since the decay process with virtual photon is
quantum mechanically indistinguishable from the decay events without photon.
On the other hand, the bremsstrahlung process could be, in principle,
distinguished from the zeroth-order process, therefore
the ${\cal M}_{BR}$ amplitude should not be added to ${\cal M}_0$, but
 the bremsstrahlung parts of the measurable quantities should be
computed by phase-space integration and spin summation of
$|{\cal M}_{BR}|^2$.

\section{Photon bremsstrahlung}

According to the Bloch-Nordsieck theorem \cite{BlochNordsieck},
bremsstrahlung photons are always present in processes where
charged particles are involved. With some finite energy resolution
of the detection system only the bremsstrahlung events with
$K>K_{min}$ photon energy can be distinguished from the no-photon
decay events. Assuming $K_{min}\sim 1 keV$ resolution energy, in
neutron decay the probabilities to get 1 and 2 photons during a
decay event are: ${\cal P}(1\gamma)\sim 0.5 \% $ and ${\cal
P}(2\gamma)\sim 0.001 \% $, respectively. Therefore, it is usually
enough to calculate the one-photon bremsstrahlung.

The photons created during the beta decay process are called
internal brems-strahlung photons. They should be distinguished
from the external brems-strahlung photons: the latter are created
during a collision process of beta electrons with  external nuclei
which are far from the beta decay point. Therefore external
bremsstrahlung is completely different from internal
bremsstrahlung.

The internal bremsstrahlung amplitude can be written as a sum of
leptonic and hadronic parts:
\begin{equation}
  {\cal M}_{BR}={\cal M}_l+{\cal M}_h,
\end{equation}
where the ${\cal M}_l$ and ${\cal M}_h$ amplitudes
 represent the diagrams with the photon
emitted by the charged lepton (here electron) and the hadrons,
respectively. The calculation of the ${\cal M}_l$ part is simply
QED, so it is always reliable. The ${\cal M}_h$ hadronic part
contains strong interaction effects, so for energetic beta decays
it is not easy to calculate reliably. This could be, however, an
advantage since for example by the measurement of the radiative
pion decay $\pi \to e\bar\nu \gamma$ one can obtain important
information about the strong interaction, and it is possible also
to restrict the weak tensor coupling constant. We mention that in
radiative pion and kaon decays the ${\cal M}_l$ amplitude together
with the point-like hadron model approximation part of ${\cal
M}_h$ is usually called inner bremsstrahlung (IB), and the
remaining is called structure-dependent contribution (SD)
 (see Ref. \cite{Bryman}, page 166).

In the case of the beta decays with small decay energy, like neutron,
$\pi_{e3}$ and nuclear beta decays, the situation is different.
For example, in neutron decay the maximum photon  energy allowed by
kinematics is $K_{max}=0.78 MeV$. We know since Yukawa that about 140 MeV
energy corresponds to 1 fermi wavelength, so in our case the minimum
wavelength of the bremsstrahlung photon is much more larger than the
dimension of the nucleons:
\begin{equation}
  \lambda_{min}^{(\gamma)}\gg \,1 fm.
\end{equation}
{\bf
 Obviously, these large wavelength photons
cannot penetrate inside the nucleons, they can see mainly their charges
(and to some extent their magnetic moment). The photon bremsstrahlung
calculation of beta decays with small decay energy is almost
model independent, therefore very reliable.}

We can see also to what extent is this photon bremsstrahlung
calculation model independent. For very small energy photon
the hadronic bremsstrahlung amplitude is proportional to the zeroth order
amplitude:
\begin{equation}
  {\cal M}_h[small\; K] \approx e\frac{(p\varepsilon)}{(pk)} {\cal M}_0
\end{equation}
where $k=(K,{\bf k})$ denotes the photon four-momentum ($K$ is the
photon energy) , $\varepsilon$ its polarization vector, and $e$ is
the electron charge. We see that for small photon energy the
bremsstrahlung amplitude is proportional to $K^{-1}$. After phase
space integration one can see that the bremsstrahlung photon
spectrum has also an order-$K^{-1}$ behavior for small $K$ (see
\cite{Schopper}, pages 76-84). The coefficient of the
$K^{-1}$-term, which is dominant for low photon energy, is model
independent. Due to the Low-theorem (see  Refs.
\cite{Low,Adler,Burnett,Fearing}), which is a consequence of the
gauge invariance of QED, not only the order-$K^{-1}$ , but also
the order-$K^0$ part of ${\cal M}_h$ can be calculated model
independently, using the zeroth-order amplitude and the
electromagnetic form factors of the hadrons involved in the beta
decay (an application of the Low-theorem to hyperon semileptonic
decays can be found in Ref. \cite{Gluck97a}). Thus only the
order-$K$ part (together with higher orders) of the hadronic
bremsstrahlung amplitude is model dependent. Compared to the
dominant order-$K^{-1}$ part this is of order $(K/m_n)^2\sim
10^{-6}$ for neutron decay. We can then conclude:

{\bf It is not possible to obtain any strong interaction dynamical
information from photon bremsstrahlung measurement in neutron decay.
The bremsstrahlung here is completely determined by QED. }

 For 3-body pion and for allowed nuclear
beta decays the above factor is larger than $10^{-6}$, but is seems
very unlikely that any structure dependent part of
the  photon bremsstrahlung could be measured in these decays.

The photon bremsstrahlung calculation is theoretically simple and
reliable. On the other hand, technically it is more complicated. In order
to compute the measurable quantities, one has to evaluate  many-dimensional
integrals like
\begin{equation}
  \int \frac{d^3{\bf p}_p}{E_p}
  \int \frac{d^3{\bf p}_e}{E_e}
  \int \frac{d^3{\bf p}_\nu}{E_\nu}
  \int \frac{d^3{\bf k}}{K}
 \delta^4(p_n-p_p-p_e-p_\nu-k) \sum_{spin} |{\cal M}_{BR}|^2.
\end{equation}

The $|{\cal M}_{BR}|^2$ expression can be evaluated by symbolic algebra
programs (like REDUCE). One has to use Dirac matrix algebra, and Lorentz-index
summation have to be performed. These symbolic formulae can be checked
by numerical calculation of the amplitude.

For the integration, there exist 2 different methods. First, it is
possible to use semianalytical integration (see Refs.
\cite{Gluck90,Gluck92}). This gives very precise results, but the
calculation of many complicated analytical integrals is rather
difficult. A more simpler method is with Monte Carlo integration.
In this case the computer evaluates the complicated integrals,
much more smaller amount of human effort is necessary with this
method. The accuracy is limited by statistics, but there is no
problem to generate $10^6$ events, and this gives already the
accuracy needed for meaningful results. The method is very
flexible: the same computer program can be used in order to
calculate radiative correction to any kind of measurable quantity.
The method is especially suitable for experimental off-line data
analysis, where the various kinematic cuts, detection efficiencies
etc. require complicated modifications of the theoretical
distributions. Using the Monte Carlo method one can also perform
weighted or unweighted event generations.

The $|{\cal M}_{BR}|^2$ integrand has large peaks for small photon energy
($K\to 0$: infrared peak), and in the case of larger decay energies
for almost parallel photon and electron momenta (collinear peak).
These peaks make the simple Monte Carlo integration inefficient
(one has to generate very large number of events). This problem
can be completely solved by importance sampling: one has to generate
more events in the phase space region where the integrand is large
(see Refs.
\cite{Gluck96,Gluck97a,Gluck97b} for different importance
sampling solutions).

In order to calculate radiative correction to some quantity, one has to
include also the very small energy photons  into the integration region.
But going with the $K$ photon energy to zero, the integral goes
logarithmically to infinity (we have seen above that the bremsstrahlung
photon spectrum behaves as $K^{-1}$ for small $K$). This
infrared divergence problem can be solved by regularization. One kind
of regularization is the following:
a small photon mass $m_\gamma$ is introduced in the photon four-momentum
formulae, and then the integrals become finite, namely
$\ln m_\gamma$ terms appear after integration. The same photon
mass regularization has to be performed for the calculations of the
virtual photon contributions, too. The virtual correction results also
contain $\ln m_\gamma$ terms, but with opposite sign and same magnitude
as the bremsstrahlung corrections.
{\bf
In the sum of the bremsstrahlung and virtual corrections
the photon mass term disappears: the complete radiative correction
is free from infrared divergence! }

If the Monte Carlo method is employed for the radiative correction
computation, an expedient method is the following: the photon bremsstrahlung
integration is divided into soft and hard regions. In the soft region
the photon energy is small, smaller than some given $\omega$ soft-photon
cutoff value;
we can use here 3-body decay kinematics,
the integrals containing the regulator photon mass can be calculated
rather simply analytically. Adding this contribution to the virtual
part, the photon mass logarithm disappears, and we get the virtual-soft
part of the radiative correction. In the hard region the photon energy
is larger than the $\omega$ cutoff, we use here 4-body kinematics, but
the photon mass can be put equal to zero in these integrals (this makes
the importance sampling method easier to apply).

{\bf The hard bremsstrahlung photons change the beta decay kinematics.
It is important to take into account this fact, in order to obtain
meaningful radiative corrections results}

In Refs. \cite{YM76,GM78}
analytical formulae were published  for the radiative
correction to the electron and neutrino asymmetry and the
electron-neutrino correlation of nuclear and neutron beta decays.
 Their result is, however, not applicable for
the precise analysis of the measurements. The reason is the
following. During the radiative  correction calculation, one has
to take into account the bremsstrahlung photons, which arise in
the beta decay with small, but non-negligible probability,
together with the electron and antineutrino. These photons change
considerably the decay kinematics. If one calculates the radiative
correction to the neutrino asymmetry and the electron-neutrino
correlation
 with electron and proton
detection,
first the electron and proton momenta have to be
fixed,  and the integrations over the other particle momenta have
to be performed with the constraint of the fixed
electron and proton momenta.
This constraint makes the analytical integrations very difficult.
On the other hand, the authors of Refs. \cite{GM78,YM76} performed the
photon
bremsstrahlung integration with fixed electron and antineutrino
momenta. In this case, the
analytical integration is possible.
Unfortunately, the proton momentum now changes with the
photon momentum during the integration, by momentum conservation.
The result of this calculation would be useful only if the
antineutrino could be detected and precisely measured.  At
present, however, this is in neutron and nuclear beta decay
measurements impossible.

We mention that the inapplicability of several published radiative
correction results for the experimental analysis of semileptonic
decay measurements was emphasized already in the eighties
\cite{Toth84,TSM86} (see also Refs. \cite{TG89,Gluck90}).
\section{Virtual correction and the UV divergence}
We have seen that both the bremsstrahlung and the virtual
correction have infrared (IR) divergence, so they are separately
not observable quantities: only their sum, where the IR divergence
disappears, has any meaning to be compared with experimental
results.

In addition to IR divergence, the virtual correction has  another
and more serious divergence: for the calculation of the virtual integrals
the virtual photon four-momentum has to be integrated over the whole
 four-dimensional phase-space region, and this makes the integrals divergent
 also for large photon energies. The UV divergence is present also
 in QED virtual integrals, but there it is possible to absorb
 the UV-divergent parts into the electron mass and charge (renormalization),
thus all the UV divergent integrals become finite: we obtain finite
radiative correction results. On the other hand,
before 1971 nobody was able to find any meaningful renormalization
procedure for the weak interactions. All the radiative corrections
of the hadronic beta decays contained ultraviolet cutoffs, both in the
four-fermion and in the vector-boson theories, and it was not possible
to get rid off this cutoff by some renormalization procedure. On the other
hand, the order-$\alpha$ correction of the muon decay within the
framework of the $V-A$ model was finite, free from the UV-cutoff.
In the sixties, after the electromagnetic nucleon form factor measurements
of Hofstaedter, several famous physicists (like Feynman, K\"allen, Berman,
Sirlin) conjectured the hypothesis that perhaps the strong interaction
could solve this UV divergency problem: for large $q^2$ the
electromagnetic form factors
go rapidly to zero, and this could make the integrals finite.
In the middle of the sixties, however, it was obvious due to
 the current algebra
investigations that the strong interaction alone cannot
solve the UV divergency problem. It was necessary to develop a
renormalizable theory of weak interactions to solve this problem.
The $SU(3)_c\times SU(2)_L\times U(1)$ non-Abelian gauge theory,
with spontaneous symmetry breaking due to the Higgs mechanism
(Standard Model), was
able to provide a renormalizable and phenomenologically realistic
model for the weak interaction. It was shown by Sirlin and others in the
seventies \cite{Lee,Sirlin74,Sirlin78} that the radiative corrections
of hadronic beta decays
in the framework of the Standard Model are free from ultraviolet divergences.

\section{The weak correction}

The Feynman diagrams of the order-$\alpha$ virtual corrections can be
divided into 2 groups: non-photonic diagrams, with $Z$ and $W$ bosons
and Higgs particle
in the loops, and photonic diagrams. The photon propagator of the photonic
self-energy diagrams can be decomposed as \cite{Sirlin74,Sirlin78}
\begin{equation}
\frac{1}{k^2}=\frac{1}{k^2-M_W^2}-\frac{M_W^2}{k^2-M_W^2}
\frac{1}{k^2}
\end{equation}

Then the whole order-$\alpha$ virtual correction can be
separated to weak and photonic parts. The weak part is defined by
the non-photonic diagrams, plus the self-energy photonic diagrams
taking the first term in the above decomposition. The photonic correction
is determined by the photonic box diagram and the
self-energy photonic diagrams with the second
term in the above decomposition. One can then see that the photonic
correction integrals are UV-finite
(with the $W$ boson mass as a natural cutoff). The weak correction integrals
are not finite, but their UV-divergence disappears after electroweak
renormalization, taking also into account the asymptotic freedom of QCD
(see Refs. \cite{Sirlin74,Sirlin78}).

The final universal result from the weak part of the order-$\alpha$
radiative correction to the decay rate is:

\begin{equation}
  r_{weak}=0.02 \%
\end{equation}

So the many weak diagrams give a very small and fairly reliable
correction result. As we will
see, the photonic correction is much more larger, and unfortunately
it is difficult to calculate reliably (with small theoretical
uncertainty), due to some non-perturbative effects.

\section{The model independent (outer) photonic correction}

The photonic virtual correction has the following general properties:
it is UV finite, IR divergent, and depends on strong interaction
models.

In the correction calculation the photon energy has to be integrated
from zero to infinity. Let us divide the whole integration region
into 3 subregions:

\begin{enumerate}
  \item small: [0,200 MeV]
  \item medium: [200 MeV, 4 GeV]
  \item asymptotic: [4 GeV,$\infty$]
\end{enumerate}

The region 'small' contains completely the IR divergence, it is almost free
from strong interaction effects, and only this part is sensitive to
the momenta and energies of the beta decay particles. It is expedient
to separate this part of the correction from the others.

This separation was introduced by Sirlin in 1967 \cite{Sirlin67}.
He did not use the above arbitrary energy boundary of the region
'small', but he defined the model independent  (MI) virtual correction
by using the convective term in the point-like hadron vertex-propagator
expression.

Then we use the following definition:

{\bf MI radiative correction = MI virtual correction +
bremsstrahlung}

The MI (model independent) radiative correction has the following
important properties:
{\bf
\begin{itemize}
  \item It has no strong interaction dependence, so it is reliable;
  \item It is sensitive to the experimental details \\
    (e.g.: the photon bremsstrahlung changes the kinematics);
  \item It changes the spectrum shapes and asymmetries.
\end{itemize}
}

Due to these properties, it is important to take into account this
correction in precision experimental analysis.

First, this correction changes the beta spectrum shape
of beta decays. Sirlin gave a universal analytical formula for the
MI correction to the beta spectrum (Eq. 20 in Ref. \cite{Sirlin67}.
For the case of neutron decay this function is tabulated in Table I
of Ref. \cite{Gluck93}. One can see that the correction to the shape
is about 1 \% for neutron decay. For pion beta decay and for
allowed nuclear beta decays with higher decay energy this correction
is larger. In $E_e\to E_{emax}$ limit the correction has a
logarithmic singularity, but for beta decays with decay energy below
10 MeV this is practically not relevant.
 It is important to take into account this correction
for accurate weak magnetism, second class form factor, and Fierz
term analysis of beta spectrum shapes.

The MI correction to the proton spectrum in neutron decay
can be found in Table IV of Ref. \cite{Gluck93}. This correction is
about 0.1 \%, but it changes the $\lambda=G_A/G_V$ fitted parameter
by 0.01 (the 'aspect' project aims to measure this $\lambda$ by 0.001
precision; see Ref. \cite{Zimmerglu}).

The MI radiative correction to the recoil spectrum shapes in
${}^6 He$ and ${}^{32} Ar$ beta decays was calculated in Ref.
\cite{Gluck98}. This correction shifts the fitted electron-neutrino
correlation parameter $a$ by 0.003; this is approximately the experimental
error of the most precise $a$ value measurements of these decays.

It is important to take into account the MI correction also to higher energy
beta decays. For example, this correction changes the fitted
$\lambda=g_1/f_1$ ratio
 in  $\Lambda\to pe\nu$ decay by 0.03 (see Ref. \cite{Gluck94}),
  which is twice larger than the
 experimental error of the most precise measurement.

The order-$\alpha$ MI corrections to the total decay rates of
some beta decays are
the following \cite{Gluck97b}:
\begin{itemize}
  \item neutron decay: $r_{MI}=$ 1.50 \%
  \item $\pi_{e3}$ decay: $r_{MI}=$ 1.04 \%
  \item $O^{14}\to N^{14}$ decay: $r_{MI}=$ 1.29 \%
\end{itemize}

The order-$\alpha$ MI corrections to the various asymmetry
quantities in polarized beta decays are usually rather small
(see Refs. \cite{Sirlin67,Shann,Yokoo,Gluck92,Gluck96a,Gluck98a}).

\section{The model dependent (inner) photonic correction}

While the MI photonic correction can be visualized as integrating the
virtual photon in the small energy region, and therefore this
correction is sensitive to the external particle momenta, the
model dependent (MD) correction gets the main contribution from the large
photon energy region, therefore it is not sensitive to the particle
momenta. For example, let us assume that a virtual electron
with $p_e'$ momentum emits a virtual photon with four momentum $k$
(this photon is then absorbed by the beta decay hadrons), and
it goes into the external electron (momentum: $p_e$).
From momentum conservation at the electron-photon vertex:
$p_e'=k+p_e$. The virtual correction integrand contains the electron
propagator depending on the $p_e'$ virtual electron momentum. If
$k$ is large, the small $p_e$ momentum can be neglected in $p_e'$, therefore
the integral is practically independent of the external electron momentum
$p_e$.

We can thus conclude that:

{\bf The model dependent correction is not sensitive to the external
beta decay particle momenta and to the experimental (kinematical) details.}

Sirlin proved in Ref. \cite{Sirlin67} the following theorem:

{\bf
Neglecting terms of order
\begin{equation}
 \alpha \frac{E_e}{m_n} \ln \left( \frac{m_n}{E_e}\right)
\sim 10^{-4} - 10^{-5},
\end{equation}
the MD correction can be absorbed into the $f_1$ vector and $g_1$ axialvector
form factors.
}

We can introduce effective form factors:
\begin{equation}
  f_1'=f_1\left( 1+\frac{\alpha}{2\pi}c\right), \quad
  g_1'=g_1\left( 1+\frac{\alpha}{2\pi}d\right).
\end{equation}
The model dependent correction is then defined by 2 numbers: $c$ and $d$.
For a given beta decay one can use the same effective form factors for all
measurable quantities. The $G_V$ weak vector coupling
and the $\lambda$ parameters can then be
redefined as
\begin{equation}
  G_V=G_\mu V_{ud}f_1',\quad \lambda=g_1'/f_1'
\end{equation}
All measurable quantities for a given beta decay depend on the same
$G_V$ and $\lambda$ parameters. For example,
in neutron decay the $\lambda$ parameter can be determined
 from the electron asymmetry
$A$ and from the electron-neutrino correlation parameter $a$. Let us denote
these values as $\lambda_A$ and $\lambda_a$. Within the framework of the
Standard Model we need: $\lambda_A=\lambda_a$. The comparison of these
measured parameters provides a sensitive test of the $V-A$ structure of the
Standard Model; in the presence of scalar or tensor couplings these
parameters should not be necessary equal. What is important here: this test is
independent of the model dependent correction; never mind what are the $c$
and $d$ MD corrections, we have to get the same $\lambda$ values
for a fixed beta decay type within the
SM. Unfortunately, the MD corrections and thus the
effective form factors could be different for the
different beta decay types (see below).

The $V_{ud}$ element of the CKM matrix can be determined from the
measured $G_V$ and $G_\mu$ couplings, and from the calculated
MD correction $c$.

{\bf The precise and reliable calculation of the $c$ model dependent correction
of the vector coupling is important for the $V_{ud}$ determination, and
thus for the CKM unitarity test}.

The leading asymptotic behavior in $M_Z$ of the order-$\alpha$
radiative correction amplitude to an arbitrary semileptonic decay
is given by the universal formula \cite{Sirlin82}
\begin{equation}
  {\cal M}_{as}=\frac{\alpha}{\pi} \ln \left(\frac{M_Z}{m_p}\right)
   {\cal M}_0,
\end{equation}
where $M_Z$ and $m_p$ are the Z-boson and proton masses.

This part of the model dependent correction depends on the high
energy behavior of QCD (asymptotic freedom), so it contains no
uncertainties from non-perturbative strong interaction models; it
is therefore in fact 'model independent'. The total model
dependent correction amplitude can be written as

\begin{equation}
  {\cal M}_{MD}={\cal M}_{as}+{\cal M}_{med}.
\end{equation}

Here the ${\cal M}_{med}$ part comes from the integration over the medium
energy range, where the photon energy is around 1 GeV. In this region the
virtual photon can have a large effect to the inner structure of the hadrons,
therefore this part of the correction depends very much on strong interaction
models. Nevertheless, ${\cal M}_{med}$ does not contain  any
large logarithm, so it is expected that the asymptotic part is dominant.

For total decay rates the asymptotic part gives a correction of

\begin{equation}
  r_{as}=\frac{2\alpha}{\pi} \ln \left(\frac{M_Z}{m_p}\right)=2.1 \%
\end{equation}
(we have to take the interference between the zeroth order and the
virtual amplitudes: $r_{as}=2{\cal R}e({\cal M}_{as}^* {\cal M}_0)/
|{\cal M}_0|^2$).
This part of the MD correction is universal: it is the same for all
beta decays.

The higher order terms of this asymptotic correction was estimated
by Marciano and Sirlin \cite{Marciano} using renormalization group
analysis. The asymptotic correction with the higher order terms is

\begin{equation}
  r_{as}^{h.o.}=2.25 \%
\end{equation}

The medium energy ( non-asymptotic) model dependent correction
to the decay rate was estimated
also by Marciano and Sirlin \cite{Marciano}:

\begin{equation}
  r_{med}=0.12 \pm 0.2 \%
\end{equation}
(see also Refs. \cite{Sirlin78,Sirlin95,Jaus,Towner98,Towner02}).

It is unlikely that the real $r_{med}$ correction should be
completely different from this result. Nevertheless,
this part of the radiative correction is the least reliable, and
probably further theoretical investigations are necessary
in order to obtain a better understanding of this correction.
It should be emphasized also that the $r_{med}$ correction is
very likely beta decay type dependent: it is different in neutron decay
and in pion beta decay (in the first case, the virtual photon
interacts with nucleons, in the second case with pions: these are completely
different hadrons).


\title*{Beyond $V_{ud}$ in Neutron Decay \protect\newline}
\toctitle{Beyond $V_{ud}$ in Neutron Decay}
%
%
\titlerunning{Beyond $V_{ud}$ in Neutron Decay}
%
\author{S. Gardner}
\authorrunning{S. Gardner}
%
%
\institute{Department of Physics and Astronomy, University of Kentucky,
\\ Lexington, KY 40506-0055 USA}

\maketitle              

\begin{abstract}
Precision measurements of neutron decay observables
impact a broad array of ``new'' physics searches\index{abstract}.
I discuss how the correlation coefficients of neutron $\beta$-decay,
the neutron lifetime, and
studies of neutron radiative $\beta$-decay can
impact searches for non-V-A currents
and new sources of CP violation.
\end{abstract}

\section{Introduction}

Precision studies of nuclear $\beta$-decay
have played a crucial role in the rise of Standard Model (SM).
More recently, precision studies of
neutron $\beta$-decay have been realized as well;
the thrust of these efforts
has been the determination of the
Cabibbo-Kobayashi-Maskawa (CKM) matrix element
$V_{ud}$ to realize, in concert with the empirical values of
$V_{us}$ and $V_{ub}$, a test of the unitarity of the CKM matrix.
The value of $V_{ud}$ is extracted from the vector coupling constant
of the nucleon, $f_1$, which is determined
from the neutron-spin--electron-momentum correlation $A$ and the
neutron lifetime $\tau_n$.

In this note we consider the systematic determination of all of the
couplings of the nucleon weak current, and how such extractions
lead to SM tests beyond that of CKM unitarity~\cite{Zhang}. In this context
we discuss tests of the conserved-vector-current hypothesis (CVC)~\cite{feyngm}
as well as the search
for second-class currents (SCC)~\cite{wein},
and, generally, the search for non-V-A currents.
In addition, we consider ancillary, low-energy SM tests
accessible through neutron radiative $\beta$-decay ---
and offer perspective on how the low-energy SM tests we enumerate
complement experiments at much higher energies.

\section{Framework}

The differential decay rate for polarized neutron $\beta$-decay
in the SM is given by
\begin{eqnarray}
d^3\Gamma=\frac{1}{(2\pi)^52m_B}(\frac{d^3{\mathbf p}_p}{2E_p}
\frac{d^3{\mathbf p}_e}{2E_e}\frac{d^3{\mathbf p}_\nu}{2E_\nu})
\delta^4(p_n-p_p-p_e-p_v )\frac{1}{2}\sum_{spins}|{\cal M}|^2\,,
\end{eqnarray}
where the transition matrix element ${\cal M}$ to leading order in the
weak interaction is
\begin{eqnarray}
 {\cal M}=\frac{G_F}{\sqrt{2}}\langle p(p_p)|J^\mu(0)|{\vec n}(p_n,P) \rangle
[\bar{u}_e(p_e)\gamma_\mu(1+\gamma_5)u_\nu(p_{\nu})]\;,
\end{eqnarray}
and the most general form of the hadronic
weak current, consistent with its $V-A$ structure, is~\cite{murph}
\begin{eqnarray}
\langle p(p_p)|J^\mu(0)|{\vec n(p_n,P)}\rangle=
\bar{u}_p(p_p)(f_1(q^2) \gamma^\mu -i \frac{f_2(q^2)}{M_n}\sigma^{\mu\nu}q_\nu
+\frac{f_3(q^2)}{M_n}q^\mu \nonumber \\
+g_1(q^2)\gamma^\mu\gamma_5 - i
\frac{g_2(q^2)}{M_n}{\sigma^{\mu\nu}}\gamma_5q_\nu+
\frac{g_3(q^2)}{M_n}\gamma_5 q^\mu)u_{\vec n}(p_n,P) \,,
\end{eqnarray}
where $\sigma^{\mu\nu} = i[\gamma^\mu,\gamma^\nu]/2$,
 $q\equiv p_n-p_p$, and
$u_{\vec n}(p_n,P)\equiv ({1+\gamma_5 {\slash\!\!\!\! P}}) u_{n}(p_n)/2$
for a neutron with polarization $P$.
Defining $f_i\equiv f_i(0)$,
we adopt the ``historic'' sign convention $\lambda \equiv g_1/f_1 >0$
for consistency with earlier literature.
In the SM, under an assumption of isospin symmetry,
we note that $f_1=(1+ \Delta_R^V)V_{ud}$, where
$\Delta_R^V$ is a small, radiative correction~\cite{sirlin},
the weak magnetism
term $f_2$ is given by the isovector anomalous magnetic moment,
as per the CVC hypothesis~\cite{gell58}, and both
$f_3$  and $g_2$ are zero. The chiral properties
of QCD determine $g_3$ in terms of other low-energy constants.
Isospin is merely an approximate symmetry, so that
$f_2, f_3,$ and $g_2$ suffer corrections of ${\cal O}(R)$, where
$R\approx (M_n - M_p)/\hat{M}$ and
$\hat{M}$ is the average neutron-proton mass.

Rewriting the differential decay rate
in terms of the correlation coefficients, we find~\cite{jtw}
\begin{eqnarray}
d^3 \Gamma \propto E_e |{p}_e| (E_e^{\rm max} -
E_e)^2 \nonumber \\
&{}\hspace{-32mm}\cdot \Big[1 + a \frac{{p}_e \cdot {p}_{\nu}}{E_e
E_{\nu}} + {P}\cdot (A \frac{{p}_e}{E_e} + B
\frac{{p}_{\nu}}{E_{\nu}} + D \frac{{p}_e\times{{p}_\nu}}{E_e
E_\nu})\Big] dE_e d\Omega_e d\Omega_\nu \,.
\end{eqnarray}
Neglecting recoil-order corrections, i.e., of ${\cal O}(R)$ with
$R\equiv E_e^{\rm max}/M_n\sim 0.0014$, as well as the pseudo-T-odd
term $D$, we have
\begin{eqnarray}
a= \frac{1 -\lambda^2}{1 + 3\lambda^2} \hspace{0.8cm}
A= 2\frac{\lambda(1 -\lambda)}{1 + 3\lambda^2} \hspace{0.8cm}
B= 2\frac{\lambda(1 +\lambda)}{1 + 3\lambda^2} \,,
\end{eqnarray}
implying $1+ A - B -a=0$ and $aB-A-A^2=0$~\cite{most76}, which test the
V-A structure of the SM to the level of recoil-order terms, or
corrections of ${\cal O}(1\%)$. The relations are satisfied
within current empirical errors~\cite{pdg02gard}.
With the measurement of the neutron lifetime,
$\tau_n\propto f_1^2 + 3 g_1^2$, $f_1$ and $g_1$ are
determined independently.
In the next generation of neutron decay measurements,
the coupling constants contained within the recoil-order corrections
should become experimentally accessible.
In specific, $f_2$, $g_2$, and $f_3$ appear, though the extraction of
$f_3$ seems unlikely. Note that
$g_3$ does not appear in this order; rather, it can be determined through
$\mu$ capture on the proton. The determination
of these coupling constants
permit not only a test of CKM
unitarity, but also a test of the CVC hypothesis and of the absence
of SCC, to the level of SM isospin violation. The measurement of
$g_3$ tests the chiral structure of QCD.
Before discussing these tests in greater detail, let us consider
how we might interpret failure ---
the indirect nature of SM tests at low energies imply that
many possibilities exist. For example, the failure of CKM
unitarity could suggest the existence of a fourth generation;
however, it is also possible that the effective Fermi constant which
characterizes the weak decay of the $d$ quark is distinct from the
Fermi constant determined in $\mu$ decay.
This can occur, e.g., in models with
TeV-scale extra dimensions in which the chiral fermions of the SM
sit at different locations in a
thick brane~\cite{Arkani-Hamed:1999dc,Rizzo:2001cy,ChangNg}.
In contrast,
the failure of a CVC/SCC test in a SM framework
could be interpreted as evidence for a non-V-A current, such as
a scalar contact interaction, as generated by a scalar leptoquark,
would provide. A $\beta$-decay constraint of this ilk would be
complementary to direct searches for such particles, as once
discussed in the context of apparent, anomalous charged-current events
at HERA~\cite{Altarelli:1997qu,Babu:1997tx,Hewett:1997ba}.

\subsection{Recoil-Order Corrections}

Let us consider how the
form of $a$ and $A$ in recoil order
permit the determination of $f_2$ and $g_2$, to test the CVC
hypothesis and the absence of SCC.
Defining $x \equiv {E_l}/{E^{max}_l}$,
$\epsilon \equiv ({M_e}/{M_n})^2$,
and
$R \equiv {E^{max}_l}/{M_n}=({M^2_n+ M^2_e-M^2_p})/{2M^2_n}$,
so that $\sqrt{\epsilon}/R \leq{x}\leq 1$,
we have~\cite{svgchi}
\begin{eqnarray}
a = \frac{1-\lambda^2}{1+3\lambda^2}
+\frac{1}{(1+3\lambda^2)^2}\Bigg\{
\frac{\epsilon}{Rx} \Big[(1 - \lambda^2)(1 + 2\lambda
 + \lambda^2
 + 2\lambda \tilde g_2 + 4\lambda \tilde f_2 - 2 \tilde f_3)
\Big] \nonumber \\
 \hspace{7mm} + 4R\Big[ (1+\lambda^2)(\lambda^2 + \lambda
 + 2\lambda(\tilde f_2 + \tilde g_2))\Big]
 - Rx\Big[
3(1+3\lambda^2)^2 + 8\lambda(1+\lambda^2) \nonumber \\
\hspace{7mm}\times(1+2\tilde f_2) + 3(\lambda^2 -1)^2 \beta^2
\cos^2 \theta \Big] \Bigg\} + {\cal O}(R^2,\epsilon)\;,
\label{recoila}
\end{eqnarray}
where $\tilde f_i\equiv f_i/f_1$ and $\theta$ is the angle between
the electron and neutrino momenta in the neutron rest frame.
The expression for $A$ in
recoil order is determined by integrating over the neutrino
variables, to yield~\cite{svgchi}
\begin{eqnarray}
A = \frac{2\lambda(1 - \lambda)}{1+3\lambda^2}
+\frac{1}{(1+3\lambda^2)^2}\nonumber \\ \hspace{8mm}\cdot \Bigg\{
\frac{\epsilon}{Rx} \Big[4\lambda^2(1-\lambda)(1 + \lambda
 + 2\tilde f_2)
+ 4 \lambda(1 - \lambda)(\lambda \tilde g_2 - \tilde f_3)
\Big] \nonumber \\
 \hspace{8mm}+ R\Big[
\frac{2}{3} (1+\lambda + 2(\tilde f_2 + \tilde g_2))(3\lambda^2 +
2\lambda -1) \Big]
  + Rx\Big[
\frac{2}{3} (1+\lambda + 2\tilde f_2) \nonumber \\
\hspace{8mm}\times(1 - 5\lambda - 9 \lambda^2  -3\lambda^3) +
\frac{4}{3} \tilde g_2 (1+\lambda + 3 \lambda^2 + 3 \lambda^3)
\Big] \Bigg\} + {\cal O}(R^2,\epsilon)\;. \label{recoilA}
\end{eqnarray}
If $\tilde g_2=\tilde f_3=0$, these expressions are in agreement with
Ref.~\cite{bile60}; we agree with Ref.~\cite{holstein74} as well. The
${\cal O}(R)$ corrections, which we term $a^{(1)}$ and $A^{(1)}$,
are plotted in Fig.~\ref{recoil1}. The $\epsilon/Rx$
terms in $a$ and $A$ are comparably small, so that the
determination of $\tilde f_3$ is infeasible.
\begin{figure}[b]
\vspace{0.55cm}
\begin{center}
\includegraphics[scale=0.40]{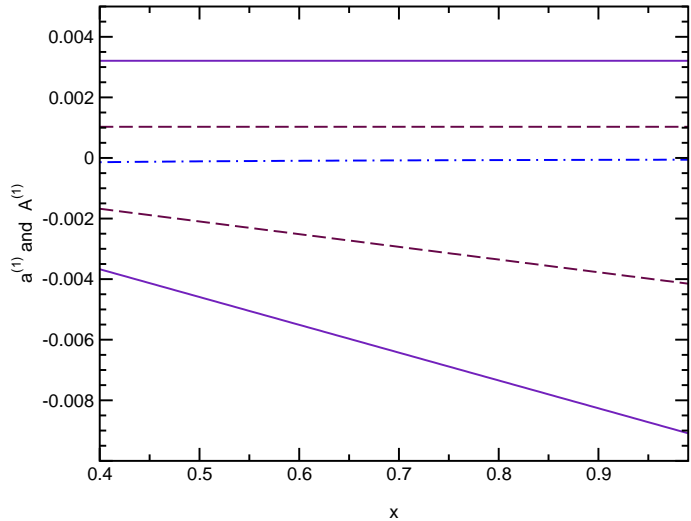}
\end{center}
\vspace{-0.4cm}
\caption[]{The leading recoil-order corrections to $a$ and $A$,
$a^{(1)}$ and $A^{(1)}$, using $\lambda=1.2670$, $\tilde g_2=0$,
$\tilde f_2=(\kappa_p  - \kappa_n)/2 = 1.8529$, and $\theta=0$.
The solid (dashed) lines indicate the $Rx$ and $R$ terms in $a$ ($A$); the
dashed-dot line indicates the $\epsilon/Rx$ term in $a$.}
\label{recoil1}
\end{figure}
To test CVC in neutron decay, that is, to determine whether $\tilde f_2$ is
given by the isovector magnetic moment~\cite{remark},
the coefficients multiplying
$Rx$ must be determined; $f_2$ and $g_2$
can then be determined independently.
A new method of measuring $a$
can determine its $x$ dependence, and a ${\cal O}(1\%)$ measurement
is possible~\cite{nista}, whereas a new experiment to measure $A$
can potentially realize ${\cal O}(0.1\%)$ accuracy~\cite{ucnA}.
A search for ``new'' physics implies that not only must the values of
$\tilde f_2$ and $\tilde g_2$ be compared to SM expectations, but
the structure of Eqs.~(\ref{recoila},\ref{recoilA})
must also be tested. To realize this, it is crucial to
determine $\tilde f_2$ and $\tilde g_2$ a plurality of ways, i.e.,
through different combinations of the $R$ and $Rx$ terms.
Were $a$ and $A$ both measured to ${\cal O}(0.1)\%$,
and $\tilde f_2$ and $\tilde g_2$ determined using the $E_e$-dependent terms,
then $\delta \tilde f_2$ would be $2.5\%$
and $\delta \tilde g_2$ would be $0.26\lambda/2$~\cite{svgchi}.
Currently the most precise nuclear test of CVC/SCC is in the mass-12 system, in
specific, through the measured difference of the
$e^\pm$ angular distributions
for purely aligned $1^+\to 0^+$ transitions,
in $^{12}B (e^-)$ and $^{12}N (e^+)$ decay to
$^{12}C$. This quantity is sensitive to both the weak magnetism and
induced tensor
terms in the nucleon weak current, though, unfortunately, the difference
in the axial charge of the two mirror transitions, induced by
isospin violation, enters as well.
In the context of the impulse
approximation, adopting the axial charge difference
$\Delta y = 0.10 \pm 0.05$
computed in Ref.~\cite{moritay}, one finds
$0.01\lambda \le 2\tilde g_2 \le 0.34 \lambda$,
if CVC is assumed~\cite{mina00}. The next generation of neutron
decay experiments can test CVC/SCC with comparable
accuracy and fewer assumptions than in the most precise nuclear
test; CVC and SCC can also be tested independently.

As $\tilde g_2$ is of ${\cal O}(R)$ in the SM,
the ability to extract $\tilde f_2$ and $\tilde g_2$ from the
recoil corrections to $a$ and $A$, as per
Eqs.~(\ref{recoila},\ref{recoilA}),
relies on the numerical magnitude
of the next-to-leading order (NLO) recoil
corrections~\cite{morita75,holstein74,Zhang,svgchi2}, i.e.,
on terms of ${\cal O}(R^2,\epsilon)$.
The uncertainties in these expressions are of particular interest.
The ${\cal O}(\alpha)$ ``outer''
radiative corrections to $a$ and $A$ must also be applied, but are
known~\cite{sirshan}. In NLO in recoil,
the momentum transfer dependence of the form factors,
as per $f_i(q^2)=f_i[1+\xi_i{q^2}/{M^2_n}+...]$,
as well as the induced pseudoscalar coupling constant $g_3$,
enter for the first time.
Only $f_1$ and $g_1$ appear in the allowed terms,
so that only the $q^2$-dependences of these form factors are
needed. However, the $q^2$-independent contributions
which emerge from the recoil expansion of the
terms in the hadronic matrix element and accompanying phase space
dominate the NLO correction, and, in turn,
the error in $g_3$ dominates the error in the NLO recoil
correction. In our analysis we use the value of $g_3$ determined
using heavy-baryon chiral perturbation theory (HBChPT)~\cite{bernard1}.
We conclude that the NLO recoil corrections, en masse, including
an estimate of the correction to $\tilde f_2$ due to isospin violation,
are sufficiently
small that the LO recoil analysis can be used to bound
$\tilde g_2$ until it is commensurate with its SM value~\cite{Zhang,svgchi2}.

\subsection{Muon Capture}

In this section we compare the
theoretical result for $g_3$
we have employed~\cite{bernard1}
with the empirical
value of $g_3$ determined in muon capture.
Both ordinary muon capture (OMC), $\mu^- p \to \nu_\mu n$,
and radiative muon capture (RMC), $\mu^- p \to \nu_\mu n \gamma$,
on the proton permit the empirical determination of $g_3$.
The kinematics of RMC
can approach the $\pi$-pole more closely and as such
is more sensitive to $g_3$. However, RMC yields a value
of $g_3$ which is roughly 1.5 times larger than the HBChPT prediction,
whereas the
value determined in OMC is marginally
consistent with theory~\cite{gorringe}.
In specific, $g_p^{\rm th} =8.44 \pm 0.23$~\cite{bernard1}, where
$g_P=-m_\mu g_3(q^2=-0.88 m_\mu^2)/\hat{M}$ and $m_\mu$ is the
muon mass, whereas an updated analysis~\cite{gorringe} of the
TRIUMF RMC experiment~\cite{wright} yields
$g_p^{\rm RMC} = 12.4 \pm 0.9 \pm 0.4$.
Extensive discussion of various
resolutions exist in
the literature, including molecular effects, (isospin-violating)
electromagnetic effects, as
well as criticisms of the
computation of hadronic matrix elements, in particular of the
inclusion of $\Delta$ degrees of freedom.
No satisfactory explanation of the OMC and RMC data has been found to
date~\cite{bernard2,ando,gorringe}; interestingly, the predicted photon
energy spectrum in RMC determined in HBChPT\cite{rmc_hbchpt}
compares favorably with
a reorganized chiral treatment in which the effects of the
$\Delta$-resonance are included explicitly~\cite{bernard2}.

\section{Neutron Radiative $\beta$-Decay}

We now turn to the discussion of the SM tests possible in neutron
radiative $\beta$-decay. We consider two distinct tests:
the determination of the weak transition form factor in
$n\to p e^- \bar \nu_e \gamma$ to test the theoretical framework
in which RMC is analyzed, as well as
the determination of a ``T-odd'' correlation, sensitive
to new sources of CP violation.

\subsection{Weak Radiative Transition Form Factor}

The photon in neutron radiative $\beta$-decay can be produced
in bremsstrahlung from either charged particle in the final state, or
it can be emitted from the effective weak vertex. The bremsstrahlung
contributions form a portion of the ``outer'' radiative corrections
in neutron $\beta$-decay~\cite{sirshan}, and the measurement of the photon
energy spectrum in neutron radiative $\beta$-decay has been suggested
as a means of testing a portion of those corrections~\cite{gaponov}.
Rather, we believe the measurement of the photon energy spectrum
offers a useful way of testing the theoretical frameworks
used to analyze RMC, be it HBChPT~\cite{rmc_hbchpt} or
reorganized HBChPT with an explicit $\Delta$ degree of
freedom, for which the $N$ and $\Delta$ splitting is presumed
small~\cite{bernard2}.
From this perspective, the bremsstrahlung
photon emitted by the charged lepton
is of lesser interest; rather the focus is on
the $VV$ and $VA\,$ 4-point functions which must be calculated
to determine the photon contribution from the hadronic side of the
decay. In the static $W^-$ approximation, appropriate to the
low-energy process we consider, the latter includes both
bremsstrahlung from the proton as well as photon emission from the effective
weak vertex. Nevertheless, the computed
4-point functions in concert with the leptonic bremsstrahlung contribution
yields a prediction of the photon energy spectrum~\cite{us};
its measurement is planned at NIST~\cite{nistg}.

\subsection{``T-odd'' Correlation}

Radiative neutron $\beta$-decay admits the possibility of a
pseudo-T-odd, P-odd correlation of the form
$p_p\cdot (p_e\times p_\gamma)$, as discussed
in  $K^+\to \pi^0 e^+ \nu_e \gamma$ decay~\cite{braguta}.
Although such a contribution
can be generated by electromagnetic final-state
interactions in the SM, it also can be generated by
new CP-violating phases in the first-generation,
charged-current interaction~\cite{gg},
paralleling the discussion of the
transverse muon polarization in $K^+ \to \mu^+ \nu_\mu \gamma$
decay~\cite{specialK}.

\section{Summary}

Precision, neutron-decay experiments are key to a rich array of
low-energy, SM tests, providing useful constraints on the
appearance of physics beyond the SM. We have considered
the systematic determination of all of the couplings
of the hadronic weak current and the SM tests such a determination
entails, focusing on those tests alternative to the extraction of
$V_{ud}$ and a test of the unitarity of the CKM matrix. We have
considered how the determination of the coupling constants contained
within the recoil corrections to the correlation coefficients
$a$ and $A$ lead to independent tests of the CVC hypothesis as
well as of the absence of SCC; such tests can be effected with
fewer assumptions and with comparable precision to the most
precise tests in nuclei. We have also considered how neutron
radiative $\beta$-decay can be used to test the theoretical
framework employed to extract the induced pseudoscalar coupling
constant $g_3$ from
RMC, as well as to search for new sources of CP violation.

\section*{Acknowledgements}
I thank the organizers for the opportunity to speak at a very enjoyable
meeting. The work presented in this talk has been done in collaboration
with Chi Zhang, whose efforts I gratefully acknowledge, and is
supported by the U.S. Department of Energy
under contract \# DE-FG02-96ER40989.

%

\def \bfgr #1{ \mbox {{\boldmath $#1$}}}

\title*{Radiative Corrections to the Neutron $\bfgr{
\beta}$-Decay Within the Standard Model
 and Their Role \\ for Precision Tests of the CKM-Unitarity}
\toctitle{Radiative Corrections to the Neutron $\bfgr{
\beta}$-Decay Within the Standard Model
 and Their Role for Precision Tests of the CKM-Unitarity}
 \titlerunning{Radiative Corrections to the Neutron $\bfgr{
 \beta}$-Decay}
\author{G.  G.  Bunatian \thanks{Email: bunat@cv.jinr.dubna.su}}
\institute{Joint Institute for Nuclear Research, 141980, Dubna,
Russia}
 \maketitle
\begin{abstract}
Radiative corrections to the neutron $\beta -$decay are calculated with
consistent allowance for electroweak interactions accordingly the
Weinberg-Salam theory. The effect of strong interactions is parameterized by
 introducing the weak nucleon transition current. The radiative
corrections to the total decay probability $W$ and to the asymmetry
coefficient of the electron momentum distribution $A$ constitute:
$\delta{W}{\approx}8{\%} \, , \, \, \delta{A}{\approx}-2{\%}$. The
accuracy attainable in the calculation proves to be $\sim{0.1{\%}}$
\end{abstract}

Nowadays, it has been well realized that careful, thorough and
all-round study of the neutron $\beta -$decay conduces to gain an
insight into the physical gist of semiweak processes and into the
elementary particle physics in general \cite{d}. In particular, the
 $CKM$ unitarity
\begin{equation}
|V_{ud}|^2 +|V_{us}|^2 +|V_{ub}|^2 =1 \label{skm}
\end{equation}
is to be verified as strictly as possible \cite{ckm1}.
 That is why a grate deal of efforts has been directed past decade
to measure, with a high accuracy, better than ${\sim}1\%$, the main
characteristics of the $\beta -$decay of free neutrons: the lifetime
$\tau$ \cite{t}, the asymmetry factors (as a neutron is polarized)
of the electron and antineutrino momentum distributions, $A$
\cite{a} and $B$ \cite{ba} respectively, the recoil proton
distribution and the electron-antineutrino correlation coefficient
$a$ \cite{p}. Further experiments are believed to come to fruition
before long.

In the report presented, the calculation of radiative corrections is
based on the electroweak Lagrangian
\begin{equation}
{\cal L}^{EW}( e , M_Z , M_W , m_f , A_{\mu} , W_{\mu}^{\pm} ,
Z_{\mu} , \psi_f , \xi =1)
\label{l1}
\end{equation}
thoroughly elaborated in several review articles and books
\cite{d,ao,h1,h2,b}, and we pursue the methods expounded in these
references. The electric charge $e=\sqrt{{\alpha}4{\pi}}$ and the
masses of particles are chosen as the bare physical input parameters
together with the physical fields $ A_{\mu} , W_{\mu}^{\pm} ,
Z_{\mu} ,
\psi_f $, the Feynman gauge, $\xi =1$, is presumed. This Lagrangian
specifies the propagators of gauge boson, quark and lepton fields, $
\; \; D^{A,W,Z}_{\mu\nu}(k^2) \, , \, \; G_f(p) \, , \; $ and
interactions between these fields
\begin{equation}
 {\cal L}^{EW}_{int} ={\cal L}^{WWZ} +{\cal L}^{WWA} +{\cal L}^{Wff}
+{\cal L}^{Zff} +{\cal L}^{Aff}, \label{l2}
\end{equation}
where $A , W , Z$ stand for electromagnetic, $W-$ and $Z-$boson
fields, and $f$ renders various kinds of fermions. In further
calculating the neutron $\beta -$decay amplitude in the one-loop
approximation, we leave out the effect of Higgs-fermion interactions
since they are of order $\sim {m_f}{/}{M_W} \ll 1$, the Higgs
coupling to fermions \cite{ao,h1,h2,b}. The multiplicative
renormalization is carried out, $g\rightarrow
g{\cdot}z_1^{W,Z}{\cdot}(z_2^{W,Z})^{-3/2} \; , \; \;
\Phi\rightarrow
\Phi{\cdot}(z_2^{\Phi})^{1/2} \; , \; \;  z=1+\delta{z}$ , accordingly the
widely-applied on-mass-shell (OMS) renormalization scheme, $g$ and
$\Phi$ represent generically couplings and fields \cite{ao,h1,h2,b}
. The original Lagrangian
(\ref{l1}) written in terms of the bare physical quantities is then
decomposed into tree and counter terms as usually \cite{ao,h1,h2,b}
\begin{eqnarray}
{\cal L}^{EW} \Rightarrow {\cal L}_{tree}^{EW}(e , M_W , M_Z , m_f
, \Phi ) \nonumber \\ \hspace{11mm}+ {\cal L}^{EW}_{ct}(e , M_W ,
M_Z , m_f , \Phi , \delta M^2_W , \delta M_Z^2 , \delta m^2_f ,
\delta z_{1,2}^{\Phi}). \label{ct}
\end{eqnarray}
Upon renormalizing, not only the ultraviolet divergencies, occurring
in the loop expansion, are absorbed into the infinite parts of the
renormalization constants, but also the finite parts of radiative
corrections are fixed. These lead to physically observable
consequences.

The neutron $\beta$-decay
\begin{equation}
n\Longrightarrow p+e^{-} +\bar\nu +\gamma \label{n}
\end{equation}
can never be reduced to the pure quark decay $u\Rightarrow d+e^{-}
+\bar\nu +\gamma$. As the nucleon is a complex composite system,
strong quark-quark interactions must be properly taken into
consideration. The total Lagrangian to describe (\ref{n}) is the sum
of electroweak and strong $qq-$interactions
\begin{eqnarray}
{\cal L}_{int}(x)={\cal L}^{EW}_{int}(x)+{\cal L}^{qq}_{str}(x) , \;
\; \; \; {\cal L}_{int}(x)\longrightarrow 0 \; \;
\; \text{when} \; \; \; x_0\longrightarrow \mp\infty \label{i1}
\end{eqnarray}
Without ${\cal L}^{qq}_{str}$, the nucleon wave function in quark
variables reads as
\begin{equation}
\Phi^q_N (P_N ,\sigma_N, x_{0} )=
\Phi^q_{0N}(P_N ,\sigma_N )
\equiv\Phi^q_N (P_N , \sigma_N , x_0{=}\pm\infty ).\label{i2}
\end{equation}
Here $P_N , {\sigma}_{N}$ stand for momentum and spin variables of
nucleons.  The $S_{str} -$matrix, caused by ${\cal L}^{qq}_{str}$,
transforms the wave function $\Phi^q_{0 N}(P_N , \sigma_N)$ of the
noninteracting quark system into the wave function of the physical
nucleon
 $\Phi^q_{N}(P_N , \sigma_N , x_0)$. So, with allowance for ${\cal
L}_{str}^{qq}$ ,\\ $\Phi^q_N (P_N , \sigma_N , x_0)={\cal
S}_{str}(x_0 , \mp\infty )\Phi^q_N (P_N , \sigma_N ,
\mp\infty )={\cal S}_{str}(x_0 , \mp\infty )\Phi^q_{0N}(P_N , \sigma_N
) \, , $ where
$$ {\cal S}_{str}(x_0 , -\infty )={\cal T}
\exp\Bigl(i\int\limits_{-\infty}^{x_0} \mbox{d}x_0 \int \mbox{d}
{\bf x} {\cal L}^{qq}_{str}(x)\Bigr) \, , \; \; \; \; \; {\cal
S}(t,t')\cdot {\cal S}(t',t_0)={\cal S}(t,t_0).$$ $\cal T$ stands
for the time-ordering operator.

The transition amplitude of (\ref{n}) is given in the general form
\begin{eqnarray}
 {\cal M}\cdot i(2\pi )^4\delta (P_n-P_p
-p_e -p_{\gamma})  = \nonumber \\ \langle \Phi_{0p}^{q \, +}(P_p
,\sigma_p ) , \psi_e^{+}(p_e ) , A(p_{\gamma})|{\cal S}_{int}
|\Phi_{0n}^q (P_n ,\sigma_n) , \psi_{\nu}(-p_{\nu})\rangle \, ,
\label{i5}
\end{eqnarray}
with $S_{int} -$matrix dictated by (\ref{i1})
\begin{eqnarray}
{\cal S}_{int} = {\cal T} \exp\Bigl(i\int
\mbox{d}^4x{\cal L}_{int}(x)\Bigr)
= {\cal T} \exp\Bigl(i\int
\mbox{d}^4x[{\cal L}_{int}^{EW}(x) + {\cal
L}_{str}^{qq}(x)]\Bigr).\label{i6}
\end{eqnarray}
Nowadays, there sees no option but to parameterize the effect of
strong interactions in treating the neutron $\beta -$decay.

At the lowest order in ${\cal L}^{EW}_{int}$, on the tree level, the
 amplitude of (\ref{n}), with allowance for ${\cal
 L}_{str}^{qq}(x)$,
\begin{eqnarray}
{\cal M}^0
=\bar u_e(p_e) \Gamma_{\alpha}^{e\nu W}u_{\nu}(-p_{\nu})
\cdot \bar U_p(P_p)\Gamma_{N \,
\beta}^{npW}U_n(P_n)\cdot D_{\alpha\beta}^W(q) \,
 ,\label{i7}\\ {\Gamma}_{\alpha}^{e\nu W}= {\Gamma}_{q \,
 \alpha}^{pnW} =
\frac{e}{2\sqrt{2}s_W}\gamma_{\alpha}(1-\gamma^5)
\, , \; \; \;
\; \Gamma_{N \, \alpha}^{np
W}=|V_{ud}|\frac{e}{2\sqrt{2}s_W}{\cal J}_{np \; \alpha}(q) \,
,\nonumber\\
q=P_n-P_p-p_e-p_{\nu} \, \; \; \; \; q^2\ll M_p^2\ll
M_W^2
\, , \; \; \; \; D_{\alpha\beta}^W
(q)=\frac{g_{\alpha\beta}}{q^2-M_W^2}
=\frac{-g_{\alpha\beta}}{M^2_W}\nonumber
\end{eqnarray}
is parameterized by introducing the nucleon weak transition current
\begin{equation}
{\cal J}^{\beta}_{np}(k)=\gamma^{\beta}+
g_{WM}\sigma^{\beta\nu}k_{\nu}-(\gamma^{\beta}g_A+g_{IP}k^{\beta})\gamma^5
. \label{i8}
\end{equation}
In (\ref{i7}), $u_l \, , \, \, U_N$ indicate Dirac spinors of
leptons and nucleons, and the usual notations $c_W{=}{M_W}{/}{M_Z}
\, , \, \, s_W^2=1-c_W^2 $ are introduced.
The Born amplitude (\ref{i7}) is written in terms of the
non-renormalized vertices and $W-$propagator which will give place
to the renormalized quantities
$
\hat\Gamma_{\alpha}^{e\nu W} ,
\hat\Gamma_{N \,\beta}^{npW} , \hat D_{\alpha\beta}^W $ in the
one-loop calculation.

The evaluation of the renormalized pure lepton vertex with the
counter terms obtained accordingly OMS gives
\begin{eqnarray}
\Gamma_{\alpha}^{e\nu W}\Rightarrow\nonumber
\\ \hat\Gamma_{\alpha}^{e\nu W}=\Gamma^{e\nu W}_{\alpha} \Bigl\{
1+\frac{\alpha}{4\pi}\Bigl(
2\ln\frac{m}{\lambda}+\ln\frac{m}{M_Z}-\frac{9}{4}+\frac{3}{s^2_W}
+\frac{6c_W^2-s_W^2}{s^4_W}\ln{c_W}\Bigr)\Bigr\}\label{i9}
\end{eqnarray}
As seen, this renormalized vertex in multiple of the
nonrenormalized one. In evaluating (\ref{i9}) and hereafter, we
neglect all the terms of the order $q{/}{M_{N,Z,S}}$,
$m_f{/}{M_{N,Z,S}}$. All the masses are taken from \cite{pdg1}.

In the one-loop approximation, the renormalized
$\hat\Gamma_{N\alpha}^{npW} -$vertex, which is due to the quark part
of ${\cal L}_{int}^{EW}$, is defined by the matrix element which
involves besides the electroweak interactions, ${\cal L}^{Zqq} ,
{\cal L}^{Wqq} , {\cal L}^{Aqq} , {\cal L}^{ZWW} , {\cal L}^{AWW}$,
the strong $qq-$interaction ${\cal L}^{qq}_{str}$ as well, via
${\cal S}_{str}$. The processes of the different kinds contribute to
$\hat\Gamma_{N\alpha}^{npW}$. In the piece of
$\hat\Gamma_{N\alpha}^{npW}$ which involves the heavy virtual boson
propagators $D^{W,Z}(k^2)$ the integration over momenta $k$ in the
loops involves, as a matter of fact, only the large values of
momenta $k^2\sim M^2_{W,Z}$. Consequently, strong $qq-$interactions
die out in intermediate states so that we deal with free quarks in
 the intermediate states in this case. In the part of
 $\hat\Gamma_{N\alpha}^{npW}$ involving a virtual photon, the photon
 propagator $D^A(k)$ is split into two pieces including large and
 small momenta
\begin{eqnarray}
D^A_{\mu\nu}(x_2-x_3)\nonumber \\ =g_{\mu\nu}
\int\frac{\mbox{d}^4k}{(2\pi)^4} \Bigl(\frac{1}{k^2-M^2_S+i0} +
\frac{-M_S^2}{(k^2-\lambda^2+i0)(k^2-M^2_S)}\Bigr)e^{-ik(x_2-x_3)}
\label{i12}\\
=D^{AS}_{\mu\nu}(x_2-x_3) + D^{Al}_{\mu\nu}(x_2-x_3) ,
 \nonumber
\end{eqnarray}
with the subsidiary matching parameter $M_S$ introduced thereby,
$M_p^2\ll M^2_S\ll M_W^2$ \cite{sh}. Quarks are free in the term
involving $D^{AS}$, the ``massive photon" propagator. The total
renormalized vertex $\hat\Gamma_{N \, \alpha}^{npW}$ is written as
\begin{equation}
\hat\Gamma_{N \, \alpha}^{npW} = \hat\Gamma_{Ns \, \alpha}^{npW} +
 \hat\Gamma_{Nl \, \alpha}^{npW}. \label{i13}
\end{equation}
The contribution $\hat\Gamma_{Ns \, \alpha}^{npW}$ from all the
processes, where ${\cal L}^{qq}_{str}$ can be ignored in the
intermediate states, proves to be
\begin{eqnarray}
\hat\Gamma^{npW}_{Ns \, \alpha}(P_n , P_p ,q) =\Gamma^{npW}_{N \, \alpha}
\Biggl\{1+\frac{\alpha}{4\pi}\Bigl(\ln\frac{M_S}{M_Z}
+ \frac{3}{s^2_W} +
\frac{6c^2_W - s^2_W}{s^4_W}\ln(c_W)\Bigr)\Biggr\} , \label{i16}
\end{eqnarray}
with the appropriate allowance for the counter terms obtained from
${\cal L}^{EW}_{int}$ , the ``massive photon" propagator $D^{AS}$
replacing $D^A$.

As $\hat\Gamma^{npW}_{Nl\alpha}$ involves the ``soft photon"
propagator $D^{Al}(k^2)$ , quarks in the intermediate state in
 $\hat\Gamma^{npW}_{Nl\alpha}$ posses small momenta and constitute
 baryonic states. Thus, $\hat\Gamma^{npW}_{Nl\alpha}$ in (\ref{i13})
 incorporates the sum over the intermediate baryonic states, the
 bound states of strong interacting quarks. In the simplified case,
 when there are only pure unexcited nucleons in the intermediate
 states with the propagators $G_N(p_N)$, and the nucleon formfactors
 are ignored, i.e. $f_{\alpha}^p =\gamma_{\alpha} \; ,
\; \;  f^n_{\alpha}=0$, we arrive at
\begin{eqnarray}
\hat\Gamma^{pnW}_{Nl \, \alpha} = {\Gamma}^{npW}_{N\alpha} \biggl(
\frac{1}{2}{\delta}{z_{0}^p}+\frac{1}{2}{\delta}{z_{0}^n} \biggr) ,
\label{i17}
\end{eqnarray}
with the finite renormalization constants of the proton and neutron
states
\begin{equation}
{\delta}{z^{p}_0}=-\frac{\alpha}{4\pi}\Bigl(2\ln\frac{M_S}{M_p} +
\frac{9}{2}-4\ln\frac{M_p}{\lambda}\Bigr) \, , \; \; \; \;
{\delta}{z^n_0}=0. \label{i18}
\end{equation}
 To realize the accuracy of this result we have estimated the
 corrections to (\ref{i17}) due to the formfactors
\begin{eqnarray}
f^{pp}_{\alpha}(k)=\Bigl(\gamma_{\alpha}+\frac{1.79}{2
M_p}k^{\beta}{\sigma}_{\alpha\beta}\Bigr)
\frac{-m^2_{\rho}}{k^2-m_{\rho}^2}
 \, , \; \; \;
  f^{nn}_{\alpha}=\Bigl(-\frac{1.93}{2
 M_n}\Bigr)\frac{-m^2_{\rho}}{k^2-m_{\rho}^2} \, , \label{i19}
\end{eqnarray}
 where $m_{\rho}$ is the $\rho -$meson mass, and due to allowance
for the $\Delta_{33}
-$isobar in the intermediate states, $$G_N (p)\longrightarrow
G_{\Delta} (p)
=\frac{\not p +M_{\Delta}}{p^2 -M_{\Delta}^2 +i0} . $$
These corrections prove to constitute no more than a few per cent to
the quantity (\ref{i17}). The relations $M_{p,n}\ll M_{W,Z} \, , \;
\; \; M^2_{p,n}\ll
 M^2_S\ll M^2_{W,Z} \, , \; \; \; m_f\ll M_{p,n} \, , \; \; \;
 q^2\ll M^2_{p,n} \, , \; \; \; M^2_{\Delta_{33}}-M^2_{p}\sim M^2_p
 $ are utilized through all the evaluations.

Thus, with the accuracy about a few per cent, the whole
renormalized $Wnp-$vertex $\hat\Gamma^{npW}_{N \, \alpha}(P_n ,
P_p ,q) = \hat\Gamma^{npW}_{Ns \, \alpha}(P_n , P_p ,q) +
\hat\Gamma^{npW}_{Nl \, \alpha}(P_n , P_p ,q)$ is
\begin{eqnarray}
 \Gamma^{npW}_{N \, \alpha}
\Biggl\{1+\frac{\alpha}{4\pi}\Bigl(\ln\frac{M_p}{M_Z}
- 2\ln\frac{\lambda}{M_p} - \frac{9}{4} + \frac{3}{s^2_W} +
\frac{6c^2_W - s^2_W}{s^4_W}\ln(c_W)\Bigr)\Biggr\} \label{i21}
\end{eqnarray}
is multiple of the nonrenormalized vertex $\Gamma_{N\alpha}^{npW}$.

 Next, the propagator of virtual $W-$boson in ${\cal M}^0$
(\ref{i7}) modifies as
\begin{eqnarray}
D^W(q)\Longrightarrow \hat D^W(q)
=\frac{1}{q^2-M_W^2
+\hat\Sigma^{W}(q^2)}\approx\Bigl(-\frac{1}{M_W^2}\Bigr)\frac{1}
{1-\frac{\hat\Sigma^{W}(0)}{M_W^2}}\label{i22}
\end{eqnarray}
which can be extracted from the $\mu-$decay analysis \cite{ms,h2}
\begin{eqnarray}
\Bigl(\frac{e}{2\sqrt{2}s_W}\Bigr)^2\hat D^W(q)=-\frac{G_{\mu}
(1-\delta_{\it v})}{\sqrt{2}}\ ,\; G_{\mu}=1.1663\cdot 10^{-5} \,
\mbox{GeV}^{-2} \, ,\; \delta_{\it v}\approx 0.006 . \nonumber
\end{eqnarray}
 Then
\begin{eqnarray}
{\cal M}^0=-\frac{G \, |V_{ud}|}{\sqrt{2}}\bigl(\bar u_e
(p_e)\gamma_{\alpha}(1-\gamma_5)u_{\nu}(-p_{\nu})\bigr)\nonumber
\\ \hspace{12mm}\cdot\bigl(\bar U_p(P_p)\gamma^{\alpha}(1-\gamma_5
g_A)U_n(P_n)\bigr) \, \label{i23}
\\ \hspace{3mm}G=G_{\mu}(1-\delta_{\it v})\nonumber
\end{eqnarray}
By now, we have considered the terms which stem from the Born
amplitude ${\cal M}^0$ (\ref{i7}) by replacing the nonrenormalized
vertices $\Gamma^{e\nu W}_{\alpha} , \Gamma_{N\alpha}^{npW}$ and
 the $W-$boson propagator by renormalized ones. The total amplitude
${\cal M}$ of (\ref{n}) also contains the part ${\cal M}_{2\gamma}$
which is of the second order in both lepton and quark electroweak
interactions, usually referred to as a contribution from the ``box
diagrams". The matrix element which defines  ${\cal M}_{2\gamma}$
incorporates the terms of different nature. The term which is due to
a photon exchange between quarks and leptons contains the photon
propagator $D^A(k^2)$ and is divided into two parts corresponding to
large and small momenta transferred by the virtual photon, the
``massive" $D^{AS}$ and ``soft" $D^{Al}$ photon propagators
respectively (\ref{i12}). The terms in  ${\cal M}_{2\gamma}$
including heavy boson electroweak interactions with quarks and
leptons, ${\cal L}^{Zqq} , {\cal L}^{Wqq} , {\cal L}^{Zee} , {\cal
L}^{Z\nu\nu} , {\cal L}^{We\nu} $, are due to the $Z-$boson exchange
between quarks and leptons. They contain the propagators $D^{W,Z}$
of virtual heavy gauge bosons. This case corresponds to large
momenta, $ \, q^2\sim M^2_{Z,W}\gg M_p^2$ , transferred from leptons
to quarks. Strong $qq-$interactions in the intermediate states can
be ignored in the terms containing the virtual heavy gauge bosons
and
 ``massive" photons. The contribution ${\cal M}_{2\gamma s}$ from
 these terms into the whole amplitude
\begin{equation}
{\cal M}_{2\gamma}+{\cal M}_{2\gamma s}+{\cal M}_{2\gamma l}
\label{im}
\end{equation}
is multiple of ${\cal M}_{0}$,
\begin{eqnarray}
{\cal M}_{2\gamma s}=-{\cal M}^0 \frac{\alpha}{4\pi}
\Biggl\{\Bigl(1+\frac{5
c^4_W}{s^4_W}\Bigr)\ln{(c_W)}-
6\ln\frac{M_W}{M_S}\Biggr\}\label{i27}
\end{eqnarray}

The amplitude ${\cal M}_{2\gamma l}$ includes the ``soft photon"
propagator $D^{Al}$ which corresponds to small momenta $q^2 < M^2_S$
transferred from leptons to quarks. Therefore the intermediate quark
system posses small momenta so that we deal with an intermediate
baryonic state $B$, a ground or excited nucleon state. Our
calculation proves that leaving out all the excited states and
presuming  for the nucleon formfactors and nucleon transition
current
\begin{equation}
f^{pp}_{\beta}=\gamma_{\beta} \, , \; \; \; f_{\beta}^{Bn}=0 \, , \;
 \; \; {\cal J}_{\alpha}^{pn}=\gamma_{\alpha}(1-\gamma^5
 g_A)\label{fj}
\end{equation}
we commit no more than a few per cent error. With this accuracy
\begin{eqnarray}
{\cal M}_{2\gamma
l}=\Bigl(\frac{e}{2\sqrt{2}s_W}\Bigr)^2|V_{ud}|\frac{1}{M_W^2}
\nonumber \\ \cdot\frac{\alpha}{4\pi} \Biggl\{\Biggl(\bar
u_e(p_e)\gamma^{\beta}(\not
p_e+m)\gamma^{\alpha}(1-\gamma^5)u_{\nu}(-p_{\nu})
\frac{1}{2\varepsilon M_p v}\nonumber\\
\times
[\ln{(x)}\ln{\frac{\lambda}{m}}-\frac{1}{4}(\ln{(x)})^2+F(1/x-1)
-{\pi^2v}{/}{\tilde v}] \label{i29}\\ -\bar
u_e(p_e)\gamma^{\beta}\gamma^{\delta}\gamma^{\alpha}
(1-\gamma^5)u_{\nu}(-p_{\nu})\frac{1}{2M_p}
\big[-\frac{p_{\delta}}{v\varepsilon}\ln{(x)}+
\delta_{0\delta}\bigl(\frac{1}{v}\ln{(x)}-2\ln{\frac{m}{M_p}})\bigr]
\Biggr)\nonumber \\ \times\bigl(\bar U_p(P_p)\gamma_{\beta}(\not
P_p+M_p)\gamma_{\alpha}
(1-\gamma^5g_A)U_n(P_n)\bigr)\nonumber\\
- \bigl(\bar u_e(p_e)\gamma^{\beta}\gamma^{\delta}\gamma^{\alpha}
(1-\gamma^5)u_{\nu}(-p_{\nu})\bigr)
\bigl(\bar U_p(P_p)\gamma_{\beta}
\gamma^{\nu}\gamma_{\alpha}(1-g_A\gamma^5)U_n(P_n)\bigr)\nonumber\\
\times g_{\delta\nu}\Bigl(\frac{3}{8}+\frac{1}{2}\bigl(
\ln{\frac{M_W}{M_p}} - \frac{M_W^2}{M_W^2-M_S^2}\ln{\frac{M_W}{M_S}}
 \bigr) - \delta_{0\delta}\frac{1}{2}\Bigr)\Biggr\}\nonumber
\end{eqnarray}
The electron energy $\varepsilon =\sqrt{m^2+p^2}\leq M_n - M_p \ll
M_p$ . $F$ - Spens-function , ${\bf v}=\frac{\bf p}{\varepsilon} \,
,$ \\ $ x={(1-v)}{/}{(1+v)} \; , \; \;
\omega_{{\nu}0}=M_n-M_p-\varepsilon \, ,$
$${\tilde v}({\varepsilon})=\frac{1}{2}\Biggl(
\sqrt{(v+\frac{m\omega_{{\nu}0}}{M_p\varepsilon} )^2 +
2v\frac{\omega_{{\nu}0}}{\varepsilon} (\frac{m}{M_p})^2} +
 \sqrt{(v-\frac{m\omega_{{\nu}0}}{M_p\varepsilon} )^2 -
2v\frac{\omega_{{\nu}0}}{\varepsilon} (\frac{m}{M_p})^2}
 \, \, \Biggr) \, .$$ As seen, this part of the total transition
 amplitude ${\cal M}$ is not multiple of the Born amplitude ${\cal
 M}_0$ (\ref{i7}).

The amplitude of the real $\gamma -$radiation with momentum $\bf k$
and polarization $\bfgr \epsilon^{(r)}$
\begin{eqnarray} {\cal M}^{(r)}_{1\gamma}(k)
=\Bigl(\frac{e}{2\sqrt{2}s_W}\Bigr)^2|V_{ud}|\bigl(\frac{-1}{M_W^2}\Bigr)
e \epsilon_a^{(r)}\nonumber  \\ \cdot \bigl(\bar
u_e(p_e)\gamma^a\frac{(\not p_e+\not k+m)}{( p_e +k)^2-m^2}
\gamma^{\lambda}(1-\gamma^5)u_{\nu}(-p_{\nu})\bigr) \times
\bigl(\bar U_p(P_p)\gamma_{\lambda}(1-g_A\gamma^5)U_n(P_n)\bigr)
 \nonumber \\ (a , r =1 , 2 , 3 ) \label{i30}
\end{eqnarray}
is also not multiple of ${\cal M}^0$ (\ref{i7}).

Absolute square of the whole transition amplitude, up to the first
$\alpha-$order,
\begin{eqnarray}
|{\cal M}|^2 = |{\cal M}^R +{\cal M}_{2\gamma l} +{\cal
M}^{(r)}_{1\gamma}|^2\approx |{\cal M}^R|^2 + |{\cal
M}^{(r)}_{1\gamma}|^2 + 2\mbox{Re}[{\cal M}^0{\cal M}_{2\gamma l}]
,\label{abs}
\end{eqnarray}
where
\begin{eqnarray}
{\cal M}^R = \bigl(\bar u_e(p_e)\hat \Gamma_{\alpha}^{e\nu\mu}
u_{\nu}(-p_{\nu})\bigr)\times\nonumber\\
\times\bigl(\bar U_p(P_p) \hat \Gamma_{\beta}^{npW} U_n(P_n)\bigr)
D^W_{\alpha\beta}(p_{\nu}+p_e)+{\cal M}_{2\gamma
s}\approx\nonumber\\
\approx{\cal
M}^0\Bigl\{1-\frac{\alpha}{4\pi}\Bigr(2\ln{\frac{M_Z}{M_p}}+
4\ln{\frac{\lambda}{m}}+\frac{9}{2}-\ln{\frac{M_p}{m}}-\label{i31}\\
 - \frac{6}{s_W^2} - 6\ln{\frac{M_Z}{M_S}}-
\frac{3+4c^2_W}{s^4_W}\ln{(c_W)}
\Bigr)\Bigr\}\nonumber
\end{eqnarray}
comprises all the terms proportional to the Born amplitude ${\cal
M}^0$ (\ref{i7}). We calculate the $\beta
-$decay probability of a polarized neutron (the polarization vector
$\bfgr
\xi$) integrated over the final proton, antineutrino and photon
momenta, and summarized over the polarizations of all the final
particles
\begin{eqnarray}
\mbox{d}{\bf W}(\varepsilon ,{\bf p}_e) =
\mbox{d}{\bf W}^R(\varepsilon ,{\bf p}_e)+
\mbox{d}{\bf W}_{1\gamma}(\varepsilon ,{\bf p}_e)+
\mbox{d}{\bf W}_{2\gamma l}(\varepsilon ,{\bf p}_e)\label{i32}
\end{eqnarray}
where $\mbox{d}{\bf W}^R , \;
\mbox{d}{\bf W}_{1\gamma} , \;
\mbox{d}{\bf W}_{2\gamma l}$ come from $|{\cal M}^R|^2 , \;  |{\cal
M}^{(r)}_{1\gamma}|^2 , \; 2\mbox{Re}[{\cal M}^0{\cal M}_{2\gamma
l}]$ , respectively.
 $ \; \; \mbox{d}{\bf W}^R(\varepsilon ,{\bf p}_e)$ is certainly
 proportional to the uncorrected, Born, decay probability
\begin{eqnarray}
\mbox{d}{\bf W}^0(\varepsilon ,{\bf
p}_e)=\frac{G^2}{2\pi^3}\varepsilon |{\bf p}_e| k_m^2
\mbox{d}\varepsilon\frac{\mbox{d}{\bf n}}{4\pi}
\bigl(1+3g_A^2+{\bfgr{v\xi}}2g_A(1-g_A)\bigr)
 \, , \label{i33}\\ {\bf n}={\bf p}_e{/}|{\bf p}_e| \, , \; \; \;
{\bfgr v}={\bf p}_e{/}\varepsilon \, , \; \; \;
k_m=M_n-M_p-\varepsilon , \nonumber
\end{eqnarray}
whereas the contribution from the real $\gamma -$radiation and from
the ``box diagrams" containing nucleon in the intermediate state,
 $\mbox{d}{\bf W}_{1\gamma}(\varepsilon ,{\bf p}_e)$ and
$\mbox{d}{\bf W}_{2\gamma l}(\varepsilon ,{\bf p}_e)$, are not
multiple of $\mbox{d}{\bf W}^0(\varepsilon ,{\bf p}_e)$. Besides the
terms proportional to $\mbox{d}{\bf W}^0(\varepsilon ,{\bf p}_e)$
they involve ones which are not.

Eventually, the electron momentum distribution in the $\beta -$decay
of a polarized neutron is written as
\begin{eqnarray}
\mbox{d}{\bf W}({\bf p}_e , \varepsilon ) =
\frac{G^2}{2\pi^3}\varepsilon |{\bf p}_e| k_m^2
\mbox{d}\varepsilon\frac{\mbox{d}{\bf n}}{4\pi}
\bigl\{ W_0(g_A , \varepsilon) + \bfgr{v\xi}
W_{\xi}(g_A , \varepsilon)\bigr\} = \nonumber\\
\mbox{d}{\bf W}^{0}({\bf p}_e , \varepsilon )\cdot [{\cal
B}(\varepsilon)+\tilde C_1(\varepsilon)] +
\frac{G^2}{2\pi^3}\varepsilon |{\bf p}_e| k_m^2
\mbox{d}\varepsilon\frac{\mbox{d}{\bf n}}{4\pi}\times \nonumber\\
\Bigl((1+3g_A^2)\tilde C'_0(\varepsilon)+2\bfgr{v\xi}g_A(1-g_A)\tilde
C'_{\xi}(\varepsilon) + C_0(g_A ,\varepsilon)+C_{\xi}(g_A
,\varepsilon)\Bigr) .\label{i34}\\
 W_0(g_A , \varepsilon)
= (1+3g_A^2)[1+\tilde C_0(\varepsilon)+{\cal
B}(\varepsilon)]+C_0(g_A , \varepsilon) , \nonumber\\ W_{\xi}(g_A ,
\varepsilon)=2g_A (1-g_A)[1+\tilde C_{\xi}(\varepsilon) +{\cal
B}(\varepsilon)] +C_{\xi}(g_A , \varepsilon)\nonumber
\end{eqnarray}
Here, all ${\cal B} , {C}_0 , \tilde{C}_0 , {C}_{\xi} ,
\tilde{C}_{\xi} $, are the cumbersome but rather plain
functions of their arguments, directly obtained from Eqs. (\ref{i9},
\ref{i21}, \ref{i22}, \ref{i27}, \ref{i29}-\ref{i32}).

The relative modification of the total decay probability ,
\begin{eqnarray} \frac{\int
\limits_{m}^{{\Delta}} \mbox{d}{\bf w} W_{0}({g_A , \varepsilon})}
{(1+3g_A^2) \int
\limits_{m}^{{\Delta}} \mbox{d}{\bf w} }-1 = \delta W \,
,\label{i35}\\ \Delta = M_n -M_p \, , \; \; \; \;
\mbox{d}{\bf w}=\varepsilon |{\bf p}_e| k_m^2
\mbox{d}\varepsilon \, \mbox{d}{\bf n} ,
\nonumber \end{eqnarray}
amounts up to $\delta W \approx 8 \% \, ( \pm \lesssim 0.3 \% )$.
The uncorrected asymmetry factor of the electron angular
distribution $A_0$ is replaced by corrected one $A({\varepsilon})$
accounting for the radiative corrections,
\begin{eqnarray}
A_{0}=\frac{2g_{A}(1-g_{A})} {1+3g_{A}^{2}}\Longrightarrow
\frac{W_{\xi }(g_A , {\varepsilon})}
{W_{0 }({g_A , \varepsilon})}= A({\varepsilon}) .
\label{i36}
\end{eqnarray}
 The difference $\delta A = A(\varepsilon )-A_0 \; $ amounts up to $
\; \delta A\approx -1.9\% \, ( \pm \lesssim 0.2\%)$. Practically,
the same values of (\ref{i35}), (\ref{i36}) were acquired in the
previous calculation \cite{i} based immediately on the effective
Lagrangian of the local $4$-fermion theory of weak interactions
\cite{d}. It is to emphasize ones again that the accuracy
attainable in the presented calculation is about a few per cent to
the obtained values of the radiative corrections, i.e. a few tenth
of per cent, $\sim 0.01\%$, to the characteristics of the neutron
$\beta -$decay, such as $W , A$. As was discussed above, the
ambiguities come from the need to allow for strong $qq
-$interactions and the nucleon compositeness, such as excited
states and formfactors, associated with intrinsic structure. In
particular, varying the parameter $M_S$ within the limits $3 M_p <
M_S < 30 M_p$ causes uncertainties $\sim 0.1\%$ in $\delta W ,
\delta A$. Introducing only the usual parameters $g_A \, , \, \,
g_{WM} \, , \, \, g_{IP}$, describing the weak nucleon transition
current, is not enough to parameterize the whole effect of strong
interactions in treating the neutron $\beta -$decay.

Apparently, (\ref{i34}) can never be transformed to an expression
multiple of (\ref{i33}), unlike the results asserted in several
calculations \cite{sha,gl1,gl2,gar} which were entailed by Ref.
\cite{sir} where the total decay probability was reduced, to all
intents and purposes, to the ``model independent" part proportional
merely to ${\cal M}^0$. That is why our results for quantities
(\ref{i35}), (\ref{i36}) differ appreciably from the results
asserted in Refs. \cite{sha,gl1,gl2,gar,sir}:
$\delta{W}_{MI}{\approx}5.4\% \, , \,
 \, \delta{A}_{MI}{\approx}0\%$. Because of the differences between
these $\delta{W}_{MI} \, , \, \,  \delta{A}_{MI} $ and our
 $\delta{W} \, , \, \, \delta{A} $, the $|V_{ud}|$ and $g_A$ values,
ascertained from experimental data processing with allowance for
$\delta{W}_{MI} \, , \, \, \delta{A}_{MI}$, would alter, when
obtained with our $\delta{W} \, , \, \, \delta{A} $. The
modifications are of the noticeable magnitude: $\delta g_A\approx
0.47\% \, , \, \, \delta |V_{ud}|{=}{-1.7\%}$. For instance, the
values $g_A=1.2739 \, , \, \, |V_{ud}|=0.9713 $ given in
\cite{a,pdg1} will be modified to $g_A{\approx}1.28 \, , \, \,
|V_{ud}|{\approx}0.96 $

So, the tenable calculation based on the electroweak theory provides
 the palpable results for radiative corrections  which must be
strictly allowed for in experimental data processing.


\title*{Breaking of Isospin Symmetry in Nuclei \\ and
Cabibbo-Kobayashi-Maskawa Unitarity }
\toctitle{Breaking of Isospin Symmetry in Nuclei \protect\newline and
Cabibbo-Kobayashi-Maskawa Unitarity}
%
%
\titlerunning{Breaking of Isospin Symmetry in Nuclei}
%
\author{H. Sagawa}
\authorrunning{H. Sagawa}
%
%
\institute{Center for Mathematical Sciences, the University of Aizu\\
 Aizu-Wakamatsu, Fukushima 965,   Japan }

\maketitle              

\begin{abstract}
We studied the effect of isospin impurity on the super-allowed
Fermi $\beta $ decay using microscopic HF and RPA (or TDA) model
taking into account CSB and CIB interactions. It is found that the
super- allowed transitions between odd-odd $J$=0 nuclei and
even-even $J$=0 nuclei are quenched because of the cancellation of
the isospin impurity effects of mother and daughter nuclei. An
implication of the calculated Fermi transition rate on the
unitarity of Cabibbo-Kobayashi-Maskawa mixing matrix is also
   discussed.
\end{abstract}

\section{Introduction}

The isospin symmetry is the first dynamical symmetry in
physics proposed by J. Heisenberg, 1932.
This hypothesis relies  entirely on the equivalence between the
p-p  and n-n two-body interactions.
Experimentally, the validity of the isospin symmetry was proved
by the finding of the isobaric analog state (IAS) in 1961 by
 J. D. Anderson and C. Wong.  On the other hand, it is known that the
two-body Coulomb interaction, the charge symmetry breaking (CSB) force and
the charge independence  breaking (CIB) force  violate this symmetry
 and induce  the impurity of isospin in the atomic nuclei.
The question is how much the isospin impurity  affects on
important or not  for  the experimental determination of the
vector coupling constant G$_V $ of nucleon $\beta $ decay.
Super-allowed Fermi $\beta $ decays have been studied intensively
for several decades in relation to the  vector coupling constant
under the conserved vector current (CVC) hypothesis. Combining the
vector coupling constant of nuclear  $\beta $ decay with that of
muon  $\beta $ decay, it is possible to determine the mixing
amplitude between ${\it u}$ and ${\it d}$ quarks in the first row
of the Cabibbo-Kobayashi-Maskawa (CKM) unitary matrix. Thus, this
amplitude , together with the  mixing amplitudes of ${\it u}$ and
${\it s}$ quarks and ${\it u}$ and ${\it b}$ quarks, provides an
opportunity to test experimentally the standard three-generation
quark model for the electro-weak interaction.

Two nuclear medium corrections have been studied for obtaining
the {\it `` nucleus\, independent\,'' ft } value\cite{OB89,THH77}. The first one is  the
``inner'' radiative correction and the second one is the effect of  isospin
non-conserving forces in nuclei.  These two corrections have been
studied intensively during last two decades and found
to be important to obtain the ``nucleus independent'' $ft$ value.
However, there is still a substantial deviation from the unitarity
in the empirical CKM matrix elements.
It has been claimed that the empirical data of Fermi decay after
subtracting the nuclear structure effects gives somewhat smaller
value than that required by the unitary condition (3 times more
than the standard deviation).  In this study\cite{SGS96}, we will
would like to pin down
 two  new  effects which have not been seriously discussed in the
previous studies  of the isospin mixings,i.e. , the isospin
impurity of the core outside the shell model space
 and the effect of the charge symmetry breaking (CSB)
and charge independence breaking (CIB) forces on the mean field potentials
\cite{Wil93,SGS95}.
  It might be interesting to see
how our model gives different results for the CKM unitarity problem
since there are  essential differences between our model and the
previous studies.
Particularly we study the Fermi $\beta $ decay in
$^{10}$C, $^{14}$O, $^{26}$Al, $^{34}$Cl, $^{38}$K, $^{42}$Sc and $^{54}$Co  nuclei
for which
the most accurate experimental ${\it ft }$ values are available.
We also report results of heavier nuclei  $^{62}$Ga, $^{66}$As, and $^{74}$Rb
for further study of CVC hypothesis, and to clarify the differences between our
model and previous calculations with different models.

\section{RPA calculations for Fermi transitions}
We have performed self-consistent
Hartree-Fock(HF)+
random phase approximation (RPA) calculations.
The
CSB and CIB interactions are taken into account  in the HF calculations.
We take even-even
nuclei (the  daughter nuclei of the $\beta $ decay except for
$^{10}$C and $^{14}$O) as the  RPA vacuum and
calculated the excited states with $J^{\pi }=0^+ $ in odd-odd nuclei.
The lowest states in the RPA spectra
are identified as the IAS of the $\beta $ decay.  We adopt the
filling approximation for the RPA vacuum in which the
particles occupy the HF single particle states from the bottom of the
potential in order and the last orbit has a partially occupied
configuration according  to the mass number. The HF+RPA calculations
are performed by using the harmonic oscillator basis.  The model space
adopted is 8 $\hbar \omega $ for $^{10}$C and $^{14}$O and 10  $\hbar \omega $
 for other nuclei.

The quenching factor $\delta _c $ for the super-allowed transition
is defined as
\begin{equation}
  | \langle J^{\pi }=0^+ T=1: \mbox{daughter}
        |T_+  |J^{\pi }=0^+ T=1 :\mbox{mother} \rangle |^2
     \equiv 2(1-\delta _c )
\end{equation}
The results are shown in Fig. \ref{fig:dc}  with and without the
CSB and CIB interactions.  We pointed out that the sum rule values
 of the super-allowed Fermi transitions are enhanced substantially
by the isospin impurity effect in nuclei with the mass $A\leq$20
\cite{SGS96}.
 Contrary to the results of the sum rule,
  the transitions are quenched substantially
except for $^{10}$C.  This is  due to the fact that the isospin mixing
of the mother state enhances the sum rule, but that of the
daughter state cancels this enhancement in the transition.
While
there is no enhancement due to the couplings to the isovector
giant monopole states,  the coupling to neighboring $ J^{\pi }$=
0$^+ $ states decrease the decay strength. This is the same mechanism
as the finding of the previous shell model
calculations\cite{OB89,THH77}.  The CSB and CIB
interactions give  20-30 \% larger quenching factors in all nuclei
as shown in Fig. \ref{fig:dc}.

\begin{figure}[hbt]
\begin{center}
\includegraphics[scale=0.45]{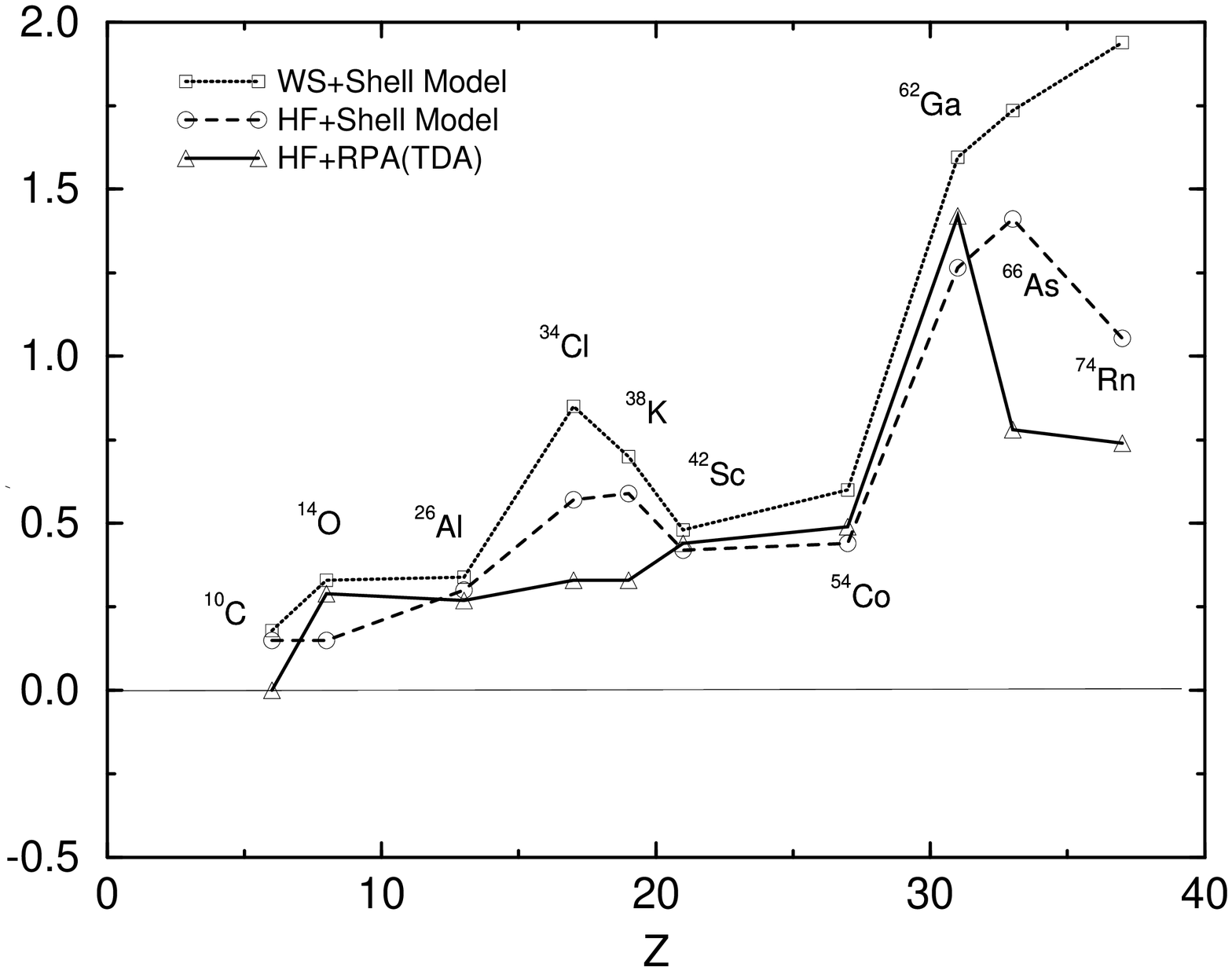}
\caption{Calculated quenching factor $\delta _c $ of the super-allowed
Fermi $\beta $ decay of seven $J$=0 odd-odd nuclei.
The solid line
with triangles  is calculated by  the HF and RPA (or TDA) with
the Skyrme interaction SG2 + the CSB and CIB interactions.
The results of HF(Woods-Saxon) + shell model denoted by the dashed(dotted)
line with circles(squares) are taken from ref. \cite{OB95}.
\label{fig:dc}}
\end{center}
\end{figure}

\section{CVC hypothesis and Cabibbo-Kobayashi-Maskawa (CKM) mixing matrix}
 Precise measurements of super-allowed $\beta $ decay between nuclei
with ($J^{\pi }=0^{+}, T=1)$ provide the most stringent probe of the
electroweak interaction and has been the subject of intensive study
for several decades. Since the axial current does not contribute to
transitions in the lowest order, the experimental $ft$-value is directly
related to the vector coupling constant $G_V $
\begin{equation}
ft = \frac{K}{G_V ^2 |M_F |^2 }
\end{equation}
where the constant $K/(\hbar c)^6 $ and the matrix $M_F$ are
defined as
\begin{eqnarray}
     K/(\hbar c)^6 &=& 2\pi ^3 ln(2)\hbar /(m_e c^2)^6  \\\nonumber
                   &=&(8120.271\pm 0.012)\times 10^{-10} GeV^{-4}\cdot s \\
    |M_F |^2 &=& | \langle J^{\pi }=0^+ T=1: \mbox{daughter} \\\nonumber
        |T_+  |J^{\pi }=0^+ T=1 :\mbox{mother} \rangle |^2  \\
     &=& 2(1-\delta _c )
\end{eqnarray}
Up to now, nine $ft$ values of $0^+ \rightarrow  0^+ $transitions have been
reported experimentally in enough  accuracy of less than
0.2 \% error to test the CVC hypothesis
; from the lightest $^{12}$C to the heaviest $^{54}$Co.
The constancy of these values is the key issue of the prediction
of CVC hypothesis.  Top of the CVC problem, the CKM mixing matrix
between $u$ and $d$ quarks ($v_{ud}$) can be determined by
comparing the decay rates for muon and nuclear Fermi $\beta $ decay.
A test of the unitarity of the matrix , made possible by the
 empirical value $v_{ud}$, is an important measure of the accuracy
for the  three generation Standard model.

For these purposes, nucleus-dependent corrections should be subtracted
 from the experimental $ft$ values.
The first is radiative corrections to the statistical rate
function $f$, denoted conventionally $\delta _R $. There is also
nucleus-independent radiative corrections  $\Delta _R ^V $.
The factor $\delta _R $ is called the ``inner'' radiative correction and
 $\Delta _R ^V $ is the ``outer'' radiative
correction including axial-vector interference terms.
The second correction is the nuclear structure factor due to the
isospin impurity.   Including all these
corrections, the ``nucleus-independent'' $Ft$ value is defined
as
\begin{equation}
Ft=ft(1+\delta _R +\Delta _R ^V )(1-\delta_c )
\end{equation}
and the matrix element $v_{ud}$ is given by
\begin{equation}
|v_{ud}|^2 =\frac{\pi^3 ln2}{Ft}\frac{\hbar ^7 }{G_F ^2 m_e ^5 c^4 }
           = \frac{2984.38(6)}{Ft}
\end{equation}
where the Fermi coupling $G_F $ is obtained from muon $\beta  $ decay.

In table 1, we list the experimental $ft$ values, the nucleus dependent
radiative correction $\delta _R $, the nuclear structure factor $\delta_c $
 and the $Ft$ values.  The outer radiative correction $\Delta _R ^V $ is
taken from ref. \cite{Tow95},
\begin{equation}
\Delta _R ^V   = (2.46 \pm 0.09) \%
\end{equation}
and added to obtain the $Ft$ values.
The  statistical factor $f$ adopted is slightly different
from the values in ref.\cite{Wil93}, but does not make any significant change
for the final results.

\begin{figure}[hbt]
\begin{center}
\includegraphics[scale=0.45]{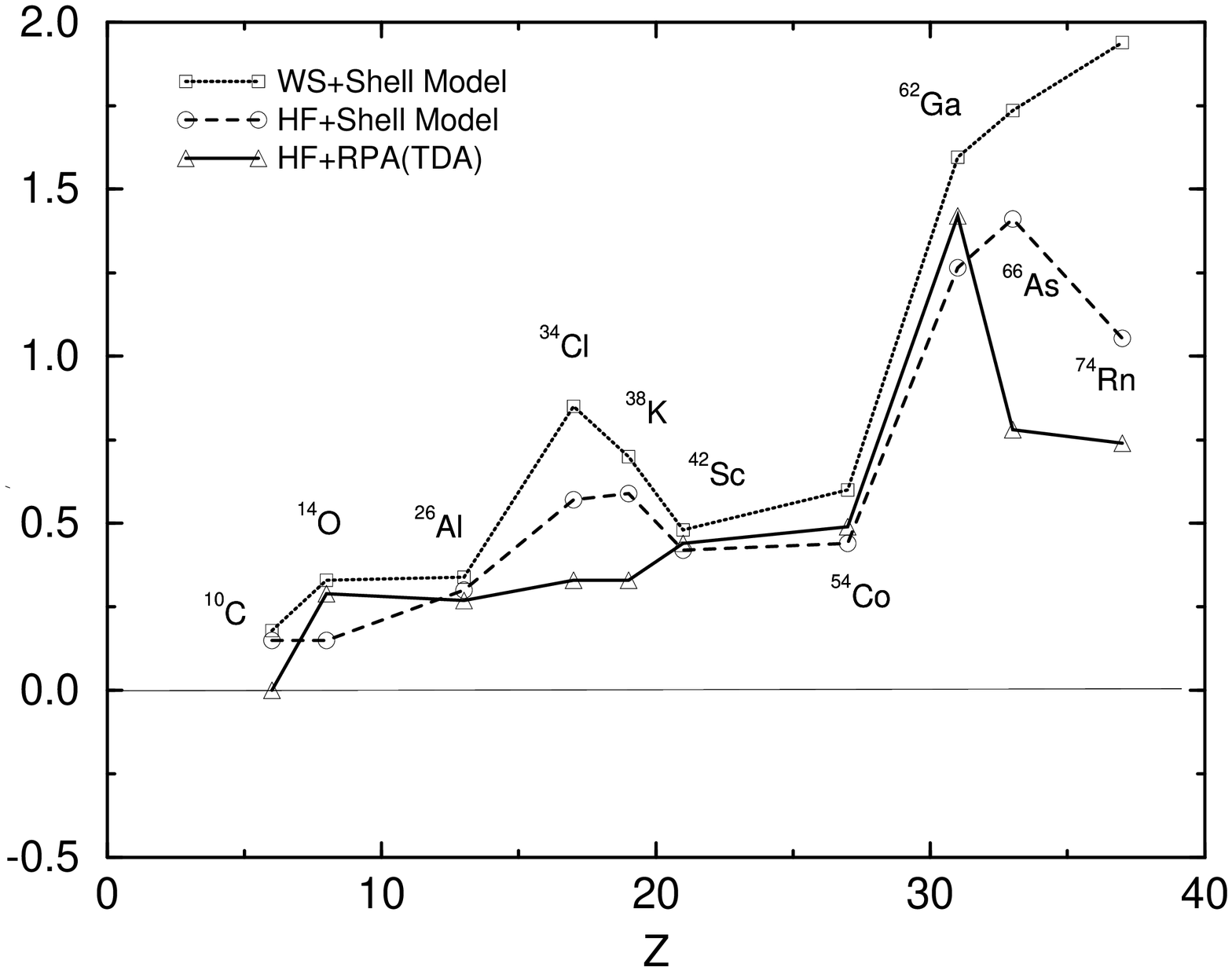}
\end{center}
\caption{Nucleus independent $Ft$ values in eq. (24) of super-allowed
Fermi transitions.  The present results are plotted by black squares,
while those of ref. \cite{Tow95} are given by black squares.
\label{fig:ft}}
\end{figure}

The average value $\bar{Ft}$ of 7 data in table 1 is
\begin{equation}
 \bar{Ft} = 3149.6 \pm 1.5
\end{equation}
The standard deviation $\sigma $ is 1.5 which could be small enough
to justify the CVC hypothesis.  There is no sigh of the Z-dependence
of $Ft$ value in Fig. 4 which was claimed in ref.\cite{Wil93}
to solve the deviation of the unitarity condition of the CKM matrix.
In ref.\cite{Tow95}, the calculated $\bar{Ft}$ of the seven data is
\begin{equation}
  \bar{Ft} = 3148.2 \pm 1.6\qquad,
\end{equation}
 while Ormand and Brown \cite{OB95} give
\begin{equation}
 \bar{Ft} = 3149.8 \pm 1.8
\end{equation}
The  average value  of our model
is surprisingly close to the previous calculations
in which the models are essentially different in the two points, i.e.,
the configuration space and the effect of CSB and CIB interactions
in the mean field.
The matrix element $v_{ud}$ is now calculated from eq. (27) to be
\begin{equation}
v_{ud}=0.9734 (2)
\end{equation}
The other two matrix elements  $v_{us}$ and $v_{ub}$ are determined
from  independent experimental information of weak decays,
\begin{eqnarray}
v_{us} &=& 0.2199 (17)  \nonumber  \\
v_{ub} &<& 0.0075 (90\% \mbox{confidence level})
\end{eqnarray}
Finally, the sum of three matrix element becomes
\begin{equation}
   |v_{ud}|^2 +|v_{us}|^2 +|v_{ub}|^2=0.9959(12)
\end{equation}
which deviates (3-4) times more than the standard deviation
$\sigma $= 0.0012
from the unitary condition.  This deviation is certainly
more than the ambiguity of nuclear structure effect since
our RPA model prediction differs in only 0.05 \% level from that
of the shell model in average and  corresponds to less than one
standard deviation.  It should be noticed that the final result in
eq. (32) relies on many small effects of the radiative corrections and
the nuclear structure. We calculated the $\delta _c $ and $Ft$ values
by using another Skyrme interaction S3.  The average $Ft$ value obtained
is 3149.7 $\pm $1.5 which is essentially identical to that of SGII in table 1.
However, the final conclusion will be easily changed by
one unknown effect  since the values discussed are always
1\% level which will be certainly discarded in most of nuclear
physics study.

\begin{table}[tbh]
\caption{Empirical $ft$ values of super-allowed Fermi transitions
and nucleus dependent corrections $\delta _R $ and $\delta _c $
The nucleus independent radiative corrections $\Delta_R ^V $=2.46
$\pm$ 0.09 ( \% )
are included
in the $Ft$ values in eq. (24).  The data of $ft, \delta _R $ and
$\Delta_R ^V $ are taken from ref. \cite{Tow95}.}
\vspace{-1.0 mm}
\begin{center}

\begin{tabular}{l|cccc} \hline
 SG2 &    & & & \\
\hline
  A& $ft (s)$ & $\delta _{R}$ (\% )&  $\delta _{c} (\% )$ &
   $ Ft (s)  $\\\hline
$^{10}$C& 3040.1 (50) & 1.30 (4)& $-$0.01 & 3154.7 (60) \\
 $^{14}$O& 3038.1 (18) & 1.26 (5) & 0.17 & 3145.8 (37)\\
 $^{26}$Al& 3035.8 (17) & 1.45 (2) & 0.27 & 3146.0 (34)\\
 $^{34}$Cl& 3048.4 (19)& 1.33 (3)& 0.33 & 3153.5 (36)\\
 $^{38}$K& 3047.9 (26) & 1.33 (4)& 0.33 & 3153.0 (41)\\
 $^{42}$Sc& 3045.1 (14) & 1.47 (5)& 0.44 & 3150.8 (36)\\
 $^{54}$Co& 3045.8 (11)& 1.39 (7)& 0.49 & 3147.6  (38)\\\hline
          &        &      & ave. & 3149.6 (15) \\\hline
 \end{tabular}
\end{center}
\end{table}

\section{Summary}
We studied the effect of isospin impurity on the super-allowed Fermi
$\beta $-decay using the HF and RPA (or TDA) model.  The Skyrme force
SGII is adopted for both HF and RPA calculations. The CSB and
CIB interactions are also taken into account in the HF calculations
for the first time.  The super-allowed Fermi transition probabilities
of light nuclei $^{10}$C and $^{14}$O are shown to be quenched less than
0.2 \% , while those of sd shell and pf-shell nuclei are quenched
up to 1\%.  These calculated values are close to those obtained
in the literatures with the HF and shell model calculations.
It is interesting to notice that the shell model configuration space
is 1 $\hbar \omega $ while, in our RPA calculations, we took into account
up to 12 $\hbar \omega $ configuration in the harmonic oscillator basis.
Another difference is the pairing which is properly taken into account
in the shell model , but not in the HF + RPA calculations.
We can notice in Fig. 2 that the quenching factor $\delta_c $ of the shell model
is somewhat larger than those of our results in nuclei at the middle of
the shells because of strong correlations in open-shell nuclei.

We estimate the so-called ``nucleus independent'' $Ft$ value taking into
account both the nuclear structure and the radiation effects.
The average $Ft$ value is obtained as
\begin{equation}
 {\bar Ft}=3149.6 (15)   \nonumber
\end{equation}
which is very close to the ones reported by Chalk River group
recently.  It is shown that the values $Ft$ are rather
Z-independent and no sign of the quadratic Z-dependence which is
suggested by ref.\cite{Wil93}. Our result shows that the average
$Ft$ value is still larger than the requested value to satisfy
strictly the unitary condition of Cabibbo-Kobayashi-Maskawa matrix
 within 0.1 \% level, although two new effects ,
the core polarization effects
and the CSB and CIB interactions, are taken into account for the first time
in our study.



\newcommand\cO{\mathcal{O}}
\title*{$g_A$ on the Lattice}
\toctitle{$g_A$ on the Lattice}
%
%
\titlerunning{$g_A$ on the Lattice}
%
\author{M. G\"ockeler\inst{ {1}\,{2}}, R. Horsley\inst{3}, D. Pleiter\inst {4}, P.E.L. Rakow\inst {5}, A.
Sch\"afer\inst{2}, \\ \and G. Schierholz\inst{{4}\,{6}}}
\tocauthor{M. G\"ockeler, R. Horsley, D. Pleiter, P.E.L. Rakow, A.
Sch\"afer\inst{2}, \\G. Schierholz}
\authorrunning{M. G\"ockeler et al.}
%
%
\institute{Institut f\"ur Theoretische Physik, Universit\"at
Leipzig, D-04109 Leipzig, Germany \and Institut f\"ur Theoretische
Physik, Universit\"at Regensburg, D-93040 Regensburg, Germany \and
School of Physics, University of Edinburgh, Edinburgh EH9 3JZ, UK
\and John von Neumann-Institut f\"ur Computing NIC / DESY Zeuthen,
D-15738 Zeuthen, Germany \and
Theoretical Physics Division, Department of Mathematical Sciences,
University of Liverpool, Liverpool L69 3BX, UK \and Deutsches
Elektronensynchrotron DESY, D-22603 Hamburg, Germany }

\maketitle              

\begin{abstract}
We describe the techniques used in lattice evaluations of hadronic matrix
elements like the neutron decay constant $g_A$. Recent results for $g_A$
are presented and the influence of the finite quark mass and the finite
volume on the determination of $g_A$ is briefly discussed.
\end{abstract}

\section{What do we want to compute?}

Lattice evaluations of $g_A$ make use of nucleon matrix elements
of the axial vector current. For the quark flavors $q=u,d$ we have
\begin{equation}
 \langle \mbox{proton},p,s | \bar{q} \gamma_\mu \gamma_5 q |
 \mbox{proton},p,s \rangle = 2 \Delta q \cdot s_\mu \,,
\end{equation}
where $p$ denotes the momentum of the proton and $s$ is its spin
vector. In parton model language, $\Delta q$ is the fraction of
the proton spin carried by the quarks of flavor $q$. Assuming
perfect isospin symmetry we can write
\begin{eqnarray} \label{gadef}
 \langle \mbox{proton},p,s |
   \bar{u} \gamma_\mu \gamma_5 u - \bar{d} \gamma_\mu \gamma_5 d |
 \mbox{proton},p,s \rangle
   \nonumber \\ \hspace{-3.3mm}= \langle \mbox{proton},p,s | \bar{u} \gamma_\mu \gamma_5 d |
 \mbox{neutron},p,s \rangle = 2 g_A \cdot s_\mu
\end{eqnarray}
and hence $g_A = \Delta u - \Delta d$.

\section{What can we compute on the lattice?}

The basic observables in lattice QCD are Euclidean $n$-point correlation
functions. Since space-time has been discretised (with lattice spacing $a$)
the path integral has become a high-dimensional integral
over a discrete set of field variables. As the (Grassmann valued) quark
fields appear bilinearly in the action, they can be integrated out
analytically leaving behind the determinant of the lattice Dirac operator
and products of quark propagators. The remaining integrals over the
gluon fields can then be evaluated by Monte Carlo methods.
In the quenched approximation, which will be employed throughout
most of this paper, the determinant of the Dirac operator is replaced
by 1. This approximation saves a lot of computer time, but it
is hardly possible to estimate its accuracy.

Let us briefly sketch how hadronic matrix elements can be extracted from
ratios of three-point functions over two-point functions.
First, one has to choose suitable interpolating fields
for the particle to be studied. For a proton with momentum
$\vec{p}$ one may take
\begin{equation}
  B_\alpha (t,\vec{p})  =
\sum_{x;x_4=t}
\mathrm e^{- \mathrm i \vec{p}\cdot \vec{x} }
\epsilon_{i j k} u^i_\alpha (x)
 u^j_\beta (x) (C \gamma_5)_{\beta \gamma} d^k_\gamma (x)
\end{equation}
and the corresponding $\bar{B}$, where $i$, $j$, \ldots are color
indices, $\alpha$, $\beta$, \ldots are Dirac indices and $C$ is
the charge conjugation matrix.

As the time extent $T$ of our lattice tends to $\infty$, the
two-point correlation function becomes the vacuum expectation value of the
corresponding Hilbert space operators with the Euclidean evolution operator
${\mathrm e}^{- H t}$ in between, i.e.\ we have, omitting Dirac indices
and momenta for simplicity:
\begin{equation}
  \langle B(t) \bar{B}(0) \rangle
   \stackrel{T \to \infty}{=}
  \langle 0 |  B {\mathrm e}^{- H t}  \bar{B} | 0 \rangle \,.
\end{equation}
If in addition the time $t$ gets large, the ground state
$|\mbox{proton} \rangle$ of
the proton will dominate the sum over intermediate states between $B$
and $\bar{B}$, and the two-point function will decay exponentially with
a decay rate given by the proton energy $E_{\mathrm{prot}}$:
\begin{equation}
  \langle B(t) \bar{B}(0) \rangle
   \stackrel{T \to \infty}{=}
  \langle 0 |  B {\mathrm e}^{- H t}  \bar{B} | 0 \rangle
   \stackrel{t \to \infty}{=}
 \langle 0 |  B | \mbox{proton} \rangle {\mathrm e}^{- E_{\mathrm{prot}} t}
       \langle \mbox{proton} | \bar{B} | 0 \rangle + \cdots
\end{equation}
Of course, if the momentum vanishes, we have
$E_{\mathrm{prot}} = m_{\mathrm{prot}}$, the proton mass.

Similarly we have for a three-point function with the operator $\cO$ whose
matrix elements we want to calculate:
\begin{eqnarray}
  \langle B(t) \cO (\tau) \bar{B}(0) \rangle \nonumber \\
   \stackrel{T \to \infty}{=}
    \langle 0 |  B {\mathrm e}^{- H (t-\tau)} \cO
       {\mathrm e}^{- H \tau} \bar{B} | 0 \rangle \nonumber
\\
  {}  =  \langle 0 |  B | \mbox{proton} \rangle
        {\mathrm e}^{- E_ {\mathrm{prot}}(t - \tau) }
      \langle \mbox{proton} | \cO| \mbox{proton} \rangle
       {\mathrm e}^{-E_{\mathrm{prot}} \tau}
      \langle \mbox{proton} | \bar{B} | 0 \rangle
      + \cdots \nonumber
\\
  {}  =  \langle 0 |  B | \mbox{proton} \rangle {\mathrm e}
              ^{- E_{\mathrm{prot}} t }
       \langle \mbox{proton} | \bar{B} | 0 \rangle
       \langle \mbox{proton} | \cO| \mbox{proton} \rangle
      + \cdots
\end{eqnarray}
if $t > \tau > 0$. Hence the ratio
\begin{equation}
 R \equiv
 \frac{ \langle B(t) \cO (\tau) \bar{B}(0) \rangle}
      { \langle B(t) \bar{B}(0) \rangle}
  = \langle \mbox{proton} | \cO| \mbox{proton} \rangle + \cdots
\end{equation}
will be independent of the times $\tau$ and $t$, if all time differences
are so large that excited states can be neglected, and then $R$ yields
the desired matrix element.

The proton three-point function for a two-quark operator contains quark-line
connected as well as quark-line disconnected pieces. In the quark-line
connected contributions the operator is inserted in one of the quark lines
of the nucleon propagator, while in the disconnected pieces
the operator is attached to an additional closed quark line which
communicates with the valence quarks in the proton only via gluon exchange.
In the limit of exact isospin invariance considered in this paper,
the disconnected contributions of the $u$ quarks and the $d$ quarks
cancel in the case of non-singlet two-quark operators.
Fortunately, the operator needed for the evaluation of $g_A$ in
Eq.(\ref{gadef}) is of this type so that we do not have to cope with the
disconnected contributions, which are very hard to compute.

\section{Chiral symmetry}

Chiral symmetry plays an important role in hadronic physics.
Unfortunately, it is not straightforward to implement it on the
lattice. ``Traditional'' formulations, like (improved) Wilson
fermions break chiral symmetry explicitly at finite lattice
spacing $a$. This has the consequence that the axial vector
current has to be renormalized and chiral symmetry is only
restored in the continuum limit $a \to 0$. However, the last years
have seen a remarkable progress in this field. We have now
``chirally symmetric'' lattice formulations of the Dirac operator
based on solutions of the Ginsparg-Wilson relation. These enjoy a
lattice version of chiral symmetry even at finite lattice spacing
such that physical consequences of chiral symmetry (e.g.\ Ward
identities) hold already at finite $a$. In particular, there is an
axial vector current which is not renormalized.

However, there is a price to be paid for these nice properties:
Chirally symmetric lattice fermions need considerably more computer time
than the ``traditional'' formulations. Therefore phenomenologically
interesting results obtained with lattice fermions of this kind are only
slowly beginning to appear.

A related problem is the chiral extrapolation: In the foreseeable future
it will not be possible to perform simulations at the physical values
of the masses of the $u$ and $d$ quarks. Hence results obtained at
higher masses have to be extrapolated to the physical mass values. This
extrapolation is, of course, the more reliable the smaller the masses in
the simulation are. Since Ginsparg-Wilson fermions allow us to work with
considerably lighter quarks than most other lattice fermions, their
use will improve the quality of the results also in this respect.

\section{Results}

In Fig.\ref{fig.roger} taken from the review \cite{lat02} we show
our own results (QCDSF and UKQCD collaborations) obtained with
quenched and unquenched $O(a)$-improved Wilson fermions
as well as results from the LHPC and SESAM collaborations \cite{LHPC}
who work with quenched and unquenched unimproved Wilson fermions.
The $g_A$ values from the simulations have been extrapolated
linearly in the quark mass to the chiral limit and are plotted versus
$a^2$ in units of the ``force scale''
$r_0$ whose phenomenological value is $\approx 0.5 \, \mbox{fm}$.
Although the agreement between the various simulations is rather good
(within the partially quite large statistical errors), the value
obtained by a simple-minded continuum extrapolation is considerably
smaller than the experimental value. Unquenching does not seem to
have a big effect, which may be due to the rather large quark masses
in the unquenched simulations.
\begin{figure}[b]
\begin{center}
\includegraphics[width=8.0cm]{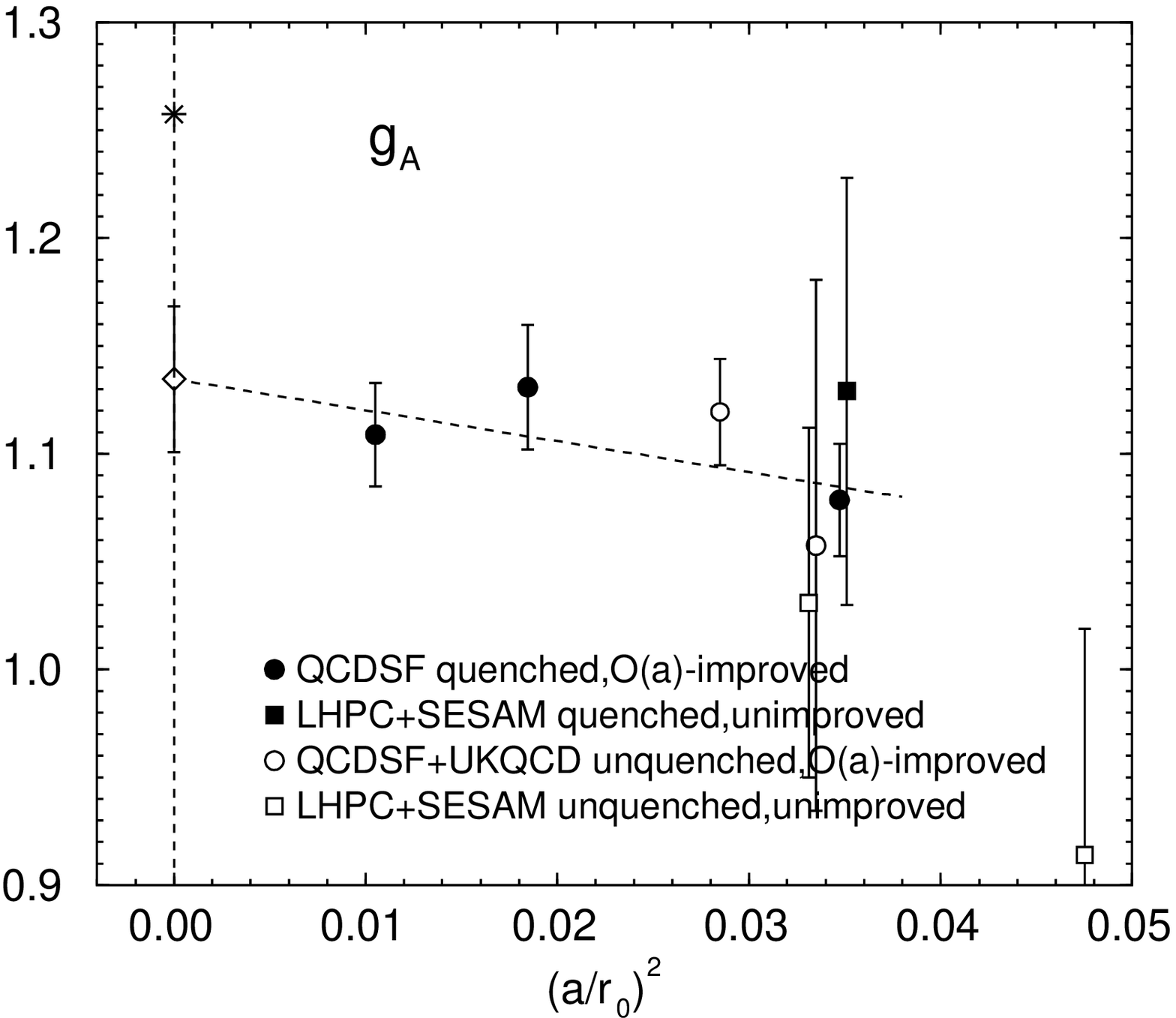}
\end{center}
\caption[]{$g_A$ from quenched and unquenched simulations versus $a^2$
  in units of $r_0 \approx 0.5 \, \mbox{fm}$. The experimental value is
  indicated by the asterisk.}
\label{fig.roger}
\end{figure}

Could the discrepancy between the simulations and experiment be caused
by the chiral extrapolation? The data at finite masses which are behind
the results displayed in Fig.\ref{fig.roger} do not show any deviation from
linearity when plotted versus the quark mass. This is exemplified in
Fig.~\ref{fig.gaq_m} where the quenched QCDSF data with lattice
artefacts subtracted are plotted versus the square of the pseudoscalar mass
in units of $r_0^{-1}$. But will this linearity
persist down to the physical mass? At small masses one has to worry
about finite size effects, and simulations with unimproved Wilson fermions
which we are performing show indeed some indications of such effects,
but no significant deviation from linearity yet.
\begin{figure}[t]
\begin{center}
\includegraphics[width=9.0cm]{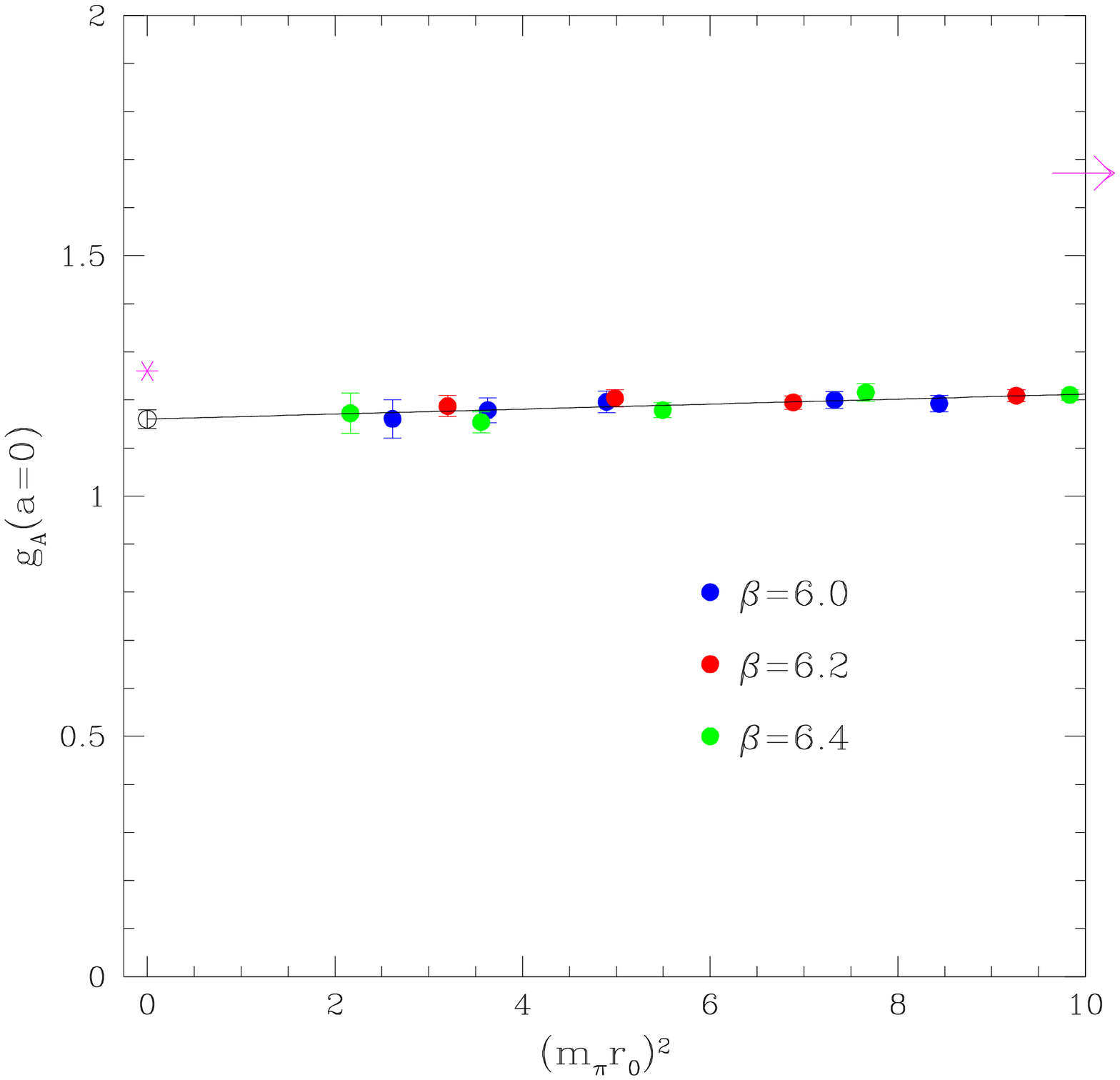}
\end{center}
\caption[]{QCDSF results for $g_A$ from quenched improved
  Wilson fermions plotted
  versus the square of the pseudoscalar mass (proportional to the
  quark mass). The line shows a linear chiral extrapolation.
  The experimental value is indicated by the asterisk.}
\label{fig.gaq_m}
\end{figure}

Clearer evidence for finite size effects in $g_A$ has been found
by the RBC collaboration \cite{RBC} in quenched simulations with
domain wall fermions (an approximate realization of
Ginsparg-Wilson fermions). For a more theoretical discussion of
the volume dependence of $g_A$ we refer to the recent papers by
Jaffe and Cohen \cite{jaffe}. Hopefully, the influence of the
finite volume will soon be better understood leading also to a
clearer picture of the quark mass dependence. Recent results from
chiral perturbation theory \cite{chpt} shed more light on the
problem of the chiral extrapolation. These developments should
eventually enable us to increase the reliability of the lattice
computations of $g_A$.


\def\Title#1{\begin{center} {\Large {\bf #1} } \end{center}}

\title*{CLEO-c and CESR-c: Allowing Quark Flavor Physics to Reach Its
Full Potential}
\toctitle{CLEO-c and CESR-c: Allowing Quark Flavor Physics \protect\newline to
Reach Its Full Potential}
%
%
\titlerunning{CLEO-c and CESR-c}
%
\author{I. Shipsey}
\authorrunning{I. Shipsey}
%
%
\institute{Department of Physics\\
Purdue University\\
West Lafayette, IN 47907, U.S.A.}

\maketitle

\begin{abstract}

We report on the physics potential of a proposed conversion of the
CESR machine and the CLEO detector to a charm and QCD factory:
``CLEO-c and CESR-c'' that will make crucial contributions to
quark flavor physics this decade, and may offer our best hope for
mastering non-perturbative QCD, which is essential if we are to
understand strongly coupled sectors in the new physics that lies
beyond the Standard Model. Of particular relevance to this
workshop CLEO-c will make a precise measurement of $V_{cd}$ that
can be combined with the beautiful measurements of $V_{ud}$
discussed elsewhere in these proceedings to test of the unitarity
of the first column of the CKM matrix.

\end{abstract}


\section{Executive Summary}

The goals of quark flavor physics are:
to test the consistency of the Standard Model (SM)
description of quark mixing and CP violation,
to search for evidence of new physics, and to sort between new physics
scenarios initially uncovered at the LHC. This will require a
range of measurements in the quark flavor changing sector of the SM
at the per cent level. These measurements will come from a variety
of experiments including
BABAR and Belle and their upgrades,
full exploitation of the facilities at Fermilab (CDF/D0/BTeV)
and at the LHC (CMS/ATLAS/LHC-b), and experiments in rare kaon decays.

However, the window to new physics that quark flavor physics can
provide, has a curtain drawn across it. The curtain represents hadronic
uncertainty. The study of weak interaction
phenomena, and the extraction of quark mixing matrix parameters remain
limited by our capacity to deal with non-perturbative strong
interaction dynamics.  Techniques such as lattice QCD (LQCD) directly address
strongly coupled theories and have the potential to eventually determine
our progress in many areas of particle physics. Recent advances in LQCD have
produced a wide variety of calculations of non-perturbative quantities
with accuracies in the 10-20\% level for systems involving one or two heavy
quark such as  $B$ and $D$  mesons, and $\Psi$ and $\Upsilon$
quarkonia. The techniques needed to reduce uncertainties
to 1-2\% precision exist, but the path to
higher precision is hampered by the absence of accurate charm
data against which to test and calibrate the new theoretical techniques.

To meet this challenge the CLEO collaboration has proposed to operate
CLEO and CESR  as a charm and QCD factory at charm threshold where the
experimental conditions are optimal. In a three year focused program
CLEO-c will obtain charm data samples
one to two orders of magnitide larger than any previous experiment operating
in this energy range, and with a detector that is significantly more powerful
than any previous detector to operate at charm threshold. CLEO-c has
the potential
to provide a unique and crucial validation of LQCD with accuracies of 1-2\%.

If LQCD is validated, CLEO-c data will lead to a dramatic improvement
in our knowledge of the quark couplings in
the charm sector. In addition CLEO-c validation of lattice calculations,
combined with
B factory, Tevatron, and LHC data will allow a significant improvement in our
knowledge of quark couplings in the beauty sector. The impact CLEO-c will have
on our knowledge of the CKM matrix makes the experiment an essential step
in the quest to understand the origin of CP violation and quark mixing.
CLEO-c allows quark flavor physics to reach its full potential,
by enabling the heavy flavor community to draw back the curtain
of hadronic uncertainty, and thereby
see clearly through the window to the new physics that lies beyond the SM.
Of equal importance, CLEO-c allows us to significantly advance our
understanding and control over strongly-coupled, non-perturbatyive
quantum field theories in general. An understanding of strongly coupled
theories will be a crucial element in helping to interpret new
phenomena at the high energy frontier.

\begin{figure}
\begin{center}
\epsfig{figure=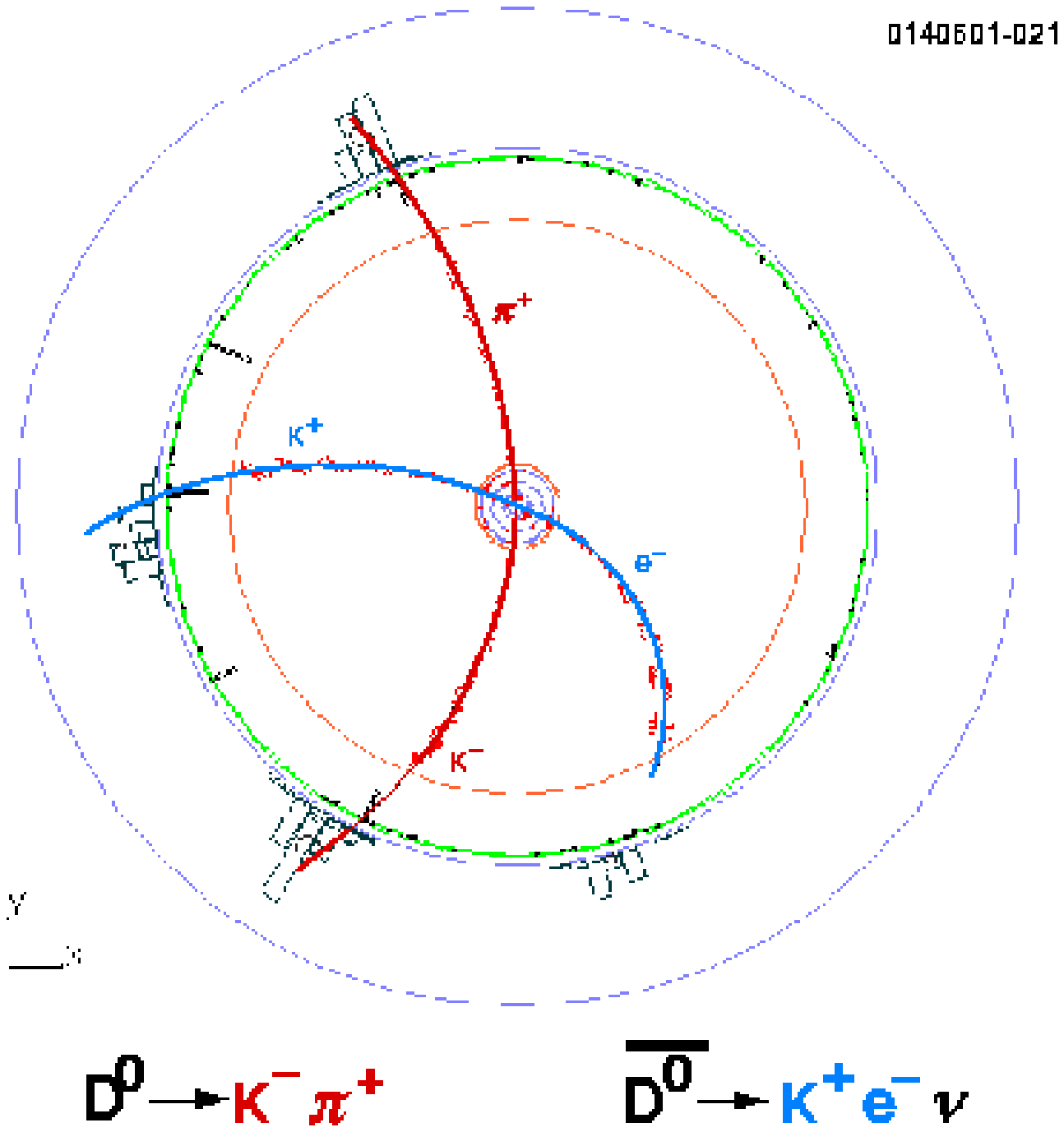,width=6.5cm,height=7cm}
\caption {A doubly tagged  event at the  $\psi(3770)$.}
\label{fig:cleoc_event}
\end{center}
\end{figure}

\section{Introduction}
\begin{figure}
\begin{center}
\epsfig{figure=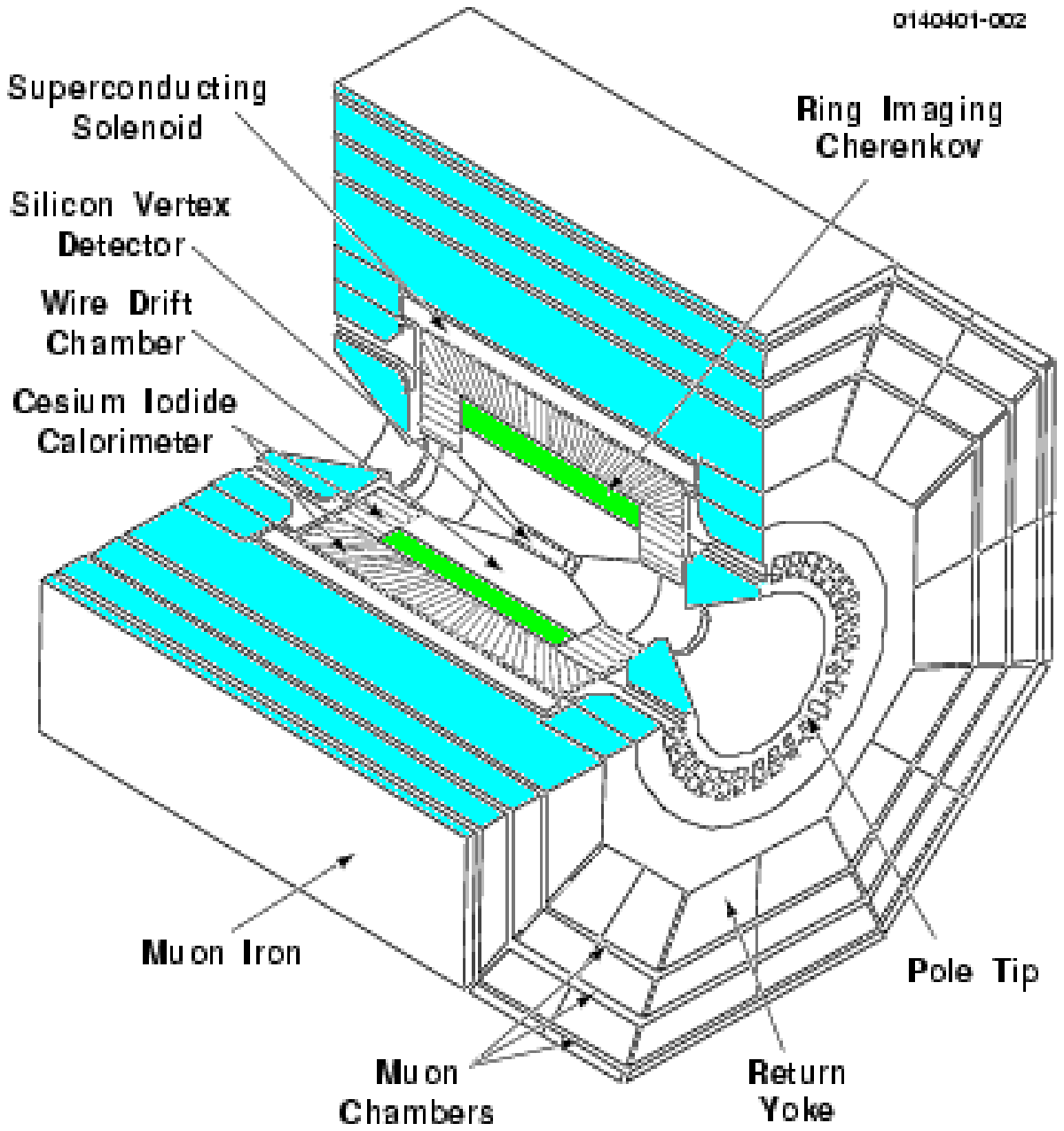,width=8.5cm,height=8.5cm}
\caption {The CLEO III detector.}
\label{fig:cleo3_det}
\end{center}
\end{figure}
For many years, the CLEO experiment at the Cornell Electron Storage
Ring, CESR, operating on the $\Upsilon$(4S) resonance, has provided most of the
world's information about the $B_{d}$ and $B_{u}$ mesons.
At the same time, CLEO, using the copious continuum pair production at the
$\Upsilon$(4S) resonance
has been a leader in the study of charm and $\tau$ physics. Now that the
asymmetric B factories have achieved high luminosity, CLEO is
uniquely positioned  to advance
the knowledge of quark flavor physics by carrying out several measurements
near charm threshold, at center of mass energies in the 3.5-5.0 GeV region.
These measurements address crucial topics which benefit from the high
luminosity and experimental constraints which exist near threshold but
have not been carried out at existing charm factories because the luminosity
has been too low, or have been carried out previously with meager statistics.
They include:

\begin{enumerate}
\item Charm Decay constants $f_D, f_{D_s}$
\item Charm Absolute Branching Fractions
\item Semileptonic decay form factors
\item Direct determination of $V_{cd}$ \& $V_{cs}$
\item QCD studies including: \\
Charmonium and bottomonium spectroscopy \\
Glueball and exotic searches \\
Measurement of R between 3 and 5 GeV, via scans \\
Measurement of R between 1 and 3 GeV, via ISR
\item Search for new physics via charm mixing, CP violation
and rare decays
\item $\tau$ decay physics
\end{enumerate}

The CLEO detector can carry out this program with only minimal
modifications. The CLEO-c project is described at length in
\cite{cleo-c} - \cite{maravin}. A very modest upgrade to the
storage ring is required to achieve the required luminosity.
Below, we summarize the advantages of running at charm threshold,
the minor modifications required to optimize the detector,
examples of key analyses, a description of the proposed run plan,
and a summary of the physics impact of the program.

\begin{table}[t]
\begin{center}
\caption{Summary of CLEO-c charm decay measurements.}
\label{tab:charm}
\begin{tabular}{cccccc}
\hline
Topic & Reaction & Energy & $L $        & current     & CLEO-c \\
      &          & (MeV)  & $(fb^{-1})$ & sensitivity & sensitivity \\
\hline
\multicolumn{1}{c}{Decay constant} &
\multicolumn{5}{c}{}\\
\hline
$f_D$ & $D^+\to\mu^+\nu$ & 3770 & 3 & UL & 2.3\% \\
\hline
$f_{D_s}$ & $D_{s}^+\to\mu^+\nu$ & 4140 & 3 & 14\% & 1.9\% \\
\hline
$f_{D_s}$ & $D_{s}^+\to\mu^+\nu$ & 4140 & 3 & 33\% & 1.6\% \\
\hline
\multicolumn{2}{c}{Absolute Branching Fractions} &
\multicolumn{4}{c}{}\\
\hline
\multicolumn{2}{c}{$Br(D^0 \to K\pi)$} & 3770 & 3 & 2.4\% & 0.6\% \\
\hline
\multicolumn{2}{c}{$Br(D^+ \to K\pi\pi)$} & 3770 & 3 & 7.2\% & 0.7\% \\
\hline
\multicolumn{2}{c}{$Br(D_s^+\to\phi\pi)$} & 4140 & 3 & 25\% & 1.9\% \\
\hline
\multicolumn{2}{c}{$Br(\Lambda_c\to pK\pi)$} & 4600 & 1 & 26\% & 4\% \\
\hline

\end{tabular}

\end{center}

\end{table}

\subsection {Advantages of running at charm threshold}

The B factories, running at the $\Upsilon$(4S) will have produced 500 million
charm pairs by 2005. However, there are significant advantages of running at
charm threshold:

\begin{enumerate}
\item Charm events produced at threshold are extremely clean.
\item Double tag events, which are key to making absolute branching fraction
measurements, are pristine.
\item Signal/Background is optimum at threshold.
\item Neutrino reconstruction is clean.
\item Quantum coherence aids $D$ mixing and CP violation studies.
\end{enumerate}

These advantages are dramatically illustrated in Fig. 1, which shows a picture of a simulated and fully reconstructed
$\psi(3770)\to D\bar{D}$ event.

\subsection {The CLEO-III Detector : Performance, Modifications and issues}

The CLEO III detector, shown in Figure~\ref{fig:cleo3_det}, consists of
a new silicon tracker, a new drift chamber,
and a Ring Imaging Cherenkov Counter (RICH), together  with the
CLEO II/II.V magnet, electromagnetic calorimeter and muon chambers.
The upgraded detector was installed and commissioned during the Fall of 1999
and Spring of 2000.
Subsequently operation has been very reliable (see below for a caveat)
and a very high quality data set has been obtained.
To give an idea of the power of the CLEO III detector
in Figure~\ref{fig:rich} (left plot) the beam
constrained mass for the Cabibbo allowed decay $B\to D\pi$ and
the Cabibbo suppressed decay $B\to DK$
with and without RICH information is shown.

The latter decay was extremely difficult to observe in
CLEO II/II.V which did not have a RICH detector.
In the right plot of Figure~\ref{fig:rich}
the penguin dominated decay $B \to K\pi$ is shown.
This, and other rare $B$ decay modes are observed in CLEO III
with branching ratios consistent with those found in CLEO
II/II.V, and are also in agreement with recent Belle and BABAR results.
Figure~\ref{fig:rich} is a demonstration that CLEO III performs very well
indeed.

\begin{figure*}[t]
\begin{center}
\epsfig{figure=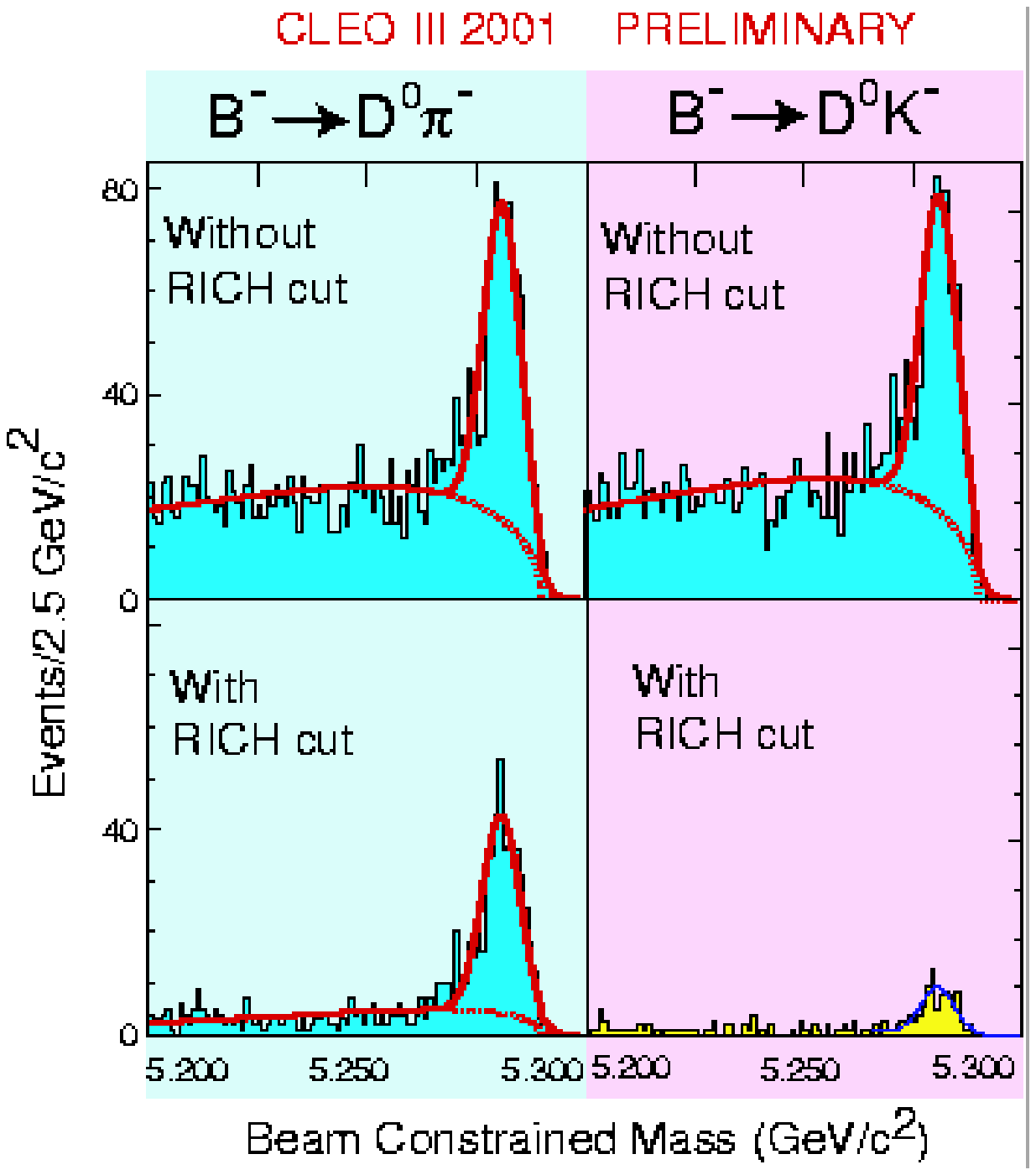,width=6.8cm,height=6.8cm}
\epsfig{figure=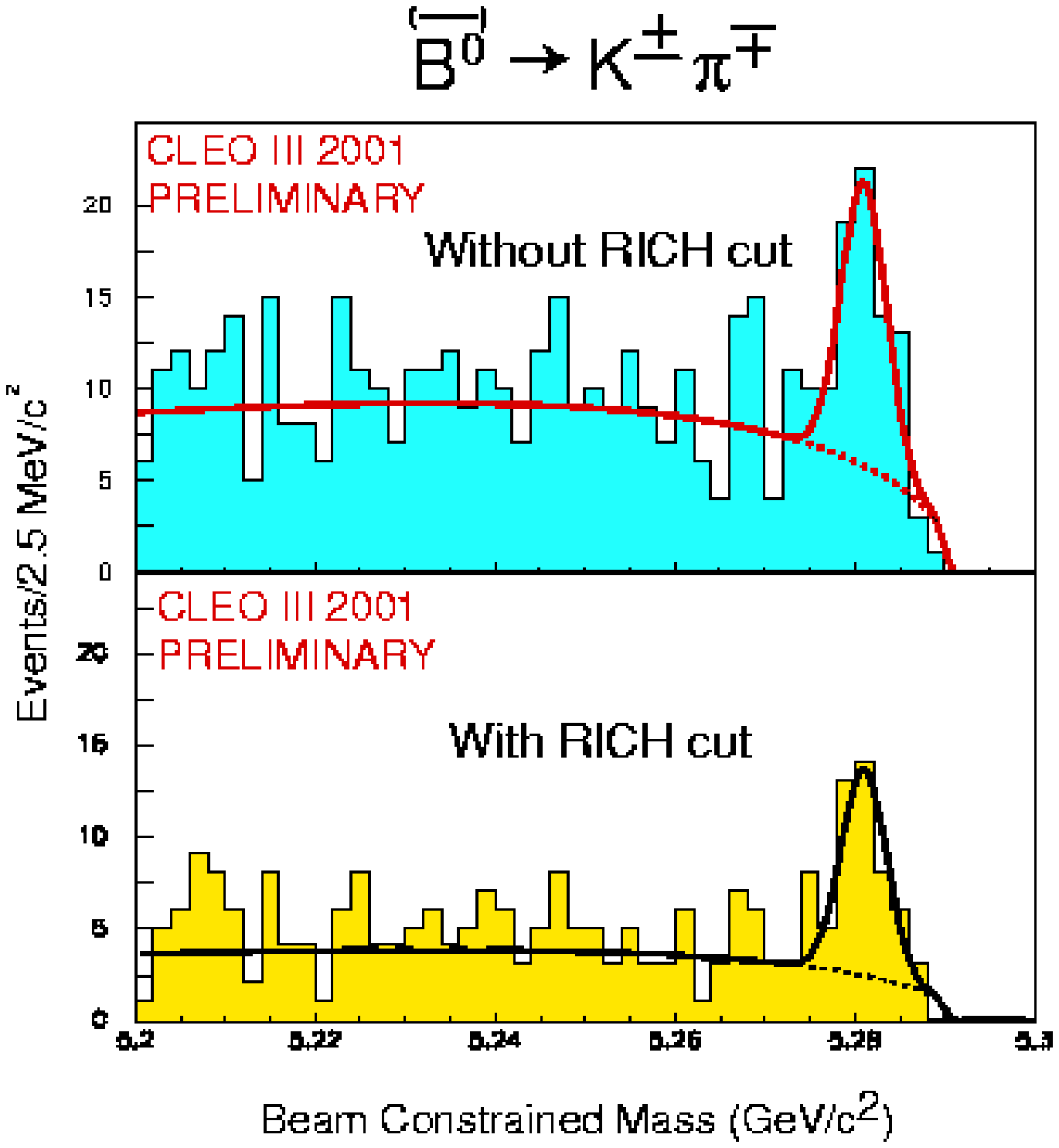,width=6.8cm,height=6.8cm}
\caption
{(Top) Beam constrained mass for the Cabibbo allowed decay $B\to
D\pi$  and the Cabibbo suppressed decay $B\to DK$ with and without
RICH information. The latter decay was extremely difficult to
observe in CLEO II/II.V which did not have a RICH detector.
(Bottom) The penguin dominated decay $B \to K\pi$. Both of these
modes are observed in CLEO III with branching ratios consistent
with those found in CLEO II/II.V.} \label{fig:rich}
\end{center}
\end{figure*}

Unfortunately, there is one detector subsystem that is not performing well.
The CLEO III silicon has experienced an unexpected and unexplained loss of
efficiency. The silicon detector
will be replaced with a wire vertex chamber for CLEO-c.
We note that if one was to design a charm factory detector from
scratch the tracking would be entirely gas based to ensure
that the detector material was kept to a minimum.
CLEO-c simulations indicate that a simple six layer stereo tracker
inserted into the CLEO III
drift chamber, as a silicon detector replacement, would provide a system
with superior momentum resolution
compared to the current CLEO III tracking system.

Due to machine issues we plan to lower the solenoid field
strength to 1.0 T from 1.5 T. All other
parts of the detector do not require modification.
The dE/dx and Ring Imaging Cerenkov counters
are expected to work well over the CLEO-c momentum range.
The electromagnetic calorimeter
works well and has fewer photons to deal with at 3-5 GeV than at 10 GeV.
Triggers will work as before.
Minor upgrades may be required of the Data Acquisition system to
handle peak data transfer rates. The conclusion is that, with the addition of
the
replacement wire chamber, CLEO is expected to work well in the 3-5 GeV
energy range at the expected rates.

\subsection {Machine Conversion}

Electron positron colliders are designed to operate optimally within
a relatively narrow
energy range. As the energy is reduced below design, there is a
significant reduction in synchrotron radiation, which is the primary means
of cooling the beam. In consequence, the luminosity drops, roughly as the
beam energy to the fourth power. Without modification to the machine,
CESR performance in the 3-5 GeV
energy range would be modest,
well below $10^{31} {\rm cm}^{-2} {\rm s}^{-1}$.
CESR conversion to CESR-c requires 18 m of wiggler magnets, to increase
transverse cooling,
at a cost of $\sim$ \$4M. With the wigglers installed,
CESR-c is expected to
achieve a luminosity in the range $2-4 \times 10^{32} {\rm cm}^{-2}
{\rm s}^{-1}$
where the lower (higher) luminosity corresponds to $\sqrt{s} = 3.1 (4.1)
{\rm GeV}$.

\subsection {Examples of analyses with CLEO-c}

The main targets for the CKM physics program at CLEO-c are
absolute branching ratio
measurements of hadronic, leptonic and semileptonic decays.
The first of these provides an absolute
scale for all charm and hence all beauty decays.
The second measures decay constants and the third
measures form factors and, in combination with theory, allows
the determination of $V_{cd}$ and $V_{cs}$.

\subsubsection {Absolute branching ratios}

The key idea is to reconstruct a $D$ meson in any hadronic mode.
This, then, constitutes the tag. Figure~\ref{fig:d0kpi}
shows tags in the mode $D\to K\pi$. Note the y axis is a log scale.
Tag modes are very clean. The signal to background ratio is $\sim$ 5000/1
for the example shown.
Since $\psi(3770) \to D\bar{D}$, reconstruction of a second D meson in a
tagged event to a final state X, corrected by the efficiency which is very
well known, and divided by
the number of D tags, also very well known, is a measure of
the absolute branching ratio
$Br(D\to X)$. Figure~\ref{fig:br_dkpipi} shows the $K^{-}\pi^{+}\pi^{+}$
signal from doubly tagged events. It is approximately background free.
The simplicity of $\psi(3770) \to D\bar{D}$
events combined with the absence of background allows the determination of
absolute branching ratios with extremely small systematic errors.
This is a key advantage of running at threshold.

\begin{figure}
\begin{center}
\epsfig{figure=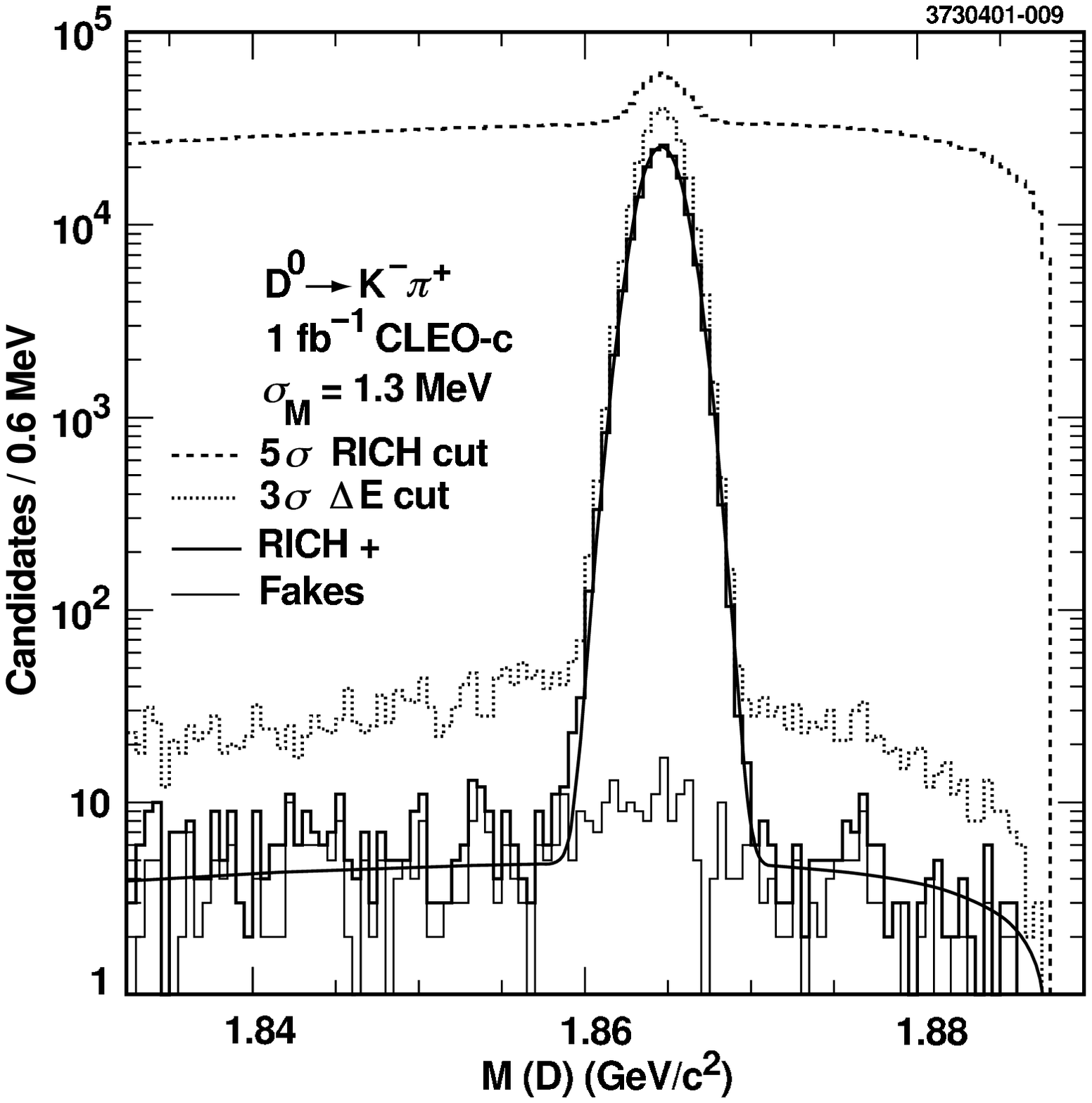,width=9cm,height=9cm}
\caption
{$K\pi$ invariant mass in $\psi(3770)\to D\bar{D}$ events
showing a strikingly clean signal for
$D\to K\pi$. The y axis is a logarithmic scale. The signal to background ratio
is $\sim$ 5000/1.}
\label{fig:d0kpi}
\end{center}
\end{figure}

\begin{figure}
\begin{center}
\epsfig{figure=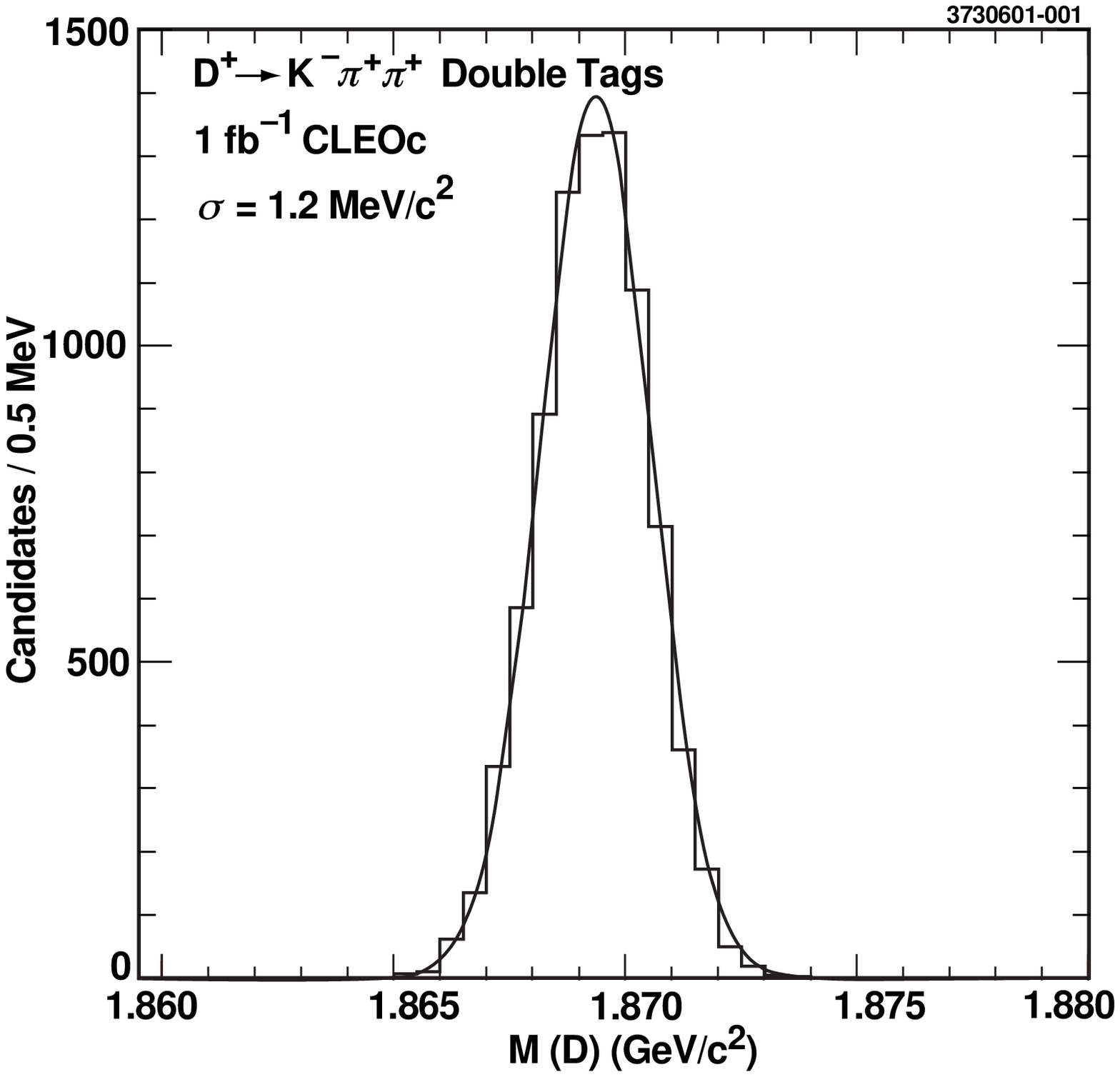,width=9cm,height=9cm}
\caption
{$K\pi\pi$ invariant mass in $\psi(3770)\to D\bar{D}$
events where the other D in the event has
already been reconstructed.
A clean signal for $D\to K\pi\pi$ is observed and the absolute
branching ratio $Br(D\to K\pi\pi)$ is measured by counting events
in the peak.}
\label{fig:br_dkpipi}
\end{center}
\end{figure}

\subsubsection {Leptonic decay $D_s\to\mu\nu$}

This is a crucial measurement because it provides information which
can be used to extract the weak decay constant, $f_{D_{s}}$. The
constraints provided by running at threshold are critical to extracting
the signal.

The analysis procedure is as follows:
\begin{enumerate}
\item Fully reconstruct one $D_s$, this is the tag.
\item Require one additional charged track and no additional photons.
\item Compute the missing mass squared ($m_{\nu}^2$) which  peaks at zero for
a decay where only a neutrino is unobserved.
\end{enumerate}

The missing mass resolution, which  is of order $\sim m_{\pi^0}$,
is sufficient to reject the backgrounds to this process as shown
in Fig.~\ref{fig:munu_pienu}. There is no need to identify muons,
which helps reduce the systematic error. One can inspect the
single prong to make sure it is not an electron. This provides a
check of the background level since the leptonic decay to an
electron is severely helicity-suppressed and no signal is expected
in this mode.

\begin{figure*}
\begin{center}
\epsfig{figure=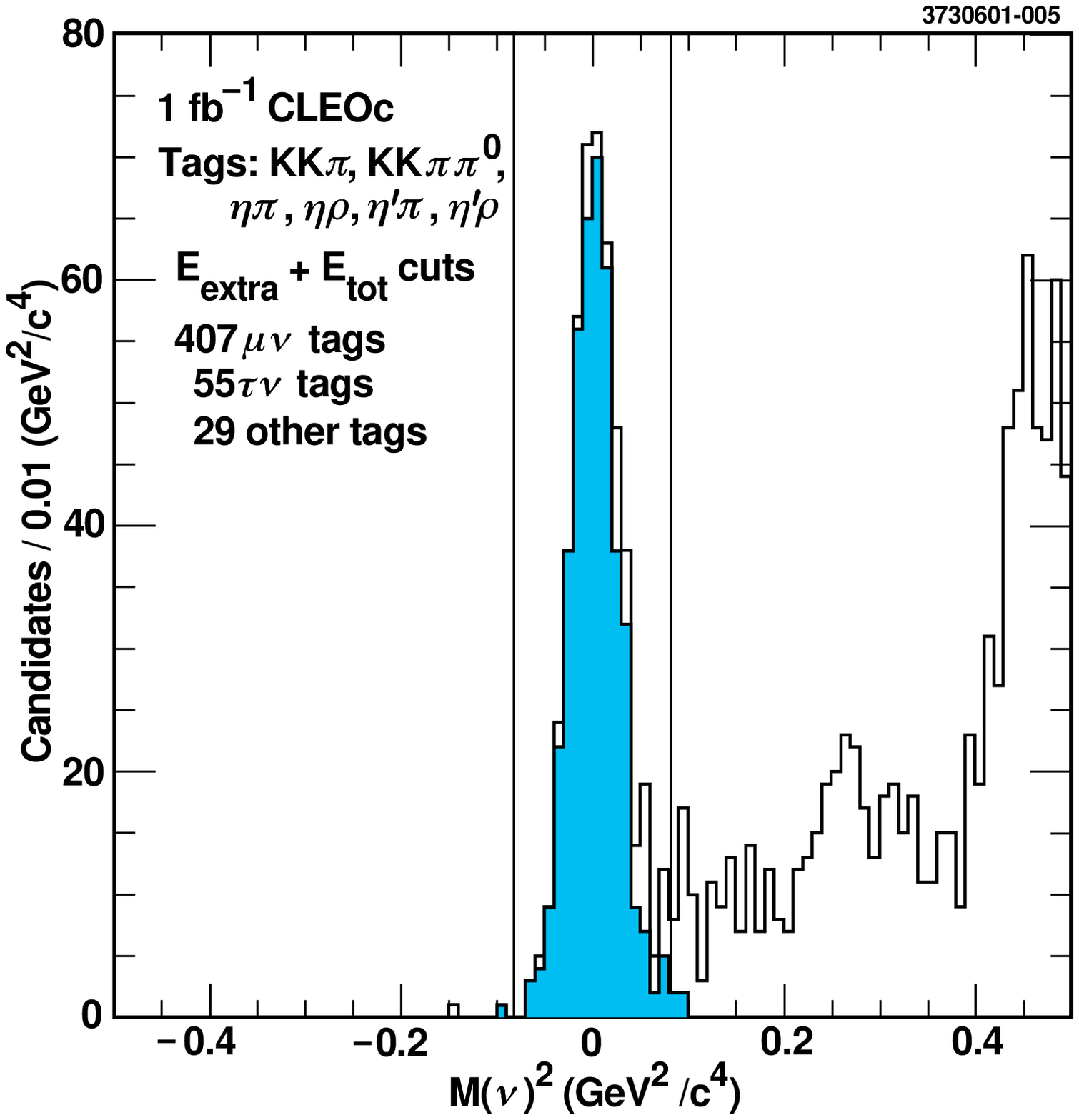,width=6cm,height=6cm}
\epsfig{figure=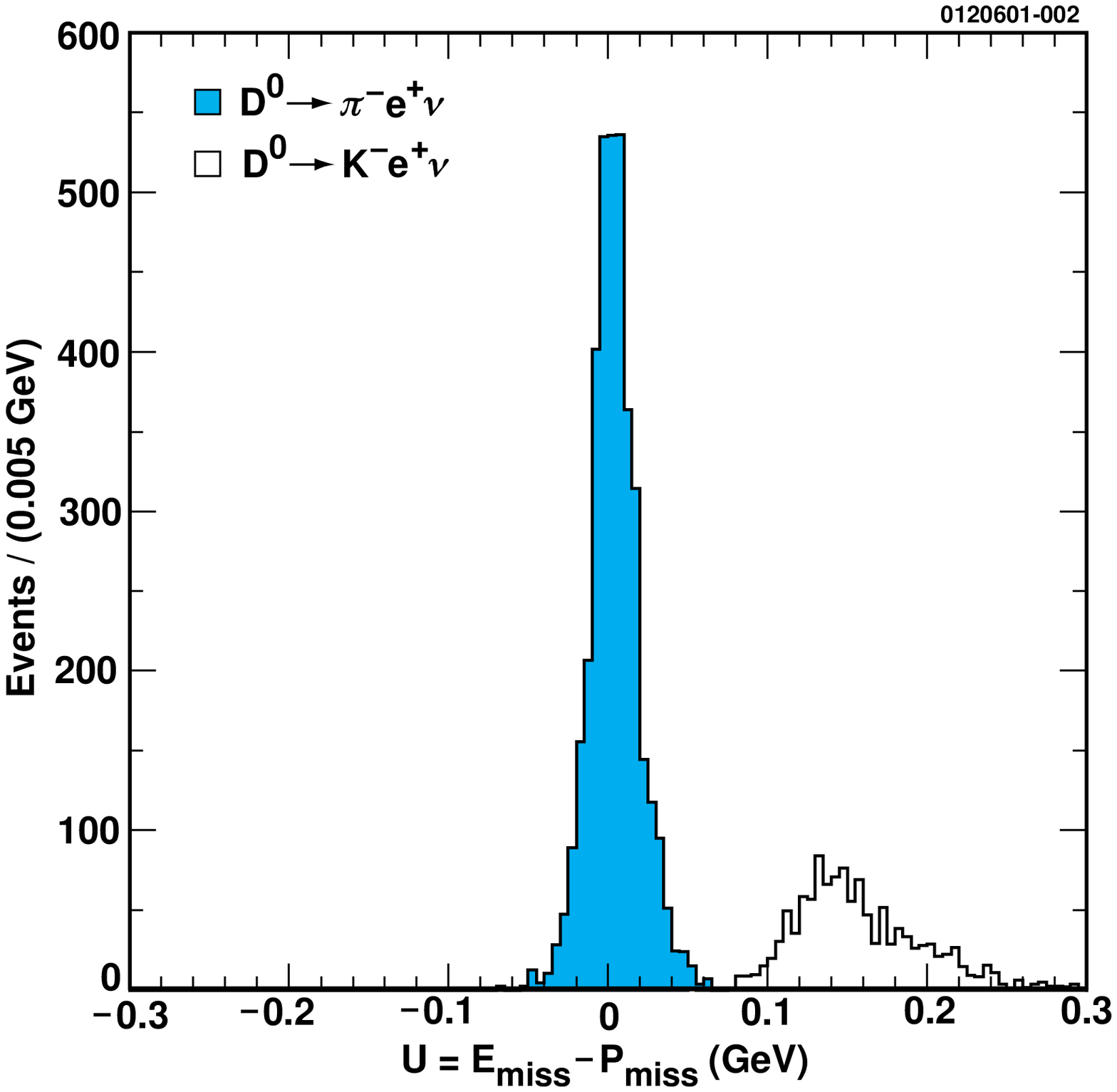,width=6cm,height=6cm}
\caption
{(Left) Missing mass squared for $D_{s} \bar{D_{s}}$ tagged pairs produced
at $\sqrt{s}=4100$ MeV. Events due to the decay $D_s\to\mu\nu$ are shaded.
(Right) The difference between the missing energy and missing
momentum in $\psi(3770)\to D\bar{D}$ tagged events for
the Cabibbo suppressed decay $D\to \pi \ell\nu$ (shaded).
The unshaded histogram arises from the ten times more copiously produced
Cabibbo allowed
transition $D\to K \ell\nu$ where the $K$ is outside the fiducial volume of
the RICH.}
\label{fig:munu_pienu}
\end{center}
\end{figure*}

\subsubsection {Semileptonic decay $D\to \pi e^+\nu$}

The analysis procedure is as follows:
\begin{enumerate}
\item Fully reconstruct one D, this constitutes the tag.
\item Identify one electron and one hadronic track.
\item Calculate the variable, $U = E_{miss} - P_{miss}$, which
peaks at zero when only a neutrino has escaped detection, which is the case
for semileptonic decays.
\end{enumerate}
Using the above procedure results in the right plot of
Figure~\ref{fig:munu_pienu}.
With CLEO-c for the first time it will become possible to make
precise branching ratio and
absolute form factor measurements of every charm meson
semileptonic pseudoscalar to
pseudoscalar  and pseudoscalar to vector transition.
This will be a lattice validation data set without
equal. Figure~\ref{fig:error_br} shows the current precision
with which the absolute semileptonic branching ratios
of charm particles are known, and the precision attainable
with CLEO-c.

\subsection {Run Plan}

CLEO-c must run at various center of mass energies to
achieve its physics goals. The ``run plan'' currently used
to calculate the physics reach is given below. This plan assumes CESR-c
achieves design luminosity.  Item 1 is prior
to machine conversion, while the remaining items are post machine conversion.

\begin{enumerate}
\item 2002 : $\Upsilon$'s --  1-2 $fb^{-1}$ each at
$\Upsilon(1S),\Upsilon(2S),\Upsilon(3S)$ \\
Spectroscopy, electromagnetic transition matrix elements, the leptonic width.
$\Gamma_{ee}$, and searches for the yet to be discovered
$h_b, \eta_b$ with 10-20 times the existing world's data sample.
As of July 2002,
most of this data has been collected.
\item 2003 : $\psi(3770)$ -- 3 $fb^{-1}$ \\
30 million events, 6 million tagged D decays (310 times MARK III)
\item 2004 : 4100 MeV -- 3 $fb^{-1}$ \\
1.5 million $D_{s}D_{s}$ events, 0.3 million tagged $D_s$ decays
(480 times MARK III, 130 times BES)
\item 2005 : $J/\psi$ -- 1 $fb^{-1}$ \\
1 billion $J/\psi$ decays (170 times MARK III, 20 times BES II)
\end{enumerate}

\subsection{Physics Reach of CLEO-c}

Tables~\ref{tab:charm}, \ref{tab:ckm} , and
\ref{tab:comparisonship}, and Figures~\ref{fig:error_br} and
\ref{fig:comparison} summarize the CLEO-c measurements of charm
weak decays, and compare the precision obtainable with CLEO-c to
the expected precision at BABAR which expects to have recorded
about 500 million charm pairs by 2005. While BABAR data allows
improvement in the precision with which these quantities can be
measured,  CLEO-c clearly achieves far greater precision for many
measurements. The reason for this is the ability to measure
absolute branching ratios by tagging, and the absence of
background at threshold. For charm quantities where CLEO-c is not
dominant, it will remain comparable in sensitivity, and
complementary in technique, to the B factories. Also shown in
Table~\ref{tab:comparisonship} is a summary of the data set size
for CLEO-c and BES II at the $J/\psi$ and $\psi'$, and the
precision with which R, the ratio of the $e^{+}e^{-}$ annihilation
cross section into hadrons to mu pairs, can be measured. The
CLEO-c data sets are over an order of magnitude larger, the
precision with which R is measured is a factor of three higher, in
addition the CLEO detector is vastly superior to the BES II
detector.

Taken together the CLEO-c datasets at the $J/\psi$ and $\psi'$ will be
qualitatively and quantitatively superior to any previous dataset
in the charmonium sector thereby
providing discovery potential for glueballs and exotics without equal.

\begin{figure}
\begin{center}
\epsfig{figure=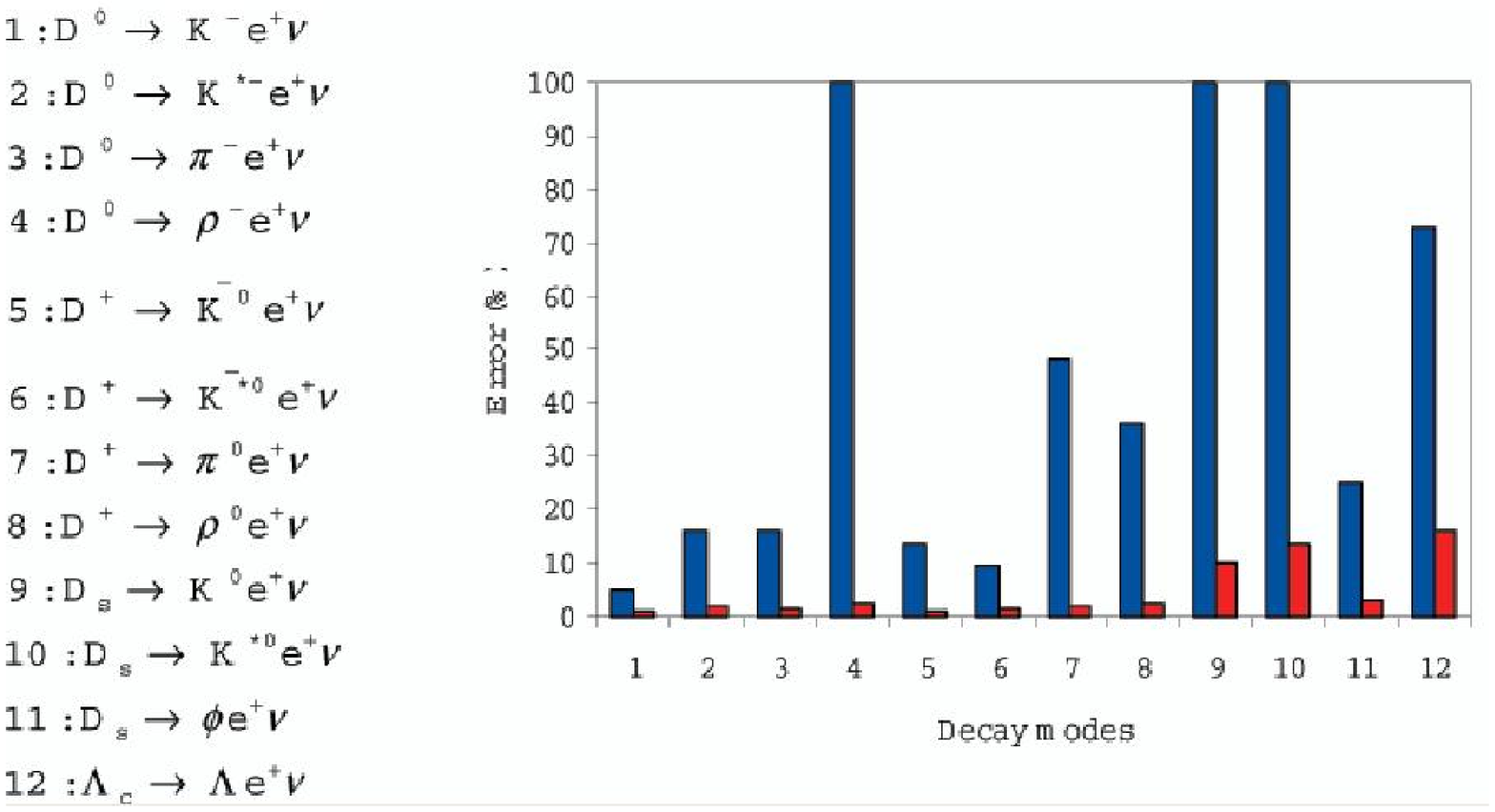,height=6cm,angle=0} \caption {Absolute
branching ratio current precision from the PDG (left entry) and
precision attainable at CLEO-c (right entry ) for twelve
semileptonic charm decays.} \label{fig:error_br}
\end{center}
\end{figure}

\begin{table}
\begin{center}
\caption{Summary of direct CKM reach with CLEO-c}
\label{tab:ckm}
\begin{tabular}{cccccc}
\hline
Topic & Reaction & Energy & $L $        & current     & CLEO-c \\

      &          & (MeV)  & $(fb^{-1})$ & sensitivity & sensitivity \\
\hline
$V_{cs}$ & $D^0\to K\ell^+\nu$ & 3770 & 3 & 16\% & 1.6\% \\
\hline
$V_{cd}$ & $D^0\to\pi\ell^+\nu$ & 3770 & 3 & 7\% & 1.7\% \\
\hline
\end{tabular}
\end{center}
\end{table}

\begin{table}
\begin{center}
\caption{Comparision of CLEO-c reach to BABAR and BES}
\label{tab:comparisonship}
\begin{tabular}{cccccc}
\hline
Quantity & CLEO-c & BaBar  & Quantity & CLEO-c & BES-II \\
\hline
$f_D$ & 2.3\% & 10-20\%    & \#$J/\psi$ & $10^9$ & $5\times 10^7$ \\
\hline
$f_{D_s}$ & 1.7\% & 5-10\% & $\psi'$   & $10^8$ & $3.9\times 10^6$ \\
\hline
$Br(D^0 \to K\pi)$ & 0.7\% & 2-3\% & 4.14 GeV & $1 fb^{-1}$ &$23 pb^{-1}$\\
\hline
$Br(D^+ \to K\pi\pi)$ & 1.9\% & 3-5\%  & 3-5 R Scan & 2\%  & 6.6\% \\
\hline
$Br(D_s^+\to\phi\pi)$ & 1.3\% & 5-10\% & \multicolumn{3}{c}{} \\
\hline
\end{tabular}
\end{center}
\end{table}

\begin{figure}
\begin{center}
\epsfig{figure=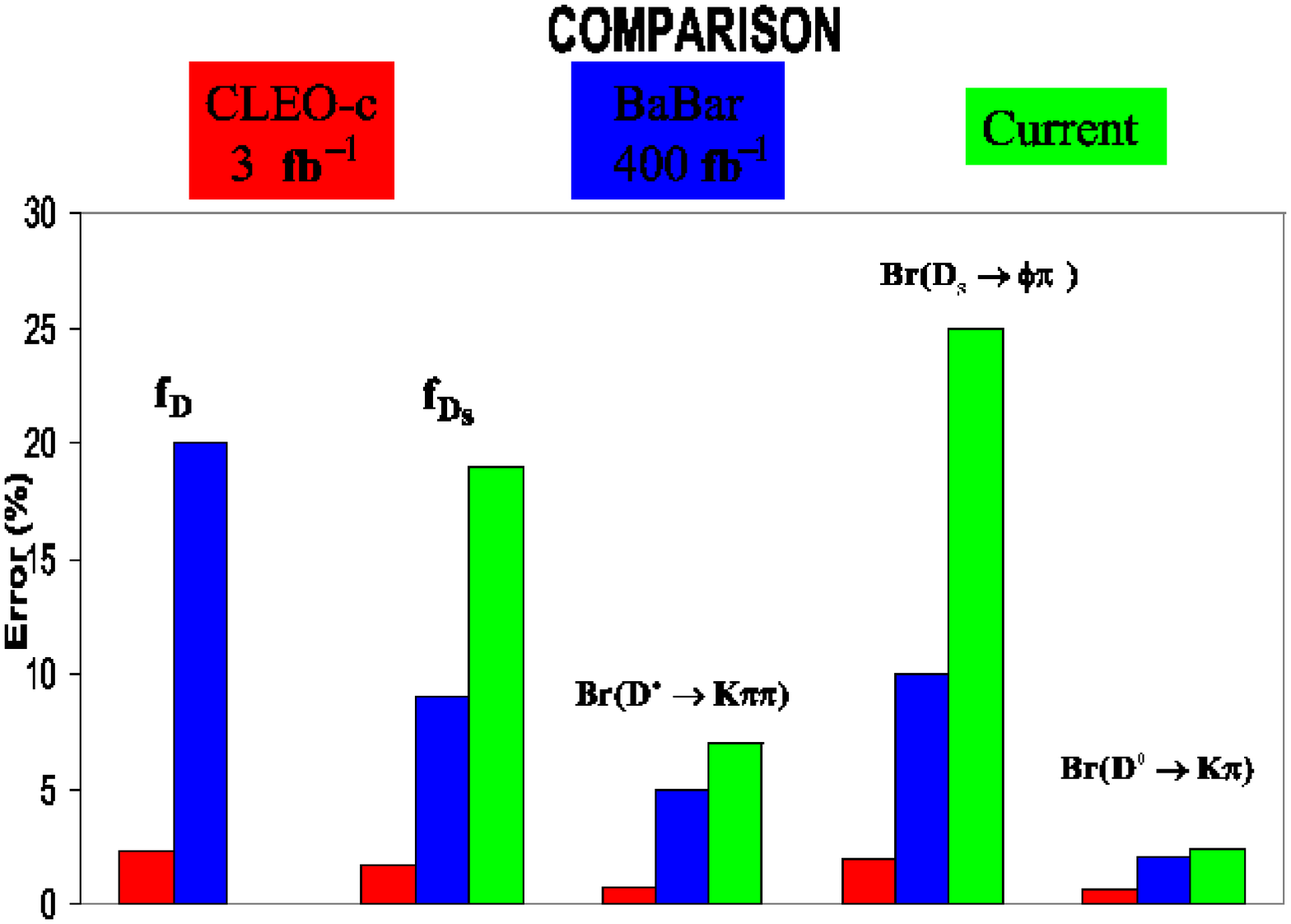,width=8cm,height=6cm}
\caption
{Comparison of CLEO-c (left) BABAR (center) and PDG2001 (right)
for the charm meson decay constants and three important charm meson
hadronic decay branching ratios.}
\label{fig:comparison}
\end{center}
\end{figure}

\subsection{CLEO-c Physics Impact}

CLEO-c will provide crucial validation of Lattice QCD, which  will
be able to  calculate with accuracies of 1-2\%. The CLEO-c decay
constant and semileptonic data will provide a ``golden'', and
timely test while CLEO-c QCD and charmonium data provide
additional benchmarks. CLEO-c will provide dramatically improved
knowledge of absolute charm branching fractions which are now
contributing significant errors to measurements involving b's in a
timely fashion. CLEO-c will significantly improve knowledge of CKM
matrix elements which are now not very well known. $V_{cd}$ and
$V_{cs}$ will be determined directly by CLEO-c data and LQCD, or
other theoretical techniques. $V_{cb}, V_{ub}, V_{td}$ and
$V_{ts}$ will be determined with enormously improved precision
using B factory and Tevatron data, once the CLEO-c program of
lattice validation is complete. Table~\ref{tab:ckm_summary}
provides a summary of the situation. CLEO-c data alone will also
allow new tests of the unitarity of the CKM matrix. The unitarity
of the second row of the CKM matrix will be probed at the 3\%
level.  Of particular relevance to this workshop CLEO-c will make
a measurement of $V_{cd}$ to better than 2\% precision. This
measurement can be combined with the beautiful measurements of
$V_{ud}$ discussed elsewhere in these proceedings to test of the
unitarity of the first column of the CKM matrix to a precision
similar to that with which the first row is now known.  CLEO-c
data will also test unitarity by measuring the ratio of the long
sides of the squashed $cu$ triangle to 1.3\%.

Finally the potential to observe new forms of matter; glueballs,
hybrids, etc. in $J/\psi$ decays and new
physics through sensitivity to charm mixing, CP violation,
and rare decays provides a discovery
component to the program.

\begin{table}
\begin{center}
\caption{Current knowledge of CKM matrix elements (row one). Knowledge of
CKM matrix elements after CLEO-c (row two).
The improvement in the precision with which
$V_{cd}$ and $V_{cs}$ are known is attainable with
CLEO-c data combined with Lattice QCD.
The improvement in precision with which $V_{cb}$, $V_{ub}$, $V_{td}$, and
$V_{ts}$ are known is obtained from CLEO-c validated Lattice
QCD calculations and B factory and Tevatron
data.}
\vspace{0.1in}
\label{tab:ckm_summary}
\begin{tabular}{cccccc}
\hline
$V_{cd}$ & $V_{cs}$ & $V_{cb}$ & $V_{ub}$ & $V_{td}$ & $V_{ts}$ \\
\hline
7\% & 16\% & 5\% & 25\% & 36\% & 39\% \\
\hline
1.7\% & 1.6\% & 3\% & 5\% & 5\% & 5\% \\
\hline
\end{tabular}
\end{center}
\end{table}

\bigskip


I would like to thank my CLEO colleagues for providing the
opportunity to represent the collaboration at this conference. It
is a privilege to be part of the CLEO collaboration. I thank
Ikaros Bigi, Gustavo Burdman, Andreas Kronfeld, Peter Lepage,
Zoltan Ligeti and Matthias Neubert for valuable discussions.
Finally, I thank Hans Abele and his support team for the superb
organization of this conference.



\def\Journal#1#2#3#4{{#1} {\bf #2}, #3 (#4)}

\def\NCA{\em Nuovo Cimento}
\def\NIM{\em Nucl. Instrum. Methods}
\def\NIMA{{\em Nucl. Instrum. Methods} A}
\def\NPB{{\em Nucl. Phys.} B}
\def\PLB{{\em Phys. Lett.}  B}
\def\PRD{{\em Phys. Rev.} D}
\def\ZPC{{\em Z. Phys.} C}

\def\st{\scriptstyle}
\def\sst{\scriptscriptstyle}
\def\mco{\multicolumn}
\def\epp{\epsilon^{\prime}}
\def\vep{\varepsilon}
\def\ra{\rightarrow}
\def\ppg{\pi^+\pi^-\gamma}
\def\vp{{\bf p}}
\def\ko{K^0}
\def\kb{\bar{K^0}}
\def\al{\alpha}
\def\ab{\bar{\alpha}}
\def\be{\begin{equation}}
\def\ee{\end{equation}}
\def\bea{\begin{eqnarray}}
\def\eea{\end{eqnarray}}
\def\CPbar{\hbox{{\rm CP}\hskip-1.80em{/}}}

\def\DAF{DA$\Phi$NE }
\def\Repsp{Re($\epsilon'/\epsilon $)}
\def\epsp{$\epsilon'/\epsilon$}
\def\ppm{$\pi^{+}\pi^{-}$}
\def\p00{$\pi^{0}\pi^{0}$}
\def\KL{K$^{0}_{L}$}
\def\KS{K$^{0}_{S}$}
\def\KP{K$^{+}$}
\def\KM{K$^{-}$}
\def\KKbar{$K^{0}\overline{K^{0}}$}
\def\K3{$K_{l3}$}
\def\K4{$K_{l4}$}
\def\Kppo{K$^{\pm} \rightarrow \pi^{\pm} \pi^{0}$}
\def\Kmunu{K$^{\pm} \rightarrow \mu^{\pm}\bar{\nu}(\nu)$}
\def\Kppoo{K$^{\pm} \rightarrow \pi^{\pm} \pi^{0}\pi^{0}$}
\def\lum{cm$^{-2}$s$^{-1}$}
\def\K00{K$^{0}_{S} \rightarrow \pi^{0}\pi^{0}$}
\def\Kpm{K$^{0}_{S} \rightarrow \pi^{+}\pi^{-}$}
\def\Kpmg{K$^{0}_{S} \rightarrow \pi^+\pi^-(\gamma)$}
\def\Lzz{K$^{0}_{L} \rightarrow \pi^{0}\pi^{0}$}
\def\Lpm{K$^{0}_{L} \rightarrow \pi^{+}\pi^{-}$}
\def\L3p{K$^{0}_{L} \rightarrow \pi^{+}\pi^{-}\pi^{0}$}
\def\Len{K$^{0}_{L} \rightarrow \pi^{\pm}e^{\mp}\bar{\nu}(\nu)$}
\def\Sep{K$^{0}_{S} \rightarrow \pi^{\pm}e^{\mp}\bar{\nu}(\nu)$}
\def\nnn{K$^{0}_{L} \rightarrow \pi^{0}\pi^{0}\pi^{0}$}
\def\Ps0{P$_{S0}$}
\def\Pl1{P$_{L1}$}
\def\epm{e$^{+}$e$^{-}$}

\title*{
Perspectives on Measuring $V_{us}$ at KLOE }
\toctitle{Perspectives on Measuring $V_{us}$ at KLOE }
%
%
\titlerunning{Perspectives on Measuring $V_{us}$ at KLOE}
%
%
\author{
The KLOE Collaboration\thanks{ A.~Aloisio, F.~Ambrosino,
A.~Antonelli, M.~Antonelli, C.~Bacci, G.~Bencivenni,
S.~Bertolucci, C.~Bini, C.~Bloise, V.~Bocci, F.~Bossi,
P.~Branchini, S.~A.~Bulychjov, R.~Caloi, P.~Campana, G.~Capon,
G.~Carboni, M.~Casarsa, V.~Casavola, G.~Cataldi, F.~Ceradini,
F.~Cervelli, F.~Cevenini, G.~Chiefari, P.~Ciambrone, S.~Conetti,
E.~De~Lucia, P.~De~Simone, G.~De~Zorzi, S.~Dell'Agnello, A.~Denig,
A.~Di~Domenico, C.~Di~Donato, S.~Di~Falco, B.~Di~Micco, A.~Doria,
M.~Dreucci, O.~Erriquez, A.~Farilla, G.~Felici, A.~Ferrari,
M.~L.~Ferrer, G.~Finocchiaro, C.~Forti, A.~Franceschi,
P.~Franzini, C.~Gatti, P.~Gauzzi, S.~Giovannella, E.~Gorini,
E.~Graziani, S.~W.~Han, M.~Incagli, W.~Kluge, V.~Kulikov,
F.~Lacava, G.~Lanfranchi, J.~Lee-Franzini, D.~Leone, F.~Lu,
M.~Martemianov, M.~Matsyuk, W.~Mei, L.~Merola, R.~Messi,
S.~Miscetti, M.~Moulson, S.~M\"uller, F.~Murtas, M.~Napolitano,
A.~Nedosekin, F.~Nguyen, M.~Palutan, E.~Pasqualucci,
L.~Passalacqua, A.~Passeri, V.~Patera, E.~Petrolo, L.~Pontecorvo,
M.~Primavera, F.~Ruggieri, P.~Santangelo, E.~Santovetti,
G.~Saracino, R.~D.~Schamberger, B.~Sciascia, A.~Sciubba, F.~Scuri,
I.~Sfiligoi, A.~Sibidanov, T.~Spadaro, E.~Spiriti, G.~L.~Tong,
L.~Tortora, P.~Valente, B.~Valeriani, G.~Venanzoni, S.~Veneziano,
A.~Ventura, S.~Ventura, R.~Versacci, G.~W.~Yu. } presented by E. De Lucia}
\authorrunning{The KLOE Collaboration}
%
%
\institute{
Sezione INFN di Roma
Universit\`a ``La  Sapienza'', \\ P.le A.Moro n.2, 00185-Roma, Italy}
\maketitle              
\begin{abstract}

  The KLOE experiment has been running since April 1999 at
  the DA$\Phi$NE e$^{+}$-e$^{-}$ collider at a center of mass energy equal
  to the $\phi$-meson mass. The luminosity integrated up to September 2002
  is $\sim 500~~pb^{-1}$. Perspectives on the measurement of the $V_{us}$
  CKM-matrix element with the KLOE detector,using both charged and neutral
  kaon semileptonic decays, are presented.

\end{abstract}

\section*{The KLOE experiment at \DAF}
The KLOE\cite{kp} detector at \DAF\cite{dafne}, the Frascati $\phi$-factory,
started data taking in April 1999. The \DAF\ e$^{+}$-e$^{-}$ collider
operates at the center of mass of the $\phi$(1020) meson  producing  almost
monochromatic \KS\ and \KL\ pairs, \KP\ and \KM\  pairs and all other $\phi$ decay
products.
Moreover at \DAF\  kaons are produced with a $\sim 110-125$ MeV/c momentum and their decay
lengths are $\lambda_{S} \sim 0.6$ cm, $\lambda_{L} \sim 340$  cm and
$\lambda_{\pm} \sim 90$ cm.
The unique feature of a $\phi$-factory is the {\it tagging}: the detection
of a  long-lived neutral kaon guarantees the presence of a \KS\ of given
momentum and direction
and viceversa, the same holds for charged kaons.

  During 2002 data taking \DAF\  reached a peak luminosity
of $\simeq 8 \times 10^{31}$\lum\ .

The integrated luminosity is $\sim 500~pb^{-1}$  for a total
number of $1.5\times 10^{6}~~K^{+}K^{-}$ pairs per $pb^{-1}$ and
$10^{6}~~K_{L}K_{S}$ pairs per $pb^{-1}$.

The KLOE detector\cite{kp,tp}
consists of a large cylindrical drift chamber surrounded by a hermetic electromagnetic
calorimeter. A superconducting  coil and an iron yoke surrounding the whole detector
provide a 0.52 T magnetic field.

The drift chamber\cite{dpap} (DC) with 4 m diameter
and  3.3 m length, has a total number of  52140 wires, arranged in 12582 cells
distributed over 58 concentric layers and with an all-stereo geometry.
In order to maximize transparency to photons and reduce \KL\ regeneration, the
mechanical support of the drift chamber  is made of carbon fiber-epoxy composite and
the operating gas mixture is  90 \% helium - 10 \% isobutane.
The achieved position resolution is  $\sigma_{r \phi}\sim 150 \mu m$ and
$\sigma_{z} \sim 2 mm$ while vertices are reconstructed with a resolution
$\sigma_{v} \sim 3 mm$. The momentum resolution is
$\sigma(p_{\perp})/p_{\perp} \sim 0.4\%$.

The electromagnetic calorimeter\cite{tp,epap} (EmC) is a lead-scintillating
fiber sampling calorimeter made of 88 modules, divided into a barrel
section and two
C-shaped end-caps, ensuring 98\% coverage
of the solid angle. The read-out of the modules is performed on both ends
with  $\sim 4.4 \times 4.4 cm^{2}$ granularity  for a total of  4880
photomultipliers.
The calorimeter has to detect with very high efficiency
photons down to 20 MeV energy and to accurately measure
their energy and time of flight.
The achieved energy resolution is  5.7$\%$/$\sqrt{E(GeV)}$,
with a  linearity in energy response better
than 1$\%$ above 80 MeV and better than 4$\%$ between 20 to 80 MeV.
The time resolution is  $\sigma_{t}$=(54/$\sqrt(E(GeV) \oplus$ 50) ps.

\section*{Measuring $V_{us}$}

The most accurate test of the unitarity condition of the
CKM-matrix is provided by the measurements of  the $V_{us}$,
$V_{ud}$ and $V_{ub}$ matrix elements. Present experimental values
indicate a 2.2$\sigma$ deviation from unitarity in the CKM matrix.
Therefore further efforts to reduce the uncertainties in the
determination of the $V_{us}$ element, besides those of $V_{ud}$
and $V_{ub}$, are needed.

  The present value of $V_{us}$ is \cite{pdg02delucia}:
\begin{equation}
V_{us}=0.2196\pm0.0026~~~(\Delta V_{us}/V_{us}=1.18\%) \nonumber
\end{equation}

  The $V_{us}$ CKM-matrix element can be measured using
the semileptonic decays of both charged and neutral kaons, in
particular $K_{e3}$ decays provide the most accurate value.

The almost monochromatic beams of \KS \KL\ and \KP \KM produced at
\DAF makes KLOE one of the most promising detectors to perform
these measurements.

The partial decay width
$\Gamma(K_{l3})$ of the kaon semileptonic decays is given by:
\begin{equation}
\Gamma(K_{l3})= \frac{G^{2}_{F} m^{5}_{K}}{192 \pi^3} \cdot S_{EW} \cdot
  C^{2}_{K}\cdot |f^{K \pi}_{+}(0) \cdot V_{us}|^{2} \cdot I_{K}(m_K
  ,m_\pi ,m_l , f^{K \pi}_{+,0}(q^2)) \cdot (1 + \delta_{K}) \nonumber
\end{equation}
%
%

where $G_{F}$ is the Fermi coupling constant obtained from $\mu$ decays,
$I_{K}$ is the phase space integral,
$f^{K \pi}_{+}(q^2)$ and $f^{K \pi}_{0}(q^2)$ are the form factors of the
strangeness changing hadronic vector current, function of the squared
momentum transfer $q$. These form factors incorporate the isospin breaking
corrections and the second order SU(3) breaking effects arising from
\textit{s-d, u} quark mass difference.The radiative corrections are
factorized into a short-distance electroweak term $S_{EW}$, and a
long-distance model-dependent QED correction $\delta_{K}$
\cite{cirigliano,calderonlopez}.

  The observable is the quantity $|f^{K \pi}_{+}(0)\cdot
V_{us}|$ so in order to extract $V_{us}$ we need to know the SU(2)
and SU(3)-flavor symmetry breaking and the radiative corrections.

These corrections are different for neutral and charged kaons (and
for $K_{e3}$ and $K_{\mu 3}$) therefore it is important to perform
the measurement in both cases.

The uncertainty of the present value of $V_{us}$ is dominated by
the theoretical knowledge of $f^{K  \pi}_{+}(0)$, contributing
with a 0.8$\%$ factor. This means that any improvement in the
experimental accuracy on the $K_{l3}$ decay properties has to be
accompanied by an improvement in the calculation of form factors
at $q^2 = 0$. However, more precise measurements of the partial
decay width and of the $q^2$ dependence of the vector form factor
$f^{K  \pi}_{+}(q^2)$ are very useful in understanding the
long-distance radiative corrections and possible effects on
$V_{us}$ arising from nonlinear terms that could be present in
$f^{K \pi}_{+}(q^2)$.
\newline Thus the set of the four kaon
semileptonic decays is fundamental to make a consistency check of the
measurements and of the different corrections applied to the decay rates.

At KLOE we have the possibility of measuring the full set of kaon
semileptonic decays using the same detector and exploiting
the tagging feature.

  The present accuracy on the partial
decay widths ($\Gamma(l3)$) can be improved:
\begin{itemize}
\item measuring the absolute branching ratios for semileptonic decays of charged and
  neutral decays
\item measuring directly the partial decay width.
\end{itemize}

The traditional method to obtain the partial decay width uses the
measurement of the branching ratio ($BR(l3)$) and of the
total decay width, coming from the lifetime ($\tau_K$). The uncertainty on $\Gamma(l3)$
is given by the propagation of the errors on $BR(l3)$ and $\tau_K$.

At KLOE the direct measurement of $\Gamma(l3)$ can be performed counting:
\begin{itemize}
\item the number of produced kaons given by the tag ($N_{K}$)
\item the number of semileptonic decays in a given decay region
  ($\Delta N_{l3} / \Delta t$)
\end{itemize}
and using the expression:
\begin{equation}
\Gamma(l3) = (\Delta N_{l3} / \Delta t) / N_{K}.
\label{dirmeas}
\end{equation}

The total decay width enters as a second order correction in eq. \ref{dirmeas}
 therefore its contribution to the uncertainty of $\Gamma(l3)$ is reduced,
 by a factor $\sim 5$, with respect to the traditional $\Gamma(l3)$ measurement.

  This is a unique feature of KLOE arising from the \textit{tagging}.

\begin{figure}[t]
\begin{center}
\includegraphics[width=7cm]{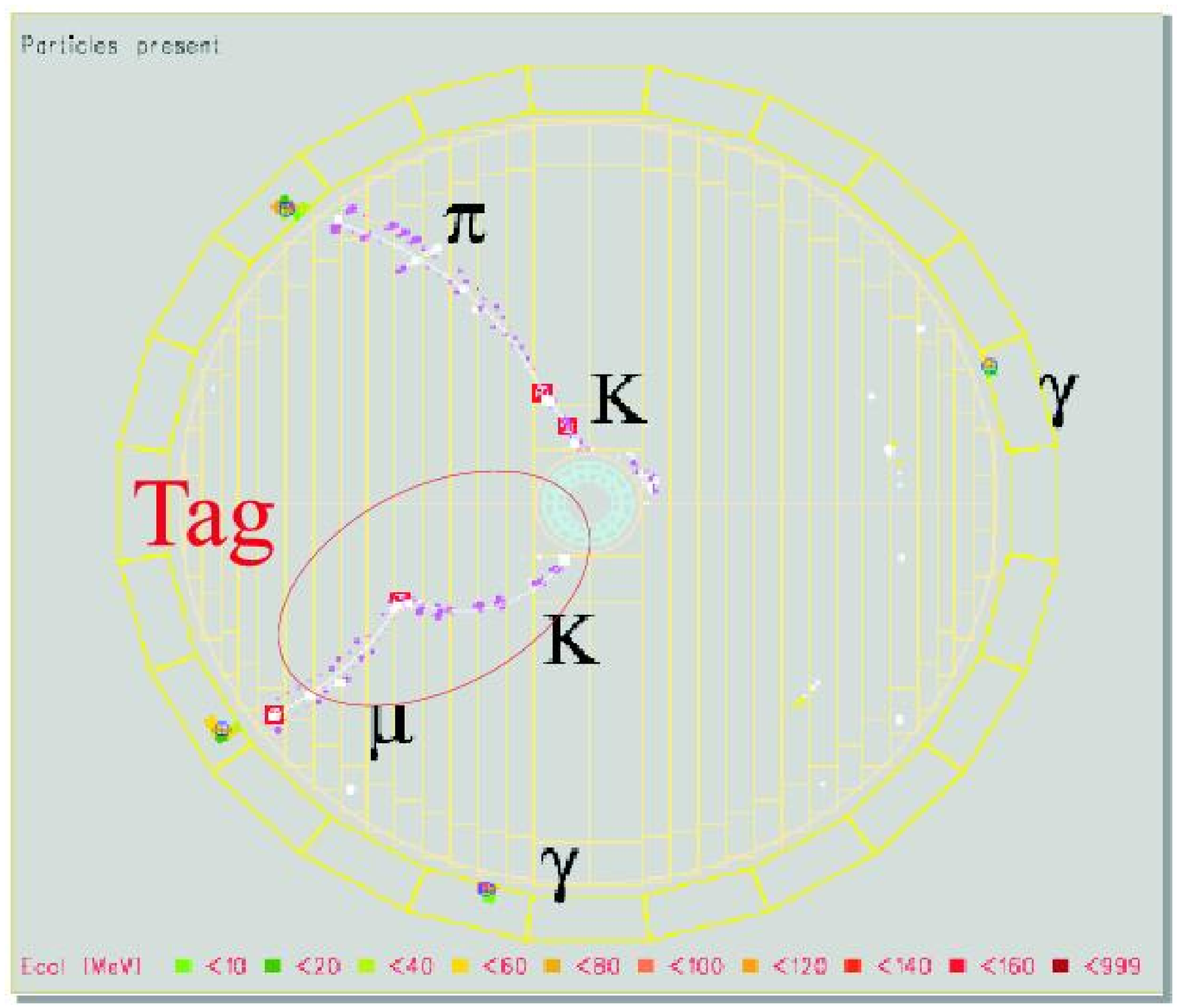}
\end{center}
\caption[]{Display showing a \KP \KM event selected exploiting the tagging.}
\label{tag}
\end{figure}

As previously said, the tagging allows us to select samples of \KP \KM and
of \KS \KL. Thus the strategy for the selection of $K_{l3}$ decays is: tag
with one kaon of the pair and look for the desired semileptonic decay of the other.

The very clean signature of the decays \Kppo\ and \Kmunu\ is
exploited to tag charged kaons, while \Kpm\ decays tag the \KL
neutral kaon (fig. \ref{tag}). Only drift chamber information is
used in both cases. The tagging efficiencies can be estimated
directly from data. For charged kaons we use the redundant
calorimetric information of the \Kmunu\ events while for the
tagging provided by \Kpm\, we use the sample of \KL\ interacting
in the Electromagnetic Calorimeter.

With the statistics of $\sim 500 \textrm{pb}^{-1}$ we can reach a
level of $O(0.07 \%)$ accuracy on tagging efficiencies.

The sample of $K^{\pm}_{l3}$ events is selected asking for a tag on one
side and on the other for a vertex
in the drift chamber and one $\pi^{0}$ in the electromagnetic
calorimeter. The time of flight information separates
charged pions from electrons, exploiting the excellent timing resolution of
the detector. 
From preliminary studies we expect the number of $K^{\pm}_{e3}$
decays to be $N_{K^{\pm}_{e3}} \simeq 2000 / \textrm{pb}^{-1}$, which means
a total number of $N_{K^{\pm}_{e3}} \simeq 10^{6}$, after analysis cuts. Most of the
selection efficiencies can be evaluated directly from data using control
samples, a method already used in the measurement of
$\Gamma (K_{S} \to\pi^{+}\pi^{-} (\gamma))/ \Gamma (K_{S}\to\pi^{0}\pi^{0})$
\cite{ksrapp}. The momentum range of the lepton in $K^{\pm}_{l3}$ is
covered by \Kppo\, \Kmunu\ and \Kppoo\ decays (fig. \ref{effi}). The energy
range of the most
energetic cluster of the $\pi^{0}$ from the semileptonic decay is covered by
 \Kppo\ and \Kppoo\ decays (fig. \ref{effi}).

\begin{figure}[t]
\begin{center}
\includegraphics[width=10.cm]{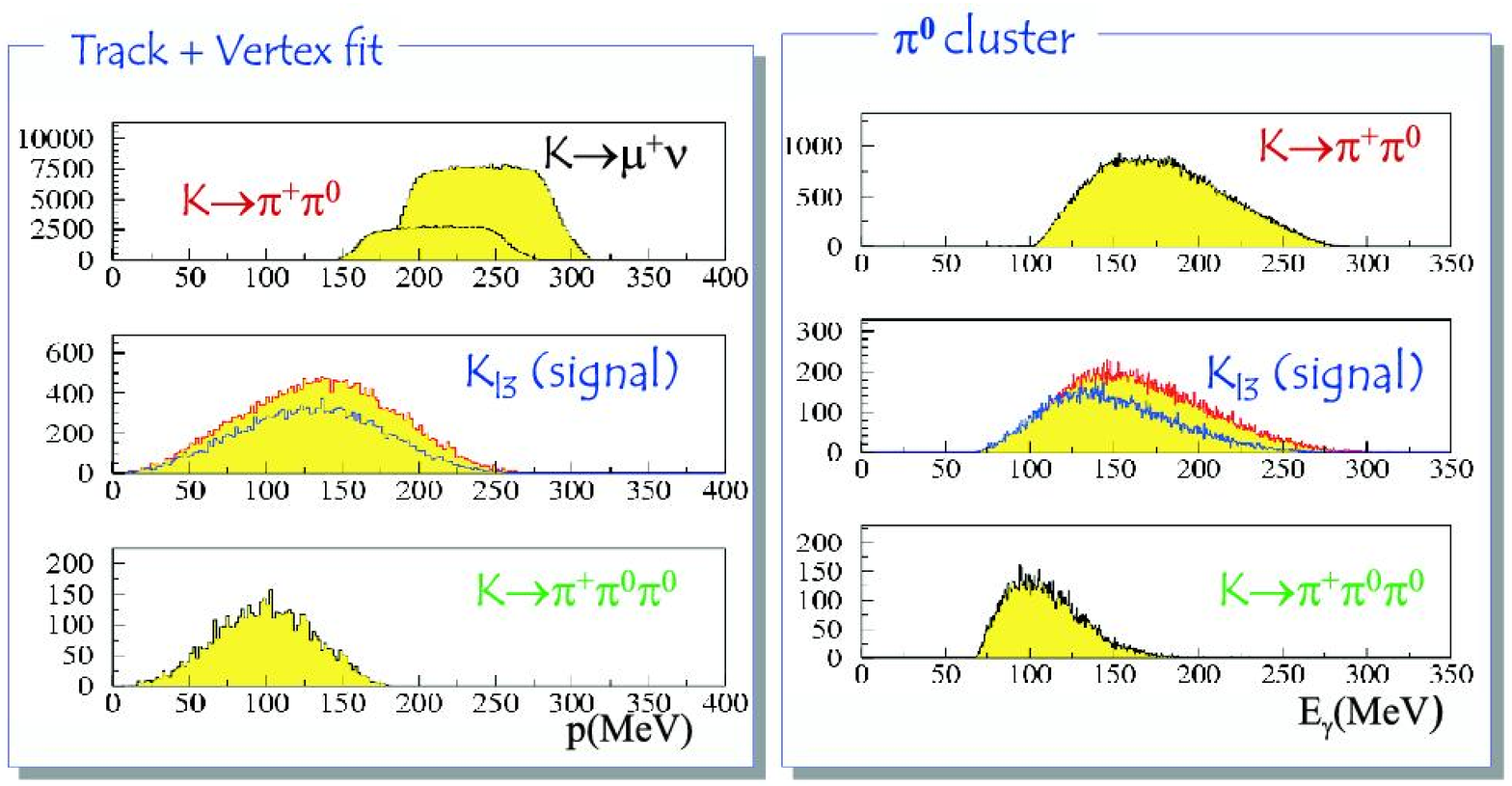}
\end{center}
\caption[]{Distribution of the momentum of the lepton (left) and of
the energy of the most energetic cluster of the $\pi^{0}$ (right)
for $K^{\pm}_{l3}$ decays and for the control samples \Kppo\, \Kmunu\ and
\Kppoo\
  decays.}
\label{effi}
\end{figure}

The sample of $K^{0L}_{l3}$ events is selected asking for a tag on one side
and on the other asking for a vertex in the
drift chamber fiducial volume, applying a cut on the invariant mass. The
time of flight information separates pions from muons and electrons.
The total number of semielectronic \KL\
decays is $N_{K^{0L}_{e3}} \sim 1.5 \times 10^{6}$, after analysis cuts.
This
procedure has been already used in the measurement of the branching ratio
of \Sep\ decay \cite{kssemil}.

Moreover at KLOE we can improve the actual measurements on the vector form
factor slope $\lambda_{+}$, using both charged and neutral $K_{e3}$ decays.
With the present data sample we can reach a statistical accuracy of
$O(10^{-4})$ for $\lambda^{+}$ from $K^{0}_{e3}$ and of $O(10^{-3})$ using
$K^{\pm}_{e3}$, values which are competitive with the ones of the actual
measurements.

\section*{Conclusions}

The KLOE experiment can improve the actual situation
of the $V_{us}$ CKM-matrix element measuring, with the same detector, the
absolute
branching ratios both for
charged and neutral kaon semileptonic decays and measuring directly the
partial decay widths.
Moreover the accuracies that can be reached are beyond the present
theoretical ones.
Therefore KLOE can give a unique contribution in understanding the
SU(2) and SU(3)$_{F}$ symmetry breaking effects and the radiative
corrections applied to the decay rates in order to determine the
value of $V_{us}$.


\title*{An Ultracold Neutron Facility at PSI\protect\newline}
\toctitle{An Ultracold Neutron Facility at PSI}
%
%
\titlerunning{An Ultracold Neutron Facility at PSI}
%
\author{M. Daum for the PSI-PNPI-ILL-Cracow-IMEP UCN Collaboration}
\tocauthor{M. Daum}
\authorrunning{M. Daum}
%
%
\institute{PSI, Paul-Scherrer-Institut, CH 5232 Villigen-PSI, Switzerland}

\maketitle              

\begin{abstract}
At PSI, we build a new type of ultracold neutron (UCN) source, based on the spallation process. The essential elements of the new source are a pulsed proton beam with a high intensity
(I$_p \geq$~2mA) and a very low duty cycle (1\,\%), a heavy element spallation target
and a moderator consisting of solid deuterium kept at a temperature of about 6\,K. Recent
experimental studies of the production of ultracold neutrons in solid deuterium open prospects for densities of 3000 ultracold neutrons per cm$^3$.
\end{abstract}

\section{The New UCN Source}
A new type of ultra-cold neutron source (SUNS, Spallation
Ultra-cold Neutron Source) based on the spallation process is under construction at PSI.
A detailed description of
the source parameters can be found in Ref.\,\cite{tm}. The essential elements of SUNS are a pulsed
proton beam with highest intensity (I$_p \geq $ 2\,mA) and a low duty cycle ($\sim$\,1\,\%),
a heavy-element spallation target, and a large moderator and converter system consisting
of about 4\,m$^3$ of heavy water at room temperature and 30 dm$^3$ of solid deuterium (SD$_2$)
kept at a low temperature ($\sim$\,6\,K) for the production of ultra-cold neutrons.
Operating the UCN source in a pulsed mode will allow maintaining the SD$_2$ at very low
temperatures despite of the large temporary heat load deposited in the spallation target.
The proton beam is directed onto the neutron production target for a few seconds only,
a time long enough to fill the intermediate UCN storage vessel ($\sim$\,2\,m$^3$).
A high neutron density is generated in the moderator assembly. The neutrons are
thermalized in the D$_2$O, further cooled in the SD$_2$ and finally, some of them are
down-scattered into the ultra-cold neutron range (T$_{\mbox{kin}} \leq 250$\,neV).

About when equilibrium between the produced and re-absorbed
UCNs is achieved in the storage
volume, the SD$_2$ moderator is separated from the latter by a reflective shutter,
in order to prevent re-absorption of the UCNs in the cold moderator.
The proton beam
is turned off, and the UCNs are transferred from the storage volume to the EDM apparatus.
The filling of the source storage volume is repeated after
about 800\,s, i.e., as soon as the UCN density in the storage volume has dropped significantly.
The source layout, pulse duration, and storage volume are optimized for a dedicated EDM
spectrometer of about 0.2\,m$^3$ volume.

Monte Carlo calculations\cite{tm} show that at this source, an average UCN density of
$\sim$\,3$\cdot 10^3$ UCN/cm$^3$ can be delivered to the experiments. This is about two
orders of magnitude more than in the present experiments at the reactors in the
Institute Laue Langevin, ILL, Grenoble and in the St.\ Petersburg Nuclear Physics Institute,
PNPI, Gatchina.
This average UCN density
corresponds to a pulsed proton beam current I$_p$ = 2 mA (600 MeV) with a pulse duration
of 8 seconds and a 1\,\% duty cycle.
The target material assumed for this calculation is lead with a filling
factor (Pb/D$_2$O ratio) of 0.5.

\begin{figure}[t]
\begin{center}

\includegraphics[width=11cm]{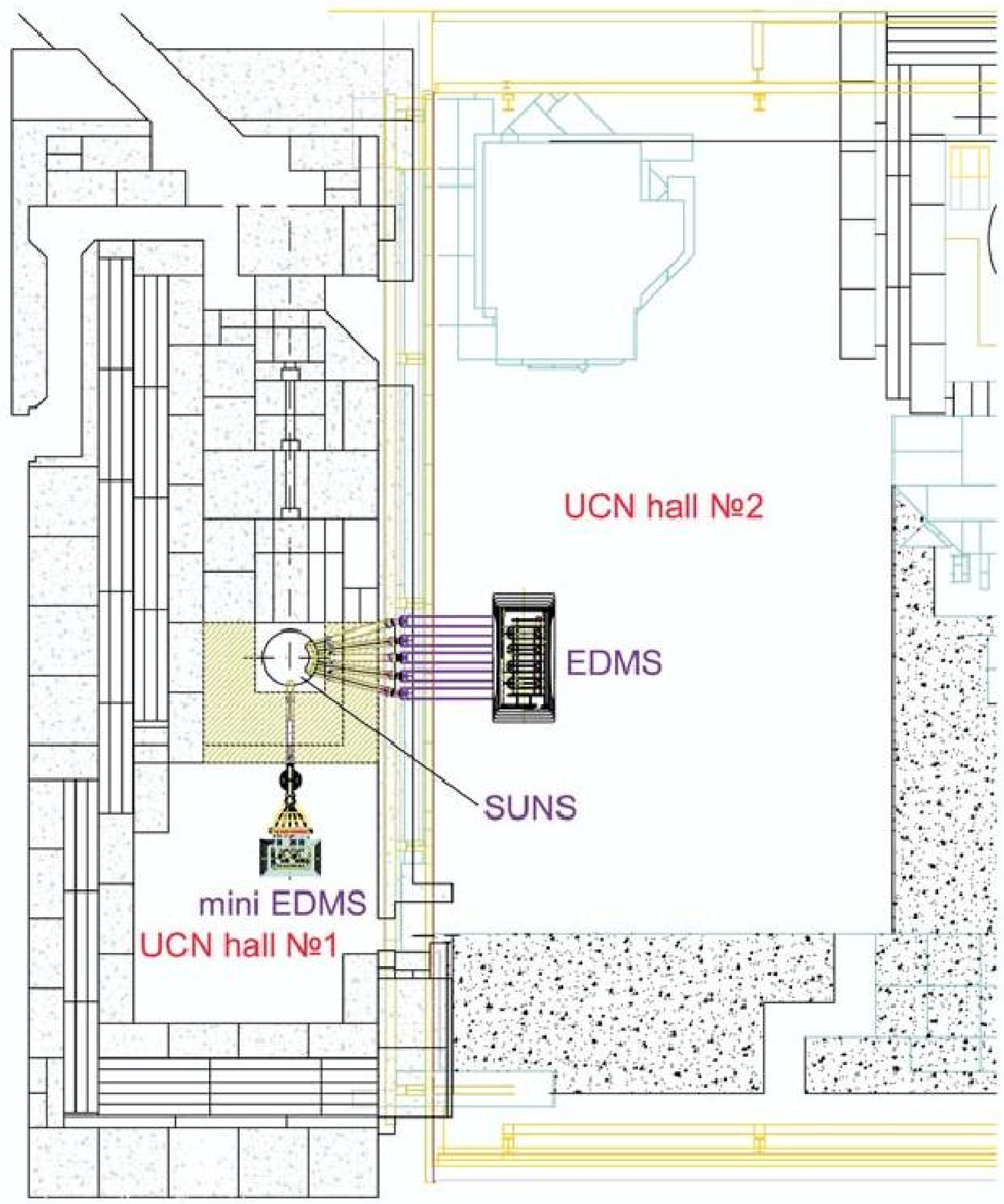}


\end{center}
\caption{Layout of SUNS and EDM spectrometers in the different experimental areas at PSI.
}
\label{areal1}
\end{figure}

The new UCN facility at PSI will be built in two steps. The layout
of SUNS and the connected experimental areas are shown in
Fig.\,\ref{areal1}. In 2006, the source is expected to be
operational and a first experimental area will be available for
experiments with UCNs (mini EDMS, UCN hall No.1,
Fig.\,\ref{areal1}). At a later stage ($\sim$2007-2008), an
extended area (EDMS, UCN hall No.2, Fig.\,\ref{areal1}) will
complete the experimental installations at this facility. Then, a
second step of an EDM Experiment with an EDM spectrometer of
highly increased volume ($\sim$0.2\,m$^3$) as well as various
other fundamental experiments can be envisaged.

\title*{Three-fold Correlation in the Interaction \\ of Polarized Neutrons
with Polarized Nuclei \\ and the Method of Oscillating
Fields\protect\newline}
\toctitle{Three-fold Correlation in the Interaction of Polarized Neutrons
with Polarized Nuclei and the Method of Oscillating
Fields }
\titlerunning{Three-fold Correlation in the Interaction of Polarized Neutrons}
\author{A.V.Aldushchenkov, A.I.Kovalev, V.V.Lukashevich}
\authorrunning{A.V.Aldushchenkov}
\institute{S.-Petersburg Nuclear Physic Institute, 188350 Gatchina, Russia}
\maketitle
\begin{abstract}
  A two coil resonance method is discussed for the measurement of possible P,
T-violation effects in the interaction of low-energy neutrons with
polarized nuclei. It is shown that a neutron phase depending
asymmetry has a direct connection with the T-odd amplitude and can
be a measure of breaking T-invariance.
\end{abstract}

  There is an intriguing problem of searching for breaking T-invariance in the low
energy physics. One of the possible variants of such searching has
been under discussion nearly for recent 20 years. This is a study
of the slow neutrons interaction  with $^{139}$La. An interest to
this nucleus is stipulated by the fact that closely lying
s,p-resonances created in a compound state by passing neutrons
could enhance the T-violation interaction on 5-6 orders.

P,T--odd interaction in $^{139}$La is like the interaction of an
axial neutron dipole moment with an electric field but it seems
more attractive because of large enhancement factor. T--odd,
P--even interaction is phenomenological and is not followed by the
extension of standard model, though such an interaction is studied
carefully in five-fold correlation \cite{journ1}.

  Traditionally, the amplitude of interaction of neutrons with a nucleus is
written as
\begin{displaymath}
        f=A+p_{t}B(\vec s\vec I)+C(\vec s\vec k)+p_tD(\vec s[\vec k\vec I]),
\end{displaymath}
\noindent
where I - spin nucleus, $p_{t}$- target polarization, s - neutron spin and
k - its wave vector.

  The experimental measurement subject is a value T--violation amplitude $D$. The
hierarchy spin depended forces is such as $B \gg C \gg D$, so it
is difficult to separate the small P,T-odd effect from the large
background effects of strong, electromagnetic and helicity
depended neutron-nucleus interactions.

  There are several  proposals how to measure the value of $D$ \cite{journ2}. However,
according to analysis \cite{journ2} and \cite{journ3}, some of
them are practically unattainable, while some others are wrong.
So, new presentations are required to resolve a problem under
discussion.

  In the given work the possibility of T-breaking amplitude measurement  is
discussed on the basis of a high sensitive method of magnetic
exact, two coil method by Ramsey \cite{journ4}.

  A neutron spin  is parallel to the magnetic field H and a nuclear polarization
at the entrance of the first coil. After a coil the spin is turned on $\frac{\pi}{2}$.
Then, a neutron goes through the target and the second coil, in which the spin
is turned on $\frac{\pi}{2}$ once more, and an analyzer measures its
polarization.

  A result of polarized neutron interaction with the efficient pseudomagnetic,
weak P-odd and T-violation fields in the polarized target is
convenient to describe within the framework of the density matrix
formalism. This enables one to define an experimental strategy.

  Let us use a standard presentation for the density matrix
\begin{displaymath}
            \rho=\frac{1}{2}[1+(\vec p\vec \sigma )],
\end{displaymath}
\noindent
where $\vec p$ is the neutron polarization vector and $\vec \sigma$ are the
Pauli matrices.
For neutrons, polarized along axis z, the matrix of density is diagonal and
the components of polarization vector are $p_{z}=p_{0}$, $p_{x}=p_{y}=0$.
The passing of neutrons through the first coil with the radio-frequency field,
target and second coil in terms of the density matrix is described as
follows

\begin{equation}
\label{eq1}
\rho=U_{2}U_{int}U_{1}\rho_{0}U_{1}^{\dagger}U_{int}^{\dagger}U_{2}^{\dagger}.
\end{equation}

In this expression $U_{1}$, $U_{2}$are the time evolution
operators for the first and second coils, accordingly. $U_{int}$   is
the evolution operator, describing a transformation of density
matrix by the interaction in a target. $\rho_{0}$ is an initial
density matrix. On condition that the field frequency in coils is equal to
the frequency of Larmor precession in the external magnetic field and that
the spin turning angle is equal to $\frac{\pi}{2}$ in each of the coils,
the Hamiltonian of a neutron in the coils in the rotating coordinate system
has a simple type
\begin{displaymath}
     H_{1}=H_{2}=\frac{\pi}{4}(\vec n\vec \sigma). \\
\end{displaymath}
Here $\vec n$ is a single vector toward the magnetic field. For
specified conditions, this vector has following components:
$n_{z}=0$, $n_{x}=\cos\delta$, $n_{y}=\sin\delta$,

where $\delta$ is a random phase, with which a neutron flies in the first coil.

After passing first coil, a density matrix becomes equal to
\begin{equation}
\label{eq2}
\rho_{1}=\frac{1}{2}[1+p_{0}(\sigma_{x}\sin\delta-\sigma_{y}\cos\delta)].
\end{equation}
That is to say, neutron spins "lie" in horizontal plane and a spin direction
is defined by the initial phase $\delta$.
  For the density matrix evolution of neutrons in the polarized target
$^{139}$La we will use the results of work \cite{journ3}, in which
general solution for the density matrix

\begin{equation}
\label{eq3}
\rho_{int}=U_{int}\rho_{1}U_{int}^{\dagger}
\end{equation}

\noindent
is given with a free matrix $\rho_{1}$.

  Let us suppose that a vector of target polarization and an external magnetic
field are directed along the axis z and neutrons are moved along the axis y.
Then, using ({\ref{eq3}}), ({\ref{eq2}}) and ({\ref{eq1}}) we find expression for
matrix on the device exit, which in the expanding form is

\begin{equation}
\label{eq4}
\rho=\frac{1}{2}[N_{0}+(\vec{P}\vec{\sigma})],
\end{equation}

\noindent where
\begin{eqnarray*}
P_{z}=-p_{x}^{'}\sin\delta_{1}+p_{y}^{'}\cos\delta_{1}, \\
P_{x}=\cos\delta_{1}(p_{x}^{'}\cos\delta_{1}+p_{y}^{'}\sin\delta_{1})+p_{z}\sin\delta_{1}
\\
P_{y}=\sin\delta_{1}(p_{x}^{'}\cos\delta_{1}+p_{y}^{'}\sin\delta_{1})-p_{z}\cos\delta_{1}.
\end{eqnarray*}
The phase $\delta_{1}$contains a constant difference $\epsilon$
between the radio frequency
phases of the two coils, so that $\delta_{1}=\delta+\epsilon$.\\
In (\ref{eq4}) $p_{i}^{'}$ are projections of the polarization
vector after the target, which are defined in \cite{journ3}. From
this work we also take the determination of efficient fields. The
vector $\vec b^{'}=\frac{\sin{qt}}{q}\vec{b}$, where
$q=\sqrt[2]{\vec b\vec b}$ and the vector $\vec{b}$ is a resulting
vector of the three fields, which a neutron interacts with and
$qt$ is the efficient angle of neutron spin rotation around this
vector.

In the chosen coordinate system values $b_{i}^{'}$ are connected with efficient
fields in a target as follows
\begin{itemize}
\item the T violation field $b_{x}=p_{t}\frac{D}{2}$,
\item the field of weak interaction $b_{y}=\frac{C}{2}$,
\item and the so called pseudomagnetic field  $b_{z}=p_{t}\frac{B}{2}$.
\end{itemize}
Here $p_{t}$ is the target polarization factor.

  A general form of the density matrix ({\ref{eq4}}) allows us to construct
 three asymmetries.
The first is an asymmetry depended on the neutron phase, the second -- on
the neutron polarization and the third - on a degree of the target polarization.
However, under detailed consideration, two last asymmetries do not allow to
select T- violation amplitude D, so we will give an analysis of the asymmetry
of neutron polarization $\eta _{\delta }$ depending on a neutron phase.
The final result for $\eta _{\delta }$ is

\begin{equation}
\label{eq5}
\eta _{\delta}=\frac{\rho_{11}(\delta)-\rho_{11}(-\delta)}{\rho_{11}(\delta)+
\rho_{11}(-\delta)}=\frac{E}{F},
\end{equation}

\noindent where $\rho_{11}$ is the matrix element of the density
matrix ($\ref{eq4}$).\begin{eqnarray*}
E=4\sin{\delta}[(p_{0}+\cos{\epsilon})Im(b_{y}^{'}b_{z}^{'\ast})+
(p_{0}-\cos{\epsilon})Im(b_{x}^{'\ast}\cos{qt})\nonumber
\\ +2p_{0}\cos{\delta}\cos{\epsilon} Re(b_{x}^{'}b_{y}^{'\ast})] + 4\sin{\delta}\sin{\epsilon}[p_{0}\cos{\delta}
(|b_{y}^{'}|^{2}-|b_{x}^{'}|^{2})\\-Im(b_{x}^{'}b_{z}^{'\ast}-b_{y}^{'\ast}\cos{qt})]\\
F=2[(1+p_{0}\cos{\epsilon}\cos{2\delta})|b_{x}^{'}|^{2}+
(1-p_{0}\cos{\epsilon}\cos{2\delta)}|b_{y}^{'}|^{2}+(1+
p_{0}\cos{\epsilon})|b_{z}^{'}|^{2}\\+(1-p_{0}\cos\epsilon)
|\cos{qt}|^{2}]+4\cos{\delta}(p_{0}+\cos{\epsilon})Im(b_{x}^{'}b_{z}^{'\ast})
\\-4\cos{\delta}(p_{0}-\cos{\epsilon})Im(b_{y}^{'\ast}\cos{qt})\\+4\sin{\epsilon}
[\cos2{\delta}Re(b_{x}^{'}b_{y}^{'\ast}+
p_{0}Re(b_{z}^{'\ast}\cos{qt})
 -\cos{\delta}(Im(b_{y}^{'\ast}b_{z}^{'}+Im(b_{x}^{'\ast}\cos{qt}))].
\end{eqnarray*}

  We can choose the instrumental phase $\epsilon$ equal to $\pi$ and  take
$p_{0}=1$ because the polarization degree of neutron beams is
approximately equal to 1 in real experiments. Under these
conditions the expression ({\ref{eq5}}) is significantly
simplified

\begin{equation}
\label{eq6}
\eta_{\delta}=\frac{2\sin{\delta}[Im(b_{x}^{'\ast}\cos{qt})-\cos{\delta}Re(b_{x}^{'}b_{y}^{'\ast})]}
{\sin^{2}{\delta}|b_{x}^{'}|^{2}+\cos^{2}{\delta}|b_{y}^{'}|^{2}+|\cos{qt}|^{2}-
2\cos{\delta}Im(b_{y}^{'}\cos{qt})}.
\end{equation}

  Now this is a critical point of the analysis. In ({\ref{eq6}}) for the quantity
$q$ we can take into account only the large term corresponding to the pseudomagnetic
 field then
\begin{displaymath}
     q=ReB+iImB.
\end{displaymath}
Let us re-write ({\ref{eq6}}) in terms of efficient fields saving only main values and
neglecting the second term in the numerator ({\ref{eq6}}) as far as it has the
second order of smallness.

  As it is pointed out by Abragam \cite{mono2} $ImB\sim10^{-3}ReB$. It allows us not to include
terms with $ImB$. After these remarks we have

\begin{equation}
\label{eq7} \eta_{\delta}\approx
p_{t}\frac{4\sin{\delta}ImD}{{\omega}ctg{\frac{{\omega}t}{2}}+4\cos{\delta}ImC},
\end{equation}

\noindent
where $\omega=ReB$ is the frequency of neutron spin precession in pseudomagnetic field.

 We can evaluate this frequency from pseudomagnetic moment $^{139}$La
which was measured in \cite{journ5}. The pseudomagnetic field in
crystal LaAlO$_{3}$(Nd$^{3+}$) is found less than 1kG at the
target polarization 50\%. The spin makes several turns in this
field for the flight time ($10^{-6}$ sec) through the target with
length 1cm. We can notice that such level of polarization in
crystal LaAlO$_{3}$(Nd$^{3+}$) was reached in the work
\cite{journ6}.

  The P,T-odd effect shown by Eq. ({\ref{eq7}}) disappears in the
case when the spin makes the integer number of turns and the
effect is maximum when the complete precession angle is different
on ${\pi}$ from the integer number of turns. The second
opportunity may be realized by the corresponding choice of the
target length for neutrons with p-wave resonance energy 0.734eV.
From the expression ({\ref{eq7}}) follows that the phase analysis
have not to be done near phase angle $\delta =0,{\pi}$.

 As a result, it is possible to conclude that the experimental found asymmetry
({\ref{eq7}}) can be a measure of breaking T-invariance at the interaction of slow
neutrons with nuclei.

 In the experimental plan the measurement of the $\eta_{\delta}$ value means
 a synchronization moment of a neutron registration in the counter (after passing
an analyzer) with the phase of radio frequency field and a selection of events
 with phases distinguishing on $\pi$. More exactly, neutrons with the energy of p-wave
resonance $E_{n}=0.734 eV$ are registered as function of phase
$\delta $. On the basis of the existing technique, the accuracy of
phase angle measurements can be order of one degree \cite{journ7}.


\title*{Trine -- A New Limit on Time Reversal Invariance Violation in Neutron $\beta$-Decay}
\toctitle{Trine -- A New Limit on Time Reversal Invariance Violation in Neutron $\beta$-Decay}
%
%
\titlerunning{Trine -- A New Limit}
%
\author{T.~Soldner\inst{1}\and L.~Beck\inst{2}\and C.~Plonka\inst{3}\and K.~Schreckenbach\inst{3}\and
 O. Zimmer\inst{4}}
\authorrunning{T.~Soldner et. al.}
%
%
\institute{Institut Laue Langevin, BP 156, F-38042 Grenoble Cedex
9, France\and Beschleunigerlabor von LMU und TU M{\"u}nchen, Am
Coulombwall, D-85747 Arching, Germany\and Physik-Department E21,
TU M{\"u}nchen, James-Franck-Str., D-85747 Arching, Germany\and
Physik-Department E18, TU M{\"u}nchen, James-Franck-Str., D-85747
Arching, Germany}

\maketitle              

\begin{abstract}
In neutron beta decay, the triple correlation between the neutron spin
and the momenta of electron and antineutrino ($D$ coefficient) tests for
a violation of time reversal invariance beyond the Standard model
mechanism of CP violation. We present a new
preliminary limit for this correlation which was obtained by the Trine
experiment:
\mbox{$D_{\rm prel.}=(-3.1\pm6.2^{\rm stat}\pm4.7^{\rm syst}
\pm4.7^{\rm syststat})\cdot10^{-4}$}.
\end{abstract}

\section{Introduction}

To create the baryon antibaryon asymmetry in the universe from a symmetric
start, a baryon number, C and CP violating process outside thermal equilibrium
is required \cite{sakharov1967}. CP violation was discovered in the decay of
neutral kaons \cite{christenson1964}. This type of CP violation is implemented
in the Standard model of particle physics via a free phase in the
quark mixing matrix \cite{kobayashi1973} but seems
to be insufficient to explain the observed baryon asymmetry \cite{herczeg2001}.

Extensions of the Standard model like SUperSYmmetric models or Grand Unified
Theories open new channels for CP violation which may be
observed in low energy particle physics like in electric dipole
moments (EDMs) or in the neutron beta decay. Especially the neutron EDM
is a sensitive test for physics beyond the Standard
model and restricts the parameter space for many alternative
models \cite{ellis1989}. The decay, however, namely the triple correlation $D$
of the spin of the decaying neutron and the momenta of electron and
antineutrino, is more sensitive for CP violation via leptoquarks
\cite{herczeg2001} which appear naturally in GUTs.

The differential decay probability of the neutron can
be written as \cite{jackson1957a}:
\begin{eqnarray}\label{eqn:JacksonFormel}
  \frac{{\rm d}W}{{\rm d}E_{\rm e}{\rm d}\Omega_{\rm e}
    {\rm d}\Omega_{\bar\nu}} &=&
    g G_{\rm E}(E_{\rm e})
    \left\{1+a\frac{\vec p_{\rm e}\vec p_{\bar\nu}}{E_{\rm e}E_{\bar\nu}} +
    b\frac{m_{\rm e}}{E_{\rm e}} +\right.\\
    &&\left.\frac{\vec\sigma_{\rm n}}{\sigma_{\rm n}}\left(
    A\frac{\vec p_{\rm e}}{E_{\rm e}} +
    B\frac{\vec p_{\bar\nu}}{E_{\bar\nu}} +
    D\frac{\vec p_{\rm e}\times\vec p_{\bar\nu}}{E_{\rm e}E_{\bar\nu}}\right)
    \right\}\nonumber
\end{eqnarray}
Here, $g$ is a normalization constant, $G_{\rm E}$ the electron spectrum,
$\vec\sigma_{\rm n}$ the neutron spin, $E_i$
the energy, $\vec p_i$ the momentum, and ${\rm d}\Omega_i$ the
solid angle of electron e and antineutrino $\bar\nu$, respectively.
The coefficients $a$, $b$, $A$, $B$, and $D$ describe the
correlations between the decay products.

Eq. (\ref{eqn:JacksonFormel}) assumes only Lorentz invariance but no discrete
symmetries like parity P, charge conjugation C, or time reversal T.
Indeed, the coefficients $A$ and $B$ are P and C violating and
nonzero ($A=-0.1162(13)$, $B=0.983(4)$ \cite{pdg2002}). In the V--A-theory
$A$ or $a$ are used to determine the ratio $|\lambda|:=|g_{\rm A}/g_{\rm V}|$
of the axial vector and the vector coupling constant
($b\equiv0$ in V--A-theory). Together with the neutron life time the
absolute values of the coupling constants can be determined. For a
precise measurement of the phase of $\lambda$, however, the $D$ coefficient
is required. A phase $\ne 0,\pi$, i.e. $D\ne0$, would indicate T violation
(and according to the CPT theorem CP violation). Up
to now, no evidence for a deviation of $D$ from 0 was found (world average
$D=-0.6(1.0)\cdot10^{-3}$ \cite{pdg2002}). The Standard model prediction
is $D<10^{-12}$. Any value above the final state effect level
($D_{\rm FS}\approx10^{-5}$ for neutrons) would indicate new physics.
For leptoquark models, this experimental range is not excluded by
measurements of alternative parameters (like, e.g., EDMs)
\cite{herczeg2001}.

\section{\boldmath Principle of a $D$ measurement\label{sec:principle}}

To measure $D$ in neutron decay, electron and proton (which can replace
the antineutrino for slow neutrons) have to be detected dependent on the neutron
spin. Integrating (\ref{eqn:JacksonFormel}) over the acceptance of
electron detector $i$ and proton detector $j$ gives the
count rate $\dot N^{ij}$ of the detector combination
${\rm e}^i{\rm p}^j$:
\begin{equation}
  \dot N^{ij} =
    \epsilon_{\rm e}^i\epsilon_{\rm p}^j
    \left\{K_1^{ij} + a K_a^{ij} + b K_b^{ij} + \vec P\left( A\vec K_A^{ij} +
    B \vec K_B^{ij} + D \vec K_D^{ij}\right)\right\}.
\end{equation}
Here, $\epsilon_{\rm e}^i$ ($\epsilon_{\rm p}^j$) describes the detector
efficiency of electron (proton) detector $i$ ($j$). $K_\eta^{ij}$ are apparatus
constants that describe the sensitivity of the
apparatus versus the coefficient $\eta\in\{1,a,b,A,B,D\}$, e.g.
\begin{displaymath}
  K_1^{ij}\propto\left\langle\int_{{\rm e}^i{\rm p}^j}
    G_{\rm E}(E_{\rm e}){\rm d}E_{\rm e}
    {\rm d}\Omega_{\rm e}{\rm d}\Omega_{\bar\nu}\right\rangle _V
    \end{displaymath}
$\quad\mbox{or}\quad$
     \begin{displaymath}
 \vec K_D^{ij}\propto\left\langle\int_{{\rm e}^i{\rm p}^j}
    G_{\rm E}(E_{\rm e})
    \frac{\vec p_{\rm e}\times\vec p_{\bar\nu}}{E_{\rm e}E_{\bar\nu}}
    {\rm d}E_{\rm e}{\rm d}\Omega_{\rm e}
    {\rm d}\Omega_{\bar\nu}\right\rangle _V.
\end{displaymath}
$\langle\ldots\rangle_V$ represents the average over the decay volume.
$\vec P$ is the neutron polarization. Modifications are necessary for
inhomogeneous $\epsilon$ or $\vec P$. The $K_\eta$ can be determined by
Monte Carlo simulations. The quotient
\begin{equation}\label{eqn:ExperimentellAsymmetrie}
  \alpha^{ij} :=
    \frac{\dot N^{ij}_\uparrow -
    \dot N^{ij}_\downarrow}
    {\dot N^{ij}_\uparrow +
    \dot N^{ij}_\downarrow}=
    \vec P
    \left(A\vec\kappa_A^{ij}+B\vec\kappa_B^{ij}+D\vec\kappa_D^{ij}\right)
\end{equation}
with
$\vec\kappa_\eta^{ij}=\vec K_\eta^{ij}/(K_1^{ij} + a K_a^{ij} + b K_b^{ij})$
is independent on detector efficiencies.

Since $D\ll A,B$ the influence of the parity violating coefficients $A$ and
$B$ has to be suppressed carefully. Therefore, the detector and the decay
volume should have two common perpendicular mirror planes ($x$-$z$ and
$y$-$z$ planes in Fig.~\ref{fig:DPrototype} (a) which shows the simplest
implementation). For such detector, $A$ and $B$ are suppressed to first order:
\begin{figure}
  \centerline{
    (a)~~~~\parbox[c]{3cm}{\centerline{
      \includegraphics[width=30mm]{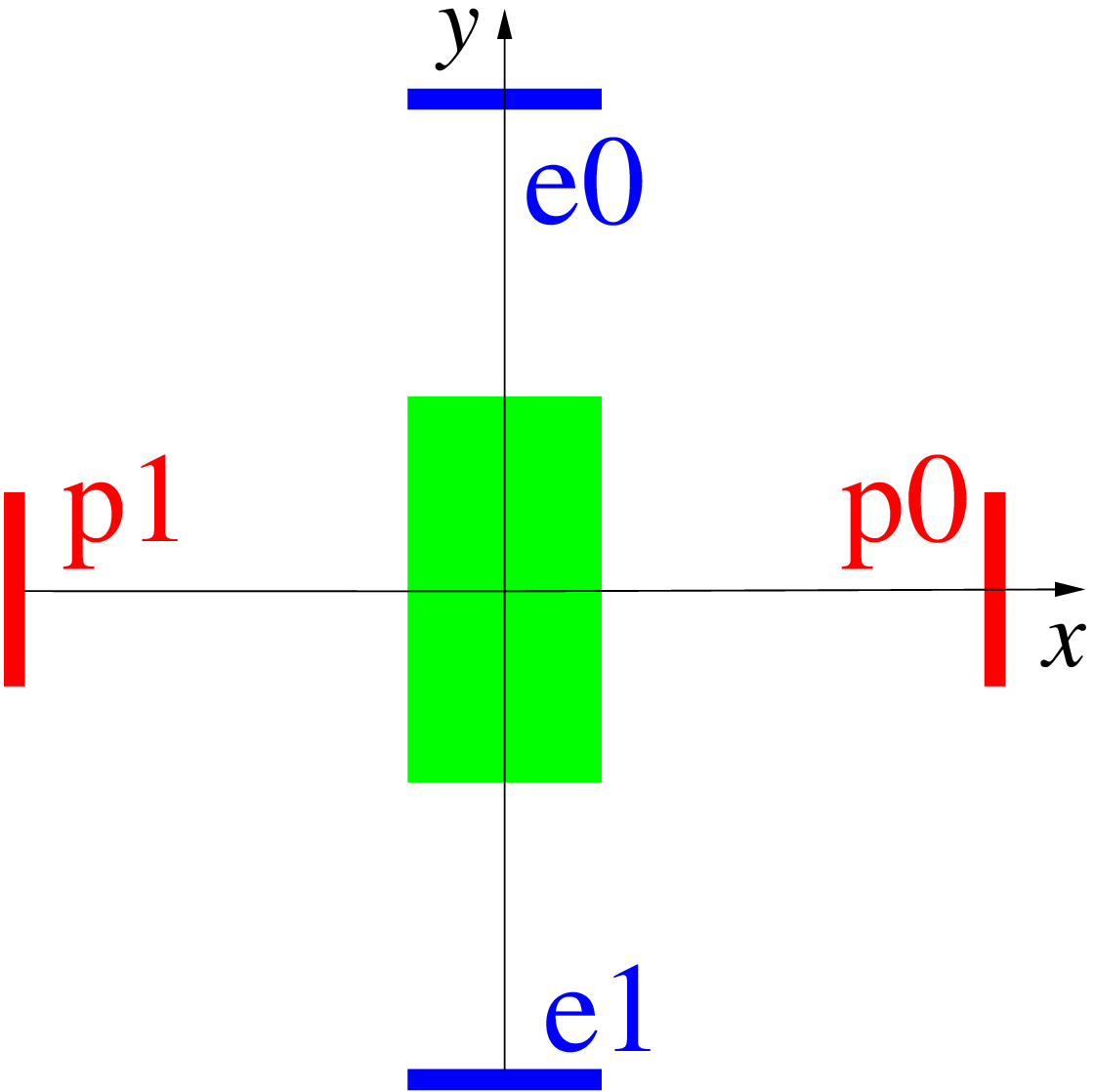}}}\hfill
    (b)~~~~\parbox[c]{6.5cm}{\includegraphics[width=6.5cm]{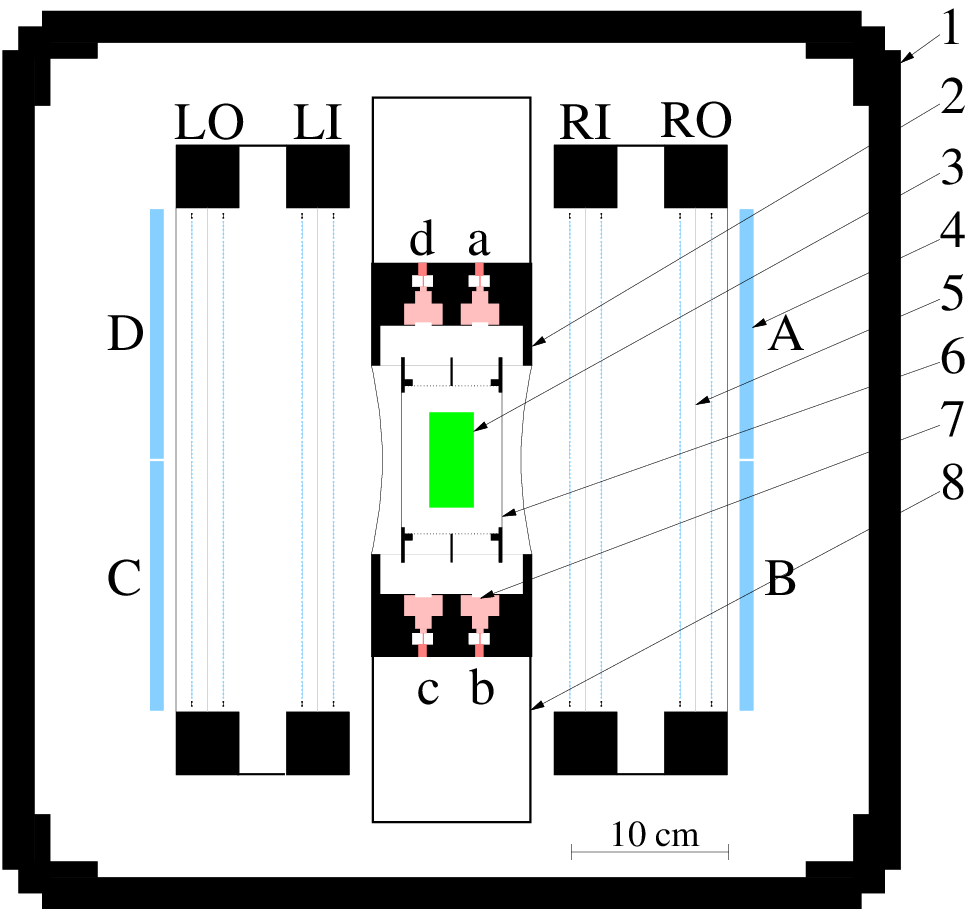}}}
  \caption{(a) Simplest symmetric detector for $D$,
    \label{fig:DPrototype}
    (b) Cross section of the Trine detector:
    1 -- outer chamber (counting
    gas), 2 -- inner vacuum chamber, 3 -- neutron beam, 4 -- plastic
    scintillator, 5 -- wire chamber, 6 -- electrode for proton acceleration,
    7 -- PIN diode, 8 -- housing for PIN preamplifier.\label{abb:TrineCross}
    For both detectors, the polarizations points in $z$ direction perpendicular
    to the plane of the drawing.}
\end{figure}
\begin{equation}\label{eqn:alphaD}
  4 P_z\kappa_{D,z} D = \alpha^{00}-\alpha^{01}-\alpha^{10}+\alpha^{11}
    =:\alpha_D.
\end{equation}
This bases on the different symmetry properties of
$\vec\kappa_A\propto \vec p_{\rm e}$, $\vec\kappa_B\propto \vec p_{\bar\nu}$,
and $\vec\kappa_D\propto \vec p_{\rm e}\times\vec p_{\bar\nu}$.
The detector is insensitive to a beam divergence
and to deviations of the polarization from $z$ axis. However,
deviations from the mirror symmetries are sources for systematic
errors. Whereas (\ref{eqn:alphaD}) suppresses the influences of the parity
violating coefficients one can define asymmetries that enhance this influence
and allow to investigate imperfections of the set-up:
\begin{eqnarray}
  \alpha_x:=&\alpha^{00}+\alpha^{01}-\alpha^{10}-\alpha^{11}&=
    4P_x(A\kappa_{A,x}+B\kappa_{B,x})+4P_yD\kappa_{D,y}\label{eqn:alphax}\\
  \alpha_y:=&\alpha^{00}-\alpha^{01}-\alpha^{10}+\alpha^{11}&=
    4P_y(A\kappa_{A,y}+B\kappa_{B,y})+4P_xD\kappa_{D,x}\label{eqn:alphay}\\
  \alpha_z:=&\alpha^{00}+\alpha^{01}+\alpha^{10}+\alpha^{11}&=
    4P_z(A\kappa_{A,z}+B\kappa_{B,z}).\label{eqn:alphaz}
\end{eqnarray}
The index of these combined asymmetries indicates the component of the
polarization the asymmetry is sensitive to
(cf. Fig.~\ref{fig:DPrototype} (a)). In principle,
(\ref{eqn:alphax})-(\ref{eqn:alphaz})
allow to derive the full polarization vector
from the measured combined asymmetries, using
the values for $A$ and $B$ from literature.

A further reduction of the sensitivity to the coefficients $A$ and $B$
can be obtained by optimizing the angle $\varphi$ between electron and proton
detector. This sensitivity can be described by
$\kappa_A(\varphi)/\kappa_D(\varphi)$
and $\kappa_B(\varphi)/\kappa_D(\varphi)$ and has a minimum at slightly
obtuse angles of about 120$^\circ$, depending on the specific detector
dimensions \cite{soldner2000a}.

The statistical sensitivity of a combination of electron and proton
detector is determined by the angular correlation
between electron and proton and the dependence $\kappa_D=\kappa_D(\varphi)$
and has its maximum at about 135$^\circ$ \cite{lising2000,soldner2000a}.

\section{The Trine Experiment}

Trine detects the electrons by 4 plastic scintillators
($560\times 158\times 8.5$ mm$^3$) in coincidence
with multi wire proportional chambers and the protons after
acceleration in a focusing electrostatic field by special PIN
diodes with thin entrance windows (diameter of active area 10 mm,
25 nm dead layer; see \cite{beck1997} for the performance).
Fig.~\ref{abb:TrineCross}~(b) shows a cross-section of the detector.
The detector consists of 16 such planes which use the same four
scintillators and wire chambers (Fig.~\ref{abb:trinetop}).
Only the central detector planes 03--14 and plane 16 were equipped with
PIN diodes. Data analysis used the 12 central planes
to avoid edge effects at the ends of the high voltage electrode.
Plane 16 served to investigate these effects.
\begin{figure}
  \vspace{2mm}
  \centerline{\includegraphics[width=90mm]{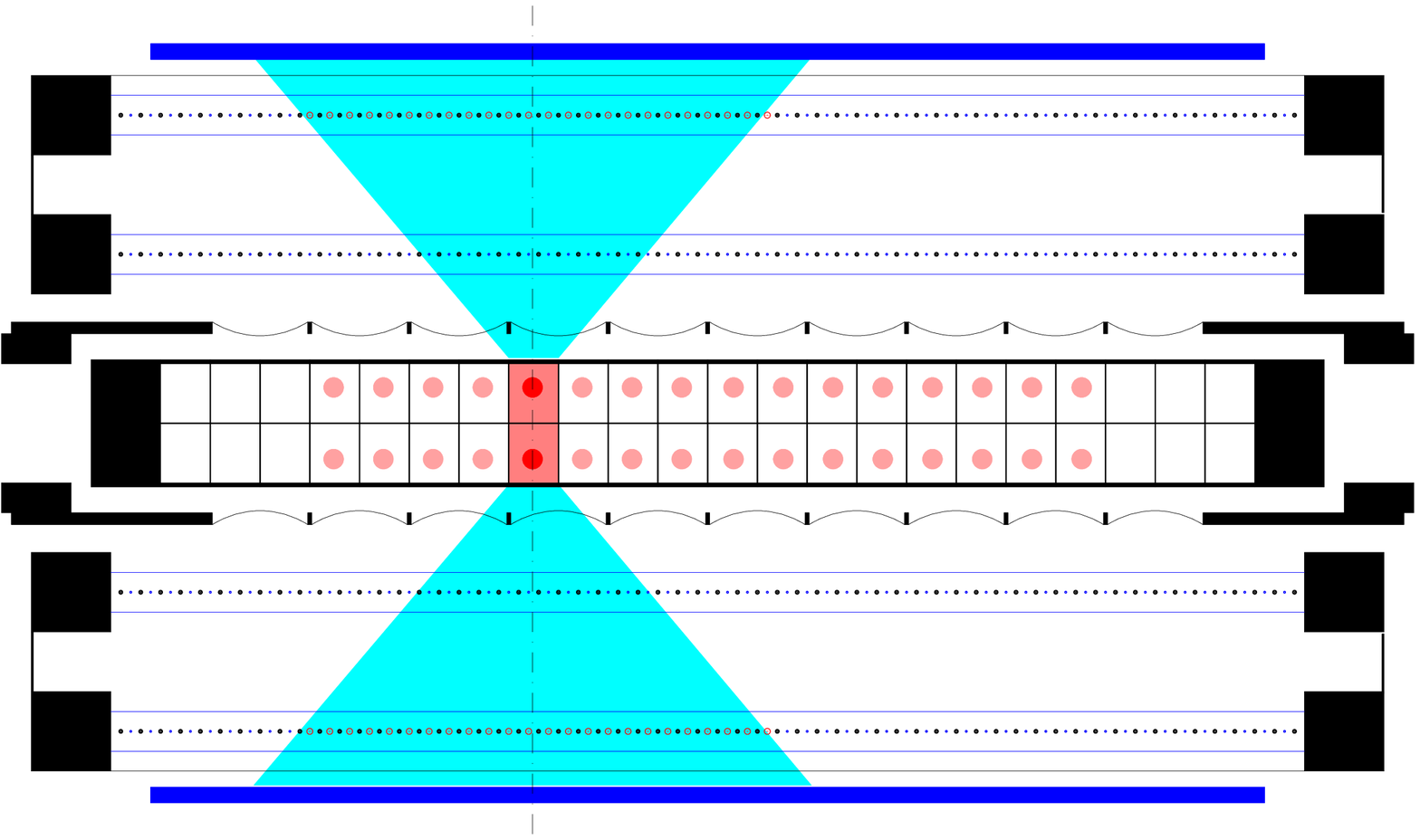}}
  \caption{Top view of the detector, symmetrization: The electron
    has to pass the wire chamber in a range symmetric to the
    PIN diode plane hit by the proton.\label{abb:trinetop}}
\end{figure}

In each plane, four groups of
detector combinations exist defined by the enclosed angle between
electron and proton detector: 50$^\circ$, 82$^\circ$, 98$^\circ$, and
130$^\circ$. Each group fulfills the symmetries requested in section
\ref{sec:principle} (Fig.~\ref{fig:DPrototype}).

The experiment was carried out at the ILL cold neutron
beam facility PF1. The beam polarization of $P=0.974(26)$ was created by a
focusing polarizer. The neutron spin was flipped every 3 s by a resonance
flipper.
An octagonal long coil (length 180 cm, diagonal 96 cm, correction coils at the
ends), surrounded by a mu metal tube to shield the earth magnetic field,
created the longitudinal spin holding field of 140 $\mu$T in the detector
region.
The field deviation $B_\perp/B_z$ from the $z$ axis was smaller than
$5\cdot10^{-3}$.

The neutron beam profile was measured at the beginning, the center and the end
of the decay volume ($z=-15, 0, 15$ cm respectively) using gold foils
which were exposed to the neutron beam and than scanned with an image
plate \cite{soldner2000a}. The profile is slightly inhomogeneous in $y$
direction (Fig.~\ref{abb:cuts}), caused by an inhomogeneous transmission of the
focusing polarizer.
\begin{figure}[b]
  \centerline{\parbox[b]{75mm}{
    \includegraphics[width=75mm]{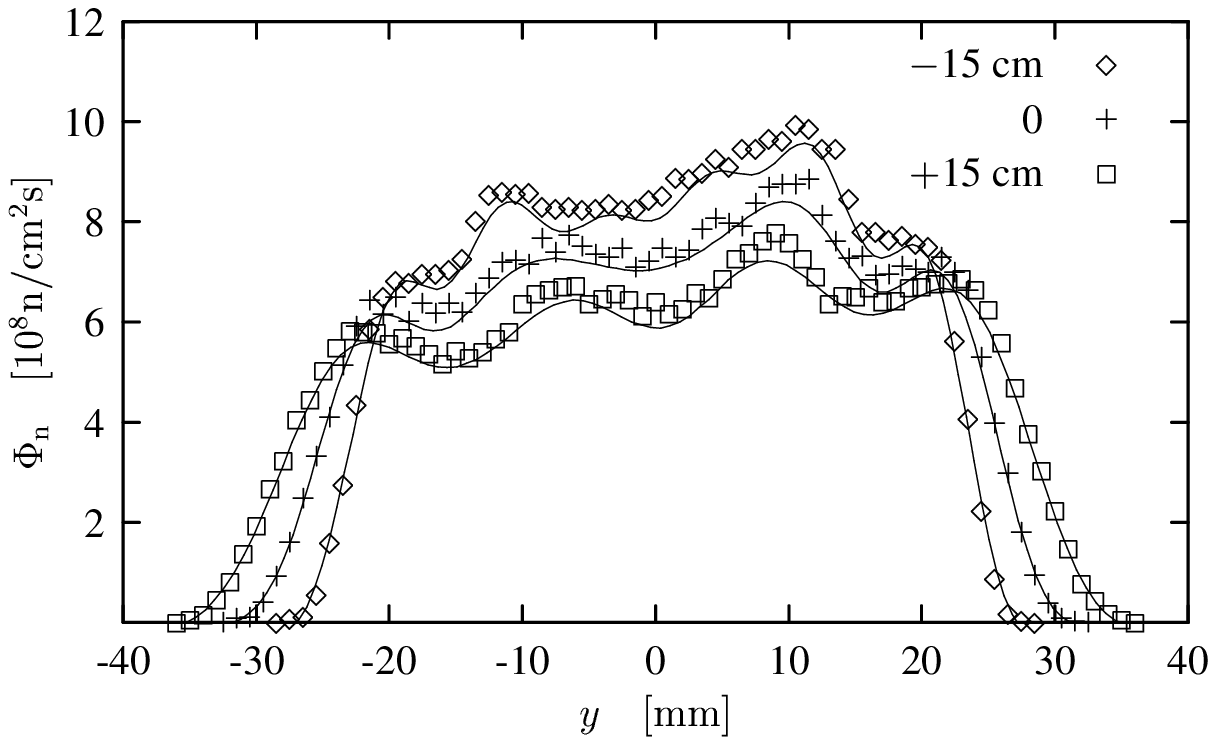}}\hfill
    \parbox[b]{45mm}{\caption{\label{abb:cuts}Cuts of the beam profile (capture
    flux) in the decay volume in $y$ direction. The solid lines correspond to
    2 dimensional Fourier expansions of the data. The decay rate from this
    flux is about $10^3$/s in the detector volume, resulting in a count rate of
    about 10/s (due to solid angle and electron-proton correlation).}}}
\end{figure}

Data acquisition required the coincidence of a scintillator and the
corresponding outer wire chamber. Thus, the trigger rate for events to store
was kept low. Events without a wire chamber signal contributed a dead time of
only 1.2~\%. For each event, the analog signals of all scintillators, the
numbers of the wires
hit in all wire chambers, the number(s) and analog value(s) of the PIN diode(s)
hit in the 10 $\mu$s after the second trigger, and the proton time of flight
(TOF) between trigger and the first PIN diode hit were
registered by a VME based acquisition system. The dead time per stored event
was 30 $\mu$s, resulting in an overall dead time of 3.3~\%. The VME bus was
read out synchronously with the
spin flip. Incomplete events (i.e. events without proton signal within the
10 $\mu$s) were sorted out
by software. Only every 16th incomplete event was saved for control
purposes. Monitor data like neutron flux, count rates of single detectors,
high voltages of the electrode and the wire chambers were stored for each spin
interval.

From the 100 days available at PF1 about 25 days in the first and
40 days in the second reactor cycle were used to collect
statistics and 10 days of the second cycle for systematic tests.
During the measurement, the scintillators were recalibrated every
10 days but only small adjustments were needed. Data from the
first cycle suffered from high voltage problems and are not
analysed yet. In the following we present the analysis of the data
from the second cycle.

\section{Data Analysis}

\subsection{Selection of events}

Spin intervals with unusual values of monitor signals together with the
following three intervals and spin intervals where one VME module
lost a trigger were removed (approx. 4~\%). Only complete events with exactly
one triggering PIN diode were used. A threshold of 150~keV was applied to the
electron signal by software (hardware threshold was about 115~keV). The
stability of the detectors was verified by an automatic generation of
software cuts to the PIN analog spectra and allowed to sum the data to
10 days samples, corresponding to the period of scintillator recalibrations.

\begin{figure}
  \centerline{\parbox[b]{75mm}{
    \includegraphics[width=75mm]{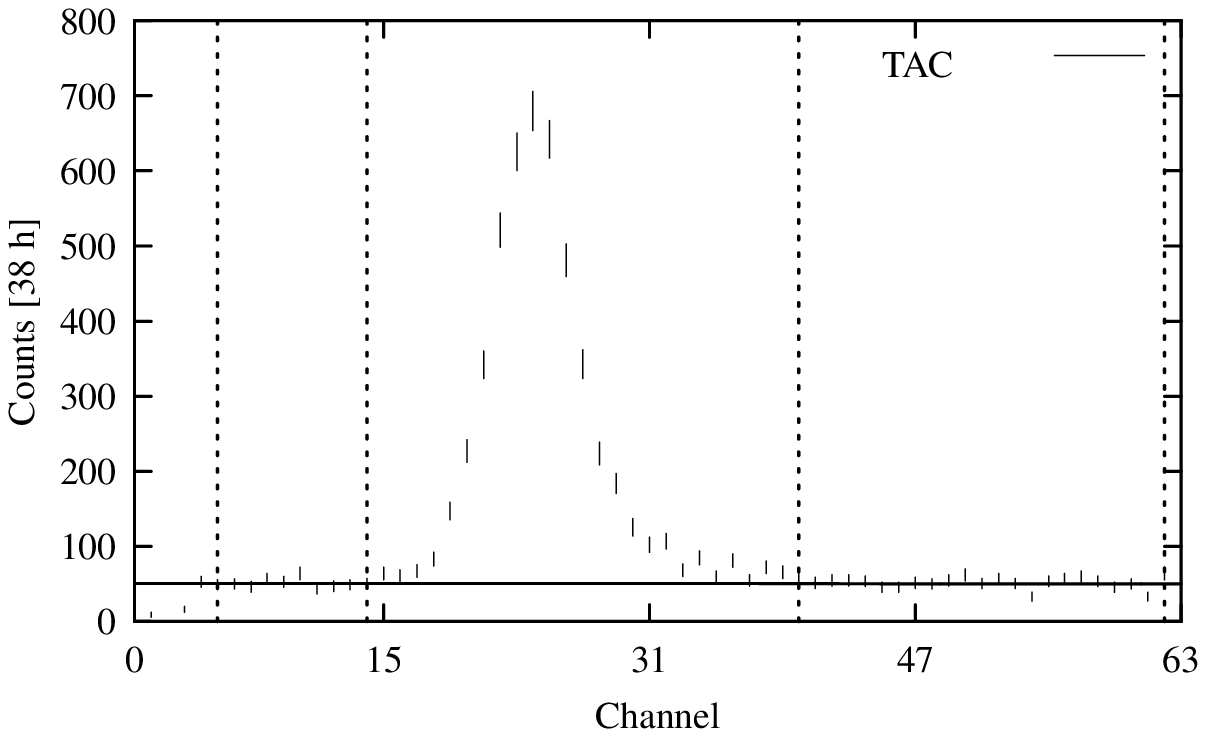}}\hfill
    \parbox[b]{45mm}{\caption{\label{abb:TOF}Typical TOF spectrum (1 PIN diode
    and 1 scintillator for 98$^\circ$). The dashed lines indicate the
    ranges for the background fit, the solid line the fit result.}}}
\end{figure}
Individual TOF spectra were calculated for all detector combination in each
sample using the events that fulfill the software cuts. The background of the
TOF spectra was fit by an exponential in a fixed range before and after the
coincidence peak (Fig.~\ref{abb:TOF}). This shape of the background
follows from the data acquisition which stopped the TOF measurement with the
first proton signal. As a further consequence, the background
behind the peak is suppressed compared to that in front of it.
The $\chi^2$ analysis showed perfect agreement between the exponential and
the data for separate fits of the two fit ranges
but a systematic increase to $\chi^2/{\rm ndf}=1.26$ (averaged over all
individual spectra) for a common fit of the
ranges. To account for this the error of the background was scaled by a factor
1.124, but anyway the effect is very small due to the excellent signal to
background ratio of 23 (averaged over the detector combinations used).
The thus obtained peak areas were normalized with the neutron monitor counts
of the particular spin to account for fluctuations
caused by upstream experiments.

\subsection{Selection of Detector Combinations}

The measured count rates of the detector combinations 50$^\circ$ and
82$^\circ$ were higher than expected from the Monte Carlo simulations.
These combinations are more sensitive to systematic effects due to
the small particle energies caused by kinematics. This increases the
scattering for electrons (e.g. by the counting gas). The low energy protons
may be disturbed by a small penetration of the electrostatic field into
the electrode. Furthermore, the sensitivity to $A$ and $B$ coefficient is
larger for angles below 90$^\circ$ than for slightly obtuse angles,
and the contribution of small angle combinations to the statistics can be
neglected (see section \ref{sec:principle} or \cite{soldner2000a}).
Therefore, only the larger angle combinations (98$^\circ$ and 130$^\circ$)
were used in the analysis.

\subsection{Detector ``Symmetrization''\label{subsec:symmetrisation}}

The single asymmetry $\alpha^{ij}$ of a detector combination close
to an end of the decay volume is high due to the spatial asymmetry
of this combination in $z$ direction, resulting in a sensitivity
to $A$ and $B$ (Fig.~\ref{fig:alphaz}, top). This sensitivity
cancels by calculating the combined asymmetry $\alpha_D$ (section
\ref{sec:principle}). However, for a real detector, effects like
inhomogeneous detector efficiencies result in an incomplete
cancellation which can fake $D\ne 0$.
\begin{figure}[t]
  \begin{center}
  \includegraphics[width=90mm]{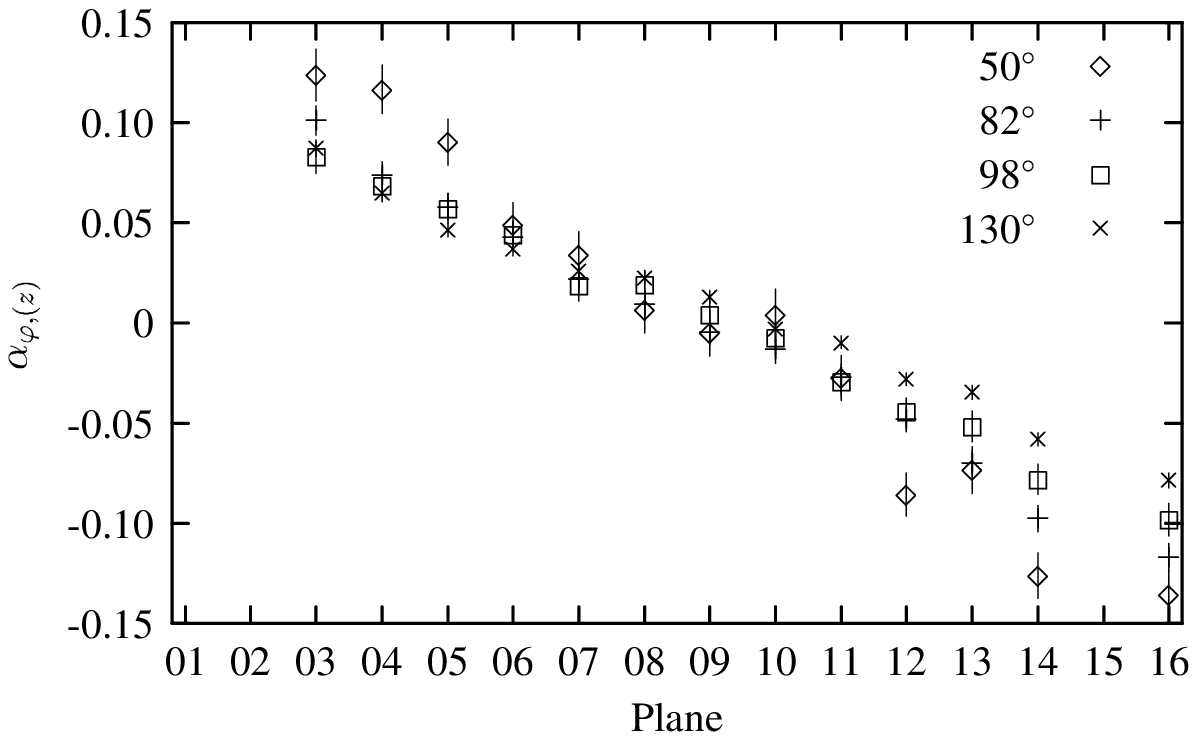}\hfill
    \includegraphics[width=90mm]{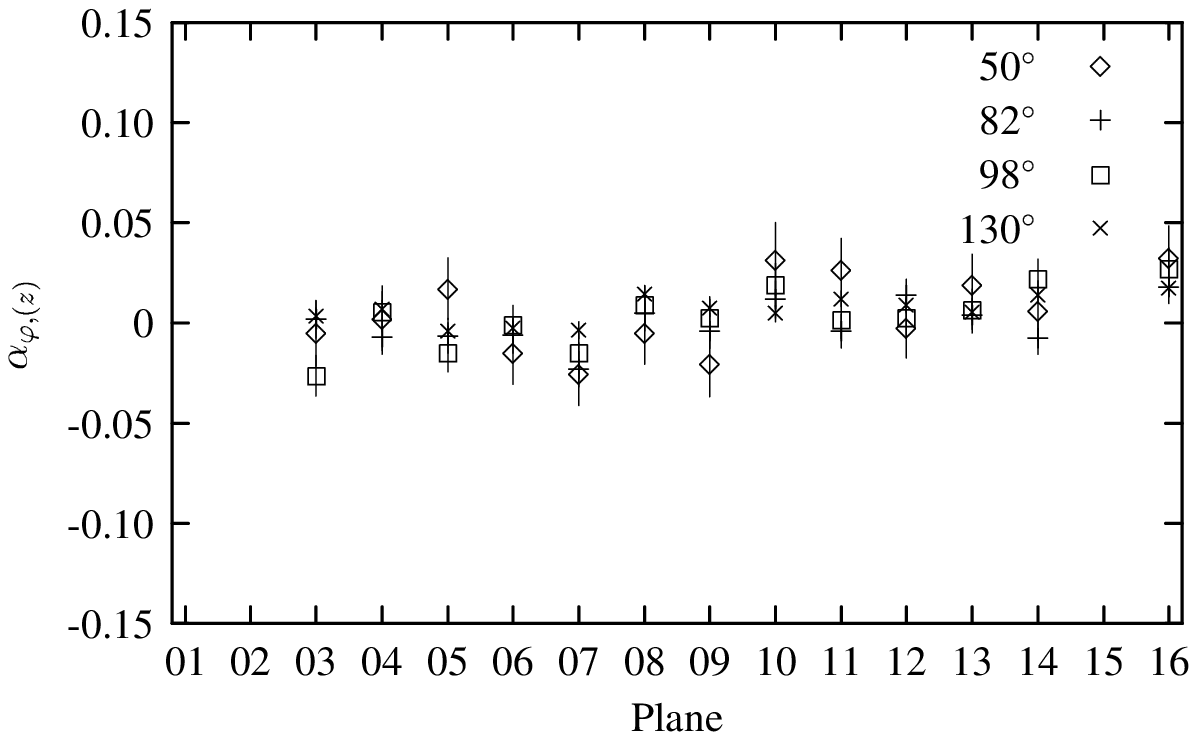}
    \end{center}
  \caption{Combined asymmetries $\alpha_z\approx 4\alpha$ as function of the
    detector plane. Top: full scintillator, Bottom: symmetrized detector with
    $\pm 10$ cm per plane.\label{fig:alphaz}}
\end{figure}
The spatial resolution of the wire chamber was used to suppress
this sensitivity already in the initial asymmetries by selecting a
symmetric electron detector range for each detector plane (see
Fig.~\ref{abb:trinetop}). The resulting asymmetries $\alpha_z$ are
plotted in Fig.~\ref{fig:alphaz} (bottom). The size of the range
was selected such that the variations of $\alpha_D$ between the
different planes were consistent with statistical variations. For
$\pm 10$ cm a $\chi^2$ of 10.4 (12.2) for 11 degrees of freedom
was found for 98$^\circ$ (130$^\circ$). The slightly higher
$\chi^2$ for 130$^\circ$ was taken into account as a systematic
error of $2.2\cdot 10^{-4}$. The change of the $\alpha_D$ values
for different sizes of the wire chamber range was not fully
compatible with statistics. Although this is expected since the
range serves to suppress systematic effects it was considered as a
contribution to the systematic error of $1.0\cdot10^{-4}$ by
comparing the $D$ values for different wire range sizes.

\subsection{Influence of the Beam Profile\label{subsec:beamprofile}}

The influence of the beam profile was investigated with test measurements
where one half or one quarter of the polarizer exit were closed to
increase the beam shift (center of mass shifted by $\Delta y=7.2$ mm for
the 3/4 beam compared to 1 mm for the full beam).
The results $D_{\rm 3/4}$ for the both detector combinations used were
consistent with 0 but were used to limit the systematic error caused
by the inhomogeneous beam profile:
$\delta_{\rm shift} D=16(13)\cdot 10^{-4}$ ($2.4(5.0)\cdot 10^{-4}$)
for 98$^\circ$ (130$^\circ$) (statistical error given).
A more precise calculation of this
systematic error by Monte Carlo simulations is in progress and will
replace the present estimation in the final result.

\subsection{Results and Outlook}

During the second cycle, $30\cdot10^6$ events were collected with the
unshifted beam. $13.8\cdot 10^6$ events fulfilled the symmetry condition
(wire chamber range). The preliminary result is
$D=(-3.1\pm 6.2^{\rm stat}\pm 4.7^{\rm syst}\pm4.7^{\rm syststat})\cdot10^{-4}$.
Syststat indicates the statistical error of the systematic error
determined with the partially covered beam and will
not enter into the final result after the Monte Carlo simulations
(section \ref{subsec:beamprofile}). The systematic error
consists of the contributions given in sections \ref{subsec:symmetrisation}
and \ref{subsec:beamprofile} and those from
the uncertainties of the apparatus constants ($0.3\cdot10^{-4}$)
and polarization ($0.08\cdot10^{-4}$).

\begin{table}[t]
  \caption{Comparison of the latest $D$ measurements.}
  \renewcommand{\arraystretch}{1.4}
  \setlength\tabcolsep{5pt}
  \centerline{
  \begin{tabular}{lrr@{}lr@{}lr@{}lr@{}l@{}l@{}l}\hline
    \rule{0pt}{14pt}& Year & \multicolumn{2}{l}{$P$} &
     \multicolumn{2}{l}{Events} &  \multicolumn{2}{l}{Sig/BG} &
     \multicolumn{4}{l}{$D$}\\
  &&\multicolumn{2}{l}{$[\%]$}&\multicolumn{2}{l}{$[10^6]$} &  &&
    \multicolumn{4}{l}{$[10^{-3}]$} \\[2pt]\hline
  \cite{steinberg1976}\rule{0pt}{14pt}&1976& 70&(7) &6&& 4&&
    --1&.1&$\pm$1.7\\
  \cite{erozolimskii1978}              &1978& 68&(3), 65(1) & 2&.5&2&.2&
      2&.2&$\pm$3.0\\
  emiT \cite{lising2000}                   &2000& 96&(2) &~15&&  2&.5&
    --0&.6&$\pm$1.2&$\pm$0.5\\
  Trine                       &2000& 97&.4(2.4) &30&/13.8&23&&
    --0&.31&$\pm$0.62&$\pm$0.47$\pm$0.47\\[2pt]\hline
  \end{tabular}}
  \label{tab:comparison}
\end{table}
Table \ref{tab:comparison} compares the last $D$ measurements. The result of
the Trine measurement profits from the suppression of systematic effects
using the spatial resolution of the wire chambers and the high segmentation
with 12 used detector planes. Because of the signal to background ratio of 23
the statistics of the neutron beam could be used completely.

Improved measurements of emiT (Trine) are in progress (preparation).
The world average for $D$ may reach a precision in the very interesting
lower $10^{-4}$ range within one year.

\section*{Acknowledgments}

We are indebted to Prof.~P.~Liaud, Dr.~A.~Bussi{\`e}re, and Dr.~R.~Kossakowski
for their help in the preparation of the experiment and for valuable
discussions. We acknowledge the support of the Institut Laue Langevin.
This work was supported by BMBF (grants 06 TM366 and 06TM879) and
is associated to SFB 375 of the DFG.

\title*{Search for Time Reversal Violating Effects,
\protect\newline R- and N-Correlations in the Decay of Free
Neutrons}
\titlerunning{Search for TRV Effects in n-Decay}
\toctitle{Search for Time Reversal Violating Effects,
\protect\newline R- and N-Correlations in the Decay of Free
Neutrons}
%
%
%

\author{
  K.~Bodek\inst{1}\and
  G.~Ban\inst{7}\and
  M.~Beck\inst{4}\and
  A.~Bia{\l}ek\inst{8}\and
  T.~Bry\'s\inst{{1}, {3}}\and
  A.~Czarnecki\inst{9}\and
  W.~Fetscher\inst{2}\and
  P.~Gorel\inst{{3},{7}}\and
  K.~Kirch\inst{3}\and
  St.~Kistryn\inst{1}\and
  A.~Kozela\inst{{2}, {8}}\and
  A.~Lindroth\inst{4}\and
  O.~Naviliat-Cuncic\inst{7}\and
  J.~Pulut\inst{{1}, {3}, {4}}\and
  A.~Serebrov\inst{6}\and
  N.~Severijns\inst{4}\and
  E.~Stephan\inst{5}\and
  J.~Zejma\inst{1}}
\authorrunning{K.~Bodek et al.}
%
%
\institute{ Institute of Physics, Jagellonian University, Cracow,
Poland \and Institute of Particle Physics, ETH,
Z\"urich,Switzerland \and Paul Scherrer Institute, Villigen,
Switzerland \and Catholic University, Leuven, Belgium \and
University of Silesia, Katowice, Poland\and St. Petersburg Nuclear
Physics Institute, Gatchina, Russia\and Laboratoire de Physique
Corpusculaire, Caen, France\and Institute of Nuclear Physics,
Cracow, Poland\and University of Alberta, Edmonton, Canada }

\maketitle              

\begin{abstract}
An experiment aiming at the simultaneous determination of the two
transversal polarization components of electrons emitted in the
decay of free, polarized neutrons is underway at the Paul Scherrer
Institute, Villigen, Switzerland. A non-zero value of $R$ due to
the polarization component, which is perpendicular to the plane
spanned by the spin of the decaying neutron and the electron
momentum, would signal a violation of time reversal symmetry and
thus physics beyond the Standard Model (SM).  The value of $N$,
given by the transverse polarization component within that plane,
is expected to be finite.  The measurement of $N$ both probes the
SM and serves as an important systematic check of the apparatus
for the $R$-measurement.  Using the Mott scattering polarimetry
technique, the anticipated accuracy of $5\times 10^{-3}$ should be
achieved within a few months of data taking.
\end{abstract}

\section*{Introduction}

According to well known theoretical conjectures, supported by
experimental observations, the combined charge conjugation and
parity symmetry ($\EuScript{CP}$) and time reversal symmetry
($\EuScript{T}$) are closely related by the
$\EuScript{CPT}$-theorem.  There are two unambiguous pieces of
evidence for $\EuScript{CP}$- and $\EuScript{T}$-violation: the
forbidden decay modes of neutral $K$ and $B$ mesons and the excess
of the baryonic matter over antimatter in the present Universe.
However, the $\EuScript{CP}$-violation found in kaon decays, and
incorporated into the SM via the quark mixing mechanism, is too
weak to explain the excess of baryons over antibaryons. Therefore,
cosmology provides a hint for the existence of an unknown source
of $\EuScript{T}$-violation, which is not included in the SM.

The SM predictions of $\EuScript{T}$-violation, originating from
the quark mixing scheme, for systems built up of $u$ and $d$
quarks, are by 7 to 10 orders of magnitude lower than the
experimental accuracies available at present.  This applies to
determinations of the $\EuScript{T}$-violating electric dipole
moments as well as to $\EuScript{T}$-violating correlations in
decay or scattering processes.  With such a strong suppression of
the SM contribution these experiments are regarded as important
searches for \textit{``Physics beyond the Standard Model.''}  New
time reversal violating phenomena may be generated by e.g. the
exchange of multiplets of Higgs bosons, leptoquarks, right handed
bosons, or by the presence of the $\theta$ term in the QCD
interactions. These exotic particles or phenomena do not
contribute to the $V$-$A$ form of the weak interaction which is
embedded into the SM. However, they may generate scalar $S$ or
tensor $T$ variants of the weak interaction or a phase different
from 0 or $\pi$ between the vector $V$ and axial-vector $A$
coupling constants. It is a general presumption that time reversal
phenomena are caused by tiny admixture of exotic interaction
terms.  Therefore, weak decays provide a favorable testing ground
in a search for such feeble forces \cite{BOEH95,HERC95}.  Physics
with very slow, polarized neutrons has a great potential in this
respect. Our experiment looks after small deviations from the SM
in two observables that have never before been addressed
experimentally in neutron decay.

\section*{Angular correlations in $\beta$-decay}

Direct, i.e. first-order access to the $\EuScript{T}$-violating
part of the weak interaction coupling constants is provided for by
measurements of directional correlations between the spins and
momenta of particles or nuclei involved in the decay process.  The
lowest order $\EuScript{T}$-violating combination of spins and
momenta appears in the form of the mixed triple product.  From the
experimentally accessible quantities, four triple products can be
formed (following the notation of \cite{JACK57a,JACK57b,EBEL57}):
\begin{center}
  \begin{tabular}{llll}
    $R$-correlation~ & ($\EuScript{T}$-odd, & $\EuScript{P}$-odd) &
      ~~$\vec{J}\cdot(\vec{p}\times\vec{\sigma})$, \\
    $D$-correlation~ & ($\EuScript{T}$-odd, & $\EuScript{P}$-even) &
      ~~$\vec{J}\cdot(\vec{P}\times\vec{p})$ =
      $\vec{J}\cdot(\vec{p}\times\vec{p}_{\nu})$, \\
    $V$-correlation~ & ($\EuScript{T}$-odd, & $\EuScript{P}$-odd) &
      ~~$\vec{J}\cdot(\vec{P}\times\vec{\sigma})$, \\
    $L$-correlation~ & ($\EuScript{T}$-odd, & $\EuScript{P}$-even) &
      ~~$\vec{P}\cdot(\vec{p}\times\vec{\sigma})$,
  \end{tabular}
\end{center} where $\vec{J}$ is the spin of the parent system,
$\vec{\sigma}$, $\vec{p}$ are the spin and momentum of the
detected lepton, $\vec{P}$ denotes the momentum of the recoil
system and $\vec{p}_{\nu}$ stands for the momentum of the
unobserved neutrino.  Either one of the first two correlations
listed above was measured in the weak decays of the muon, the
neutron, kaons, hyperons, several nuclei and in the decay of
polarized $Z^0$ \cite{ABEK95}.  The only system for which both $D$
and $R$ have been determined is $^{19}$Ne \cite{SCHN83}.  The
latter two correlations, which require two difficult measurements
simultaneously, were not addressed experimentally yet.

For our discussion, the relevant terms in the formula for the
decay rate $W$ for a semileptonic transition from an oriented
sample of nuclei or particles with vector polarization $\vec{J}$
can be written as \cite{JACK57a,JACK57b}:
\begin{eqnarray}
  W\!\!\propto\!\!\left[ 1 +
    A\,\frac{\vec{J}\cdot\vec{p}}{E} +
    B\,\frac{\vec{J}\cdot\vec{p_{\nu}}}{E_{\nu}} +
    D\,\frac{\vec{J}\cdot(\vec{p}\times\vec{p_{\nu}})}{EE_{\nu}}+
    R\,\frac{\vec{J}\cdot(\vec{p}\times\vec{\sigma})}{E} +
    N\;\vec{J}\cdot\vec{\sigma} + \!\cdot\cdot\cdot \right],
    \nonumber
\end{eqnarray}
where $E$, $E_{\nu}$ are the total energies of emitted leptons,
and $A$ and $B$ are the usual decay asymmetry parameters arising
from parity violation for the charged lepton and the neutrino,
respectively. The correlation parameter $N$ is also connected to
the charged lepton's spin, however, it is
$\EuScript{T}$-conserving. $N$ was rarely considered in
$\beta$-decay discussions so far.

\subsection*{The $R$-correlation}

Our interest is focused on the $R$-correlation, the time reversal
violating observable, which has not yet been measured for the free
neutron decay.  The physical interpretation is straightforward:
the numerical value of the $R$-coefficient represents the
transverse component of the electron polarization which is
contained in the plane perpendicular to the neutron spin axis.  In
contrast to $D$, which is sensitive primarily to the complex terms
in the vector/axial-vector interference, the $\EuScript{P}$-odd,
$\EuScript{T}$-odd $R$-observable may disclose the exotic scalar
or tensor interaction terms.  The explicit expression for the
$R$-amplitude, in terms of Fermi and Gamow-Teller matrix elements
$M_F$, $M_{GT}$ and weak interaction coupling constants $C_i$
($i=S,V,A,T$), is given by \cite{JACK57b}.  For neutron decay, we
obtain:
\begin{eqnarray}
  R \;&=&\; \frac{\Im{\left[ (C_V^*+2C_A^*)(C_T+C'_T)+C_A^*(C_S+C'_S)
  \right]}}{|C_V|^2+3|C_A|^2} \;+\; R_{FSI}
  \;\end{eqnarray}\begin{eqnarray}\hspace{-30mm}=\; 0.28 \cdot S \,+\, 0.33 \cdot T \;+\; R_{FSI},
  \nonumber
\end{eqnarray}
where $S \equiv \Im{\left[ (C_S+C'_S)/C_A \right]}$ and $T \equiv
\Im{\left[ (C_T+C'_T)/C_A \right]}$ and $M_F=1$,
$M_{GT}=\sqrt{3}$, $C_V = {C'}_V = \Re{C_V} = 1$, $C_A = {C'}_A =
\Re{C_A} = -1.26$, and $|C_S|, |{C'}_S|, |C_T|, |{C'}_T| \ll 1$
were assumed.  While the lowest order expression of $R$ vanishes
for the SM, the value including final-state interaction becomes
finite:
\begin{eqnarray}
 R_{\rm FSI, SM} \;&=&\;  \frac{\alpha Z m}{p} \cdot A_{\rm SM}.
 \nonumber
\end{eqnarray}
With $A = -0.1189(8)$ \cite{REIC00}, this implies $R_{\rm SM}
\approx 0.001$ due to FSI-effects, which is beyond the scope of
this experiment, though the value of this correction is known with
the absolute precision of $10^{-5}$~\cite{VOGE83}\footnotemark[1].
The exclusion plot in the $S-T$ plane, including the results from
Refs.~\cite{SCHN83,SROM96} and from electron-neutrino angular
correlations in the decay of $^{33}$Ar~\cite{ADEL93} and
$^{32}$Ar~\cite{ADEL99} is shown in Fig.~\ref{FIG1}.  We note that
the neutron experiment, with an accuracy of 0.005 in the
$R$-coefficient, has a potential either to determine finite values
of the $\EuScript{T}$-violating charged current scalar couplings
or to bring a significant improvement in their upper limit.
\footnotetext[1]{With present input data the precision may reach
$5\times 10^{-6}$}
\begin{figure}[t]
  \begin{center}
    \includegraphics[height=7.0cm,angle=180]{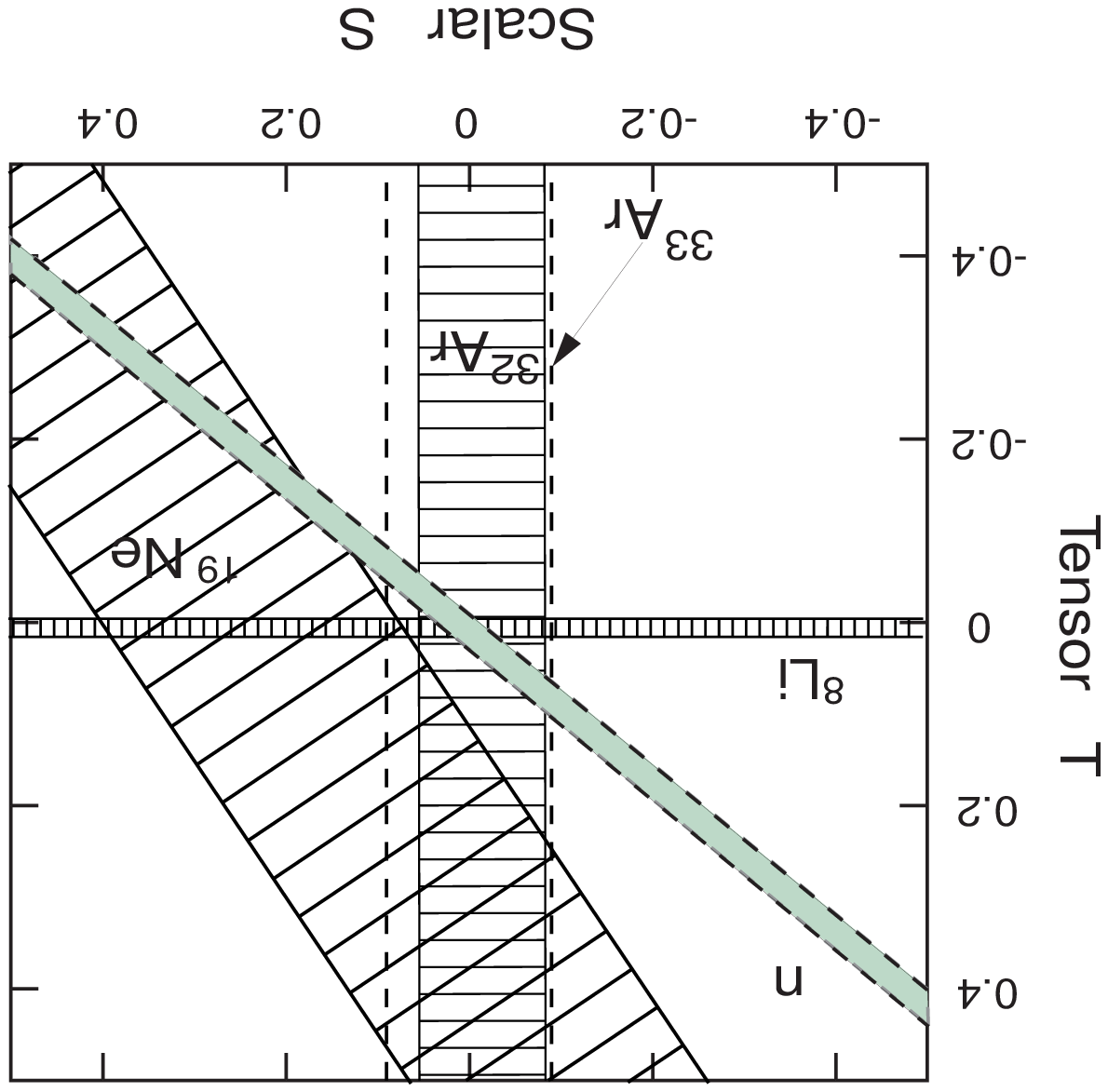}
  \end{center}
\vspace{3pt} \caption{\label{FIG1}
           Results from the experiments testing the scalar and tensor
           weak interaction.  The bands indicate $\pm 1\sigma$ limits.
           Constraints from the study of the $R$-correlation in the
           free neutron decay with an accuracy of $\pm 0.005$ are
           attached.  This prediction is arbitrarily fixed at
           $S,T$ = 0.}
\end{figure}

\subsection*{The $N$-correlation}

According to our knowledge this correlation has not been measured
directly in nuclear or neutron decay before. As for the
$R$-correlation, $N$ can be determined by measuring the neutron
polarization, and the momentum and transverse polarization of the
emitted electron. The experimental apparatus capable of measuring
$R$ will in a natural way measure $N$ simultaneously. The
numerical value of the $N$-coefficient multiplied by
sin$\,\theta_e$, $\theta_e$ being the electron emission angle with
respect to the neutron spin direction, represents the transverse
component of the electron polarization which is contained in the
plane spanned by the neutron polarization and the electron
momentum. $N$ conserves $\EuScript{T}$ and is given in
Ref.~\cite{JACK57b}.  We note that the Standard Model value of $N$
scales with the decay asymmetry $A$, corresponding to:
\begin{eqnarray}
  N_{\rm SM} \;=\; -\;\frac{m}{E}\; A_{\rm SM}
  \;=\; \frac{m}{E} \;\frac{2(\lambda^2+\lambda)}{1+3\lambda^2}
  \;\approx\; + 0.119\;\frac{m}{E}, \nonumber
\end{eqnarray}
where $\lambda$ denotes the ratio $C_A/C_V$.  This neutron decay
experiment aims at an absolute sensitivity of 0.5\% which
translates into a measurement of $N$ at the 5\% (relative) level.
Because $A$ has been measured to the 1\% level one can not expect
any progress in the determination of the Standard Model $\lambda$
from the measurement of $N$. However, the $N$ measurement will
test additional couplings in the general expression with its 0.5\%
sensitivity. This is the same sensitivity as of the current best
$B$ \cite{SERE98} experiment and thus the measurement of $N$ will
provide an independent check on the involved scalar and tensor
couplings.

\section*{Experiment}

The main challenge of the experiment is the measurement of the
polarization of the low energy electrons (end-point energy of
783~keV in neutron decay).  Large angle Mott scattering is
sensitive to the transverse component of the electron polarization
and the analyzing power reaches exceptionally high values of -0.4
to -0.5 as shown in Fig.  2a.  Such a high analyzing power,
together with the large polarization of the cold neutron beam
($\sim 95$\%) provides an unprecedented sensitivity for spin
observables. However, for neutron decay, the difficulty arises
from relatively weak decay source (10$^3$ -- 10$^4$~s$^{-1}$).
This should be considered in the context of high background
generated by slow neutrons captured in the neighborhood of the
experiment.
\begin{figure}[b]
 \centering

   \includegraphics[height=11.cm]{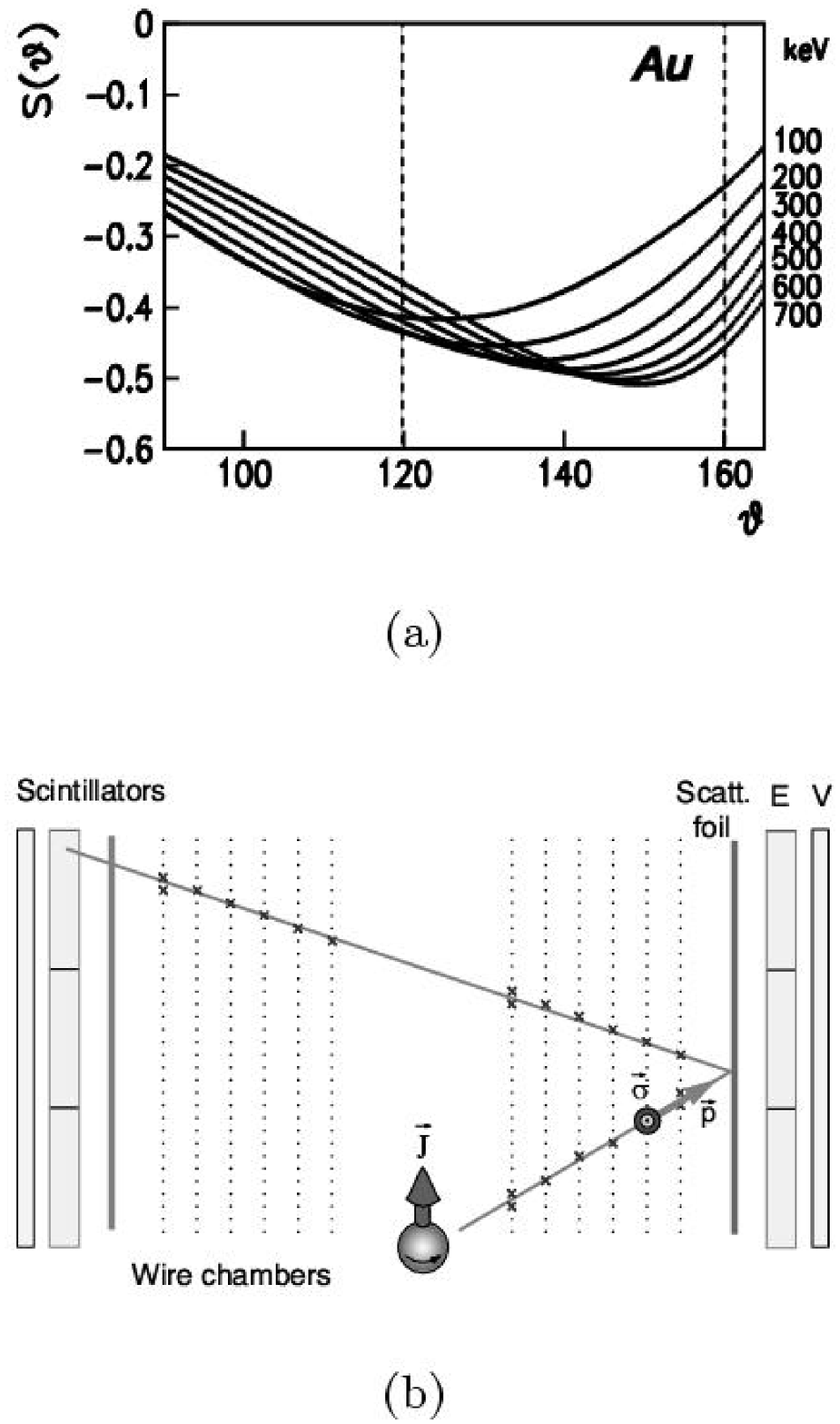}

 \caption{(a) Mott scattering analyzing power $S(\vartheta)$ for
 $^{197}$Au obtained by interpolation of the Sherman tables
 \cite{SHER56}.  (b) The principle of the experiment.  The electrons
 are backscattered from the spin analyzer Au foil.  From the electron
 tracks, the scattering angle and the Mott scattering asymmetry can
 be determined.}
\end{figure}

The principle of the measurement is sketched in Fig. 2b. The
electron emitted from a polarized neutron and scattered from an
analyzing foil is tracked by a system of two multiwire gas
chambers and stops in the plastic scintillator. In this way, all
the angular and energy information necessary to determine the
momentum of the electron and the Mott scattering asymmetry is
provided.

For the vertically oriented neutron spin in a simultaneous
measurement of $R$ and $N$ one of the correlations will produce an
up-down asymmetry while the other leads to a forward-backward
asymmetry. Turning the neutron polarization by 90$^\circ$ between
the beam axis and the vertical axis interchanges the relation of
the correlations to the observable asymmetries. One can now fully
appreciate the measurement of $N$ as an aid for the
$R$-measurement: because the deviation of $N$ from its SM-value is
expected to be small (as implied by the good knowledge of the
asymmetry parameters $A$ and $B$) it provides an ideal,
positive-effect calibration of the apparatus.
\begin{figure}[t]
 \centering
   \includegraphics[height=10.5cm]{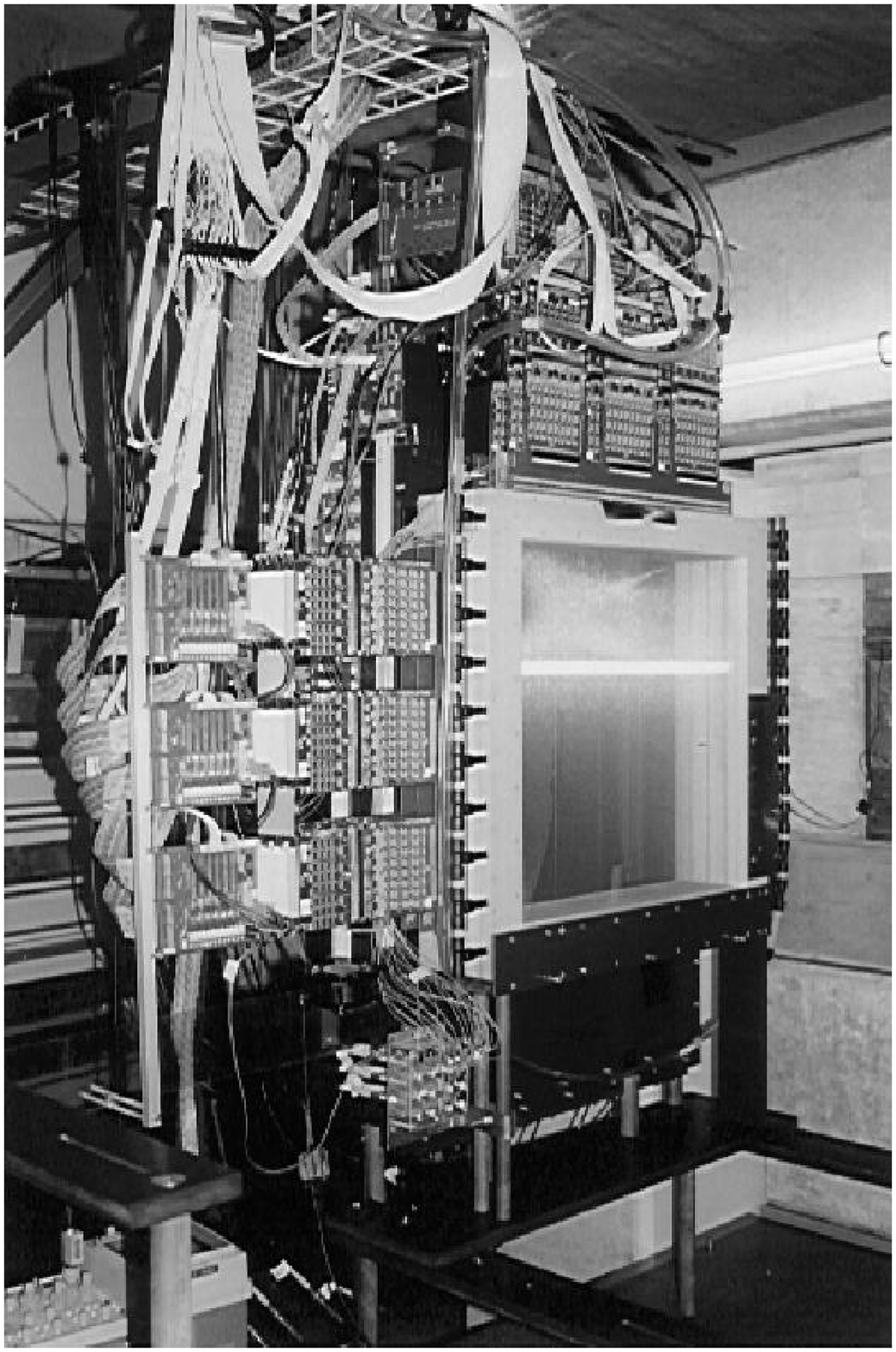}
  \caption{Full size MWPC-1 placed inside the experimental bunker.
  The neutron decay chamber is disassembled.}
\end{figure}
\begin{figure}[b]
 \centering
 \includegraphics[height=5.5cm]{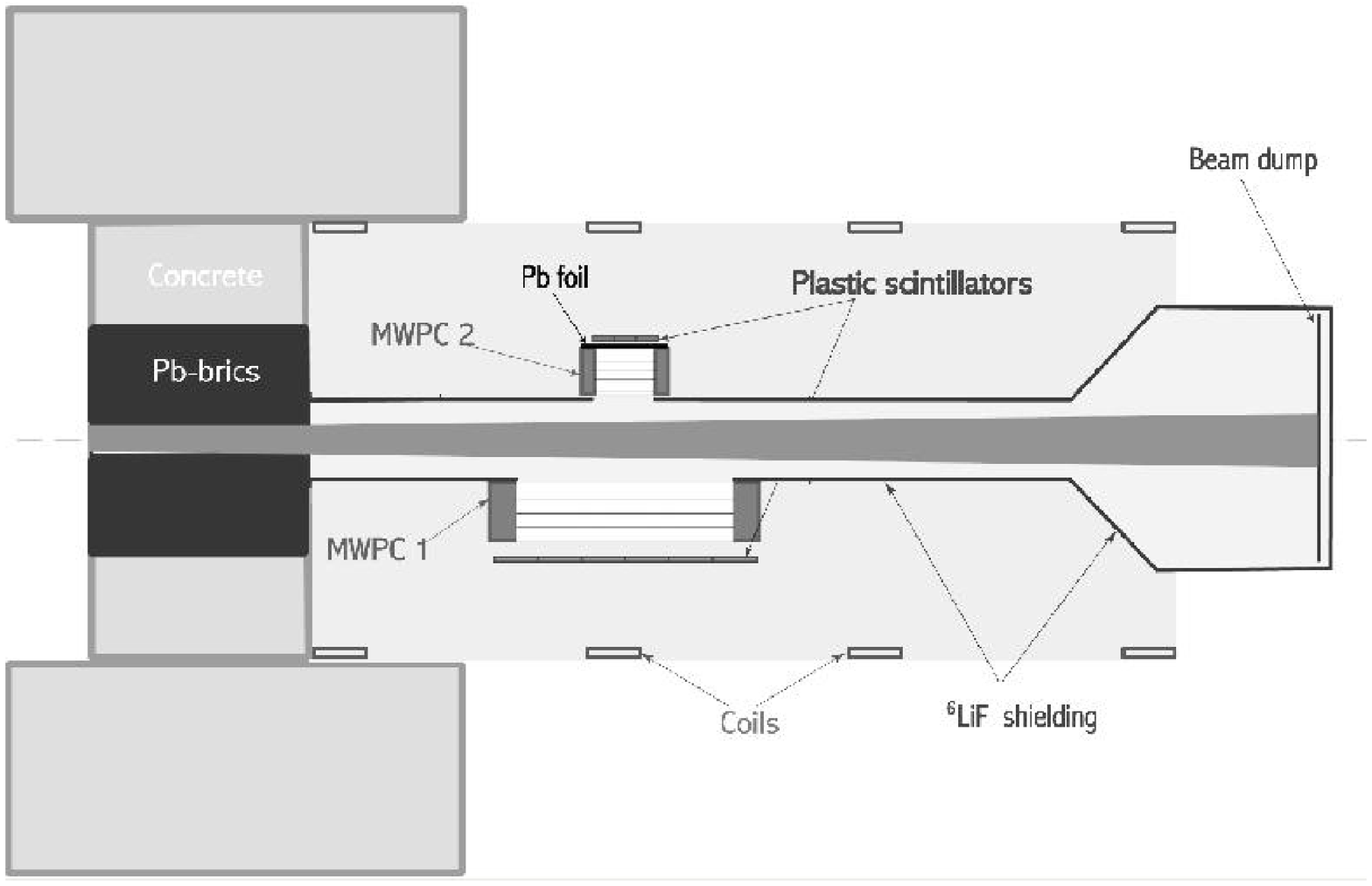}
  \caption{Layout of the
  experimental setup where the missing MWPC-2 has been replaced by the
  prototype detector.  A 30 mg/cm$^2$ thick Pb foil is used as the Mott
  target.}
\end{figure}
\begin{figure}[t]
 \centering
 \subfigure[\label{SB1}]{
   \includegraphics[height=4.2cm]{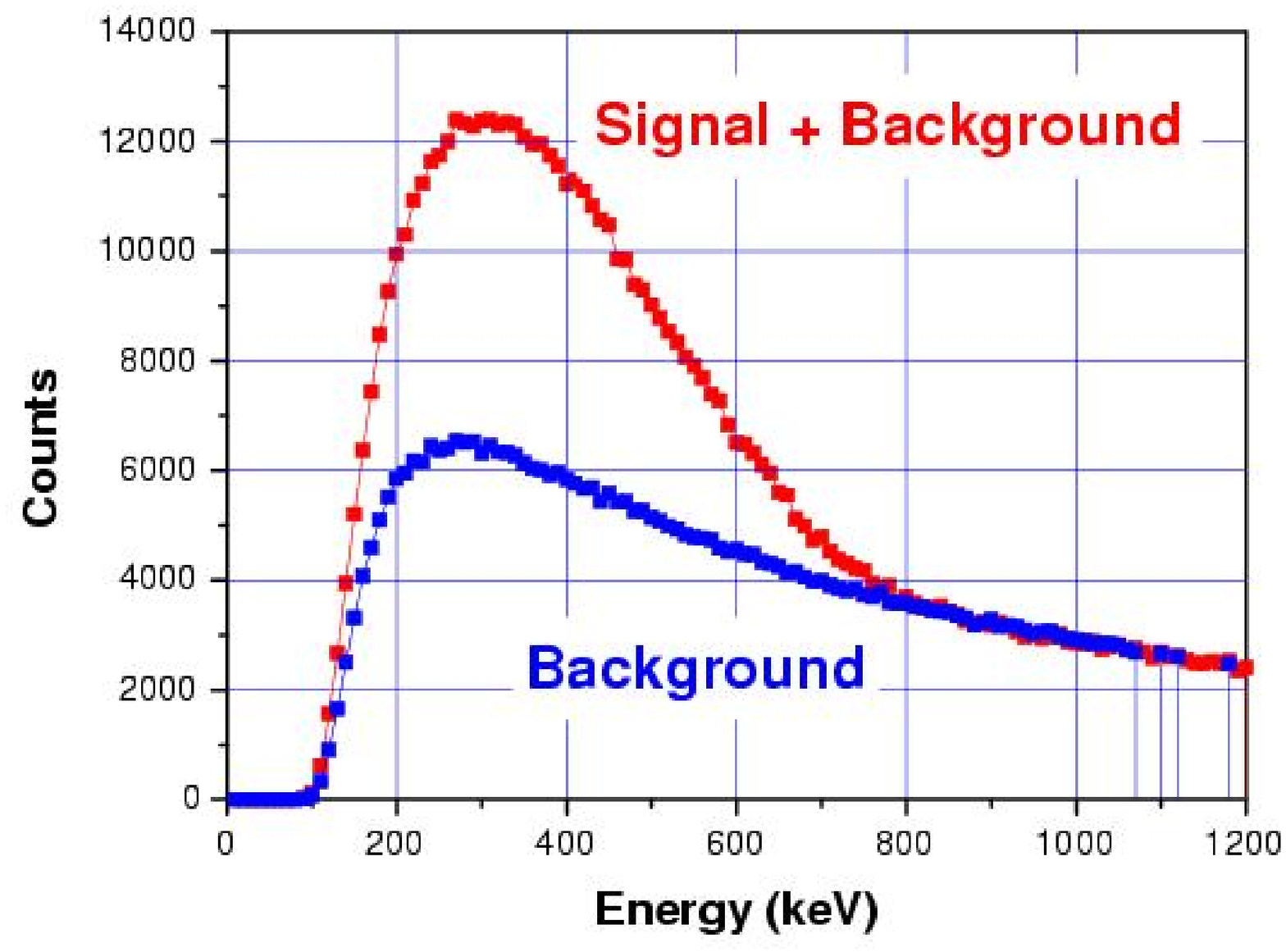}}
 \subfigure[\label{BETA1}]{
   \includegraphics[height=4.4cm]{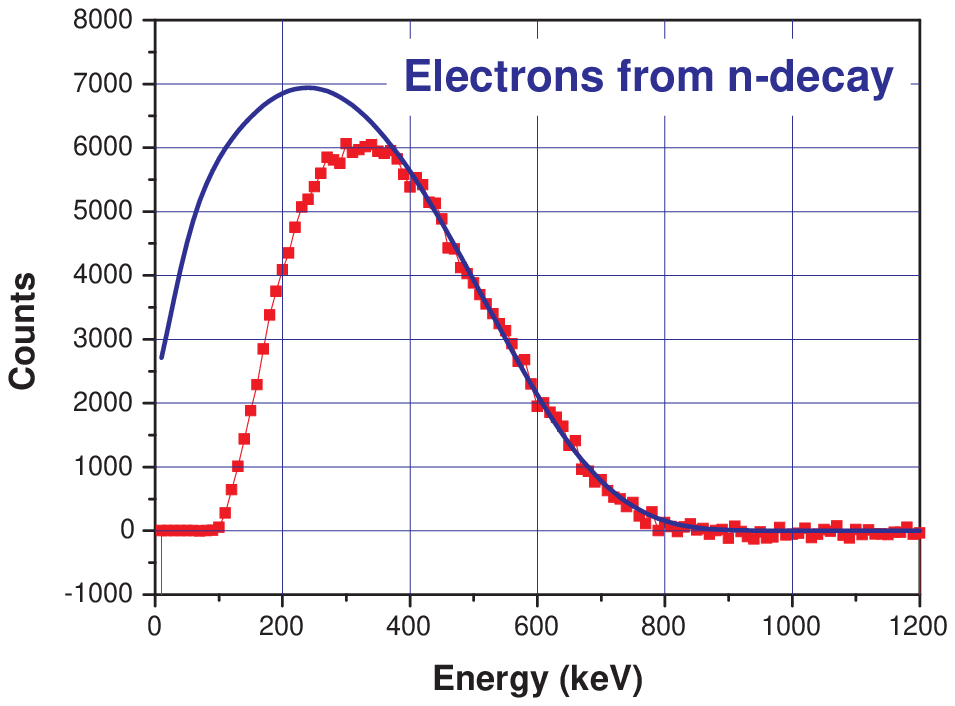}}\hspace{0.5cm}
 \subfigure[\label{VTRACK}]{
   \includegraphics[width=12.0cm]{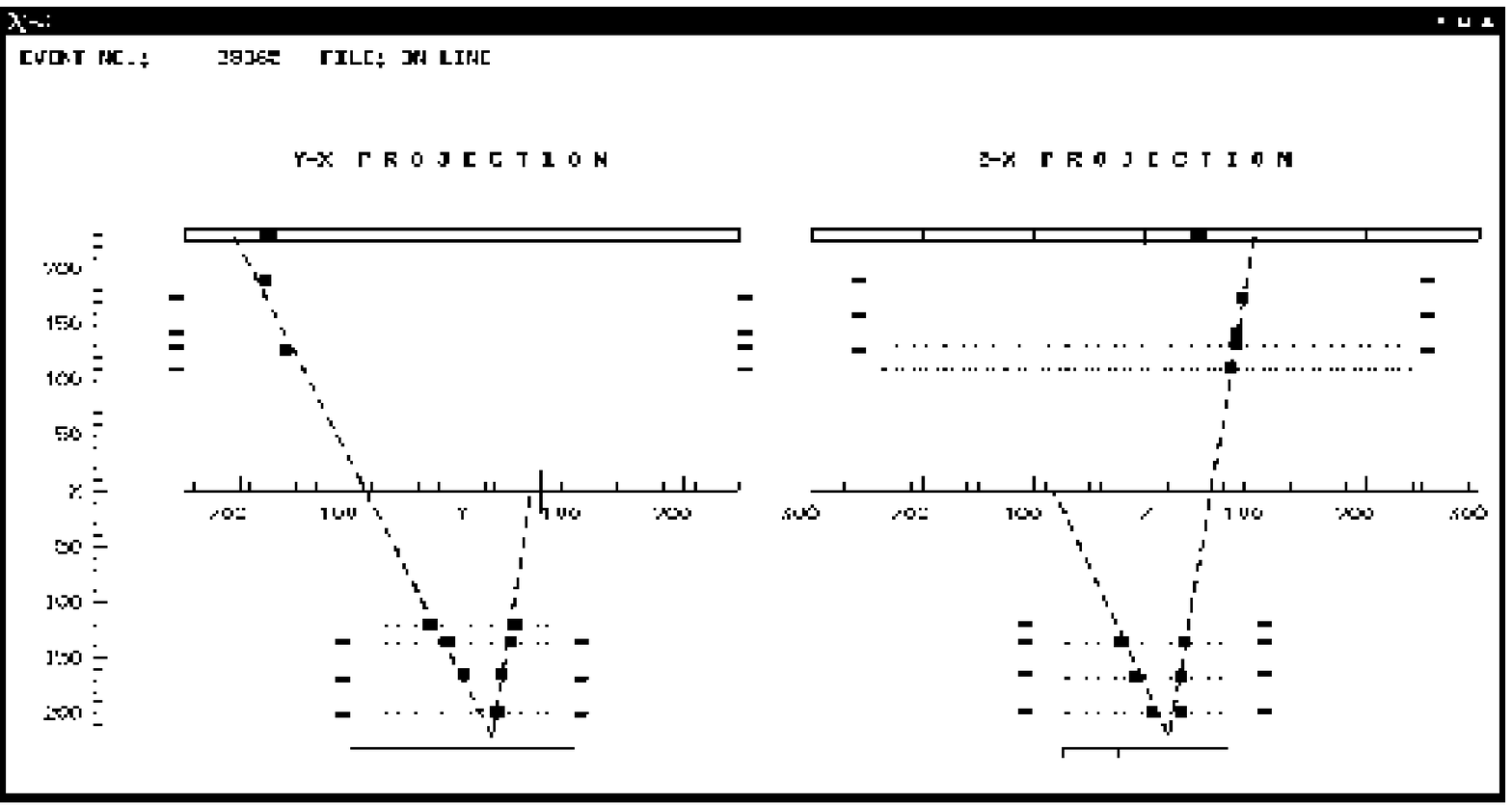}}
  \caption{(a) Energy spectra of electrons reconstructed as coming from
  the neutron beam region (``Signal+Background'') and originating in walls
  (``Background''), respectively.
  (b) Experimental energy spectrum after background subtraction.
  The solid line represents the theoretical $\beta$ spectrum from neutron
  decay.  (c) On-line display of a the Mott scattering event.}
\end{figure}
For fundamental physics experiments, a dedicated polarized cold
neutron facility has been constructed at the spallation source
SINQ.  Its description can be found in Ref.~\cite{NEUN00}.  New
measurements give the following performance parameters at the
place of the experimental setup:
\begin{center}
  \begin{tabular}{lrl}
    Flux:~~ & ($2.46 \pm 0.04$) &
    ~~$\times 10^8\;\textrm{(cm}^2\cdot\textrm{s}\cdot\textrm{mA)}^{-1}$ \\
    Equivalent thermal flux:~~ & ($6.49 \pm 0.10$) &
    ~~$\times 10^8\;\textrm{(cm}^2\cdot\textrm{s}\cdot\textrm{mA)}^{-1}$ \\
    Total intensity:~~ & ($1.48 \pm 0.03$) &
    ~~$\times 10^{10}\;\textrm{(s}\cdot\textrm{mA)}^{-1}$ \\
    Density:~~ & ($2.95 \pm 0.04$) &
    ~~$\times 10^3\;\textrm{(cm}^3\cdot\textrm{mA)}^{-1}$ \\
    Mean polarization:~~ & ($95.05 \pm 0.09$) & ~~\% \\
    Horizontal divergence:~~ & 0.014 & ~~rad \\
    Vertical divergence:~~ & 0.011 & ~~rad
  \end{tabular}
\end{center}
An efficient detector providing good rejection of undesired events
is of primary importance.  The key method of selecting the true
events, where the electron emitted in the neutron decay was
scattered from the analyzing foil, is based on the electron
identification via energy spectrum and the reconstruction of the
scattering vertex: the measured energy (corrected for losses) must
not exceed the end-point energy of 783~keV and the reconstructed
scattering vertex must be located on the analyzing foil.

These two conditions governed the laboratory development of
prototype detectors and optimization studies for the experimental
environment around the decay source: the part of the neutron beam
viewed by detectors.  The detector should have low mass and should
be constructed of low Z materials.  This leads to the concept of a
gas detector with all electrodes of thin wire grids and gas
mixture based on helium.  Also the neutron beam must be
transported in pure helium in front of the thin window (Mylar, 2
$\mu$m).  The results of the laboratory investigations of the
prototype MWPC are described in detail in Ref.~\cite{BODE01}. The
experience made by testing this detector in the real environment
influenced by neutron beam led to a construction of the full size
detectors with an active area of $50 \times 50\;\textrm{cm}^2$.
Now the first one of two MWPCs (see Fig. 3) is systematically
investigated in the setup shown schematically in Fig. 4. The
missing second MWPC has been replaced by the laboratory prototype.
In this way, the trigger and the readout electronics can be
reliably tested, too.

Sample data taken with the present setup are promising: the
electrons originating from neutron decay are clearly identified as
can be seen in Figs.~\ref{SB1} and \ref{BETA1}.  The missing part
of the experimental $\beta$ spectrum at low energies is due to
absorption effects in gas and the energy threshold of the
scintillation detectors.  Fig.~\ref{VTRACK} shows the on-line
display of an example Mott scattering event.

It is planned that the experiment should start data taking in
summer 2003 and within a few months should collect enough data for
the anticipated accuracy of $5 \times 10^{-3}$ for the $R$- and
$N$-correlation parameters in the decay of free neutrons.

\section*{Acknowledgments}

Polish authors kindly acknowledge partial financing of the project
by the Polish Committee for Scientific Research under the grant
No. 1450/P03/2002/22.


 \title*{The Measurement of Neutron Decay Parameters with the
    Spectrometer PERKEO II}
  \toctitle{The Measurement of Neutron Decay Parameters \protect\newline with the
    Spectrometer PERKEO II}
  \titlerunning{The Measurement of Neutron Decay Parameters}
  \author{M.B. Kreuz$^{1,2}$}
\tocauthor{M.B. Kreuz}
  \authorrunning{M.B. Kreuz}
  \institute{$^{1}$Institut Laue-Langevin, Grenoble, France\protect\linebreak
             $^{2}$Universit\"at Heidelberg, Germany}

  \maketitle

  \begin{abstract}
    The decay of free neutrons is a simple system to study the weak
    interaction in detail and to search for physics beyond the
    Standard Model. In particular, the beta asymmetry A is well suited
    to test the unitarity condition of the quark mixing CKM matrix.
    The neutrino asymmetry B is sensitive to the existence of
    right handed currents.\\
    The spectrometer PERKEO has measured the correlations A and B.
    We developed a technique which allows us to register both
    electron and proton in the same detector. This gives us access to
    a sensitive and systematically clean method for a B measurement.
  \end{abstract}

  \section{Introduction}
  \label{intro}
  Particle physics tries to examine structure and strength of particle
  interactions and particle properties. Traditionally, it is associated
  with high energy physics working in the energy range of presently
  $GeV$ and $TeV$. In contrast, our experiment was done with cold
  neutrons whose energies are even much lower than those of ordinary
  gas molecules at the other end of the energy spectrum.\\
  In the Standard Model of elementary particle interactions the
  fundamental fermionic constituents of matter are leptons and quarks
  found in three generations. In the decay of free neutrons
  a d-quark couples to a u-quark and the electron to an
  electron-antineutrino via a W-Boson exchange. This is a simple
  process where all particles of the first family are
  present. It allows to derive precise information about the first
  generation of the Standard Model and to access a number of
  interesting questions in particle physics, for example
  \begin{itemize}
  \item
    the ratio of the coupling constants $\lambda=\frac{g_{A}}{g_{V}}$
    (beta asymmetry $A$)
  \item
    the quark mixing and the unitarity of the CKM matrix ($A$ and
    neutron lifetime $\tau$)
  \item
    the universality of the electroweak interaction ($A$ and $\tau$)
  \item
    the origin of parity violation and the proposed existence of right
    handed currents (neutrino asymmetry $B$)
  \item
    a violation of time reversal invariance beyond the Standard Model
    (triple correlations D and R \cite{torsten})
  \end{itemize}
  In all these fields important progress has been made in the last few
  years.

  \section{The Spectrometer PERKEO}
  \label{perkeo}

  The spectrometer PERKEO II \cite{reich99} \cite{reich00} was build
  to access several of these
  coefficients, in particular the beta asymmetry $A$ and the neutrino
  asymmetry $B$, describing the correlation between the neutron spin
  and the momenta of the electron and the neutrino respectively.
  For $A$, it is sufficient to detect the decay electrons with respect
  to the neutron spin. For $B$ both the electron and the proton have
  to be traced to be able to deduce the neutrino direction.

  We use the measurement scheme shown in Fig.\ref{fig1}.
  A spin polarized, high flux cold neutron beam passes the
  spectrometer perpendicular to a strong magnetic field of
  1T, created by large superconducting coils in a split pair
  configuration. Some of the neutrons will decay inside the
  spectrometer in the volume defined by a set of baffles.
  The uncharged neutrons pass the magnetic
  field without deviation, but the charged decay products -- i.e. the
  electron and the proton -- gyrate along the magnetic field
  lines and can be detected in the two detectors left and right of the
  beam. By using the magnetic field many problems of beta
  decay experiments are suppressed:
  the solid angle is well known to be $2\times2\pi$. In addition the
  count rate is rather high because all particles reach
  the detector. By this the signal to background ratio improves
  accordingly.
\begin{figure}[b]
    \begin{center}
      \includegraphics[width=0.9\textwidth]{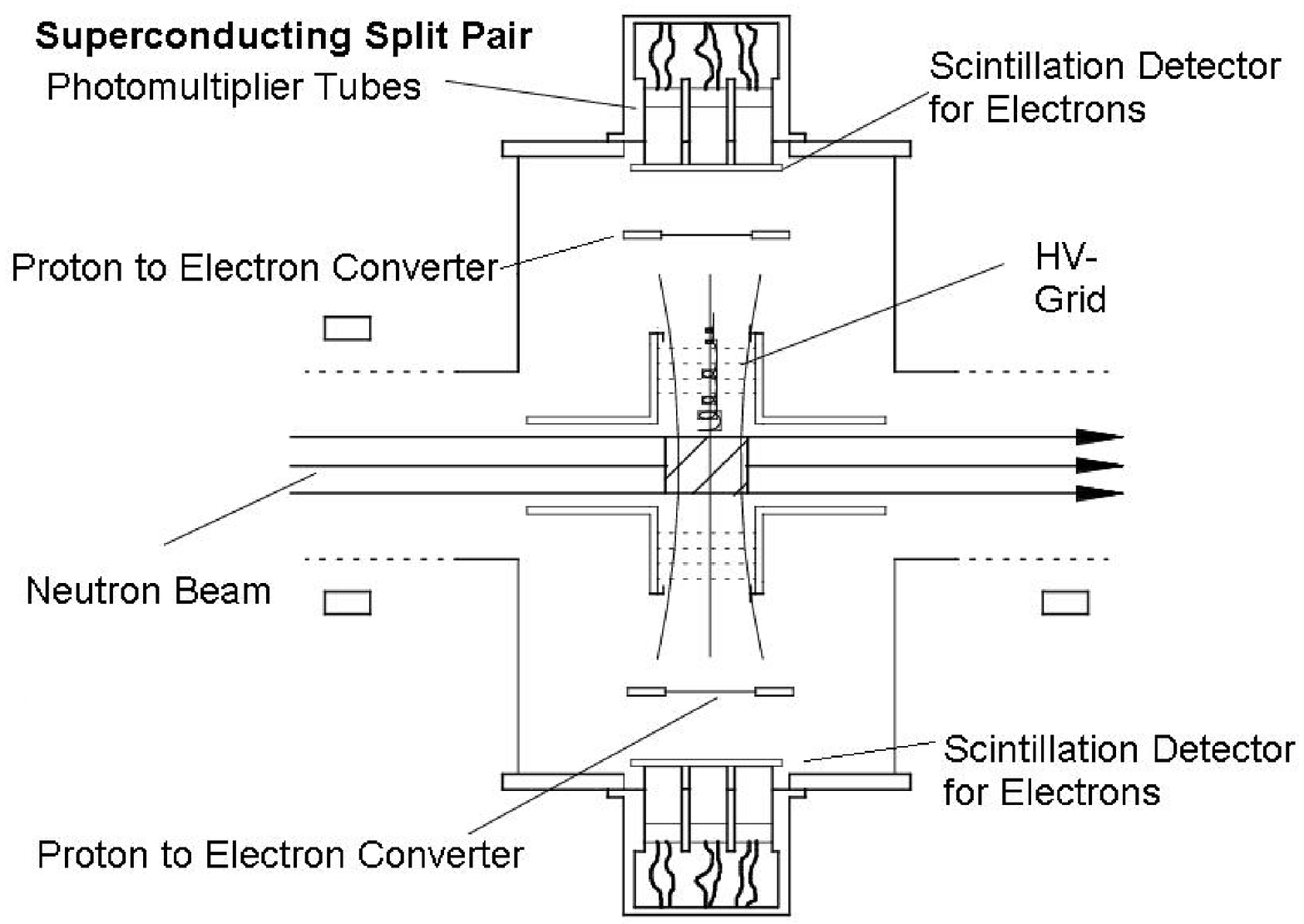}
    \end{center}
    \caption{Layout of the spectrometer PERKEO for the measurement of
      the correlation $B$. The electron and the
      proton from the neutron decay are guided by a $1T$ magnetic field to
      the two combined electron proton detectors.}
    \label{fig1}
  \end{figure}
  For the electron detection, plastic scintillators with
  photomultiplier readout are used. This is a fast and efficient
  combination. The threshold for electrons is 60keV. For the $A$
  measurement count rates of about 300s$^{-1}$ have been
  obtained. In addition false effects from electron backscattering --
  a serious source of systematic error in beta spectroscopy -- are
  suppressed by the magnetic field.

  The detection of the protons is a challenge due to their low energy
  ($<750$eV). For systematic reasons we also wanted to be able to
  detect both particles in both hemispheres (see section
  \ref{neutrino}). To this aim, the protons are converted
  into electrons via a negatively charged thin carbon foil: they are
  accelerated by the negative potential towards the foil where they
  create secondary electrons, which can be detected in the
  electron detector \cite{stratowa}. Since the electric potential is
  considerably lower than the electron energies observable in the
  experiment, the electrons will pass the foils unhindered.

  \section{The Beta Asymmetry A}
  \label{beta}
  As already mentioned for the determination of the beta asymmetry
  $A$ only the electrons have to be detected. In our experiment, the
  experimental asymmetry in two detectors is defined by
  \begin{equation}
    A_{exp}:=\frac{N^{\uparrow\uparrow}-N^{\downarrow\uparrow}}
    {N^{\uparrow\uparrow}+N^{\downarrow\uparrow}}
    =APf\frac{v}{c}
  \end{equation}
  with the polarization $P$, the spin flip efficiency $f$, the electron
  velocity $v$ and the count rates parallel and antiparallel to the
  neutron spin $N^{\uparrow\uparrow}$ and
  $N^{\downarrow\uparrow}$.

  $A$ is very sensitive to the ratio of the axial vector to vector
  coupling strength $\lambda=\frac{g_{A}}{g_{V}}$
  of the weak interaction. In the Standard Model $A$ and $\lambda$ are
  linked by a simple formula:
  \begin{equation}
    A=-2\frac{|\lambda|^{2}-Re(\lambda)}{1+3|\lambda|^{2}}
  \end{equation}
  From our experiment we get $A=-0.1189(7)$ and
  $\lambda=-1.274(2)$. The main experimental errors in this experiment
  have been the statistical error ($0.45\%$) and the determination of
  the neutron spin polarization ($0.3\%$). Since the last measurement,
  important improvements in both fields have been made. First, the
  neutron flux at the new ballistic supermirror guide at the
  instrument PF1B at the ILL is a factor of $3$ higher compared with
  the previous experimental zone \cite{harry}. Second, a new arrangement
  of two supermirror polarizers allows to achieve an unprecedented
  degree of polarization of more than $99.5\%$ and its determination
  at the $0.1\%$ level \cite{sascha}.

  From $\lambda$ and the neutron lifetime $\tau$ \cite{pdgkreuz} $g_{A}$
  and $g_{V}$
  can be determined. They are important parameters for many problems
  in astrophysics, e.g. for primordial nucleosynthesis. Furthermore,
  they can be used to test the Standard Model: According to
  the conserved vector current hypothesis (CVC) the value of $C_{V}$
  is not changed by QCD effects in the nuclear medium, which has been
  checked for many nuclei. On the other hand, $g_{A}$ depends on the
  nuclear medium and has to be measured.
  The value of $g_{V}$ can also be used to determine the first element
  $V_{ud}$ of the quark mixing matrix or CKM matrix via
  \begin{equation}
    V_{ud}=\frac{g_{V}}{G_{F}}
  \end{equation}
  with the Fermi coupling constant $G_{F}$.
  Together with the particle data group values for $V_{us}$
  and $V_{ub}$, the unitarity of the CKM matrix required by the Standard
  Model can be tested for the first row. In a unitary matrix the squared
  sum of all rows and columns has to be equal to one. For our value of
  $V_{ud}$ the test reads
  \begin{equation}
    |V_{ud}|^{2}+|V_{us}|^{2}+|V_{ub}|^{2}=0.9917(29)
  \end{equation}
  which gives a $2.7$ sigma deviation from the expected value
  \cite{abele}. The
  particle data group uses the world average $V_{ud}=0.9728(12)$ from
  neutron decay, leading to a unitarity gap of only $2.2$ sigma
  \cite{pdgkreuz}.
  Using data from nuclear beta decay, one still finds a $2.3$ sigma
  deviation \cite{hardy} from unity. New measurements are required to
  determine
  whether a discrepancy between reality and the Standard Model has
  been found.

  \section{The Neutrino Asymmetry B}
  \label{neutrino}
  It is necessary to detect both electron and proton to be able to
  trace the neutrino and to measure the neutrino asymmetry $B$. In our
  setup with a combined electron-proton detector on both sides there
  are two possibilities to define an observable asymmetry: electron
  and proton in the same or in opposite hemispheres. We can
  define the following asymmetries:
  \begin{equation}
    \label{basym}
    B_{exp,1}=\frac{N^{\uparrow\uparrow\uparrow}-
      N^{\downarrow\uparrow\uparrow}}
    {N^{\uparrow\uparrow\uparrow}+N^{\downarrow\uparrow\uparrow}}
  \end{equation}
  \begin{equation}
    B_{exp,2}=\frac{N^{\uparrow\uparrow\downarrow}-
      N^{\downarrow\uparrow\downarrow}}
    {N^{\uparrow\uparrow\downarrow}+N^{\downarrow\uparrow\downarrow}}
  \end{equation}
  The arrows indicate the direction of the neutron spin and the
  hemisphere direction of the electron and proton respectively.
  \begin{figure}[tbp]
    \begin{center}
        \includegraphics[width=8cm]{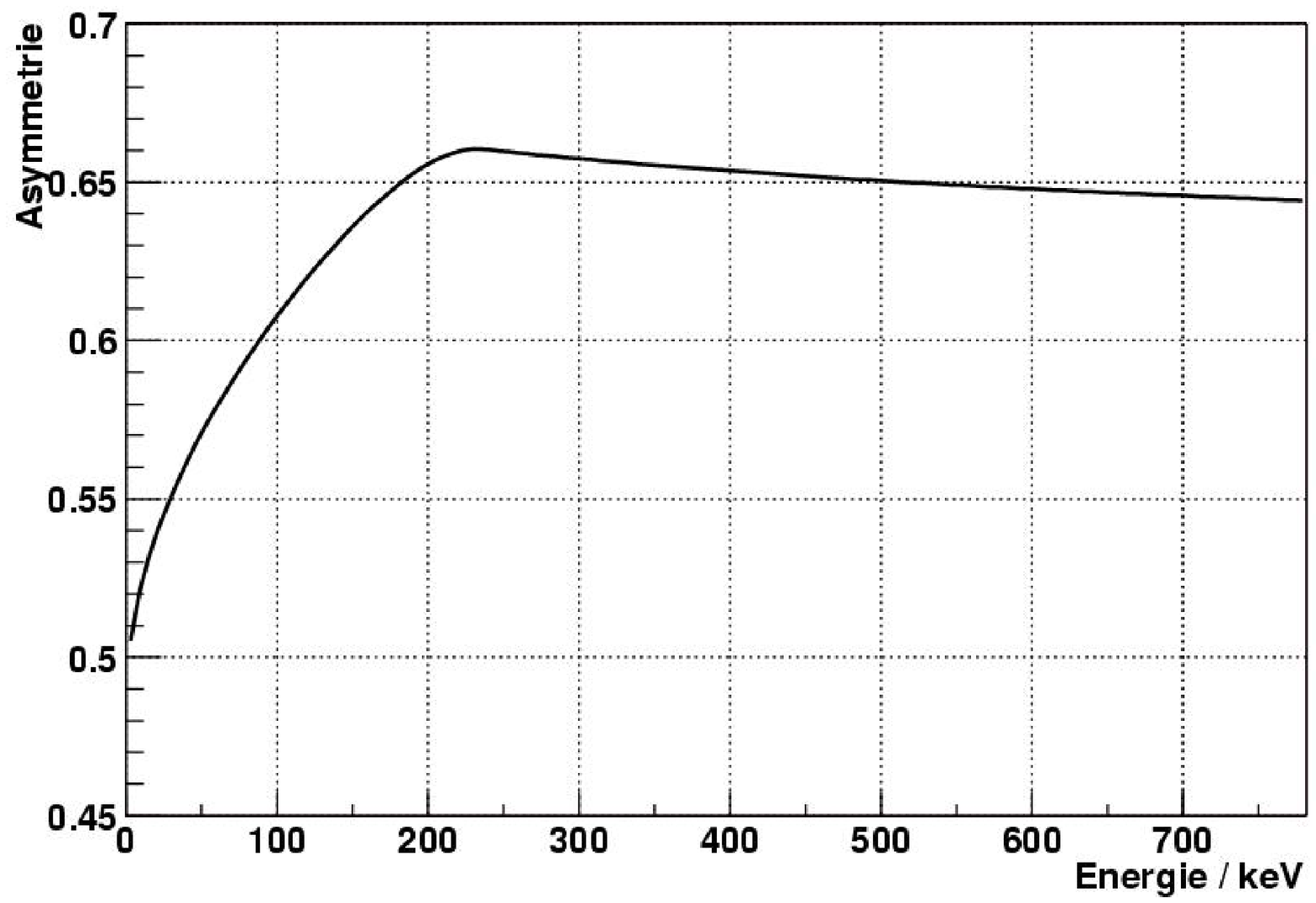}   
        \includegraphics[width=8cm]{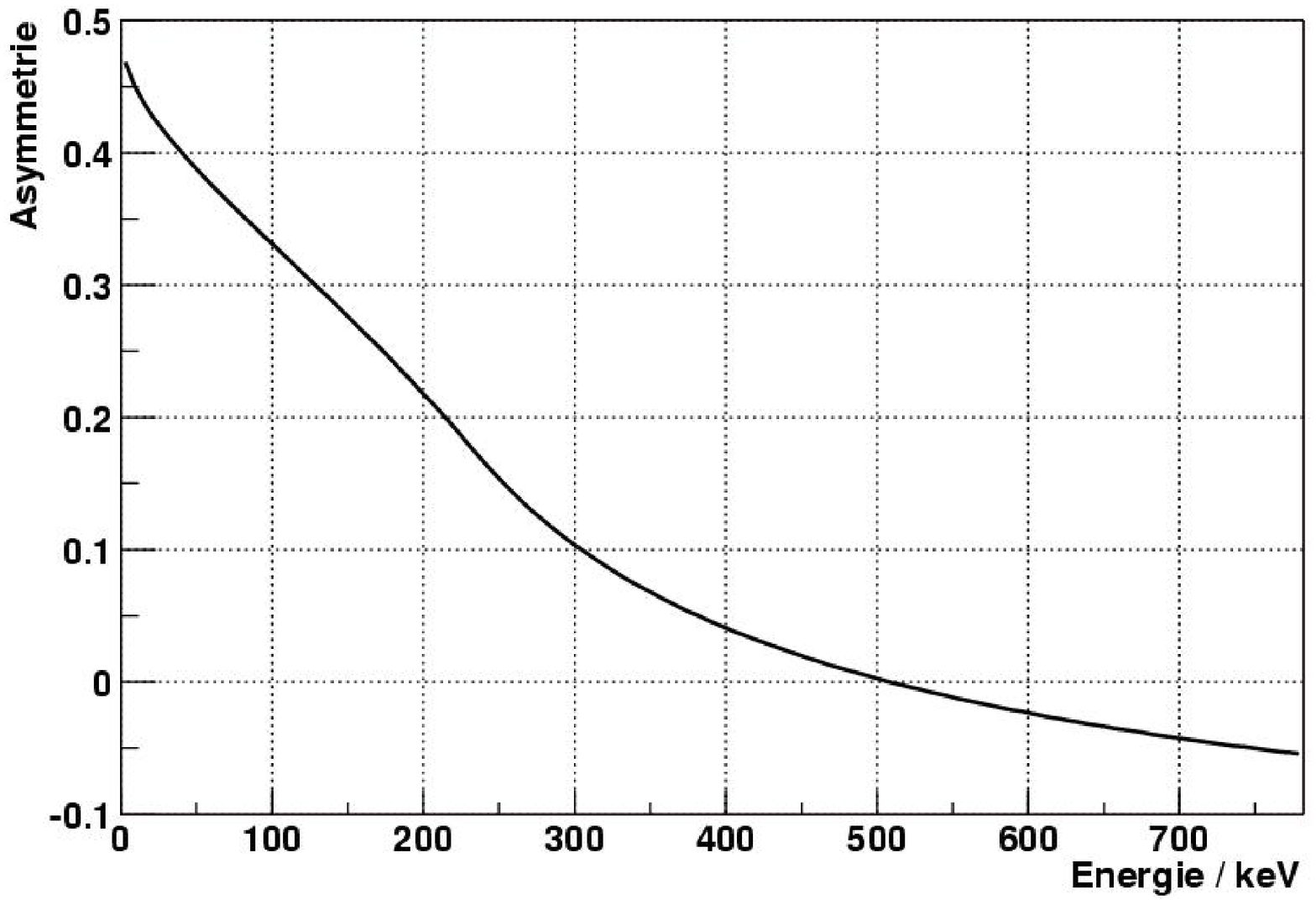}   
    \end{center}
    \caption{Electron-proton asymmetry in dependence of the electron
      energy for both possible configurations. Top: both
      particles in the same hemisphere. Bottom: in opposite
      hemispheres.}
    \label{fig2}
  \end{figure}
  The resulting dependence of the asymmetry from the electron energy
  can be seen in Fig.\ref{fig2}. From a systematic point of view it is
  much better to use the same hemisphere: If electron and proton are
  detected in the same detector the direction of the neutrino is well
  defined. Therefore the asymmetry is almost independent from the
  electron energy and thus insensitive to the detector calibration
  and resolution. In addition the influence of other asymmetry
  coefficients is suppressed and a high statistical sensitivity for
  $B$ is achieved over the complete energy range.
  In the case of opposite hemispheres the direction of the neutrino is
  not well defined and the asymmetry depends strongly on the electron
  energy. Furthermore, the result depends on a precise knowledge of
  the beta asymmetry parameter $A$ and the
  The highest sensitivity to $B$ is in the low energy part of
  the spectrum where the spectroscopy is most difficult because of
  background and threshold effects. Considering these systematics
  an evaluation of the asymmetry of equation\ref{basym} is strongly
  preferable, but any measurement with PERKEO will automatically allow
  both.

  In contrast to $A$, $B$ is not very sensitive to the ratio of the
  coupling
  strengths $\lambda$. But since $B$ measures directly the
  antineutrino asymmetry, it is rather sensitive to a hypothetical
  admixture of right handed components to the weak interaction which
  is purely left handed in the Standard Model. It can be used to test
  left right symmetric models. In these models parity is not violated
  for high energies and even in our low energy world it should not be
  maximally violated. Small remnants of the right handed currents can
  still be found, suppressed by the much higher mass of the
  corresponding W-Boson. The manifest left right symmetric models are
  described by three
  parameters: the mixing angle $\zeta$ between the weak W-eigenstates
  and the mass eigenstates, the ratio $\delta$ of the square of the
  boson masses and the ratio $\lambda$ of the coupling constants.
  Using $A$ and $B$ from the experiment and $\tau$ from the PDG an
  exclusion plot in the $\delta-\zeta$-plane can be
  drawn. The Standard Model
  ($\delta=0$,
  $\zeta=0$) lies within the  allowed region at $90\%$ confidence
  level. Please note that these are not the best limits
  reached so far by high energy physics and astronomy, but due to the
  high sensitivity of B, a slight improvement of the precision will
  make them competitive.
  The evaluation of our 2001 measurement of the neutrino asymmetry is
  still in progress. It should be finished at the end of 2003.

  \section{Summary}
  \label{summary}
  Within our group two different correlation coefficients in the decay
  of free neutrons have been measured and precision tests of the
  Standard Model have been performed. $V_{ud}$ the first element of
  the CKM matrix has been derived from neutron decay experiments in
  such a way that a unitarity test of the CKM matrix can be performed
  based solely on particle physics data. With this value we find a
  2.7 sigma deviation from unitarity.






\title*{Summary Talk }
\toctitle{Summary Talk}
%
%
\titlerunning{Summary Talk}
%
\author{N. Severijns}
\authorrunning{N. Severijns}
%
%
\institute{Katholieke Universiteit Leuven, Celestijnenlaan 200 D,
B-3001 Leuven, Belgium}

\maketitle              

\begin{abstract}
A brief overview is given of the most important results of the
workshop, with special emphasis on the possibilities for future
advances in the field of weak interaction physics offered by
experiments in neutron decay.
\end{abstract}

\section{Unitarity}
At the current level of precision the unitarity test based on the
first row of the CKM quark mixing matrix depends only on the
$V_{ud}$ and $V_{us}$ matrix elements.

The $V_{ud}$ matrix element is determined in beta decay processes.
At their respective precisions the results from the superallowed
$0^+ \rightarrow 0^+$ pure Fermi transitions (i.e. $V_{ud}$ =
0.9740(5) \cite{hardy02}), from the lifetime $\tau$ and the
electron asymmetry parameter $A$ in neutron decay (i.e. $V_{ud}$ =
0.9740(13)\cite{abele02}) and from the pion beta decay branch
(i.e. $V_{ud}$ = 0.9771(56) \cite{pocanic02}) agree with each
other and point to a 2 $\sigma$ to 2.5 $\sigma$ deviation from
unitarity.

The highest precision at present is reached by the superallowed
$0^+ \rightarrow 0^+$ transitions. The error bar is in this case
dominated by the corrections. Although theorists believe that both
the radiative as well as the nuclear structure related corrections
are well under control, the isospin corrections are rather
frequently questioned. At several facilities experimental programs
are therefore ongoing to acquire experimental information that
should allow one to check the reliability of these corrections.

Due to important recent progress in the experimental precision for
both the lifetime $\tau$ and the electron asymmetry parameter $A$
in neutron decay the precision of the $V_{ud}$ value from neutron
decay is now almost at the same level of precision as the one from
the $0^+ \rightarrow 0^+$ transitions. A number of new
measurements is moreover being prepared or well advanced already.
In addition, important cross checks will be provided by new
measurements of the electron-neutrino correlation coefficient $a$
(with about an order of magnitude higher precision) as well as by
a combined measurement of the electron asymmetry parameter $A$,
the neutrino asymmetry parameter $B$ and the electron-neutrino
correlation coefficient $a$ in neutron decay.

In pion decay, reaching a high precision for $V_{ud}$ is hampered
by the low beta decay branching ratio of order $10^{-8}$ only.
However, the analysis of a currently available large data set from
a recent experiment at PSI will significantly improve the value
mentioned above. It is finally important to note that the error on
the $V_{ud}$ values obtained from neutron decay and from pion beta
decay are still dominated by the experimental error bars. New
measurements that allow one to further reduce these should
therefore be strongly supported.

The $V_{us}$ matrix element is determined most precisely in Kaon
decay. As it turns out, the presently accepted value for $V_{us}$
is based on experiments that were carried out about 25 years ago.
Recent results obtained in $K_{e3}$ decay at Brookhaven indicate
that the branching ratios on which the value for $V_{us}$ is based
may be off by several standard deviations. If so, it is not
excluded that this might solve the unitarity problem. New efforts
that are being planned in this respect \cite{renk, delucia} should
clarify this.

It is interesting to note that measurements planned at the CLEO-c
facility that is constructed to do charm physics at threshold,
which in general yields clean signals, will soon allow one to test
unitarity also for other rows and columns of the CKM matrix. Most
important will be the determination of the $V_{cd}$ matrix element
with an anticipated precision of about 1.7\% \cite{shipsey}, which
would yield a unitarity test for the first column with an almost
identical precision to that obtained now for the first row. This
will indicate whether the current deviation from unitarity for the
first row is more likely to be due to a problem with $V_{ud}$ or
with $V_{us}$. In addition, the determination of the $V_{cs}$
matrix element with an about 1.6\% precision \cite{shipsey} would
allow to test unitarity also for the second row and the second
column, although with somewhat less precision.

\section{Search for new weak interactions}
Apart from providing important input information to test
unitarity, correlation experiments in neutron decay also provide
very important tests for the presence of scalar, tensor or
right-handed weak interactions that are not included in the
Standard Model and would be mediated by gauge bosons other than
the W- and Z-bosons or by leptoquarks. The potential of neutron
decay experiments in this respect was demonstrated several years
ago already by Mostovoi et al. \cite{mostovoi00}, and this type of
physics interpretation of correlations in neutron decay will
certainly gain in importance as the experimental precision
increases further. At that point it will also become important to
include in the analysis of the data the Fierz interference term
$b$ (which is zero in the Standard Model) as well as radiative and
recoil corrections, which are now usually neglected. Measurements
of the energy dependence of the recoil corrections to the electron
asymmetry parameter $A$ and the electron-neutrino correlation
coefficient $a$ would lead to improved CVC and second class
current tests.

\section{Time reversal violation}
In recent years a lot of progress was made in searches for the
neutron electric dipole moment. The current limit for the neutron
EDM has ruled out already a number of models and is close to the
discovery limit for other models. Large efforts are currently
ongoing at several places to further improve on this, mainly based
on the design of powerful ultra cold neutron facilities. With the
planned and ongoing developments of new UCN sources that will
provide unprecedented amounts of neutrons, an improvement of two
orders of magnitude seems realistic, thereby improving the
sensitivity for neutron EDM experiments to about 2 x $10^{-28}$ e
cm. These efforts should be strongly supported.

Other ongoing searches for T-violation in neutron decay, i.e.
determinations of the D- and R-triple correlation coefficients,
should be pursued as well as these can be interpreted with much
fewer theoretical uncertainties than the EDM, or they are
complementary.

\section{Technical advances}
Recent new ideas for alternative measurement principles and new
set-ups for lifetime and correlation measurements with both cold
and ultra cold neutrons have already lead and will still lead to
reduced systematic errors and increased statistics. Especially
important in this respect are the successful new developments with
respect to neutron polarimetry. Using supermirrors a precision of
0.5\% is now routinely available, while polarized $^{3}He$
polarimeters even allow one to limit the uncertainty on the
neutron polarization to as low as 0.1\%. Further, several
facilities providing cold and ultra cold neutrons will provide
both higher intensity beams as well as pulsed beams, thereby
allowing for significant experimental improvements.

\section{Conclusion}
The field of fundamental research in neutron decay is very active.
In the past decade significant progress was made with respect to
the obtainable experimental precision, both due to increased
neutron beam intensities and the development of improved neutron
polarimetry techniques. With the presently ongoing technical
developments, the large amount of new ideas for improved
experimental set-ups, and the new approaches for the determination
of several observables in neutron decay that are currently
available and being developed, the neutron will also in the next
decade remain an important laboratory for fundamental weak
interaction physics.

\title*{List of Participants\protect\newline}

\toctitle{List of Participants} \vspace*{1.5cm}
\titlerunning{List of Participants}
\author{}
 \institute{}
\authorrunning{List of Participants}
%
%

\maketitle              

 

\begin{flushleft}
\vspace{-1.5cm}
\vspace{4pt}\emph{\scshape{H. Abele}}, Physikalisches Institut, Universit\"at
Heidelberg, \\ Philosophenweg 12, 69120 Heidelberg, Germany

\vspace{4pt}\emph{\scshape{A. Alduschenkov}}, Petersburg Nuclear Physics Institute,
Orlova Roscha, \\ 188350 Gatchina, Russia

\vspace{4pt}\emph{\scshape{A. Ali}}, DESY, Notkestra\ss{}e 85,\\ 22607 Hamburg, Germany

\vspace{4pt}\emph{\scshape{S. Bae\ss{}ler}}, Institut f\"ur Physik, Universit\"at Mainz,
\\ Staudingerweg 7, 55128 Mainz, Germany

\vspace{4pt}\emph{\scshape{E. Barberio}}, CERN , EP Division Group HC, Bld. 40, Office
3-C32, \\ 1211 Geneva 23, Switzerland

\vspace{4pt}\emph{\scshape{P. Barker}}, Physics Department, University of Auckland,
Private Bag, \\ 92019 Auckland, New Zealand

\vspace{4pt}\emph{\scshape{K. Bodek}}, Paul Scherrer Institut, 5232 Villigen, PSI,
Switzerland

\vspace{4pt}\emph{\scshape{B. B\"ohm}}, Physikalisches Institut, Universit\"at
Heidelberg, \\ Philosophenweg 12, 69120 Heidelberg, Germany

\vspace{4pt}\emph{\scshape{L. Bondarenko}}, Kurchatov Institute, \\ Kurchatov Sq. 1, 123182
Moscow, Russia

\vspace{4pt}\emph{\scshape{M. Brehm}}, Physikalisches Institut, Universit\"at Heidelberg,
\\Philosophenweg 12, 69120 Heidelberg, Germany

\vspace{4pt}\emph{\scshape{H. Bruhns}}, Physikalisches Institut, Universit\"at
Heidelberg, \\ Philosophenweg 12, 69120 Heidelberg, Germany

\vspace{4pt}\emph{\scshape{G.G. Bunatian}}, Joint Institute for Nuclear Research, \\ 141980
Dubna, Moscow Region, Russia

\vspace{4pt}\emph{\scshape{J. Byrne}}, Physics and Astronomy Subject Group, University of
Sussex,\\ Brighton, BN1 9QH, United Kingdom

\vspace{4pt}\emph{\scshape{M. Daum}}, Paul Scherrer Institut, 5232 Villigen, PSI,
Switzerland

\vspace{4pt}\emph{\scshape{D. Dubbers}}, Physikalisches Institut, Universit\"at
Heidelberg, \\ Philosophenweg 12, 69120 Heidelberg, Germany

\vspace{4pt}\emph{\scshape{S. Gardner}}, Department of Physics and Astronomy University
of Kentucky, \\ Lexington, KY 40506-0055, USA

\vspace{4pt}\emph{\scshape{P. Geltenbort}}, Institut Laue-Langevin, \\ B.P. 156, 6 rue Jules
Horowitz, 38042 Grenoble, France

\vspace{4pt}\emph{\scshape{F. Gl\"uck}}, Institut f\"ur Physik, Universit\"at Mainz,
\\ Staudingerweg 7, 55128 Mainz, Germany

\vspace{4pt}\emph{\scshape{M. G\"ockeler}}, Institut f\"ur Theoretische Physik, Universit\"at Regensburg, \\ 93040 Regensburg, Germany

\vspace{4pt}\emph{\scshape{R. Golub}}, Hahn-Meitner-Institut, \\ Glienickerstra\ss{}e 100, 14109
Berlin, Germany

\vspace{4pt}\emph{\scshape{J. Hardy}}, Cyclotron Institute, Texas A\&M University
College, \\ Station, TX 77843, USA

\vspace{4pt}\emph{\scshape{J. Hartmann}}, Technische Universit\"at M\"unchen,
Physik-Department E18, \\ James-Franck-Stra\ss{}e, 85748 Garching,
Germany

\vspace{4pt}\emph{\scshape{W. Heil}}, Institut f\"ur Physik, Universit\"at Mainz,
\\ Staudingerweg 7, 55128 Mainz, Germany

\vspace{4pt}\emph{\scshape{P. Huffman}}, Harvard University, National Institute of
Standards and Technology, Bld. 235, Room A159, Gaithersburg, MD
20899 USA

\vspace{4pt}\emph{\scshape{M. Jamin}}, Institut f\"ur Theoretische Physik Universit\"at
Heidelberg, \\ Philosophenweg 16, 69120 Heidelberg, Germany

\vspace{4pt}\emph{\scshape{K.-P. Jungmann}} Rijksuniversiteit Groningen, Kernfysisch
Versneller Instituut (KVI), \\ Zernikelaan 25, 9747 AA Groningen,
Netherlands

\vspace{4pt}\emph{\scshape{M. Klein}}, Physikalisches Institut, Universit\"at Heidelberg,
\\ Philosophenweg 12, 69120 Heidelberg, Germany

\vspace{4pt}\emph{\scshape{E. Korobkina}}, Hahn-Meitner-Institut, \\ Glienickerstra\ss{}e 100,
14109 Berlin, Germany

\vspace{4pt}\emph{\scshape{A. Kozela}}, Institut f\"ur Teilchenphysik, \\ ETH H\"onggerberg,
HPK, 8093 Z\"urich, Switzerland

\vspace{4pt}\emph{\scshape{M. Kreuz}}, Institut Laue-Langevin, \\ B.P. 156, 6, rue Jules
Horowitz, 38042 Grenoble, France

\vspace{4pt}\emph{\scshape{E. De Lucia}}, I.N.F.N. University of Rome "La Sapienza", \\ P.
le A. Moro n. 2, 00185 Rome, Italy

\vspace{4pt}\emph{\scshape{B. M\"arkisch}}, Physikalisches Institut, Universit\"at
Heidelberg, \\ Philosophenweg 12, 69120 Heidelberg, Germany

\vspace{4pt}\emph{\scshape{W. J. Marciano}}, Brookhaven National Laboratory, \\ Upton, New
York 11973, USA

\vspace{4pt}\emph{\scshape{D. Mund}}, Physikalisches Institut, Universit\"at Heidelberg,
\\ Philosophenweg 12, 69120 Heidelberg, Germany

\vspace{4pt}\emph{\scshape{O. Nachtmann}}, Institut f\"ur Theoretische Physik,
\\ Universit\"at Heidelberg, Philosophenweg 16, 69120 Heidelberg,
Germany

\vspace{8pt}\emph{\scshape{C. Plonka}}, Technische Universit\"at M\"unchen, Physik
Department E21, \\ 85747 Garching, Germany

\vspace{4pt}\emph{\scshape{D. Pocanic}}, Department of Physics, University of Virginia,
\\ PO Box 400714, Charlottesville, VA 22904-4714, USA

\vspace{4pt}\emph{\scshape{H. Rauch}}, Atominstitut der \"Osterreichischen
Universit\"aten, \\ Stadionallee 2, 1020 Wien, Austria

\vspace{4pt}\emph{\scshape{B. Renk}}, Institut f\"ur Physik, Universit\"at Mainz,
\\ Staudingerweg 7, 55128 Mainz, Germany

\vspace{4pt}\emph{\scshape{S. Ritt}}, Paul Scherrer Institut, 5232 Villigen, PSI,
Switzerland

\vspace{4pt}\emph{\scshape{H. Sagawa}}, Center for Mathematical Sciences, University of
Aizu, \\ Aizu-Wakamatsu Fukushima 965, Japan

\vspace{4pt}\emph{\scshape{K. Schindler}}, Physikalisches Institut Universit\"at
Heidelberg, \\ Philosophenweg 12, 69120 Heidelberg, Germany

\vspace{4pt}\emph{\scshape{C. Schmidt}}, Physikalisches Institut, Universit\"at
Heidelberg, \\ Philosophenweg 12, 69120 Heidelberg, Germany

\vspace{4pt}\emph{\scshape{U. Schmidt}}, Physikalisches Institut, Universit\"at
Heidelberg, \\ Philosophenweg 12, 69120 Heidelberg, Germany

\vspace{4pt}\emph{\scshape{F. Schwab}}, Institut f\"ur Theoretische Physik, Universit\"at
Heidelberg, \\ Philosophenweg 16, 69120 Heidelberg, Germany

\vspace{4pt}\emph{\scshape{A. Serebrov}}, Petersburg Nuclear Physics Institute, \\ Gatchina,
188350 Leningrad District, Russia

\vspace{4pt}\emph{\scshape{N. Severijns}}, Instituut voor Kern- en Stralingsfysica,
Katholieke Universiteit Leuven, Celestijnenlaan 200 D, 3001
Leuven, Belgium

\vspace{4pt}\emph{\scshape{I. Shipsey}}, Purdue University, Department of Physics, \\ 1396
Physics Building West Lafayette, IN 47907, USA

\vspace{4pt}\emph{\scshape{H.-W. Siebert}}, Physikalisches Institut, Universit\"at
Heidelberg, \\ Philosophenweg 12, 69120 Heidelberg, Germany

\vspace{4pt}\emph{\scshape{W. Snow}}, Indiana University/IUCF, \\ 2401 Milo B. Sampson Ln
Bloomington, IN 47408, USA

\vspace{4pt}\emph{\scshape{T. Soldner}}, Institut Laue-Langevin, \\ BP 156, 6, rue Jules
Horowitz, 38042 Grenoble, Cedex 9, France

\vspace{4pt}\emph{\scshape{A. Vicini}}, Dipartimento di Fisica, Universita' degli Studi
di Milano,\\  Via Celoria 16, I-20133 Milano, Italia

\vspace{4pt}\emph{\scshape{F. E. Wietfeldt}}, Physics Department, Tulane University, \\ 2001
Percival Stern New Orleans, LA 70118, USA

\vspace{4pt}\emph{\scshape{B. Yerozolimsky}}, 104, Stowe Court, \\ Andover, MA 01810, USA

\vspace{4pt}\emph{\scshape{A.R. Young}},  Physics Department, NC State University, \\ 312
Coxhall, Box 8202, Raleigh, NC 27695, USA

\vspace{4pt}\emph{\scshape{O. Zimmer}}, TU M\"unchen, Physik-Department E18,
\\ James-Franck-Stra\ss{}e, 85748 Garching, Germany

\end{flushleft}

%



\begin{thebibliography}{99}
\bibitem{Soldner1}T. Soldner, Recent progress in neutron polarization and its analysis, ref. \cite{HADM}.
\end{thebibliography}
\begin{thebibliography}{8.}
\bibitem{remark}
We have implicitly assumed that
the radiative corrections to $f_1$ and $f_2$ are equal; the
latter have never been calculated.
\end{thebibliography}
\end{document}